%% file: ms.tex
\documentclass[a4paper, 11pt, twoside, openright]{book}

\include{preamble}

\begin{document}

\notesoff

\frontmatter

\begin{titlepage}
\thispagestyle{empty}
\begin{center}
\vspace*{2cm}

{\LARGE \bf Quantum Spectral Curve  \\ \vspace{0.4cm} for the $\bm{\eta}$-deformed $\bm{\ads}$ superstring}

\vfill

{\Large Dissertation \\
\medskip
zur Erlangung des Doktorgrades \\
an der Fakult\"at f\"ur Mathematik, \\
Informatik und Naturwissenschaften \\
\medskip
Fachbereich Physik \\
der Universit\"at Hamburg \\}

\vspace{2cm} 

{\large vorgelegt von\\
\medskip
Rob Klabbers}

\vspace{2cm}

\large{Hamburg \\
\medskip
2017 \\ }

\vfill

\bigskip
\bigskip
\medskip
\end{center}
\end{titlepage}

\newpage
\thispagestyle{empty}

\vspace*{\fill}


\noindent Tag der Disputation: 24.~Januar 2018
\newline Folgende Gutachter empfehlen die Annahme der Dissertation:
\newline Prof.~Dr.~G.E.~Arutyunov
\newline Prof.~Dr.~J.~Teschner

\large
\newpage
\thispagestyle{empty}

\newpage 

\begin{flushright}
\vspace*{3cm}
To my grandfather Jacques
\end{flushright}

\newpage
\vspace*{\fill}
$\,$\\
\section*{Abstract}

Being able to solve an interacting quantum field theory exactly is by itself an exciting prospect, as having full control allows for the precise study of phenomena described by the theory. In the context of the AdS/CFT correspondence, which hypothesises a duality between certain string and gauge theories, planar $\mN =4$ super Yang-Mills theory shows signs of great tractability due to integrability. The spectral problem of finding the scaling dimensions of local operators or equivalently of the string energies of string states of its AdS/CFT-dual superstring theory on the $\ads$ background turned out to be very tractable. After over a decade of research it was found that the spectral problem can be solved very efficiently through a set of functional equations known as the quantum spectral curve. Although it allowed for a detailed study of the aforementioned theory, many open questions regarding the wider applicability and underlying mechanisms of the quantum spectral curve exist and are worth studying. 

In this thesis we discuss how one can derive the quantum spectral curve for the $\eta$-deformed $\ads$ superstring, an integrable deformation of the $\ads$ superstring with quantum group symmetry. This model can be viewed as a trigonometric version of the $\ads$ superstring, like the \textsc{xxz} spin chain is a trigonometric version of the  \textsc{xxx} spin chain. Our derivation starts from the ground-state thermodynamic Bethe ansatz equations and discusses the construction of both the undeformed and the $\eta$-deformed quantum spectral curve. We reformulate it first as an analytic $Y$-system, and map this to an analytic $T$-system which upon suitable gauge fixing leads to a $\Pf\mu$ system -- the quantum spectral curve. We then discuss constraints on the asymptotics of this system to single out particular excited states. At the spectral level the $\eta$-deformed string and its quantum spectral curve interpolate between the $\ads$ superstring and a superstring on ``mirror'' $\ads$, reflecting a more general relationship between the spectral and thermodynamic data of the $\eta$-deformed string.



\thispagestyle{empty}
\clearpage

\newpage
\vspace*{\fill}
$\,$\\
\thispagestyle{empty}
\section*{Zusammenfassung}
\selectlanguage{ngerman}
Die M"oglichkeit eine exakte L"osung einer wechselwirkenden Quantenfeldtheorie zu finden ist isoliert betrachtet bereits eine interessante Aussicht, da sie uns unbeschr"ankte Kontrolle liefert. Sie erm"oglicht es Ph"anomene, die von der Theorie beschrieben werden, sehr pr"azise zu analysieren. Im Kontext der AdS/CFT-Korrespondenz, die eine Dualit"at zwischen bestimmten Eich- und Stringtheorien beinhaltet, wurden Theorien untersucht, "uber die man eine au\ss ergew"ohnlich hohe Kontrolle hat, weshalb ein enormes Forschungsinteresse am Finden exakter L"osungen besteht. Insbesondere wurde festgestellt, dass die sogenannte planare $\mN =4$ Super Yang-Mills Theorie, und ihre AdS/CFT-duale Superstringtheorie auf dem $\ads$-Hintergrund eine integrable Struktur besitzen. Speziell das Spektralproblem, welches die Suche nach Skalendimensionen von lokalen Operatoren oder "aquivalent die Suche nach Stringenergien der Stringzust"ande beinhaltet, ist einfach zu handhaben. Nach mehr als einem Jahrzehnt Forschung wurde herausgefunden, dass das Spektralproblem sehr effizient mit Hilfe eines funktionalen Gleichungssystems, der sogenannten \emph{Quantum Spectral Curve} oder \emph{Quantenspektralkurve}, gel"ost werden kann. Es erm"oglicht weitere detaillierte Forschung zu den zuvor genannten Theorien. Gleichzeitig sind viele Fragen bez"uglich der weiteren Verwendung und der zugrunde liegenden Mechanismen der Quantenspektralkurve noch unbeantwortet und es wert weiter erforscht zu werden. 

In dieser Dissertation analysieren wir, wie man die Quantenspektralkurve der $\eta$-de-formierten $\ads$-Superstringtheorie, welche eine integrable Deformation der $\ads$-Superstringtheorie mit Quantengruppensymmetrie ist, herleiten kann. Dieses Modell kann wie eine Trigonometrisierung des $\ads$-Superstrings betrachtet werden, vergleichbar mit dem Verh"altnis zwischen den Heisenberg \textsc{xxz}- und \textsc{xxx}-Spinketten. Unsere Herleitung beginnt mit den \emph{Thermodynamischen-Bethe-Ansatz}-Gleichungen f"ur den Grundzustand und behandelt die Konstruktion der beiden Quantenspektralkurven. Wir formulieren es zuerst in ein analytisches $Y$-System und anschlie\ss end in ein analytisches $T$-System um. Nach der Festlegung von einer passenden Eichung verwandeln wir das $T$-System in ein $\Pf\mu$-System, welches die Quantenspektralkurve darstellt. Dann behandeln wir die notwendigen Einschr"ankungen des asymptotischen Verhaltens des Systems, um bestimmte angeregte Zust"ande zu beschreiben. Aus Sicht des Spektrums interpolieren die $\eta$-deformierte Stringtheorie und ihre Quantenspektralkurve zwischen der $\ads$-Superstringtheorie und einer Superstringtheorie auf dem ``\emph{mirror}" $\ads$. Damit wird auf eine allgemeinere Relation zwischen den spektralen und thermodynamischen Daten der $\eta$-deformierten Superstringtheorie hingewiesen. 
 

\thispagestyle{empty}
\clearpage
\selectlanguage{english}


\newpage
\vspace*{\fill}
\noindent \textbf{This thesis is based on the publication:} 
\begin{itemize}
\item[\cite{Klabbers:2017vtw}] 
Klabbers, Rob and van Tongeren, Stijn J., \emph{Quantum Spectral Curve for the eta-deformed AdS$_5\times$S$^5$ superstring}, Nucl.Phys. B925, 2017, 252, \\\href{https://arxiv.org/abs/1708.02894}{\color{blue} arxiv:1708.02894 [hep-th]}.
\end{itemize}
 
\noindent \textbf{Other publications by the author:}
\begin{itemize}
\item[\cite{Arutyunov:2017dti}]
Arutyunov, Gleb and Frolov, Sergey and Klabbers, Rob and Savin, Sergei, \emph{Towards 4-point correlation functions of 1/2-BPS-operators from supergravity}, JHEP 1704, 2017, 5,  \href{https://arxiv.org/abs/1701.00998}{\color{blue} arxiv:1701.00998 [hep-th]}. 
\item[\cite{Klabbers:2016cdd}]
Klabbers, Rob, \emph{Thermodynamics of Inozemtsev's elliptic spin chain}, Nucl.Phys. B907, 2016, 77, \href{https://arxiv.org/abs/1602.05133}{\color{blue} arxiv:1602.05133 [math-ph]}. 
\end{itemize}

\thispagestyle{empty}

\thispagestyle{empty}
\clearpage
\thispagestyle{empty}

\cleardoublepage
\pagenumbering{gobble}
\tableofcontents
\cleardoublepage
\pagenumbering{arabic}


\mainmatter

\part{Introduction}
\chapter{Introduction}
\label{chap:introduction}
\input{introduction}

\chapter{Background on the $\eta$-deformed model}
\label{chap:etadeformed}
\input{background}

\part{Constructing the $\boldmath{\eta}$-deformed QSC}
\label{part:qsc}

\chapter{Introduction}
\label{chap:etaQSC}
\input{etaQSC}

\chapter{Analytic $Y$-system}
\label{chap:Ysystem}
\input{Ysystem}

\chapter{Analytic $T$-system}
\label{chap:Tsystem}
\input{Tsystem}

\chapter{Quantum spectral curve}
\label{chap:QSC}
\input{QSC}

\chapter{Solution-state correspondence}
\label{chap:solutionstate}
\input{solutionstate}

\chapter{Conclusions and future work}
\label{chap:conclusions}
\input{conclusions}

\begin{appendices}
\chapter{$x$ functions}
\label{app:xfunctions}
\input{xfunctions}

\chapter{Definitions and conventions}
\label{app:Definitions}
\input{definitionsandconventions}

\chapter{Dressing-phase simplification}
\label{app:dressing}
\input{dressingphase}

\chapter{$Y$-system computations}
\label{app:Ysystemcomputations}
\input{Ysystemcomputations}

\chapter{Two lemmas}
\label{app:constrainingtrigonometricpolynomials}
\input{constrainingtrigpol}
\end{appendices}

\bibliographystyle{nb}
\bibliography{bibliography,mypapers}
\addcontentsline{toc}{chapter}{References}
\chapter*{Acknowledgements}

First and foremost I gladly thank you, Gleb, for your faith in me and all the trouble you go through to support me in my scientific endeavours. I really enjoy our discussions about physics, teaching and life in general. 

Working together really is a great and important aspect of science. I would like to thank Stijn for all the things I have learnt from him: your explanations and comments have helped me very much by connecting various scattered pieces of knowledge. Also observing your approach to science was very helpful. Jules, I really value our collaboration, it inspires and drives me. Your time in Hamburg was a lot of fun, as were our hikes in Les Houches. In particular thanks for introducing me to bouldering. 
Sergei, sharing an office with you is very enjoyable, as is disentangling the incredible mess that is supergravity together. Sergey, thanks for our collaboration and your insights on that last topic. 

Thanks to all the members of the theory department at DESY for the friendly atmosphere; in particular Rutger for sharing his incredibly positive outlook on life and Joerg for being my co-supervisor. Special thanks to you, Elizabeth, for all your help over the last years and the many laughs we shared. Also thanks to Stefano for sharing his experience with me in the early stages and Jan-Peter for his feedback on the abstract. 

It has been great to have the opportunity to also work in close proximity to mathematicians in the Geomatikum: I would like to thank the RTG for its support and all its members for our discussions during seminars and over coffee. I also thank Lorenz: our discussions were particularly helpful for my work and I thoroughly enjoyed talking with you. 

It has been great to travel around to meet with peers: I would like to thank all the people I have met during schools and conferences for the great time we have had and the big inspiration they were to me. I especially want to thank Matthias and Vladimir for their hospitality in Berlin and Paris.

I would like to thank all the people who have made living in Hamburg such a fun experience: in particular to Yannick, my bouldering crew and Thibaud, also for feedback on the introduction of this thesis. Ich m\"{o}chte mich insbesondere bedanken bei Lutz und Michi: ihr habt es mir leicht gemacht mich in Hamburg wohl zu f\"{u}hlen und mir viel geholfen in den ersten Monaten. Lutz, auch vielen Dank f\"{u}r deine ausf\"{u}hrliche Kommentar zu der Zusammenfassung. Ich danke auch Hanna, mit der ich immer mein Essen teilen kann. 

Lieve vrienden en familie, bedankt voor alles: voor tennis, bandrepetities, de gast-vrijheid, de gezellige avonden, het lekkere eten en al het andere. Bedankt As, Luuk, Dirk, Janneke and Jac. Willem, wat onze vriendschap voor me betekent gaat veel verder dan dit proefschrift. Bedankt daarvoor. 

Dear Dashunya, you are an enrichment to my life. I have thoroughly enjoyed our journey so far and cannot wait to see where it takes us next. Thanks for everything. 

Bedankt opa, voor al je moeite gedurende bijna twee decennia om me te laten zien hoe geweldig intrigerend wetenschap is. 

Lieve pap en mam, voor jullie onvoorwaardelijke steun kan ik jullie nooit genoeg bedanken.









\newpage
\vspace*{\fill}
\noindent \textbf{Eidesstattliche Erkl\"arung}
\medskip
\newline
Hiermit erkl\"are ich an Eides statt, dass ich die vorliegende Dissertationsschrift selbst verfasst und keine anderen als die angegebenen Quellen und Hilfsmittel benutzt habe.
\vspace{2cm}
\newline 
Hamburg, den 1. Dezember 2017 $\qquad \qquad \qquad$ Unterschrift
\thispagestyle{empty}
\clearpage
\end{document}

%% file: preamble.tex
\usepackage{amssymb}
\usepackage{amsmath,amsbsy}

\usepackage[english,ngerman]{babel}

\numberwithin{equation}{section}
\usepackage[titletoc,toc,title]{appendix}
\usepackage{mathtools}
\usepackage{mathrsfs}
\usepackage{fullpage}
\usepackage{float}
\usepackage{graphicx}
\usepackage[small,skip=0pt]{caption} 
\usepackage{subcaption}

\usepackage{slashed} 

\usepackage{bm}

\usepackage{tikz}

\newcommand{\Z}{\mathbb{Z}}
\newcommand{\ket}[1]{\mathopen| #1 \mathclose\rangle}

\newcommand{\bd}{\begin{displaymath}}
\newcommand{\ed}{\end{displaymath}}
\newcommand{\be}{\begin{equation}}
\newcommand{\ee}{\end{equation}}
\newcommand{\bea}{\begin{eqnarray}}
\newcommand{\eea}{\end{eqnarray}}
\newcommand{\C}{\mathbb{C}}
\newcommand{\R}{\mathbb{R}}
\newcommand{\N}{\mathbb{N}}
\newcommand{\e}{\epsilon}

\newcommand{\rai}{\rightarrow \infty}
\newcommand{\nn}{\nonumber \\}
\newcommand{\ssum}[1]{\sum_{#1 =1}^{\infty}}

\newcommand{\bfB}{\textbf{B}}
\newcommand{\bfH}{\textbf{H}}

\newcommand{\mK}{\mathcal{K}}
\newcommand{\slashconv}{\hat{\slashed{\star}}}

\newcommand{\ad}{c}
\newcommand{\s}{\mathbf{s}}
\newcommand{\Pf}{\mathbf{P}}
\newcommand{\Pft}{\tilde{\Pf}}
\newcommand{\Qf}{\mathbf{Q}}
\newcommand{\Qft}{\tilde{\Qf}}
\newcommand{\Qm}{\mathcal{Q}}

\newcommand{\bR}{\mathbf{R}}
\newcommand{\bL}{\mathbf{L}}
\newcommand{\bQ}{\mathbf{Q}}
\newcommand{\bH}{\mathbf{H}}
\newcommand{\bC}{\mathbf{C}}
\newcommand{\mA}{\mathcal{A}}
\newcommand{\mAp}{\mathcal{A}^{\text{p}}}

\newcommand{\mJ}{\mathbb{J}}

\newcommand{\hbT}{\hat{\mathbb{T}}}
\newcommand{\mbT}{\mathbb{T}}
\newcommand{\bfT}{\textbf{T}}
\newcommand{\JT}{\mathscr{T}}
\newcommand{\mcT}{\mathcal{T}}
\newcommand{\hmcT}{\hat{\mathcal{T}}}

\newcommand{\mF}{\mathcal{F}}
\newcommand{\hG}{\hat{G}}
\newcommand{\mN}{\mathcal{N}}

\newcommand{\bO}[1]{\mathcal{O}\left(#1\right)}
\newcommand{\ii}{i}

\def \mut {\tilde{\mu}}
\def \omegat {\tilde{\omega}}
\def\ads{{\rm AdS}_5\times {\rm S}^5}

\DeclareMathOperator{\arcsec}{arcsec}

\DeclareMathOperator{\arcsinh}{arcsinh}

\newcommand{\factornumbering}[2]{\ensuremath{ \underbrace{ #1}_{\tiny
\begin{tikzpicture}[baseline=-2mm]
\node[circle, draw] {#2};
\end{tikzpicture}
} } }

\makeatletter
\DeclareRobustCommand\widecheck[1]{{\mathpalette\@widecheck{#1}}}
\def\@widecheck#1#2{%
    \setbox\z@\hbox{\m@th$#1#2$}%
    \setbox\tw@\hbox{\m@th$#1%
       \widehat{%
          \vrule\@width\z@\@height\ht\z@
          \vrule\@height\z@\@width\wd\z@}$}%
    \dp\tw@-\ht\z@
    \@tempdima\ht\z@ \advance\@tempdima2\ht\tw@ \divide\@tempdima\thr@@
    \setbox\tw@\hbox{%
       \raise\@tempdima\hbox{\scalebox{1}[-1]{\lower\@tempdima\box
\tw@}}}%
    {\ooalign{\box\tw@ \cr \box\z@}}}
\makeatother
\usepackage{scalerel}
\usepackage{stackengine,wasysym}

\newcommand\reallywidetilde[1]{\ThisStyle{%
  \setbox0=\hbox{$\SavedStyle#1$}%
  \stackengine{-.1\LMpt}{$\SavedStyle#1$}{%
    \stretchto{\scaleto{\SavedStyle\mkern.2mu\AC}{.5150\wd0}}{.3\ht0}%
  }{O}{c}{F}{T}{S}%
}}

\newcommand{\cD}{\hat{\Delta}}
\newcommand{\ewD}{\hat{\Delta}_e}

\usepackage{marginnote}
\gdef\notesoff%
{\gdef\note##1{\mbox{}\marginnote{%
\color{white}    ##1}}}

\newcommand{\cwanalyticcontinuation}[2] {
#1_{\hspace{-2pt}\begin{tikzpicture}
\draw[->] (0.15,-0.03) arc (350:0:0.15cm);
\node at (0,0) {\scalebox{0.75}{$#2$}};
\end{tikzpicture}
}
}

\newcommand{\ccwanalyticcontinuation}[2] {
#1_{\hspace{-2pt}\begin{tikzpicture}
\draw[<-] (0.15,-0.03) arc (350:0:0.15cm);
\node at (0,0) {\scalebox{0.75}{$#2$}};
\end{tikzpicture}
}
}

\newcommand{\defeq}{:=}

\usepackage{mciteplus}

\usepackage[hidelinks,plainpages=false,hypertexnames=false,linktocpage=true, linkcolor=blue, citecolor=blue, urlcolor=blue]{hyperref}

%% file: introduction.tex
The ultimate goal of physics is to understand nature. A daunting task, considering the huge amount of phenomena one needs to understand in order to reach this goal, ranging all the way from the interactions of elementary particles on ultra-short distance scales to colliding galaxies on ultra-long distance scales.
Nevertheless physicists have succeeded in explaining many phenomena through a framework of laws, the most prominent ones being the frameworks of quantum field theory and general relativity. Although these frameworks are very successful, they do contain unsolved puzzles that theoretical physicists have been breaking their heads over for almost a century. Moreover, for some phenomena appearing in nature these frameworks do not seem to be very natural and careful study is extremely hard. Colour confinement in quantum chromodynamics forms an excellent example: although captured in the framework of quantum field theory colour confinement is hard to study theoretically, since it requires a good understanding of the theory at strong coupling. One way to understand colour confinement better is to find a simpler model that exhibits confinement yet is easier to study than quantum chromodynamics, such as two- and three-dimensional abelian gauge theories or even certain coupled spin chains of spin-$1/2$ particles \cite{Lake:2009pi}. Indeed, studying a phenomenon in relative isolation -- that is, without many of the complications presented to us by nature -- can help a great deal to discover its origins. One of the driving forces of mathematical physics is to find and study such simple models in order to uncover the underlying mechanisms of nature. A particularly important role is reserved for models which are not only easy to describe, but also to a certain extent easy to \emph{solve}. These so-called \emph{tractable} models often exhibit beautiful mathematical structures as well, and it is often the presence of these structures that causes the high degree of solvability. 

Some of these structures are known as \emph{integrable} structures and can be found in various branches of physics: the oldest instances come from classical mechanics under the name of Liouville integrability. Many of the newer incarnations, in particular in quantum physics, are inspired by the seminal work of Hans Bethe \cite{bethe1931theorie} on the quantum mechanical spin chain known as the Heisenberg \textsc{xxx} spin chain \cite{heisenberg1928theorie}. They go under the name of 
\emph{quantum integrability}. Integrability even arises in realistic physical systems, such as the quantum Newton cradle \cite{kinoshita2006quantum} or the Korteweg-de-Vries equation modelling shallow water waves \cite{boussinesq1877essai,korteweg1895onthe}. The presence of integrability allows one to find far more precise results than one can usually obtain using perturbative methods and in many cases it allows one to solve the theory exactly, therefore allowing for very precise studies of ideas and concepts arising in physics. This very attractive feature is one of the motivations to find out when and why integrability arises in models in the first place and why it leads to such precise results. Understanding this better could help physicists design their toy models and thereby ease the study of complicated ideas. 

\paragraph{$\bm{\mN= 4}$ super Yang-Mills theory.} A prime example\note{Making the first thing bold} of a tractable model is the gauge theory known as $\mN=4 $ super Yang-Mills theory in four dimensions with gauge group SU$(N)$ and gauge coupling $g_{\textsc{YM}}$ (or $\mN=4$ SYM for short). Its tractability is due to the large degree of symmetry present in the theory: not only is this theory maximally supersymmetric, it is also conformal, thereby organising its excitations in superconformal multiplets. Moreover, the supersymmetry of the theory ``protects" certain quantities, meaning they do not receive quantum corrections, and also forms the key ingredient in \emph{localisation}, a method that allows for the exact computation of quantities expressed in terms of path integrals, for example certain Wilson loops of $1/8$ BPS\footnote{The abbreviation stands for Bogomol'nyi-Prasad-Sommerfield.} operators \cite{Pestun:2009nn}. Despite this high amount of symmetry, the theory is a highly non-trivial non-abelian gauge theory and being able to study any aspect of it with a large amount of precision is a rare and valuable asset for theoretical physics. The theory becomes even more manageable in what is known as the \emph{planar} limit -- sending $N\rai$ while keeping the combination $\lambda = g_{\textsc{YM}}^2 N$ known as the 't Hooft coupling finite -- because of the emergence of integrability.\footnote{It is called planar because only the Feynman diagrams that one can draw on a genus-zero surface contribute in this limit to the perturbative series of correlation functions.} 

$\mN=4$ SYM is most famous for the central role it plays in the \emph{AdS/CFT correspondence}, a correspondence between gauge and string theory that has inspired many theorists in the past two decades.\note{List of physics inspired by AdS/CFT} It was hypothesised by Maldacena \cite{Maldacena:1997re} based on the holographic principle first introduced by 't Hooft \cite{tHooft:1973alw}. It conjectures that certain string theories defined on an (asymptotically) anti-de-Sitter spacetime in $d+1$ dimensions are equivalent to particular conformal field theories in flat $d$-dimensional spacetime. The best understood example of the correspondence features $\mN=4$ SYM on the gauge theory side and the ten-dimensional type IIB string theory on an $\ads$ background on the string theory side. The most striking feature of this correspondence is the fact that it is a \emph{weak-strong} duality: the strongly coupled regime of $\mN=4$ SYM -- that is, for $\lambda \gg 1$ -- can be identified with the weak tension regime of the string theory and vice versa. Because of this it is possible to use perturbative techniques in string theory to learn more about the usually inaccessible strong coupling regime of a gauge theory and vice versa. This is a very exciting prospect and indeed has led to many remarkable results. However, the correspondence itself has to be scrutinised as well, as any hypothesis should. It is at this point that integrability -- whose presence in some strongly-coupled gauge theories was already recognised by Lipatov in \cite{Lipatov:1993yb} -- can come in handy: for most cases one cannot use the usual perturbation theory to analyse the correspondence, exactly because it is a weak/strong duality. Therefore obtaining exact results proves to be extremely useful to test the AdS/CFT correspondence and to investigate possible extensions. Therefore, even though the correspondence is believed to hold for all values of $N$, almost all of the evidence has been collected in the planar limit, which on the string theory side leads to free string theory: for large $N$ the string coupling constant $g_s$ tends to zero since it relates to the 't Hooft coupling $\lambda$ as $g_s =\lambda/4\pi N$. 

Before we continue, let us think for a second what it means to ``solve" a quantum field theory: for a generic QFT, that requires one to find the entire spectrum of elementary excitations and the $n$-point correlation functions relating them. These excitations can in principle be any gauge-invariant function of the fields, be it local operators built from fields evaluated at one point in spacetime or non-local operators such as Wilson loops. For a theory with conformal symmetry such as $\mN=4$ SYM the problem of $n$-point correlation function of local operators is simplified a lot since all these functions can be constructed out of the two- and three-point correlators. Moreover, in a convenient basis for the two-point correlators the basis elements are specified by just one coupling-dependent function, the \emph{scaling dimension}, and one can find this basis and these functions by diagonalising the \emph{dilatation operator}. For every triple of basis elements one extra function specifies the associated three-point correlator, and together with the scaling dimensions these parametrise all the correlation functions of local operators. 

\paragraph{Spectral problem.} It is in the search for scaling dimensions   -- usually dubbed the \emph{spectral problem} -- that the presence of integrability proved to be of vital importance: the gauge-invariant single-trace operators built up from a fixed amount $L$ of only two of the scalars have a very simple structure, being a trace over products of these scalars. One finds that these operators form a closed subspace under the action of the dilatation operator, implying we can represent it as a finite-dimensional matrix. As long as $L$ is not too large it is possible to find scaling dimensions by direct diagonalisation of this matrix, but these results would not be very insightful as they would not reveal any of the underlying structure. An alternative was presented by Minahan and Zarembo in \cite{Minahan:2002ve}, using the fact that these operators can be approached from a different, well-studied direction: representing the scalars by ``spin up" and ``spin down" respectively every operator can be interpreted as a state of a periodic spin chain of spin-$1/2$ particles of length $L$, where the periodicity follows from the trace. The dilatation operator in this picture becomes a spin chain hamiltonian, which in itself is not yet remarkable. However, Minahan and Zarembo found that at one loop this is not just any hamiltonian, it is the Heisenberg \textsc{xxx} spin chain. This means that at one loop the problem of finding scaling dimensions for operators of different length is integrable and can be unified by the \emph{Bethe ansatz}, implying one can write down a simple set of equations depending parametrically on $L$ known as \emph{Bethe equations} to describe the scaling dimensions. 

It was soon found this result can be extended to the entire theory at one loop \cite{Beisert:2003jj} and for the $\mathfrak{su}(2)$ subsector discussed above also to higher loops \cite{Beisert:2003tq,Serban:2004jf}, where the interaction becomes more and more long-range. Assuming that integrability should be present at all loop orders one can in fact uniquely determine the Bethe equations of the all-loop dilatation operator of $\mN=4$ SYM up to a single phase factor by bootstrapping the spin-chain $S$ matrix \cite{Beisert:2005tm,Beisert:2004hm,Beisert:2005fw}. A problem of this approach is that the interactions of these integrable spin chain hamiltonians become more and more long-range as the loop order increases, such that at a certain point the interaction length exceeds the spin chain length and wraps all the way around the spin chain. At this point the found Bethe equations no longer yield the correct result for the scaling dimensions and one has to correct for the wrapping of the interactions. Nevertheless, for operators of sufficient length $L$ these Bethe equations do in fact correctly yield scaling dimensions up to a loop order that is unachievable by normal perturbative methods. The main inspiration to deal with these wrapping corrections come from the other side of the AdS/CFT correspondence, the string theory on $\ads$. 

\paragraph{String theory.} The most commonly used formulation of the type IIB string theory on $\ads$ is in the Green-Schwarz formalism, which allows one to actually write down the action of the model in a compact form \cite{Metsaev:1998it}. In contrast, the presence of a self-dual Ramond-Ramond five-form flux makes it unclear how to follow the usual approach of Neveu-Schwarz-Ramond to construct the action. In the Green-Schwarz formalism, the string is described as a non-linear sigma model with the quotient group 
\begin{displaymath}
\frac{\text{PSU}(2,2|4)}{\text{SO}(4,1) \times \text{SO}(5)}
\end{displaymath}
as its target space. Already at this stage it was noticed that there is something special about this string theory: as defined by the Green-Schwarz action it is classically integrable, allowing for a description of the equations of motion in terms of a Lax connection \cite{Bena:2003wd}. 

After fixing the light-cone gauge to remove unphysical degrees of freedom (see \cite{Arutyunov:2004yx,Frolov:2006cc}) and obtaining the world-sheet hamiltonian one can ask how to quantise this integrable field theory. Due to the complicated form of the hamiltonian one cannot hope to quantise it directly and resorting to a different approach is more fruitful: motivated by the presence of integrability on the gauge theory side of the AdS/CFT correspondence one can try to fix the $S$ matrix of the two-dimensional conformal field theory on the world sheet assuming quantum integrability, quite similar to the approach followed for $\mN=4$ SYM. The abundance of symmetry present again constrains the $S$ matrix -- which up to a change of basis is the same as the $S$ matrix found on the gauge theory side -- up to an overall phase factor called the \emph{dressing phase} \cite{Arutyunov:2006yd,Arutyunov:2006ak}. An important thing to take into account here is that the string world sheet is a cylinder and therefore there is formally no notion of asymptotic states, precluding the normal use of $S$-matrix theory. However, by taking the circumference of the cylinder to be very large we can still proceed, and using the $S$ matrix we can construct multi-particle states and derive equations that look very much like Bethe equations to describe their energy spectrum \cite{Arutyunov:2004vx}. These equations still contain one missing element, the dressing phase. Fortunately, since we are now in the context of quantum field theory (as opposed to the quantum-mechanical setting of the spin chain) it is clearer how one can find the dressing phase: requiring the $S$ matrix to be unitary and obey a non-relativistic version of crossing symmetry\footnote{In the light-cone gauge the $\ads$ string theory is not relativistic, hence we cannot expect to impose the usual form of crossing symmetry.} leads to a set of functional equations that the dressing phase should satisfy \cite{Janik:2006dc}. Although these equations have many solutions, the solution presented in \cite{Beisert:2006ez} can be argued to be the correct one, by requiring a set of physically motivated constraints.\footnote{For a review on this subject, see \cite{Vieira:2010kb}.} In particular it was shown to agree with gauge theory computations up to four loops \cite{Bern:2006ew,Beisert:2007hz}. 

So far, attacking the spectral problem from the string theory side has only brought us a way to find the dressing phase. The problem how to deal with wrapping interactions remains and in fact finds its way into the string description: the quantum corrections to the string energy, which are the observables dual to the scaling dimensions, come in the form of virtual particles that travel all around the circle. The way to treat these virtual particles for relativistic integrable field theories has been known for a long time under the name of \emph{thermodynamic Bethe ansatz}.

\paragraph{Thermodynamic Bethe ansatz.} The name of the thermodynamic Bethe ansatz (TBA) method suggests it is meant to obtain information about the thermodynamics of physical systems. Indeed, the original application of the TBA method by Yang and Yang  to find the free energy of the Lieb-Liniger gas had exactly this purpose \cite{Yang:1968rm} and has since been applied to many models, such as the Hubbard model \cite{takahashi1972one,Korepin} and by the author to Inozemtsev's elliptic spin chain \cite{Klabbers:2016cdd}. How, then, did this method find its way into the solution to the spectral problem of $\ads$? The answer lies in an idea first proposed by Zamolodchikov \cite{Zamolodchikov:1989cf} that is based on the fact that one can compute the partition function of a QFT by Wick rotating it to depend on an imaginary-time coordinate and subsequently compactifying this new time direction on a circle with circumference $1/T$. For our two-dimensional QFT this changes the spacetime to a torus on which one can consider ``time" evolution along either of the circles. By Wick rotating further we obtain a model, dubbed the \emph{mirror model} by Arutyunov and Frolov in \cite{Arutyunov:2007tc}, where the roles of space and time have been formally interchanged. Following this through shows that to find the ground-state energy of the original model in a finite volume we can equivalently compute the free energy of the mirror model at finite temperature but in \emph{infinite} volume, such that wrapping corrections can be neglected. If in addition the mirror model is integrable, we can apply the ideas in the previous section to find its $S$ matrix and compute its free energy using the TBA method, thereby finding the ground-state energy of the original model.

The approach explained above can indeed be applied to the spectral problem of $\ads$: the mirror model was found in \cite{Arutyunov:2007tc} and a further derivation of the TBA equations followed soon \cite{Arutyunov:2009zu,Arutyunov:2009ur,Bombardelli:2009ns,Gromov:2009bc}. These equations allowed for the computation of the ground-state energy, which in fact equals zero, but more importantly allowed for the analysis of excited states through analytic continuation \cite{Dorey:1996re,bazhanov1997quantum}. Indeed, many case studies were done \cite{Bajnok:2009vm,Balog:2010xa,Arutyunov:2009ax,Gromov:2009bc,Arutyunov:2011mk,Arutyunov:2012tx} based on the excited-state TBA equations derived in \cite{Arutyunov:2009ax,Gromov:2009bc}.

\paragraph{Beyond the TBA.} For many systems the TBA equations provide the simplest form of the spectral problem. Although it is fairly straightforward to write the equations in the form of a \emph{$Y$ system} \cite{Arutyunov:2009ur,Gromov:2009tv,Bombardelli:2009ns}, a system of functional difference equations, these do not provide a true simplification: the $Y$-system equations have many solutions and selecting the correct solution can be very difficult. In the particular case of the interacting non-abelian gauge theory that is $\mN=4$ SYM it seems like a lot to ask for an even simpler form than the TBA. In some integrable field theories, however, a simplification is possible in the form of Destri-de Vega (DdV) equations, which are a \emph{finite} set of integral equations \cite{Destri:1992qk,Fioravanti:1996rz}. Furthermore, the TBA equations are difficult to treat analytically. With the TBA it is possible to compute one of the simplest excitations known as the Konishi operator up to five loops \cite{Eden:2012fe} in normal perturbative quantum field theory, the same loop order as was achieved for the TBA \cite{Arutyunov:2010gb,Bajnok:2009vm,Balog:2010xa}.

It is therefore not surprising that the search for a simpler formulation of the spectral problem did not stop at this stage:  the presence of integrability and the analytic results at the first few loop orders do give hope that it is possible to find analytic expressions for the scaling dimensions also at higher loops and the TBA equations and the associated $Y$ system are an excellent starting point for this investigation. 

In \cite{Gromov:2010km} the first steps towards further simplification were taken, proposing a general solution to the $Y$ system in terms of Wronskian determinants, reducing the number of independent functions, called $Q$ functions, to only seven. The connection to the analyticity requirements coming from the TBA equations is not clear from this construction, but was salvaged: a minimal set of analytic data was found that singles out the solution of the $Y$ system that corresponds to the solution of the ground-state TBA equations \cite{Balog:2011nm,Cavaglia:2010nm}. This data together with the $Y$ system equations form the \emph{analytic $Y$-system}. Although equivalence with the TBA equations is only established for the ground state, its generalisation to excited states seems less troublesome than for its TBA predecessor, as one only needs to allow for extra poles and zeroes in the solutions of the analytic $Y$ system, without changing the $Y$-system equations themselves. In contrast, although the contour deformation trick \cite{Arutyunov:2009ax} does allow for a similar generalisation, the TBA equations for excited states are different from the ground-state TBA-equations. Finding solutions to the $Y$ system in practice is not particularly easy though. A further simplification can be formulated based on the fact that the $Y$ system can be viewed as a gauge-invariant form of the \emph{Hirota equation}
\be
\label{eq:hirota}
T_{a,s}(u+i/2)T_{a,s}(u-i/2) = T_{a+1,s}(u)T_{a-1,s}(u)+T_{a,s+1}(u)T_{a,s-1}(u),
\ee
where the labels $a,s$ live on what is known as the $T$ hook (see fig. \ref{fig:thook}) and the $T$ functions are building blocks that one decomposes the $Y$ functions into. The extra analyticity data takes a simple form in terms of the $T$ functions and together with the Hirota equation it forms the \emph{analytic} $T$ system. This allows for the reduction of the infinite $Y$ system to a \emph{finite} set of integral equations called FiNLIE \cite{Gromov:2011cx} (see also \cite{Balog:2012zt}), which allow for the computation of scaling dimensions of up to eight loops. However, starting from the fact that the $T$ functions can be decomposed into $Q$ functions as follows from the Wronskian solution of the $Y$ system a further (and most likely final) major simplification of the spectral problem is possible: transferring the analytic properties of the $T$ functions from the FiNLIE to the language of the Wronskian determinants and its $Q$ functions simplifies them once again and the resulting system of equations takes a very simple and symmetric form known as the \emph{Quantum Spectral Curve} (QSC) \cite{Gromov:2013pga,Gromov:2014caa}. The FiNLIE equations still are non-linear integral equations, making analytical study difficult, the QSC equations on the other hand are functional equations relating different evaluations of its constituent functions on different sheets of the Riemann surface on which they are defined. Although solving equations of this type can generically also be very difficult, it is possible for the AdS/CFT spectral problem, since the solutions one needs to describe scaling dimensions have such nice properties. More precisely, one can solve these equations analytically for any state using the same universal perturbative algorithm \cite{Marboe:2014gma}, and for ``simple" states one can find their scaling dimension up to $14$ loops, a stunning result! An important reason for this unification is the fact that the way that the quantum numbers of a state under consideration are encoded in the QSC is very simple: all six quantum numbers appear on equal footing in the asymptotics of the basic functions as one sends the spectral parameter to infinity. This is in contrast with for example the TBA, where the scaling dimension follows after solving the integral equations for a set of $Y$ functions and the other five quantum numbers change the form of the equations explicitly. 

Possibly more interesting even is the fact that the QSC is supposed to be valid for any value of the coupling and could potentially lead to exact results for all-loop scaling dimensions! Until now only a numerical algorithm is known \cite{Gromov:2015wca} which computes scaling dimensions at least up to 't Hooft coupling around $1000$ with tens of digits of precision. Using data extracted from this algorithm investigations into an analytic strong coupling solution of the QSC has been initiated \cite{Hegedus:2016eop}.

The QSC has led to many interesting results: not only did it allow for the analysis of arbitrary states such as twist operators \cite{Gromov:2014bva}, it turned out to be a starting point for the study of different observables in $\mN=4$ SYM, such as the BFKL pomeron \cite{Alfimov:2014bwa}, the cusped Wilson line \cite{Gromov:2015dfa} and the quark-anti-quark potential \cite{Gromov:2016rrp}. This is remarkable, as these observables are outside of the scope of the original spectral problem. 

Its wide applicability suggests a deeper level to the QSC that is yet to be understood. \note{hmm..} One might also wonder whether the occurrence of such a drastic simplification to the spectral problem is unique to the $\ads$  case.
By now in fact there are a couple theories for which a QSC has been constructed, perhaps the most physical model of which is the Hubbard model \cite{Cavaglia:2015nta}.  More recently, the QSC for another AdS/CFT pair was constructed, namely for the type IIA string theory on AdS$_4\times$ CP$^3$, whose integrability was established long ago in \cite{Arutyunov:2008if}, and its AdS/CFT dual planar $\mN=6$ superconformal Chern-Simons theory known as ABJM theory \cite{Bombardelli:2017vhk}. This QSC resembles the $\ads$ case in many ways, most noticably in the analytic structure of its constituent functions. The presence of OSp$(4|6)$ symmetry compared to the PSU$(2,2|4)$ symmetry in the $\ads$ case at first glance changes the algebraic structure of the QSC significantly. Closer inspection shows, however, that after the proper identification of functions the algebraic structure can be made to match exactly, whereas their analytic properties seem to have been swapped compared to the $\ads$ case. 

\paragraph{Deformations.} The fact that the spectral problem for $\ads$ could be simplified to the QSC is such a great achievement, that one can also rightly ask how unique the QSC's existence is. A way to find out is is to look at deformations, alterations of the original theory continuously parametrised by a parameter such that one can see exactly what changes when the deformation is turned on. Since the QSC is a solution to the spectral problem on both sides of the AdS/CFT correspondence we can look for deformations on both the string and the gauge theory sides. Many such deformations have been found and the literature on this topic is rich, but unfortunately so too is the list of names given to them. Many deformations are reviewed in the reviews \cite{Zoubos:2010kh,vanTongeren:2013gva}. 

\subparagraph{Hopf-twisted deformations.} On the gauge theory side perhaps the most natural thing to look for are exactly marginal deformations of the lagrangian, since after all the $\mN=4$ theory is conformal. The existence of $\mN=1$ marginal deformations was proven in \cite{Leigh:1995ep} and a particular three-dimensional family of deformations, known as \emph{Leigh-Strassler deformations}, was proposed, which under a suitable condition containing the gauge coupling in fact form exactly marginal deformations. However, for the study of the QSC it seems to be necessary to restrict our attention to integrable deformations, since the construction of the QSC -- and more generally the simplification of the spectral problem -- makes very heavy use of the integrability machinery. Careful investigations have been done whether any of the members of the family discussed above preserves integrability, see for example \cite{Mansson:2008xv}. All the integrable deformations in the Leigh-Strassler family can be related to a one-parameter subfamily known as the real-$\beta$ deformation when using the notion of Hopf twisting: these change the $R$ matrix underlying the deformation by applying certain quantum Hopf algebra transformations corresponding to a quantum deformation of the SU$(3)$-$R$ symmetry. In \cite{Beisert:2005if} it was found that the real-$\beta$ deformation itself can be incorporated in this framework, rendering all the integrable deformations in the Leigh-Strassler family to be a Hopf-twisted version of $\mN=4$ SYM. 

\subparagraph{TsT-based deformations.} The existence of the AdS/CFT correspondence immediately induces two questions about the CFT deformations: is there a gravity dual for the found integrable deformations and -- more generally -- how can we describe deformations in the language of string theory? The nice framework of non-linear sigma models has helped a lot in the consideration of this question, as we will see in chapter \ref{chap:etadeformed}. Historically, however, other methods seemed more natural to consider. Indeed, it was recognised in \cite{Lunin:2005jy} that one could deform the background of the string theory by a so-called \emph{TsT transformation}, a combination of a $T$-duality transformation of one of the angle variables followed by a shift along one of the isometry directions and a subsequent $T$-duality on the first angle variable. This deformation generates a one-parameter family of theories and the case considered in \cite{Lunin:2005jy} is in fact the gravity dual of the real-$\beta$ deformation and now goes by the name \emph{Lunin-Maldacena background}. Subsequent TsT transformations are possible as well and for example give rise to the $\gamma$ deformation \cite{Frolov:2005dj}. Importantly, it turns out that TsT transformations preserve integrability \cite{Alday:2005ww}, although in general they break at least some of the supersymmetry. As was discovered recently, they are in fact special (namely abelian) cases of a larger class of deformations called \emph{Yang-Baxter deformations} \cite{Osten:2016dvf}. 

\subparagraph{Yang-Baxter deformations.} These deformations are obtained by deforming the underlying Poisson structure of the Lax formulation of the model \cite{Klimcik:2002zj,Klimcik:2008eq,Delduc:2013fga}. The input for these deformations are anti-symmetric solutions (or \emph{$r$ matrices}) of the modified classical Yang-Baxter equation, thereby allowing for a classification of these integrable deformations through classification of the solutions to this equation. Indeed, many models of this type were constructed \cite{Kawaguchi:2014qwa,Matsumoto:2015jja,vanTongeren:2015soa,Hoare:2016ibq,Hoare:2016hwh} and previously known deformations such as the real-$\beta$ deformation can be reformulated as a Yang-Baxter deformation \cite{Matsumoto:2014nra}, thereby unifying a large class of deformations. Finding or even proving the existence of a dual gauge theory is very interesting from the perspective of the AdS/CFT correspondence and can in some cases actually be done. As shown in \cite{Matsumoto:2014gwa, vanTongeren:2015uha, vanTongeren:2016eeb} a large class of Yang-Baxter deformations is dual to various noncommutative versions of supersymmetric Yang-Mills theory, where the noncommutativity is governed by the same $r$ matrix that plays a role in the deformation of the sigma model. Whether the spectral problem of these theories is tractable in any sense, however, is unclear. If possible at all, it seems that the approach taken for the undeformed case should be the most fruitful one. 

\subparagraph{Quantum group deformations.} A very important example where this turns out to be possible is the real-$q$ deformation of the $\ads$ superstring (commonly called the \emph{$\eta$ deformation}) \cite{Delduc:2013qra,Delduc:2014kha}, which follows and builds upon earlier work on deforming the principal chiral model \cite{Klimcik:2002zj,Klimcik:2008eq}. This particular deformation breaks all the supersymmetry and all the non-abelian isometries, but seems tractable still due to its integrability. Indeed, although far from trivial it was shown in \cite{Arutyunov:2013ega} that there exists an $S$ matrix that matches the tree-level bosonic $S$ matrix in the large tension limit. This $S$ matrix was already constructed in \cite{Beisert:2008tw} by considering the quantum group deformation $U_q\left(\mathfrak{psu}(2|2)\right)$ of the algebra $\mathfrak{psu}(2|2)$ which underlies the $S$ matrix in the original $\ads$ theory. This quantum group deformation comes with a complex parameter $q$, but usually only the cases for real $q$ or $q$ a root of unity are considered. The appearance of $U_q\left(\mathfrak{psu}(2|2)\right)^{\otimes 2}$ as the symmetry group for the $S$ matrix of the quantised theory is no surprise, as it was shown in \cite{Delduc:2013fga} that the classical lagrangian has $U_q\left(\mathfrak{psu}(2,2|4)\right)$ symmetry, thereby also closely following the construction of the $S$ matrix of the original $\ads$ superstring. Moreover, the appearance of quantum groups in integrability is ubiquitous, with the role it plays for the Heisenberg \textsc{xxz} spin chain as the most famous example. As we will see, the $\eta$-deformed theory plays a role very similar to the one played by the \textsc{xxz} spin chain. 

With an $S$ matrix at hand one can of course also consider the spectral problem of the deformed theory, as was done in \cite{Arutynov:2014ota} by deriving the TBA equations, setting up the possiblity to find a QSC for this deformed model. It is precisely this quest we are rapporting on in this thesis. Let us finally mention that the $S$ matrix with $U_q\left(\mathfrak{psu}(2|2)\right)^{\otimes 2}$ symmetry with $q$ a root of unity can also be matched to a classical string theory, the Pohlmeyer reduced $\ads$ superstring. This theory is a fermionic extension of a gauged Wess-Zumino-Witten model with level $k$ (see \cite{Grigoriev:2007bu,Hoare:2011wr,Hoare:2012nx} and the review \cite{vanTongeren:2013gva}). Even though its symmetry group and $S$ matrix are so closely related to the $\eta$-deformed theory, as of yet this classical theory has not been incorporated in the class of Yang-Baxter deformations. We will review the construction of the Yang-Baxter deformations in more detail for the $\eta$-deformed case in chapter \ref{chap:etadeformed}. 

\subparagraph{Even more deformations.} Other ways to deform either the $\mN=4$ gauge or string theory have been considered. One large class we have not mentioned yet are the \emph{orbifoldings}: taking a discrete subgroup of the $R$-symmetry group of the gauge theory one can define a projection of the fields dependent on this subgroup. The projected fields are less supersymmetric than their unprojected parents. For the string theory dual orbifolding can be done on the level of the background: any discrete subgroup $\Gamma$ that acts on the $\ads$ background can be used to define a string theory on the quotient $\ads/\Gamma$. See the reviews \cite{Zoubos:2010kh,vanTongeren:2013gva} for more information. 

\begin{figure}[!t]
\includegraphics[width=\textwidth]{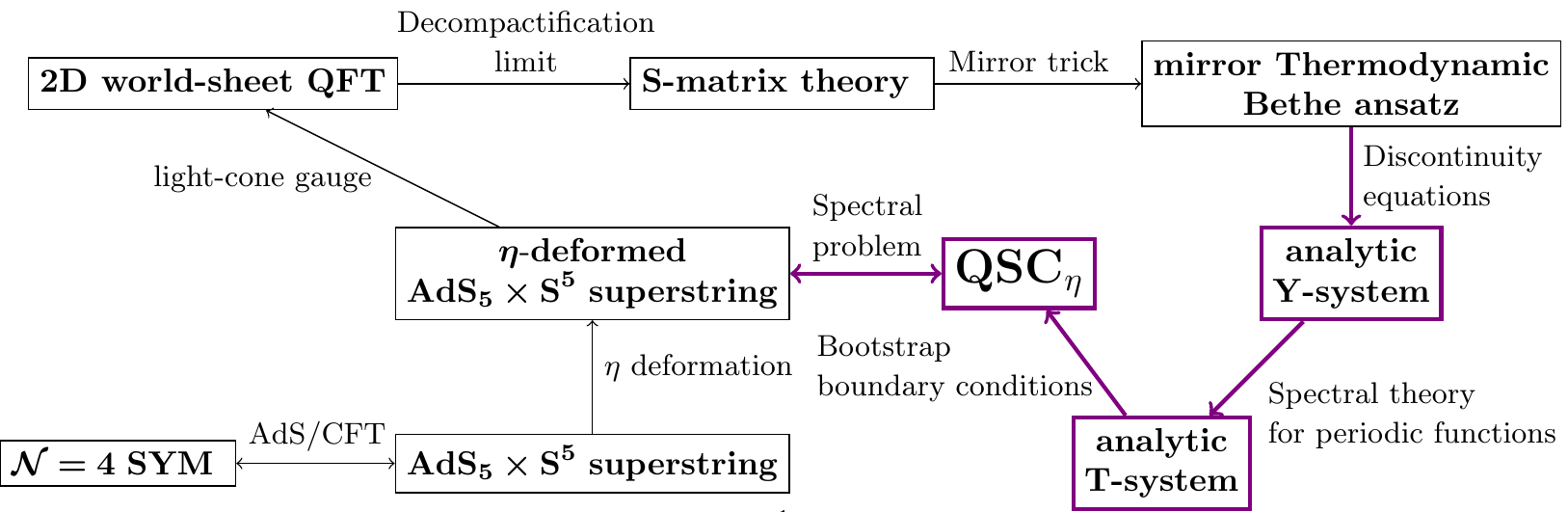}
\vspace{0.3cm}
\caption{An overview of the topics in this thesis: starting from the AdS/CFT correspondence we consider the $\eta$ deformation of the $\ads$ superstring and the steps necessary to arrive at the quantum spectral curve. The thick violet part of the diagram will be considered in Part \ref{part:qsc} of this thesis and constitutes our contribution to the subject.}
\label{fig:Overview}
\end{figure}

\paragraph{Aim of this thesis and summary.} In this thesis we  consider the spectral problem for the $\eta$-deformed $\ads$ superstring, ultimately culminating in the construction of the $\eta$-deformed quantum spectral curve, a one-parameter deformation of the quantum spectral curve constructed for the $\ads$ superstring. An overview of the necessary steps can be found in fig. \ref{fig:Overview}. 

In chapter \ref{chap:etadeformed} we consider the preliminaries necessary to undergo this quest: we first introduce the classical $\ads$ superstring and its $\eta$ deformation. We discuss how to obtain the exact $S$-matrix for the undeformed theory in the Hopf algebra formalism. This allows us to introduce the quantum group deformation corresponding to the classical $\eta$ deformation. We then bootstrap the integrable $S$ matrix with $U_q\left(\mathfrak{psu}(2|2)\right)^{\otimes 2}$ symmetry and discuss the matching of this $S$ matrix to the classical theory. We see how from this $S$ matrix one can derive the TBA equations describing the spectrum of the deformed string theory. In part \ref{part:qsc} we discuss the construction of the QSC from the TBA equations in detail. As our construction stays fairly close to the construction of the undeformed QSC, we also review this construction, commenting on the similarities and differences along the way. In this way, we aim to provide a complete overview of the construction of the undeformed QSC as well as explain how to construct the $\eta$-deformed QSC. After an introduction to this construction in chapter \ref{chap:etaQSC} we start by deriving the analytic $Y$-system in chapter \ref{chap:Ysystem}, showing how the analyticity data can be extracted from the TBA equations to supplement the $Y$-system equations. In chapter \ref{chap:Tsystem} we then derive the analytic $T$-system, by reparametrising the $Y$ functions using $T$ functions. We discuss the construction of four $T$ gauges carrying the analyticity data. In chapter \ref{chap:QSC} we introduce the final reparametrisation in the form of the $\Pf\mu$ system, the quantum spectral curve. We discuss how the analyticity data enters the $\Pf\mu$ system and how the entire $QQ$ system can be derived. In chapter \ref{chap:solutionstate} we discuss which solutions of the QSC correspond to which states in the deformed string theory. In chapter \ref{chap:conclusions} we summarise and discuss the possibilities for future work. 

\paragraph{A note on notions and notation.} We have tried to stay close to the literature in our choice of notation, which should allow for easy comparison. As this thesis builds on work by many others we have had to make some choices which conventions to stick to. 

The quantum-deformation parameter of the $\eta$-deformed model has been denoted in this thesis as $\ad$ (so $q=e^{-\ad})$, contrary to the ``$a$" used in \cite{Arutynov:2014ota}, even though the deformation has been called ``$\eta$ deformation". This may seem inconsistent, but in fact it is not: $\eta$ originates from the classical deformation of the theory, but due to renormalisation there is no reason why $\eta$ and the quantum deformation parameter $\ad$ should be equal. We use $\ad$ in favour of ``$a$" to avoid notational issues in the QSC, where ``$a$" is a common subscript. 

Since different parts of the undeformed QSC construction have been performed using different conventions, we have opted to change conventions midway through as the change is not dramatic. We discuss this in more detail in section \ref{sec:shortcuttedfunctions}. 

We have generically tried to be precise in our use of mathematical notions such as analytic, meromorphic and regular, but to avoid writing overly convoluted we do expect some interpretation from the reader: for example, we call functions $f$ satisfying $f(z+\omega) = f(z)$ for some $\omega \in \R$ \emph{real periodic} to distinguish it from $2i \ad$ periodic, not asserting anything further about the function in question. When writing \emph{deformed} we always mean $\eta$ deformed, and \emph{undeformed} always pertains to the $\ads$ superstring. Whenever relevant we add a sub- or superscript ``und" to quantities to emphasise that they are related to the undeformed theory. Finally, even though the $\eta$ deformation of the $\ads$ superstring is not a string theory itself, we do not always emphasise this in our discussions: for example, when writing ``deformed string energy" we merely want to discuss the quantity in the deformed model that is a deformation of the undeformed string energy.

 \note{More things to add here?}


%% file: background.tex
\section{Introduction}
The construction of the $\eta$-deformed quantum spectral curve is based on various lines of research: it not only builds on tremendous amount of work done to find the undeformed QSC for the canonical AdS$_5$/CFT$_4$ pair, but also on work on non-linear sigma models and representation theory of quantum algebras. In this chapter we will review all the aspects of the $\eta$-deformed model necessary to derive the QSC. Moreover, since we will compare our derivation of the $\eta$-deformed QSC with the undeformed construction it will be beneficial to have some of the background of the undeformed model available as well. 
\\[5mm]
We first discuss the construction of the $\ads$ non-linear sigma model starting from its target space PSU$(2,2|4)$ and integrability. We then introduce the $\eta$ deformation at the classical level and review some of its properties. The next step on the way to the QSC is to quantise the classical theory. This is approached in two ways: we discuss the process of light-cone gauge fixing that allows for a perturbative quantisation of the classical action. This gives rise to the perturbative $S$-matrix. We then start using integrability: we show that integrable field theories in two dimensions have very restricted scattering processes, which can be captured in the structure of Hopf algebras. We discuss in detail how to $q$ deform the Hopf algebra of the undeformed theory to end up with the centrally extended quantum group $U_q\left(\mathfrak{su}(2|2)\right)_{\text{c.e}}$ that forms the basis of the scattering theory for the $\eta$-deformed model: considering the quasi-cocommutativity condition on the level of the fundamental representation yields a set of equations that uniquely determine the exact $\eta$-deformed $S$-matrix. We then argue how one can match this with the perturbative $S$-matrix to fully fix its form, thereby defining the quantised $\eta$-deformed model. We discuss various important properties such as mirror duality. The final part of this chapter discusses the asymptotic and thermodynamic Bethe ansatz, which were used to obtain equations describing the spectrum of both the undeformed and the $\eta$-deformed model. We review these methods in general and give the resulting equations for both cases, as they form the starting point for our analysis in part \ref{part:qsc}. 
\\[5mm]
For the sake of brevity we do not discuss all these steps in full detail, but instead refer to the relevant literature. In particular the reviews \cite{Arutyunov:2009ga,Beisert:2010jr} contain the entire derivation of the undeformed TBA equations and form an excellent starting point for anyone interested in the early history of the AdS$_5$/CFT$_4$ spectral problem. All the details about obtaining the $q$-deformed TBA-equations can be found in \cite{vanTongeren:2013gva} and references therein. A recent introduction to the QSC can be found in \cite{Gromov:2017blm}.

\section{Classical $\ads$ superstring theory}
To put our explorations on a firm foundation, we now first introduce the $\ads$ superstring theory in the Green-Schwarz formalism, starting from the superconformal algebra $\mathfrak{psu}(2,2|4)$. 
\subsection{$\mathfrak{psu}(2,2|4)$}
\label{sec:psu224}
The algebra denoted as $\mathfrak{psu}(2,2|4)$ can be defined as a quotient algebra of the matrix algebra $\mathfrak{su}(2,2|4)$, which can be realised as follows: consider $\mathfrak{sl}(4|4)$ matrices written in $4 \times 4$ blocks as 
\begin{equation}
\mathbf{M} = \left( \begin{array}{c|c}
 M_1 & \theta_1 \\ \hline 
\theta_2 & M_2 
\end{array} \right),
\end{equation}
where the entries of the $M_i$ are bosonic variables and those of the $\theta_i$ fermionic Grassmann variables. $\mathfrak{sl}(4|4)$ matrices have vanishing supertrace, i.e. 
\begin{equation}
\text{str}(\mathbf{M}) \defeq \text{tr}(M_1) - \text{tr}(M_2) = 0.
\end{equation}
The algebra $\mathfrak{sl}(2,2|4)$ can be identified as the fixed-point set of the anti-linear Cartan involution  $\tau : \mathfrak{sl}(4|4)\rightarrow \mathfrak{sl}(4|4)$ defined as
\begin{equation}
\label{eq:su44reality}
\tau \left(\mathbf{M}\right) = -\mathbf{H} \mathbf{M}^{\dagger} \mathbf{H}^{-1}, \quad \text{ with } \mathbf{H} =\left( \begin{array}{c|c}
H_1 & 0 \\ \hline 0 & \mathbb{I}_4
\end{array} \right)
\quad \text{ and } H_1 = \begin{pmatrix}
\mathbb{I}_2 & 0 \\ 0 & \mathbb{I}_2
\end{pmatrix},
\end{equation}
where the adjoint of $\mathbf{M}$ is defined as $\mathbf{M}^{\dagger} = \left(\mathbf{M}^{t}\right)^*$. By analysing eqn. \eqref{eq:su44reality} we find that the allowed matrices $M_1,M_2$ span $\mathfrak{su}(2,2)$ and $\mathfrak{su}(4)$ respectively, leaving a one-parameter freedom generated by the central element $\ii \mathbb{I}_8$, which is left fixed under the Cartan involution and has vanishing supertrace. Together this forms the bosonic subalgebra $\mathfrak{su}(2,2)\oplus \mathfrak{su}(4)\oplus \mathfrak{u}(1)$. We can now define $\mathfrak{psu}(2,2|4)$ as the quotient algebra of $\mathfrak{su}(2,2|4)$ by identifying elements which differ only by a multiple of $\ii \mathbb{I}_8$, i.e. by modding out the $\mathfrak{u}(1)$ factor. 
\\[5mm]
Apart from the obvious $\Z_2$-grading due to the presence of Grassmann variables, $\mathfrak{sl}(4|4)$ also has a non-trivial $\Z_4$-grading that will play an important role. This $\Z_4$ grading is induced by the fourth-order automorphism $\Omega : \mathfrak{sl}(4|4) \rightarrow \mathfrak{sl}(4|4)$ defined by
\begin{equation}
\Omega(\mathbf{M}) = -\mathbf{K} \mathbf{M}^{st} \mathbf{K}^{-1}, \quad \text{ with } \mathbf{K} = \left( \begin{array}{c|c}
K & 0 \\ \hline 0 & K
\end{array}\right) \quad \text{ and }
K = \begin{pmatrix}
0 & -1 & 0 & 0  \\ 1 & 0 & 0 & 0 \\ 0 & 0 & 0 & -1 \\ 0 & 0 & 1 & 0
\end{pmatrix},
\end{equation}
where the supertranspose is defined as 
\begin{equation}
\label{eq:supertranspose}
\mathbf{M}^{st} = \left( \begin{array}{c|c} M_1^{t} & -\theta_2^{t} \\ \hline  
\theta_1^{t} & M_2^{t} 
\end{array} \right).
\end{equation}
Under the action of this automorphism $\mathfrak{g} = \mathfrak{sl}(4|4)$ decomposes into a direct sum 
\begin{equation}
\mathfrak{g} = \mathfrak{g}^{(0)} \oplus \mathfrak{g}^{(1)}\oplus \mathfrak{g}^{(2)} \oplus \mathfrak{g}^{(3)},
\end{equation}
where each graded subspace is defined as 
\begin{equation}
\mathfrak{g}^{(n)} = \left\{\mathbf{M} \in \mathfrak{g} \, |\, \Omega(\mathbf{M}) = \ii^n \mathbf{M} \right\}.
\end{equation}
For any $\mathbf{M}\in \mathfrak{sl}(4|4)$ its projection $\mathbf{M}^{(k)}$ onto the subspace $\mathfrak{g}^{(k)}$ is then given by
\begin{equation}
\mathbf{M}^{(k)} = P_k(\mathbf{M}) \defeq \frac{1}{4} \left( \mathbf{M} + \ii^{3k}\Omega(\mathbf{M})+ \ii^{2k}\Omega^2(\mathbf{M})+ \ii^{k}\Omega^3(\mathbf{M})\right),
\end{equation}
where we define the projectors $P_k$. Importantly, the automorphism $\Omega$ can be consistently restricted to $\mathfrak{su}(2,2|4)$, therefore inducing a decomposition of any matrix $\mathbf{M}\in \mathfrak{su}(2,2|4)$. In particular, writing now $\mathfrak{g} = \mathfrak{su}(2,2|4)$ we see that $\mathfrak{g}^{(0)}$ can be identified with the subalgebra $\mathfrak{so}(4,1)\oplus \mathfrak{so}(5)\subset \mathfrak{su}(2,2)\oplus \mathfrak{su}(4)$ and that the central element $\ii \mathbb{I}_8\in \mathfrak{g}^{(2)}$. 

\subsection{Coset description of the Green-Schwarz superstring}
These are all the elements we need to introduce the coset description of the Green-Schwarz string: we view the string as the embedding of a two-dimensional world sheet $\Sigma\cong \R \times S^1 $ with coordinates $(\tau, \sigma)$ into a target space given by the coset 
\begin{equation}
\frac{\text{PSU}(2,2|4)}{\text{SO}(4,1) \times \text{SO}(5)},
\end{equation}
which models the $\ads$ space as the bosonic subgroup SO$(2,2)\times$SU$(4)\subset \text{PSU}(2,2|4)$ is locally isomorphic to SO$(4,2)\times$ SO$(6)$.
\\[5mm]
Let $\mathbf{g}:\Sigma \rightarrow G$ with $G=  \text{SU}(2,2|4)$ and define the one-form current $A=-\mathbf{g}^{-1}d \mathbf{g}$, satisfying the zero-curvature condition
\begin{equation}
\label{eq:Aflatness}
F \defeq dA-A\wedge A = 0,\quad \text{ or in components } F_{\alpha \beta} = \partial_{\alpha} A_{\beta}-\partial_{\beta} A_{\alpha} - \left[ A_{\alpha}, A_{\beta}\right] = 0. 
\end{equation}
The $\ads$ superstring is defined by the langrangian density \cite{Metsaev:1998it}
\begin{equation}
\label{eq:adslagrangrian}
\mathcal{L} = -\frac{g}{2} \left( \gamma^{\alpha \beta}\text{str}\left( A_{\alpha}^{(2)}A_{\beta}^{(2)}\right) + \kappa \e^{\alpha \beta} \text{str}\left( A_{\alpha}^{(1)}A_{\beta}^{(3)}\right)  \right), 
\end{equation}
where $g$ is the dimensionless string tension, $\gamma^{\alpha \beta} = \sqrt{-h}h^{\alpha \beta}$ is the Weyl-invariant tensor constructed from the world-sheet metric $h^{\alpha \beta}$ and $\kappa$ is a real constant to ensure reality of $\mathcal{L}$. 
The first term in eqn. \eqref{eq:adslagrangrian} is the kinetic term, while the second is the Wess-Zumino term. The lagrangian density is invariant under the multiplication $\mathbf{g}\rightarrow \mathbf{g} \mathbf{h}$ for some element $\mathbf{h}\in \text{SO}(4,1)\times \text{SO}(5)$. It is this phenomenon that inspired the name coset model: indeed, it seems that the lagrangrian really depends only on an element from the coset 
\begin{equation}
\frac{\text{SU}(2,2|4)}{\text{SO}(4,1) \times \text{SO}(5)}.
\end{equation}
This is actually not the whole story yet: SU$(2,2|4)$ also contains the central element $\ii \mathbb{I}_8$ which generates a U$(1)$ subgroup and multiplying $\mathbf{g}$ by an element from this subgroup also leaves the lagrangian invariant, so in fact it depends only on an element from the coset
\begin{equation}
\frac{\text{PSU}(2,2|4)}{\text{SO}(4,1) \times \text{SO}(5)},
\end{equation}
as we initially wanted. An easy way to implement the modding out of the U$(1)$ subgroup is to enforce tracelessness of $A^{(2)}$. The isometry group of the lagrangian is now given by PSU$(2,2|4)$, which acts by left multiplication. The form of the lagrangian density \eqref{eq:adslagrangrian} is not the most convenient for our deforming purposes, but in order to rewrite it we should first analyse its current form a bit better. When restricting this lagrangian density to its bosonic variables only one recovers the usual Polyakov action for bosonic strings on the background $\ads$, providing a justification for its particular form in eqn. \eqref{eq:adslagrangrian}.

\paragraph{Bosonic action.} For later convenience we introduce the Polyakov action describing the bosonic part of the model above. With target-space coordinates $X^M = 
\{t,\rho,\zeta,\psi_i\} \cup \{\phi,r,\xi,\phi_i \}$ for the AdS$_5$ and the S$^5$ spaces and target-space metric $G_{MN}$ it can be written in the standard form
\begin{align}
\label{eq:bosonicPolyakov}
S^b &= -\frac{g}{2} \int d\sigma d\tau \gamma ^{\alpha \beta} \partial_{\alpha} X^M \partial_{\beta} X^N G_{MN},\nonumber \\
ds^2_{\text{AdS}_5} &= - \left(1+\rho^2\right) dt^2 + \frac{d\rho^2}{(1+\rho^2)} \nn
&+\rho^2 \left( d\zeta^2 +\cos^2 \zeta d\psi_1^2  \right)+ \rho^2 \sin^2 \zeta d\psi_2^2 \nn
ds^2_{\text{S}^5} &= \left(1-r^2\right) d\phi^2 + \frac{dr^2}{(1-r^2)} \nn
&+ r^2 \left( d\xi^2 +\cos^2 \xi d\phi_1^2  \right)+ r^2 \sin^2 \xi d\phi_2^2,
\end{align}
where $G_{MN}$ is defined by the infinitesimal line elements split into an AdS$_5$ and a S$^5$ part. 

\paragraph{Equations of motion.} Varying the lagrangian with respect to $\mathbf{g}$ gives the equation of motion
\begin{equation}
\label{eq:Aeoms}
\partial_{\alpha}\left(\gamma^{\alpha \beta} A_{\beta}^{(2)}\frac{\kappa}{2} \e^{\alpha \beta}\left(A_{\beta}^{(1)}-A_{\beta}^{(3)}  \right)  \right) -\left[ A_{\alpha}, \left(\gamma^{\alpha \beta} A_{\beta}^{(2)}\frac{\kappa}{2} \e^{\alpha \beta}\left(A_{\beta}^{(1)}-A_{\beta}^{(3)}  \right)  \right) \right] = 0.
\end{equation}
Varying the lagrangian with respect to the world-sheet metric $h^{\alpha \beta}$ leads to the Virasoro constraints
\begin{equation}
\label{eq:heoms}
\text{str}\left( A_{\alpha}^{(2)}A_{\beta}^{(2)}\right) - \frac{1}{2} \gamma_{\alpha \beta} \gamma^{\delta \rho} \text{str}\left( A_{\delta}^{(2)}A_{\rho}^{(2)}\right) = 0,
\end{equation}
where the left-hand side can be recognised as being the world-sheet stress-tensor. Its vanishing reflects the reparametrisation invariance of the action. 
\paragraph{$\kappa$ symmetry.} The presence of $\kappa$ symmetry -- a particular local fermionic symmetry -- is an essential property of the superstring theory we are considering: it is $\kappa$ symmetry that ensures that the spectrum of the theory is space-time supersymmetric and can additionally be used to gauge away half of the fermionic degrees of freedom. The action as defined by the lagrangian density \eqref{eq:adslagrangrian} does not have $\kappa$ symmetry for arbitrary $\kappa \in \R$, but only when $\kappa =\pm 1$ (see \cite{McArthur:1999dy} or the explicit treatment in \cite{Arutyunov:2009ga}). Choosing $\kappa = 1$ now allows us to rewrite the lagrangian density : introducing the orthogonal chiral projectors $P_{\pm}^{\alpha \beta} = \frac{1}{2}\left(\gamma^{\alpha \beta} \pm \e^{\alpha \beta}\right)$ we can write eqn. \eqref{eq:adslagrangrian} as
\begin{equation}
\label{eq:adslagrangrianprojectors}
\mathcal{L} = -\frac{g}{2} P_-^{\alpha \beta}\text{str}\left(A_{\alpha}d_0 A_{\beta}  \right),
\end{equation}
where we have introduced the linear combination of projectors
\begin{equation}
\label{eq:undeformedprojectors}
d_0 = P_1 + 2P_2 - P_3.
\end{equation}
As we will see, it is in this form that the $\eta$ deformation takes its most natural form. 

\paragraph{Hamiltonian formalism.}\label{sec:hamiltonianformalism} In order to understand how to derive the $\eta$ deformation we consider the $\ads$ superstring in the hamiltonian formalism \cite{Delduc:2014kha}.\footnote{The author is indebted to Gleb Arutyunov for his exposition of this topic.} We start from the loop group $\hat{G} = C\left( S^1, G\right)$ consisting of continuous maps from the circle to $G$. Its cotangent bundle $T^*\hat{G}$ comes equipped with a symplectic structure characterised by the Poisson brackets. Let $\mathbf{g} \in \hat{G}$ and $\mathbf{X} \in T\hat{G}$, then the Poisson brackets are given by
\begin{align}
\left\{ \mathbf{g}_1(\sigma)  , \mathbf{g}_2(\sigma') \right\} &= 0,  \nn
\left\{ \mathbf{X}_1(\sigma)  , \mathbf{g}_2(\sigma') \right\} &= C_{12} \mathbf{g}_2(\sigma) \delta_{\sigma \sigma'},  \nn
\left\{ \mathbf{X}_1(\sigma)  , \mathbf{X}_2(\sigma') \right\} &= \left[ C_{12}, \mathbf{X}_2(\sigma)\right] \delta_{\sigma \sigma'}, 
\end{align}
where $C_{12} \in \mathfrak{g} \otimes \mathfrak{g} $ is the quadratic tensor Casimir\footnote{We do not need its explicit form here, see appendix A of \cite{Delduc:2014kha} for more details.} for $\mathfrak{g}$ and we use the tensorial shorthands $\mathbf{M}_1 = \mathbf{M}(\sigma_1) \otimes 1$ and $\mathbf{M}_2 = 1\otimes \mathbf{M}(\sigma_2)$. The canonical fields describing our model at this level are $A$ and $\Pi$ whose embedding into $G$ and $\mathfrak{g}$ take the following familiar form (see the definition of $A$ above eqn. \eqref{eq:Aflatness}):
\begin{equation}
\label{eq:hamilformA}
\mathbf{A} = - \mathbf{g}^{-1} \partial_{\sigma} \mathbf{g} \in G, \quad \Pi = - \mathbf{g}^{-1}X\mathbf{g} \in \mathfrak{g},
\end{equation}
which obey the following Poisson brackets for its projected components
\begin{align}
\label{eq:PoissonApi}
\left\{A_1^{(i)}(\sigma)  , A_2^{(j)}(\sigma') \right\} &= 0, \nn
\left\{ A_1^{(i)}(\sigma)  , \Pi_2^{(j)}(\sigma') \right\} &= \left[ C_{12}^{(i,4-i)}, A_2^{(i+j)}(\sigma)\right] \delta_{\sigma \sigma'} - C_{12}^{(i,4-i)} \delta_{i+j,0 } \,\delta'_{\sigma \sigma'} ,  \nn
\left\{ \Pi_1^{(i)}(\sigma)  , \Pi_2^{(j)}(\sigma') \right\} &= \left[ C_{12}^{(i,4-i)},\Pi_2^{(i+j)}(\sigma)\right] \delta_{\sigma \sigma'}, 
\end{align}
where $\delta'$ is the derivative of the $\delta$ function and the counting of the projector components is mod $4$. In particular $C_{12}^{(i,j)}$ denotes projection onto $\mathfrak{g}^{(i)} \otimes\, \mathfrak{g}^{(j)}$. This Poisson structure defines the model together with a hamiltonian $H\left(A,\Pi\right)$ whose explicit form we will not need here.\footnote{See section $2$ of \cite{Delduc:2014kha}.} The integrability of this model follows due to the existence of a Lax pair $\left(L,M\right)$ (see also \cite{Bena:2003wd}):
\begin{align}
\label{eq:Lax}
L(z) &= A^{(0)} + \frac{1}{4} \left( z^{-3} + 3 z\right) A^{(1)} + \frac{1}{2} \left( z^{-2} +  z^2\right) A^{(2)} + \frac{1}{4} \left(3 z^{-1}+ z^3\right)A^{(3)} \nn
&+ \frac{1}{2} \left(1 - z^{4}\right) \Pi^{(0)} + \frac{1}{2}\left( z^{-3} - z \right) \Pi^{(1)} + \frac{1}{2}\left( z^{-2} - z^2 \right) \Pi^{(2)}+ \frac{1}{2}\left( z^{-1} - z^3 \right) \Pi^{(3)} \nn
M(z) &= A^{(0)} - \frac{1}{4} \left( z^{-3} - 3 z\right) A^{(1)}  - \frac{1}{2} \left( z^{-2} -  z^2\right) A^{(2)} - \frac{1}{4} \left(3 z^{-1}- z^3\right)A^{(3)} \nn
&+ \frac{1}{2} \left(1 - z^{4}\right) \Pi^{(0)} - \frac{1}{2}\left( z^{-3} + z \right) \Pi^{(1)} - \frac{1}{2}\left( z^{-2} + z^2 \right) \Pi^{(2)}- \frac{1}{2}\left( z^{-1} + z^3 \right) \Pi^{(3)}.
\end{align}
where $z\in \C$ is a spectral parameter.  The zero-curvature condition 
\begin{equation}
\label{eq:Lflatness}
\left[ \partial_{\sigma}  - L, \partial_{\tau} - M\right] = 0
\end{equation}
of the Lax pair is equivalent to the hamiltonian equations of motion
\begin{equation}
\label{eq:hamileoms}
\partial_{\tau} A = \left\{ A,H\right\}, \quad \partial_{\tau} \Pi = \left\{ \Pi,H\right\}
\end{equation}
as specified by our Poisson structure and hamiltonian $H$. The Lax matrix $L$ satisfies the Poisson brackets
\begin{equation}
\label{eq:PoissonLax}
\left\{ L_1(z_1, \sigma), L_2(z_2, \sigma') \right\} = \left[ \mathcal{R}_{12}, L_1(z_1,\sigma) \right] \delta_{\sigma \sigma'} - 
\left[ \mathcal{R}_{21}, L_2(z_2,\sigma') \right] \delta_{\sigma \sigma'} + \left(\mathcal{R}_{12} + \mathcal{R}_{21}\right) \delta'_{\sigma \sigma'},
\end{equation}
where we have defined
\begin{equation}
\mathcal{R}_{12}(z_1,z_2) = 2 \sum_{j=0}^3 \frac{z_1^j z_2^{4-j} C_{12}^{(j,4-j)}}{z_1^4-z_2^4} \phi_S^{-1}(z_2).
\end{equation}
The function $\phi_S$ is known as the \emph{twist function} and in this case is given by
\begin{equation}
\phi_S(z) = \frac{4 z^4}{(1-z^4)^2}, \text{ with inverse }  \phi_S^{-1}(z) = \frac{1}{4} z^{-4} -\frac{1}{2}+z^4,
\end{equation}
and it satisfies $\phi_S(\ii z) = \phi_S(z)$ compatible with the $\Z_4$ grading of $G$. The fact that we can write the Poisson bracket of $\mathcal{L}$ in the form \eqref{eq:PoissonLax} is crucial for the construction: the Poisson brackets \eqref{eq:PoissonApi} are satisfied precisely when the Poisson brackets for $\mathcal{L}$ can be written in the form \eqref{eq:PoissonLax} for some matrix $\mathcal{R}$. Moreover, the flatness of the Lax pair leads to the fact that $\mathcal{L}$ generates an infinite tower of conserved charges in the form of the eigenvalues of the monodromy 
\begin{equation}
\label{eq:Tmonodromy}
T(\tau|z) = P\overleftarrow{\exp}\left(\int_{-L/2}^{L/2} L_{\sigma}(\tau,\sigma| z) d\sigma \right),
\end{equation}
with $P\overleftarrow{\exp}$ the usual path-ordered exponential. 

\paragraph{Retrieving $\mathbf{g}$ and $\mathbf{X}$.}\label{par:retrievinggX} One particular consequence of this construction is that we can find the fields $\mathbf{g}$ and $\mathbf{X}$ by analysing the behaviour of the Lax matrix around $z=1$, which is the pole of the twist function $\phi_S$:
\begin{equation}
L(z) = A - 2(z-1) \Pi + \bO{(z-1)^2}.
\end{equation}
We introduce the gauge transformation
\begin{equation}
L^{\mathbf{g}} (z) \defeq \left( \partial_{\sigma} \mathbf{g} \right) \mathbf{g}^{-1} + \mathbf{g} L(z) \mathbf{g}^{-1}
\end{equation}
and consider the equation
\begin{equation}
\label{eq:gaugeL}
L^{\mathbf{g}}(1) = 0, 
\end{equation}
which when written explicitly reads 
\begin{equation}
\mathbf{g} A \mathbf{g}^{-1} + \partial \partial_{\sigma} \mathbf{g} \mathbf{g}^{-1} = 0. 
\end{equation}
We recognise that due to the parametrisation \eqref{eq:hamilformA} this equation is indeed satisfied. In addition we find that 
\begin{equation}
\label{eq:gaugeL2}
\frac{1}{2} \frac{d L^{\mathbf{g}}}{dz} \Big|_{z=1} = \mathbf{X}. 
\end{equation}
This observation might seem meaningless, but it will prove to be crucial as we will see in the deformed case. 

\section{$\eta$ deforming the classical $\ads$ superstring theory}
The model we are primarily interested in this thesis is the $\eta$ deformation of the $\ads$ superstring theory that we have discussed until now. Although the quantisation of this deformed model -- described by $S$-matrix theory -- was discovered first in \cite{Beisert:2008tw} by a $q$ deformation of the $\ads$ $S$ matrix, we will consider the classical string theory first. It was constructed by Delduc, Magro and Vicedo in a series of papers \cite{Delduc:2013fga,Delduc:2013qra,Delduc:2014kha} inspired by earlier work by Faddeev and Reshetikhin \cite{1986AnPhy.167..227F} and in particular the work by Klimcik \cite{Klimcik:2002zj,Klimcik:2008eq} on deforming the principal chiral model. The main strength of this approach, which can be characterised as a deformation of the Poisson structure, is that it maintains the integrability of the model manifestly during the deformation procedure. 
\paragraph{Lagrangian.} One can bypass the entire construction from the Poisson structure and directly postulate the lagrangian density for the deformed theory and a Lax pair exhibiting its integrability. For $\eta\in [0,1)$ it is given by (compare with eqn. \eqref{eq:adslagrangrianprojectors})
\begin{equation}
\label{eq:deformedlagrangian}
\mathcal{L}_{\eta} = -\frac{g}{2}\left(1+\eta^2 \right) P_-^{\alpha \beta}\text{str}\left(A_{\alpha}d_{\eta}\circ O_{\eta}^{-1} A_{\beta}  \right),
\end{equation}
where 
\begin{equation}
\label{eq:deformedd}
d_{\eta} = P_1 + \frac{2}{1-\eta^2}P_2 - P_3,\quad O_{\eta} = 1-\eta R_{\mathbf{g}}\circ d_{\eta}.
\end{equation}
The new element $R_{\mathbf{g}}$ is a special new operator responsible for the deformation. It is defined as
\begin{equation}
R_{\mathbf{g}}(\mathbf{M}) = \mathbf{g}^{-1} r\left( \mathbf{g} \mathbf{M} \mathbf{g}^{-1}\right) \mathbf{g},
\end{equation}
where $r$ is an $r$ matrix. 
\paragraph{$r$ matrix.} The deformation is governed by an $r$ matrix, being a skew-symmetric non-split solution of the \emph{modified classical Yang-Baxter equation} on $\mathfrak{g}$, i.e. it solves
\begin{equation}
\label{eq:mCYBE}
\left[ r(\mathbf{M}),r( \mathbf{N})\right]- r\left( \left[ r(\mathbf{M}), \mathbf{N}\right]+\left[ \mathbf{M},r( \mathbf{N})\right]\right) = \left[ \mathbf{M}, \mathbf{N} \right], \quad \text{ for all } \mathbf{M}, \mathbf{N} \in \mathfrak{g}.
\end{equation}
A skew-symmetric solution obeys tr$\left( \mathbf{M} r( \mathbf{N}) \right) = -$tr$\left( r(\mathbf{M})  \mathbf{N} \right)$. Such a solution has a canonical construction \cite{Klimcik:2008eq}: consider a Cartan-Weyl basis of the complexified algebra $\mathfrak{psl}(2,2|4)$ consisting of a basis $\{\mathbf{H}_i\}_{i\in I}$ of the Cartan subalgebra and a set of root vectors $\{\mathbf{E}_{\pm \alpha}\}_{\alpha\in A}$. The operator $r: \mathfrak{g} \rightarrow \mathfrak{g}$ defined by its action on this basis as
\begin{align}
\label{eq:Rmatrixaction}
r\left( \ii \mathbf{H}_k \right) &= 0, \quad \text{ for } k\in I \nonumber \\
 r\left(\frac{\ii}{\sqrt{2}} \left( \mathbf{E}_{+\alpha}+\mathbf{E}_{-\alpha} \right)\right) &= \frac{1}{\sqrt{2}} \left( \mathbf{E}_{+\alpha} - \mathbf{E}_{-\alpha} \right) \quad \text{ for } \alpha\in A,\nn
 r\left(\frac{1}{\sqrt{2}} \left( \mathbf{E}_{+\alpha} - \mathbf{E}_{-\alpha} \right)\right) &= - \frac{\ii }{\sqrt{2}} \left( \mathbf{E}_{+\alpha} + \mathbf{E}_{-\alpha} \right) \quad \text{ for } \alpha \in A
\end{align}
is skew-symmetric, solves eqn. \eqref{eq:mCYBE} and also satisfies $r^3 = -r$ by construction. 

\paragraph{Polyakov form. } The bosonic part of the action becomes (compare with eqn. \eqref{eq:bosonicPolyakov})
\begin{align}
\label{eq:etadeformedbosonicPolyakov}
S^b_{\eta} &= -\frac{1}{2}\left(\frac{1+\eta^2}{1-\eta^2} g \right) \int d\sigma d\tau \left( \gamma ^{\alpha \beta} \partial_{\alpha} X^M \partial_{\beta} X^N G_{MN}^{\eta}
-\e^{\alpha \beta} \partial_{\alpha} X^M \partial_{\beta} X^N B_{MN}\right),
\end{align}
which now has a deformed metric given by the infinitesimal line element
\begin{align}
ds^2_{(\text{AdS}_5)_{\eta}} &= - \frac{1+\rho^2}{ 1-\varkappa^2\rho^2} dt^2 + \frac{d\rho^2}{(1+\rho^2)( 1-\varkappa^2\rho^2)} \nn
&+ \frac{\rho^2}{  1+\varkappa^2 \rho^4\sin^2\zeta } \left( d\zeta^2 +\cos^2 \zeta dpsi_1^2  \right)+ \rho^2 \sin^2 \zeta d\psi_2^2, \nn
ds^2_{(\text{S}^5)_{\eta}} &= \frac{1-r^2}{ 1+\varkappa^2\rho^2} d\phi^2 + \frac{dr^2}{(1-r^2)( 1+\varkappa^2 r^2)} \nn
&+ \frac{r^2}{ 1+\varkappa^2 r^4\sin^2\xi } \left( d\xi^2 +\cos^2 \xi d\phi_1^2  \right)+ r^2 \sin^2 \xi d\phi_2^2,
\end{align}
with $\varkappa = \frac{2\eta}{1-\eta^2}$ as well as a non-trivial $B$ field \cite{Arutyunov:2013ega}
\begin{align}
B_{(\text{AdS}_5)_{\eta}} &= +\frac{\varkappa}{2} \left(\frac{\rho^4 \sin(2\zeta)}{1+\varkappa^2 \rho^4 \sin^2 \zeta} d\psi_1\wedge d\zeta + \frac{2\rho}{1-\varkappa^2 \rho^2} dt \wedge d\rho   \right),\nonumber \\
B_{(\text{S}_5)_{\eta}} &= -\frac{\varkappa}{2} \left(\frac{r^4 \sin(2\xi)}{1+\varkappa^2 r^4 \sin^2 \xi} d\phi_1\wedge d\xi + \frac{2r}{1 +\varkappa^2 r^2} dt \wedge d r   \right).
\end{align}
The geometric interpretation of the deformation on the target space is that of squashing, breaking the isometry algebra down to U$(1)^3 \times $U$(1)^3$, consistent with the breaking of all the isometries in $\mathfrak{psu}(2,2|4)$ except for the Cartan generators. Note that as $\eta\rightarrow 0$ one recovers the $\ads$ action \eqref{eq:bosonicPolyakov}. 

\paragraph{Symmetries.} This lagrangian has the following symmetries:
\begin{itemize}
\item It has $\mathfrak{psu}_q(2,2|4) \defeq U_q\left(\mathfrak{psu}(2,2|4)\right)$ symmetry with $q \in \R$. Note however that the realisation of this symmetry is far from obvious: in the undeformed case the target space manifestly carried the $\mathfrak{psu}(2,2|4)$ symmetry as its isometry group. The $q$-deformed symmetry in the present case has a more complicated realisation and is in some sense ``hidden": from the Lax formalism it is possible to derive the Poisson algebra of charges, which turns out to be (isomorphic to) $\mathfrak{psu}_q(2,2|4)$.\footnote{The fact that the $q$-deformed symmetry is not realised as an isometry group means that it remains an open question whether it is possible to construct a point particle model with this symmetry, as the point particle limit of the $\eta$-deformed model destroys this symmetry, in contrast with the undeformed case.} 
\item It contains a fermionic gauge invariance that reduces to the standard $\kappa$ symmetry as $\eta\rightarrow 0$.
\item It has conformal symmetry.
\item It is integrable by construction.
\end{itemize}
\subsection{Construction from the Poisson structure.} Now let us find out how to construct the $\eta$ deformation from the undeformed sigma model: as anticipated we consider the undeformed model in the hamiltonian formalism as discussed in section \ref{sec:hamiltonianformalism}. The dynamics of this integrable model is completely defined in terms of the Lax matrix $L$, the fields $A$ and $\Pi$ and their Poisson brackets: they give rise to the conserved charges and the equations of motion in terms of $A$ and $\Pi$. To find the embedding of this model in the group $G$ we solve the equations \eqref{eq:gaugeL} and \eqref{eq:gaugeL2} to find the dependence of $\mathbf{g}$ and $\mathbf{X}$ on $A$ and $\Pi$. The crucial observation is that the integrability of this model can be formulated solely on the level of $A$ and $\Pi$; that is, the actual form of the Poisson bracket in terms of the group and algebra elements $\mathbf{g}$ and $\mathbf{X}$ is not important here.
\\[5mm]
Therefore, the idea of the deformation as introduced in \cite{Delduc:2014kha} is to introduce a new Poisson bracket $\{\, , \,\}_{\e}$ such that the form of the Lax matrix, its Poisson relation \eqref{eq:PoissonLax} and its zero-curvature condition \eqref{eq:Lflatness} in terms of $A$ and $\Pi$ are kept fixed. Whether such a deformation exists is not immediately clear, since the deformed bracket should still be a Poisson bracket \emph{and} together with some deformed hamiltonian yield the correct equations of motion, which might in practice be too stringent of a constraint.
\\[5mm]
To deform the bracket we deform the twist function $\phi_S$ into $\phi_{\e}$. Non-degeneracy of the $\mathcal{R}$ matrix on $\mathfrak{g}$ requires $\phi_{\e}$ to satisfy still $\phi_{\e}(\ii z) = \phi_{\e}(z)$, and further well-definedness of the Poisson brackets restricts the possible deformation even further, such that we ultimately consider the new twist function
\begin{equation}
\phi_{\e}^{-1}(z) = \phi_S^{-1}(z) + \e^2,
\end{equation}
which can be inverted to give
\begin{equation}
\phi_{\e}(z) = \frac{4z^4}{\prod_{k=0}^3 \left(z-\ii^k e^{\ii \vartheta} \right)\left(z-\ii^{-k} e^{-\ii \vartheta} \right)    },
\end{equation}
where $\e = \sin 2\vartheta$. It is immediate that as $\e \rightarrow 0$ we obtain the undeformed twist function $\phi_S$. Also note that due to this deformation the double pole at $z=1$ has been split into two single poles at $z=e^{\pm \ii \vartheta}$. 
\\[5mm]
To see the effect of this deformation we first look at the Poisson relation \eqref{eq:PoissonLax}: since we do not want to change the Lax matrix or the form of this equation the deformation introduces a deformation of the bracket on the left-hand side. This kind of deformation was pioneered by Faddeev and Reshithikin in an effort to find a better description of certain models in order to quantise them \cite{1986AnPhy.167..227F}. To see the effect of the deformation in our model, we should trace it back to the level of the original variables $\mathbf{g}$ and $\mathbf{X}$. In the undeformed case we had a clear recipe (see \ref{par:retrievinggX}) to find these variables from an expansion of the Lax matrix, which however changed due to the deformation. So we should search for a generalisation of this recipe. 
\\[5mm]
The generalisation found by Delduc, Magro and Vicedo considers the complexified group $G^{\C}$ whose Iwasawa decomposition reads
\begin{equation}
G^{\C} = G \oplus K \oplus N,
\end{equation}
where the first subgroup can be identified with $G$ itself, $K$ is the non-compact Cartan subgroup and $N$ is the nilpotent piece. The Cartan involution $\tau$ that selects the real algebra $G$ out of $G^{\C}$ by $\tau\left(G\right) = G$ has $\tau\left(\mathbf{K}\right) = -\mathbf{K}$ for all $\mathbf{K} \in K$. Let us now consider the first equation \eqref{eq:gaugeL}: in the undeformed case we required the Lax matrix to vanish at $z=1$ which was a pole of the twist functions, but this is no longer a pole in the deformed case. We can instead consider the Lax matrix at the poles $e^{\pm i \vartheta}$ of the deformed twist function. Asking the Lax matrix to vanish there is too much to ask though, as the Lax matrix takes non-real values away from the real axis, such that $L^{\mathbf{g}}\left(  e^{ \pm \ii \vartheta}\right)$ can never vanish. However, we can instead require that
\begin{equation}
\label{eq:deformedLaxg}
L^{\mathbf{g}}\left(  e^{ \ii \vartheta}\right) \in K\oplus N,
\end{equation}
which is compatible with the action of the Cartan involution $\tau$:
\begin{equation}
L^{\mathbf{g}}\left(e^{- \ii \vartheta}\right) = \tau \left( L^{\mathbf{g}}\left(e^{\ii \vartheta}\right)\right) \in K\oplus N. 
\end{equation}
This can truly be regarded as a generalisation of the undeformed case since $\mathbf{0} \in K \otimes N$. Before solving this equation let us first consider how to deform the defining equation for $\mathbf{X}$ \eqref{eq:gaugeL2}: for non-zero $\e$ we directly define $\mathbf{X}$ as
\begin{equation}
\label{eq:deformedLaxX}
\mathbf{X} = \frac{\ii}{2\gamma(\e)} \left(  L^{\mathbf{g}}\left(e^{\ii \vartheta}\right) - L^{\mathbf{g}}\left(e^{-\ii \vartheta}\right)\right),
\end{equation}
where the parameter $\gamma$ is $\e$-dependent and can be fixed by demanding $A$ and $\Pi$ to satisfy the deformed Poisson bracket whereas $\mathbf{g}$ and $\mathbf{X}$ satisfy the original canonical Poisson brackets. Its explicit form is 
\begin{equation}
\gamma(\e) = - \e \sqrt{1-\e^2},
\end{equation}
which implies that the $\e\rightarrow 0$ limit of eqn. \eqref{eq:deformedLaxX} is indeed eqn. \eqref{eq:gaugeL2} as required. This limit also shows that eqn. \eqref{eq:deformedLaxX} can be regarded as a ``finite-difference derivative". The final thing we need to do is find the explicit form of the action. In order to do this we use that for every $\mathbf{X} \in \mathfrak{g}$ there is a unique $\mathbf{b}$ in the Lie algebra of $K \oplus N$ such that
\begin{equation}
\mathbf{X} = \frac{\ii}{2} \left(  \mathbf{b} - \tau\left( \mathbf{b}\right) \right). 
\end{equation}
As it turns out, the action of $r$ as defined in \eqref{eq:Rmatrixaction} can be written compactly as
\begin{equation}
r\left(\mathbf{X}\right) = r\left( \frac{\ii}{2} \left(  \mathbf{b} - \tau\left( \mathbf{b}\right) \right)\right) =  \frac{1}{2} \left(  \mathbf{b} + \tau\left( \mathbf{b}\right) \right). 
\end{equation}
Applying this to eqn. \eqref{eq:deformedLaxX} we can derive two equations to relate $A$ and $\Pi$ to $\mathbf{g}$ and $\mathbf{X}$:
\begin{equation}
L^{\mathbf{g}}\left(e^{\pm \ii \vartheta}\right) = -\mathbf{g}^{-1} \partial_{\sigma} \mathbf{g} + \gamma(\e) \mathbf{g} \left( r \mp \ii \right) \mathbf{X} \mathbf{g}^{-1},
\end{equation}
where the left-hand side has a definite dependence on $A$ and $\Pi$. Replacing $A$ and $\Pi$ by their expressions in terms of $\mathbf{g}$ and $\mathbf{X}$ in the hamiltonian gives us the deformed hamiltonian. The deformed lagrangian \eqref{eq:deformedlagrangian} can now be found by performing the inverse Legendre transform going from the deformed hamiltonian back to the langrangian picture and integrating out the dependence of $\mathbf{X}$. 

\subsection{Interesting properties}
\paragraph{Curvature singularity.} In the coordinates introduced above one sees that for $\rho = 1/\varkappa$ the metric has a singularity that is in fact a curvature singularity as the Ricci scalar has a pole there \cite{Borsato:2016hud}. This restricts the radial AdS-coordinate $0\leq \rho < 1/\varkappa$. Its origin lies in the non-invertibility of the operator $O_{\eta}$, which is necessary to write the lagrangian, but its implications are unclear. In the hamiltonian formalism there is no need for inverting $O_{\eta}$, so it seems naively that one does not see the singularity there, thereby posing a puzzling problem that is yet to be resolved. 

\paragraph{Maximal deformation limit.}\label{sec:maximaldeformationlimit} The $\eta$-deformed theory constitutes a one-parameter family, deforming the $\ads$ superstring theory more and more as $\eta$ increases from zero to one. The point $\eta=1$ cannot be accessed directly from the langrangian as it is singular in that point. The limit to $\eta=1$ is known as the \emph{maximal deformation limit} and has been studied from the hamiltonian \cite{Magrohabilitation,Delduc:2013qra} as well as the lagrangian \cite{Arutyunov:2014cra,Hoare:2014pna} perspective, but remains somewhat mysterious. The hamiltonian analysis suggests that the maximal deformation limit corresponds to an undeformed sigma model on PSU$(4,4)^*/($SO$(4,1)\times$SO$(5))$, which has a bosonic sector that corresponds to dS$^5\times$H$^5$.\footnote{For a Poisson Lie group $G$, $G^*$ is its (Poisson) dual.} The lagrangian approaches lead to a geometry which is doubly $T$-dual to dS$^5\times H^5$. The geometry obtained in \cite{Hoare:2014pna} can formally be embedded into type IIB supergravity, but its RR five-form is imaginary. The theory obtained in \cite{Arutyunov:2014cra} by taking a rescaled limit of the bosonic metric is real on the world sheet and has another important role: once light-cone gauge-fixed, this sigma model is the \emph{mirror} model of the $\ads$ superstring, the model that was central to the solution of the spectral problem for the $\ads$ superstring. In particular, one can complete this background to a full solution of type IIB supergravity. This is in sharp constrast with another possible approach: taking the original action \eqref{eq:etadeformedbosonicPolyakov} one can consider the maximal deformation limit directly on the RR five-form. It was shown in \cite{Arutyunov:2015qva} that this does not yield a solution of type IIB supergravity as we will discuss in more detail in the next paragraph. At present this discrepancy remains to be understood.
\\[5mm]
Let us mention that the fact that the $\eta$-deformed theory seems to interpolate between the $\ads$ superstring and its mirror model is very interesting. We will return to this point once we have a better understanding of the $\eta$-deformed $S$-matrix, as the notion of mirror duality will shed more light on this issue.

\paragraph{Type IIB supergravity.} In the low-energy limit the $\ads$ superstring theory is described by a type IIB supergravity theory providing a convenient framework for computations. A natural question is what happens to the low-energy limit if we turn on the $\eta$ deformation and its answer proves to be puzzling: the $\eta$-deformed background does not solve the supergravity equations of motion \cite{Arutyunov:2015qva}, despite the presence of $\kappa$ symmetry. This implies that the quantised model is not Weyl invariant, i.e. does not have conformal symmetry at the quantum level. The presence of $\kappa$ symmetry in general only implies that the background is a solution to a set of generalised supergravity equations \cite{Arutyunov:2015mqj,Wulff:2016tju}, which do not directly imply Weyl invariance. This breakdown is worth investigating: the $\eta$ deformation was introduced such as to manifestly maintain integrability and is from that perspective a ``mild" deformation. Indeed, we will see that its spectral problem is as solvable as the undeformed $\ads$ string theory. Nevertheless, the deformed model seems considerably less physical, showing that a mathematically natural deformation does not always provide a physically natural model. We will return to the question of conformality after constructing the $\eta$-deformed $S$-matrix. It is noteworthy that although the $\eta$-deformed theory has a conformal anomaly, it is (classically) $T$ dual to a conformal (non-unitary) type IIB* string model that can be consistently embedded into supergravity \cite{Hoare:2015wia}. 

\subsection{Related work}
The construction of the $\eta$ deformation of the $\ads$ string theory has sparked a widespread interest, which has led to explorations of many possible extensions and generalisations of the $\eta$-deformed theory. For example, many classical solutions have been found, see for example \cite{Arutyunov:2014cda,Banerjee:2014bca,Kameyama:2014vma,Khouchen:2015jfa,Roychowdhury:2016bsv,Hernandez:2017raj}. Also, the $\eta$ deformations for theories on the lower-dimensional AdS-spaces have been found, in particular on AdS$_2\times$S$^2$ and AdS$_3\times$S$^3$ \cite{Hoare:2015gda}, which also establishes a connection between the $\eta$-deformed $\ads$ theory and its root-of-unity cousin known as the \emph{$\lambda$-deformed} model: this model is a fermionic extension of a gauged Wess-Zumino-Witten model with constant level $k$ and was shown to have $\mathfrak{psu}_q(2,2|4)$ symmetry with $q = e^{i \pi /k}$ a root of unity \cite{Mikhailov:2007xr,Grigoriev:2007bu,Hollowood:2014qma}. 
\\[5mm]
As discussed in the introduction the natural construction of the $\eta$ deformation as discussed in this section proved a great starting point for constructing many integrable deformations of the $\ads$ string theory: solutions of the classical Yang-Baxter equation -- from our perspective a generalisation of the mYBE \eqref{eq:mCYBE} -- generate new models and studying these models has led to some interesting models \cite{vanTongeren:2015soa,vanTongeren:2015uha,Hoare:2016ibq,Hoare:2016hwh,Osten:2016dvf,vanTongeren:2016eeb}: for a particular class of these models it proved possible to construct a non-commutative dual gauge theory, where the non-commutativity is governed by a deformed Moyal product containing the $r$ matrix on which the deformation is based \cite{vanTongeren:2015uha}. 

\section{The world-sheet quantum field theory}
Having gathered all the relevant information from the classical theory we are ready to consider the quantum version of these models. Even though both the undeformed and deformed models are integrable and have a lot of symmetry it is impossible to consider quantisation directly. In order to quantise we first have to get rid of the spurious degrees of freedom by gauge fixing. 
\subsection{Light-cone gauge}
\label{sec:lightconegauge}
The presence of reparametrisation as well as $\kappa$ symmetry implies that not all the degrees of freedom of the classical string theory are physical. This is an obstruction for the quantisation of this theory and should therefore be removed, which can be done by choosing a particular gauge known as the light-cone gauge \cite{Arutyunov:2004yx,Frolov:2006cc}. We treat both the undeformed and deformed cases simultaneously here. 
\\[5mm]
The first step is to move the hamiltonian formalism: introducing conjugate momenta in the first-order formalism
\begin{equation}
p_M = \frac{\delta S^b}{\delta \partial_{\tau}X^M }
\end{equation}
we can rewrite the (bosonic)\footnote{For a clearer exposition we omit the fermions. The interested reader is referred to \cite{Arutyunov:2009ga} for a full derivation for the undeformed theory.} action in the form
\begin{equation}
S^b = \int d\sigma d\tau \left( p_M \partial_{\tau} X^M+\frac{\gamma^{\tau \sigma}}{\gamma^{\tau \tau}}C_1+\frac{1}{2 g\gamma^{\tau \tau}}C_2\right),
\end{equation}
where the $C_i$ are the Virasoro constraints, so we see that $\gamma^{\alpha \beta}$ acts as a Lagrange multiplier. The next step is to consider new coordinates built up from two of the coordinates along two commuting isometries. Concretely, we consider the time $t$ in the AdS$_5$ space and the angle $\phi$ in the compact S$^5$ space. Introducing the \emph{light-cone coordinates} 
\begin{equation}
\label{eq:lightconecoordinates}
x^+ = (1-a) t + a\phi,\quad x^- = \phi-t,
\end{equation}
depending on a real parameter $a\in [0,1]$, we find that the conjugate momentum read
\begin{equation}
p_+ = p_t +p_{\phi}, \quad p_- = -ap_t + (1-a) p_{\phi}. 
\end{equation}
The \emph{uniform light-cone gauge} can now be imposed by demanding
\begin{equation}
\label{eq:uniformlcg}
x^+ = \tau + a m \sigma, \quad p_- = 1,
\end{equation}
where $m \in \N$ signals how often $\phi$ winds around the circle. The uniformity of this gauge comes from the fact that $p_-$ does not depend on $\sigma$, causing light-cone momentum to be uniformly distributed. In this gauge, after the Virasoro constraints have been satisfied the bosonic action can be written as
\begin{equation}
S^b = \int d\sigma d\tau \left( p_{\mu} \partial_{\tau}X^{\mu} -H^b\right), \quad H^b= - p_+\left(X_{\mu},p_{\mu}\right),
\end{equation}
where $\mu$ labels only the directions transverse to the light-cone directions. Solving the first Virasoro constraint $C_1 = p_M \partial_{\sigma}X^M=0$ we find an expression for the derivative $\partial_{\sigma}x^-$, but to obtain a consistent closed-string theory we need to additionally impose that 
\begin{equation}
\int_{-L/2}^{L/2} d\sigma \partial_{\sigma}x^- = 2\pi m,
\end{equation}
which is known as the \emph{level-matching condition}. For most purposes it is convenient to restrict ourselves to the case $m=0$, which we will do from now on. 
\\[5mm]
By solving the second Virasoro constraint $C_2=0$ one finds an explicit expression for the light-cone hamiltonian. It is furthermore important to note that the light-cone directions are isometry directions and as such give rise to conserved quantities. Indeed, we find the string energy $E$ and the angular momentum $J$ along $\phi$ as
\begin{equation}
E= \int_{-L/2}^{L/2} d\sigma p_t, \quad J= \int_{-L/2}^{L/2} d\sigma p_{\phi},
\end{equation}
which in light-cone coordinates get recombined into
\begin{equation}
\label{eq:lccquantities}
-P_+ = \int_{-L/2}^{L/2} d\sigma H^b = E-J, \quad P_- = \int_{-L/2}^{L/2} d\sigma p_- = a E-(a-1)J = L,
\end{equation}
where for the last equality we used our particular gauge choice \eqref{eq:uniformlcg}. In the first equation we see how the string energy is related to the eigenvalues of the hamiltonian $H^b$, giving us a first clue how to solve the spectral problem of our string theories: all we need to do is find the spectrum of the two-dimensional quantum field theory on the closed-string world sheet defined by the hamiltonian $H^b$. The presence of highly non-linear interactions should at first instance raise doubts to whether it is possible to find the spectrum at all. The simplest choice for the gauge parameter $a$ seems to be the temporal gauge $a=0$: in this case the hamiltonian depends on $J$ only and its spectrum directly gives the world-sheet energy \cite{Arutyunov:2009ga}. Also, $L=J$, so the light-cone theory is defined on a cylinder with circumference $J$. We follow the literature in our choice of $a$ and set it to zero from now on. 
\\[5mm]
To make it easier to approach the spectrum, the next step is to consider the limit $P_-\rightarrow \infty$ of the light-cone momentum. Since by our gauge choice $L = P_- $ this has the effect of decompactifying the cylinder to a plane, hence justifying the name \emph{decompactification limit}. Note that in this limit both $E$ and $J$ become large such that their difference -- the world-sheet energy -- remains finite. Once put on the plane, studying the spectral problem of these theories becomes a lot easier: one can use the $S$-matrix formalism to study the scattering of asymptotic states.
\\[5mm]
Of course, not the full $\mathfrak{psu}(2,2|4)$ symmetry (or $\mathfrak{psu}_q(2,2|4)$ in the $\eta$-deformed case) of the original coset model survives the process described above, containing both gauge fixing and choosing a time direction by going to the hamiltonian formalism. One can check that in the undeformed case the eight bosonic and eight fermionic degrees of freedom in the resulting off-shell theory transform under the tensor product $\mathfrak{psu}(2|2)_{\text{c.e.}}^{\otimes2}$ \cite{Beisert:2005tm}, where the central charges match. We discuss this algebra in more details in section \ref{sec:uqsu22}. For the $\eta$-deformed case the result is the same with the algebra replaced by $\mathfrak{psu}_q(2|2)_{\text{c.e.}}^{\otimes2}$ with $q= e^{-\frac{2\eta}{g\left(1+\eta^2\right)}}$ \cite{Arutyunov:2013ega}. 

\paragraph{Fermions.} We have illustrated light-cone gauge fixing above by considering the bosonic part of the theories only. To obtain the $S$ matrix it is necessary, however, to have a good grasp on the fermions after gauge fixing as well. Moreover, in principle the presence of fermions might spoil the procedure described above. As it turns out, for both the undeformed as well as the $\eta$-deformed theory this is not the case, although performing the gauge fixing in full generality is very cumbersome. Luckily, to obtain the integrable $S$ matrix we only need terms quadratic in the fermions \cite{Metsaev:1998it, Grisaru:1985fv, Henneaux:1984mh} (and \cite{Arutyunov:2013ega,Borsato:2016hud} for the deformed case) and nothing of the qualitative features of the above derivation get spoiled. 
\subsection{Finding the perturbative $S$-matrix}
\label{subsec:perturbativeSmatrix}
With the hamiltonian containing all the terms quadratic in fermions we can proceed to derive the $S$ matrix describing scattering of the quantised model on the world sheet through the method of perturbative quantisation: by expanding the hamiltonian in inverse powers of the string tension $g$ keeping only the leading order and canonically quantising the result we obtain a perturbative description of the scattering processes. A crucial aspect of the light-cone gauge that allows for the application of perturbative quantisation is the fact that the kinetic terms for all the fields give rise to a canonical Poisson structure, i.e. with canonical equal-time (anti)-commutation relations for all the fields. These can then be quantised by promoting the fields to operators and replacing the Poisson bracket, which for bosons takes the form
\begin{equation}
\{\,,\,\}_{\text{P.B.}} \rightarrow -\frac{\ii}{\hbar}\left[ \, , \, \right].
\end{equation}
Concretely, for the eight bosonic (labelled by $\mu,\nu$) and the eight fermionic (labelled by $\dot{\mu},\dot{\nu}$) creation and annihilation operators we obtain the canonical (anti)-commutation relations
\begin{equation}
\left[ a^{\mu}(p,\tau), a^{\dagger}_{\nu}(p',\tau)\right] = \delta^{\mu}_{\nu}\delta\left(p-p'\right), \quad \left\{ a^{\dot{\mu}}(p,\tau), a^{\dagger}_{\dot{\nu}}(p',\tau)\right\} = \delta^{\dot{\mu}}_{\dot{\nu}}\delta\left(p-p'\right). 
\end{equation}
From this procedure we also obtain the dispersion relation $\omega(p)$. These creation and annihilation operators create states from the vacuum $\ket{\mathbf{0}}$ (suppressing the $\tau$ dependence)
\begin{align}
\label{eq:nmomentumstate}
\ket{p_1,p_2,\ldots,p_n}_{k_1,k_2,\ldots k_n} = a^{\dagger}_{k_1}(p_1)a^{\dagger}_{k_2}(p_2)\cdots a^{\dagger}_{k_n}(p_n)\ket{\mathbf{0}},
\end{align}
where we have collected the dotted and undotted indices in the index $k$, simultaneously treating bosons and fermions. The analysis of the states in eqn.  \eqref{eq:nmomentumstate} at arbitrary $\tau$ is very difficult because of the presence of interactions. Now, since we have decompactified the world-sheet cylinder we can instead consider the scattering of asymptotic states. We can define in and out operators $a_{\text{in},\text{out}}$ that evolve freely, i.e. their time evolution is governed by the free part $H_{\text{free}}$ of the hamiltonian only:
\begin{align}
\partial_{\tau} a_{\text{in},}^k (p,\tau) &= \ii \left[ H_{\text{free}}\left( a_{\text{in}}^{\dagger},a_{\text{in}}\right), a_{\text{in}}^k(p,\tau)\right], \nn
\partial_{\tau} a_{\text{out}}^k (p,\tau) &= \ii \left[ H_{\text{free}}\left( a_{\text{out}}^{\dagger},a_{\text{out}}\right), a_{\text{out}}^k(p,\tau)\right].
\end{align}
These in and out operators coincide with the original $a^k$ at $\tau =\pm \infty$, where the interactions are absent:
\begin{equation}
\lim_{\tau\rightarrow -\infty }a^k(p,\tau) = \lim_{\tau\rightarrow -\infty }a_{\text{in}}^k(p,\tau),\qquad
\lim_{\tau\rightarrow \infty }a^k(p,\tau) = \lim_{\tau\rightarrow \infty }a_{\text{out}}^k(p,\tau). 
\end{equation}
The states they create are in and out states
\begin{align}
\label{eq:inandoutstates}
\ket{p_1,p_2,\ldots,p_n}^{\text{in}}_{k_1,k_2,\ldots k_n} &= a^{\dagger}_{\text{in},k_1}(p_1)a^{\dagger}_{\text{in},k_2}(p_2)\cdots a^{\dagger}_{\text{in},k_n}(p_n)\ket{\mathbf{0}}, \nn
\ket{p_1,p_2,\ldots,p_n}^{\text{out}}_{k_1,k_2,\ldots k_n} &= a^{\dagger}_{\text{out},k_1}(p_1)a^{\dagger}_{\text{out},k_2}(p_2)\cdots a^{\dagger}_{\text{out},k_n}(p_n)\ket{\mathbf{0}}.
\end{align}
Since all the creation and annihilation operators $\left(a^{\dagger},a\right),\left(a^{\dagger}_{\text{in}},a_{\text{in}}\right)$ and $\left(a^{\dagger}_{\text{out}},a_{\text{out}}\right)$ satisfy the same canonical commutation relations, they must be related by unitary transformations
\begin{align}
a(p,\tau) &= \mathbb{U}_{\text{in}}^{\dagger}(\tau) a_{\text{in}}(p,\tau) \mathbb{U}_{\text{in}}(\tau), \nn
a(p,\tau) &= \mathbb{U}_{\text{out}}^{\dagger}(\tau) a_{\text{out}}(p,\tau) \mathbb{U}_{\text{out}}(\tau),
\end{align}
where $\mathbb{U}_{\text{in},\text{out}}$ should satisfy 
\begin{equation}
\lim_{\tau\rightarrow -\infty} \mathbb{U}_{\text{in}}(\tau) = \mathbb{I} = \lim_{\tau\rightarrow \infty} \mathbb{U}_{\text{out}}(\tau). 
\end{equation}
These operators can be constructed explicitly:
\begin{align}
\mathbb{U}_{\text{in}}(\tau) &= \mathcal{T}\exp\left( -\ii \int_{-\infty}^{\tau} d\tau' V\left(a_{\text{in}}^{\dagger}\left(\tau'\right),a_{\text{in}}\left(\tau'\right)\right)   \right), \nn
\mathbb{U}_{\text{out}}(\tau) &= \mathcal{T}\exp\left( -\ii \int_{\tau}^{\infty} d\tau' V\left(a_{\text{out}}^{\dagger}\left(\tau'\right),a_{\text{out}}\left(\tau'\right)\right)   \right),
\end{align}
where we have defined the potential $V = H-H_{\text{free}}$ and the time-ordered exponential $\mathcal{T}\exp$. Out of these operators we can build a unitary operator $\mathbb{S}$ -- known as the \emph{$S$ matrix} -- that maps out states to in states
\begin{equation}
\label{eq:defSmatrix}
\ket{p_1,p_2,\ldots,p_n}^{\text{in}}_{k_1,k_2,\ldots k_n} = \mathbb{S} \ket{p_1,p_2,\ldots,p_n}^{\text{out}}_{k_1,k_2,\ldots k_n}
\end{equation}
and is given by
\begin{equation}
\label{eq:SinUs}
\mathbb{S} = \mathbb{U}_{\text{in}}(\tau) \mathbb{U}_{\text{out}}(\tau). 
\end{equation}
The time dependence on the right-hand side is in fact only apparent: the time dependence of the in and out operators cancel each other, implying that $\mathbb{S}$ is time-independent. An important consequence of this fact is that 
\begin{equation}
\mathbb{S} = \lim_{\tau \rightarrow \infty} \mathbb{U}_{\text{in}}(\tau) = \lim_{\tau \rightarrow -\infty} \mathbb{U}_{\text{out}}(\tau).
\end{equation}
The $S$ matrix furthermore satisfies
\begin{equation}
a_{\text{in}}^{\dagger}(p,\tau) = \mathbb{S}\, a_{\text{out}}^{\dagger}(p,\tau)\,\mathbb{S}^{\dagger}, \quad a_{\text{in}}(p,\tau) = \mathbb{S}\, a_{\text{out}}(p,\tau)\, \mathbb{S}^{\dagger}, \quad \mathbb{S}\, \ket{\mathbf{0}} = \ket{\mathbf{0}}. 
\end{equation}
To find the $S$ matrix perturbatively one needs to consider the large-tension expansion of the expression \eqref{eq:SinUs}, which at leading order is given by
\begin{align}
\mathbb{S} = \mathbb{I} -\ii \int_{-\infty}^{\infty} d\tau' V\left(\tau'\right) +\ldots,
\end{align}
where $V$ expands as 
\begin{equation}
V = \frac{1}{g} H_4 +\bO{\frac{1}{g^2}},
\end{equation}
with $H_4$ being the quartic part of the hamiltonian. Therefore at leading order the $S$ matrix describes $2\rightarrow 2$ particle scattering. By continuing the expansion in $1/g$ one could find the quantum corrections, but the models we are considering allow for a more efficient continuation as we will see soon. 
\subsection{Bootstrapping the exact $S$-matrix}
For a generic QFT, finding the $S$ matrix perturbatively as discussed in the previous section is the best tool we have available to find the spectrum. For integrable field theories on the other hand we can use much more powerful techniques to obtain an \emph{exact} $S$-matrix. Unfortunately, it is very difficult to establish the integrability of the quantum model, as we cannot for example read off integrability directly from the perturbative $S$-matrix. Therefore, the only viable approach is to \emph{assume} quantum integrability, employ the special techniques and see whether the resulting $S$-matrix is consistent with the perturbative $S$-matrix \cite{Beisert:2005tm,Arutyunov:2006yd} (and \cite{Arutyunov:2013ega} in the deformed case). 

One way to obtain the exact $S$-matrix goes under the name \emph{factorised scattering theory}. The idea is to consider an abstract Hilbert space of asymptotic states that carries a representation of the Zamolodchikov-Faddeev (ZF) algebra as well as of the symmetry algebra $\mathfrak{J}$ of the model. It relies on a physical picture to understand scattering events. We will see later that for the $\eta$-deformed theory many aspects of the physical picture are in fact irrelevant: the $S$ matrix already follows uniquely from imposing compatibility with the structure of the $q$-deformed algebra. In order to appreciate this point we will first discuss the application of the ZF algebra and how one can capture most of its power in the form of Hopf algebras, although our discussion can only scrape the surface of the vast amount of literature on the topic. When we are ready to think about the $\eta$-deformed $S$-matrix we will use this machinery to obtain the $S$ matrix from purely algebraic means. 

\subsection{Integrable field theories in two dimensions}
\label{sec:integrablefieldtheories}
To understand what is so special about the theories we consider, let us analyse scattering in two-dimensional integrable theories in more generality: integrability is due to the existence of infinitely many symmetries, giving rise to an infinite set of commuting charges $\{ \mathbb{Q}_j \}$  which all mutually commute, i.e. 
\begin{equation}
\left[\mathbb{Q}_j,\mathbb{Q}_k\right] = 0. 
\end{equation}
We furthermore take a Hilbert space spanned by in and out states as before, where the label $k_i$ can now be interpreted as an abstract flavour label. Since we can diagonalise all the charges simultaneously, we can pick the basis to be such that
\begin{equation}
\mathbb{Q}_j \ket{p}_{k}^{\text{in},\text{out}} = q_j(p,k) \ket{p}_{k}^{\text{in},\text{out}}
\end{equation}
and it is straightforward to generalise this to $M$-particle states:
\begin{equation}
\mathbb{Q}_j \ket{p_1,\ldots,p_M}_{k_1,\ldots,k_M}^{\text{in},\text{out}} = \left(q_j(p_1,k_1) + \cdots + q_j(p_M,k_M) \right) \ket{p_1,\ldots,p_M}_{k_1,\ldots,k_M}^{\text{in},\text{out}}.
\end{equation}
Now, evolving an in state $\ket{p_1,\ldots,p_M}_{k_1,\ldots,k_M}^{\text{in}}$ to an out state $\ket{p_1',\ldots,p_{M'}'}_{k_1',\ldots,k_{M'}'}^{\text{out}}$ we see that conservation of the charge $\mathbb{Q}_j$ implies that
\begin{equation}
\sum_{l=1}^M q_j (p_l,k_l) = \sum_{l=1}^{M'} q_j (p_l',k_l').
\end{equation}
Since this should hold for all the infinite number of charges given a set of incoming and outgoing momenta $\{p_j\}$ and $\{p_j'\}$ respectively, we find that this can only hold if these sets are equal, i.e. 
\begin{equation}
\{ p_j\}_{1\leq j \leq M} = \{ p_j'\}_{1\leq j \leq M'}.
\end{equation}
For two-dimensional theories, this statement has far-reaching consequences: consider an $M$-particle in-state $\ket{p_1,\ldots,p_M}_{k_1,\ldots,k_M}^{\text{in}}$ at $\tau=-\infty$. In order for the particles to scatter at some finite $\tau$ the momenta better be ordered such that $p_1>p_2>\cdots>p_M$. As the particles are constrained on a line all pairs of particles will necessarily scatter at some time $\tau$. When two particles with momenta $p_i>p_j$ meet and scatter, the conservation of charges dictates that the resulting scattering state must be proportional to $\ket{p_j,p_i}_{k_2',k_1'}$, where the flavour index may or may not have changed. Assuming for now that all scattering is $2\rightarrow 2$ we can scatter all pairs of particles to finally end up with the out state $\ket{p_M,\ldots,p_1}_{k_M',\ldots,k_1'}$ at $\tau=\infty$. 
\\[5mm]
Of course, in principle one should also consider more general $M\rightarrow M$ scattering. However, a crucial point first observed in \cite{Zamolodchikov:1978xm} is that the existence of the infinite tower of charges can be used to simplify the scattering analysis immensely: each charge gives rise to unitary transformations on the momenta, effectively allowing us to rescale the momenta such that all scattering events are transformed into subsequent $2\rightarrow 2$ scattering. This effect, called \emph{factorised scattering}, is extremely powerful, as it means we can describe all scattering just by knowing the two-body $S$-matrix, but does lead to a consistency question: given a scattering event, does it matter in which way we factorise this into $2\rightarrow 2$ scattering? Let us consider the simplest case of $3\rightarrow 3$ scattering: the factorised scattering of three particles can occur in two possible ways: either particle one scatters first with particle two, or it scatters first with particle three. This means we can factorise $3\rightarrow 3$ scattering in two a priori inequivalent ways. To have a consistent scattering theory, however, we must impose that  this choice is immaterial, i.e. that both orderings lead to the same physics. This leads to an equation for the two-body $S$-matrix known as the \emph{Yang-Baxter equation}, which we will encounter in the next section. This equation forms the heart of quantum integrability. Its power is twofold: first of all it is very constraining and secondly it actually ensures consistency of \emph{all} scattering. Indeed, one can use the Yang-Baxter equation to show that any two factorisations of an $M\rightarrow M$ scattering event are equivalent. 
\subsection{Zamolodchikov-Faddeev algebra}
The ideas in the previous subsection can be captured by a special type of creation and annihilation operators $A_k^{\dagger}(p)$ and $A^k(p)$, which form the Zamolodchikov-Faddeev algebra. The $A_k^{\dagger}(p)$ create in-going or out-going particles with definite flavour and momentum from the vacuum
\begin{equation}
A^k(p) \ket{\mathbf{0}} = 0, \quad A_k^{\dagger}(p) \ket{\mathbf{0}} = \ket{p}_k^{\text{in}} = \ket{p}_k^{\text{out}}.
\end{equation}
Furthermore, implementing the conservation of the set of momenta can be done by the following action: for $p_1>p_2$
\begin{align}
\label{eq:ZFaction}
\ket{p_1,p_2}_{k_1,k_2}^{\text{in}} &= A^{\dagger}_{k_1}(p_1)A^{\dagger}_{k_2}(p_2)\ket{\mathbf{0}}, \nn
\ket{p_1,p_2}_{k_1,k_2}^{\text{out}} &= (-1)^{\e(k_1)\e(k_2)}A^{\dagger}_{k_2}(p_2)A^{\dagger}_{k_1}(p_1)\ket{\mathbf{0}},
\end{align}
where $\e(k) = 1$ if $k$ is a fermionic flavour and $\e(k)=0$ if $k$ is a bosonic flavour. The $(-1)$ prefactor therefore takes into account the exchange of fermions. We can now use the definition of the $S$ matrix in eqn. \eqref{eq:defSmatrix} to obtain the algebra relations of the ZF algebra: for $2\rightarrow 2$ scattering eqn. \eqref{eq:defSmatrix} reads in components
\begin{equation}
\ket{p_1,p_2}_{k_1,k_2}^{\text{in}} = S^{k_3,k_4}_{k_1,k_2}(p_1,p_2)\ket{p_1,p_2}_{k_3,k_4}^{\text{out}},
\end{equation}
thus $S^{k_3,k_4}_{k_1,k_2}(p_1,p_2)$ is a matrix element of $\mathbb{S}$. Combining this with the action of the ZF operators on two-particle states \eqref{eq:ZFaction} gives
\begin{equation}
\label{eq:ZFalgebra}
A^{\dagger}_{k_1}(p_1)A^{\dagger}_{k_2}(p_2) = (-1)^{\e(k_3)\e(k_4)} S^{k_3,k_4}_{k_1,k_2}(p_1,p_2)A^{\dagger}_{k_2}(p_2)A^{\dagger}_{k_1}(p_1). 
\end{equation}
We can streamline the notation by considering the two-particle states as tensor products of one-particle states, i.e. for in states
\begin{equation}
\ket{p_1,p_2}_{k_1,k_2}^{\text{in}} =\ket{p_1}_{k_1}^{\text{in}}\otimes \ket{p_2}_{k_2}^{\text{in}} \in \mathcal{V}\otimes \mathcal{V},
\end{equation}
where $\mathcal{V}$ is the module generated by the action of $A_k^{\dagger}$ on the vacuum. It is also useful to introduce the $R$ matrix $\mathbb{R}$ as follows
\begin{equation}
R_{k_1,k_2}^{k_3,k_4}(p_1,p_2) = (-1)^{\e(k_3)\e(k_4)}S^{k_3,k_4}_{k_1,k_2}(p_1,p_2),
\end{equation}
which can be compactly written as $\mathbb{R} = \Pi^g \mathbb{S}$ with $\Pi^g$ the graded permutation operator. The ZF algebra relation \eqref{eq:ZFalgebra} can now be written as
\begin{equation}
\mathbb{A}^{\dagger}_1\mathbb{A}^{\dagger}_2 = \mathbb{A}^{\dagger}_2 \mathbb{A}^{\dagger}_1 \mathbb{R}^{\dagger}_{12},
\end{equation}
where the subscripts indicate on which tensor factor(s) the operator acts. To complete the algebra, we can find similar relations also including annihilation operators
\begin{equation}
\mathbb{A}_1\mathbb{A}_2 = \mathbb{R}_{12}\mathbb{A}_2\mathbb{A}_1, \quad \mathbb{A}_1\mathbb{A}^{\dagger}_2 = \mathbb{A}_2 \R_{21}\mathbb{A}_1^{\dagger} + \delta(p_1-p-2) \mathbb{I}.
\end{equation}
Using these relations we can in principle start computing $S$-matrix elements. The approach we would like to follow here is a bit more algebraic however, as the $\eta$ deformation can be more naturally formulated in the context of Hopf algebras.
\\[5mm]
Before we move to the Hopf algebra setting, let us use the ZF algebra for some final observations: first of all the ZF algebra relations directly imply that $\R$ satisfies the braiding unitarity relation
\begin{equation}
\label{eq:braidingunitarity}
\R_{21}\R_{12} = \R_{12}\R_{21} = \mathbb{I}. 
\end{equation}
Also, we can extend the action of the ZF algebra to multi-particle states as follows
\begin{align}
\ket{p_1,\ldots,p_M}_{k_1,\ldots k_M}^{\text{in}} &= A_{k_1}(p_1)^{\dagger}\cdots A_{k_M}(p_M)^{\dagger}\ket{\mathbf{0}}, \nn
\ket{p_1,\ldots,p_M}_{k_1,\ldots k_M}^{\text{out}} &= (-1)^{\sum_{j<l} \e(k_j)\e(k_l)}A_{k_1}(p_M)^{\dagger}\cdots A_{k_M}(p_1)^{\dagger}\ket{\mathbf{0}}, 
\end{align}
where $p_1>p_2>\cdots>p_M$. For consistency, however, we have to deal with the same type of ordering problem already alluded to in the previous section: indeed, already for a three-particle state we have two ways of expressing out states in terms of in states: we can write it in one of the following ways
\begin{align}
\mathbb{A}_1^{\dagger}\mathbb{A}_2^{\dagger}\mathbb{A}_3^{\dagger} &= \mathbb{A}_3^{\dagger}\mathbb{A}_2^{\dagger}\mathbb{A}_1^{\dagger}\R_{12}\R_{13}\R_{23} \nn
\mathbb{A}_1^{\dagger}\mathbb{A}_2^{\dagger}\mathbb{A}_3^{\dagger} &= \mathbb{A}_3^{\dagger}\mathbb{A}_2^{\dagger}\mathbb{A}_1^{\dagger}\R_{23}\R_{13}\R_{12},
\end{align}
which are equivalent if the $R$ matrix satisfies the Yang-Baxter equation
\begin{equation}
\label{eq:YangBaxterR}
\R_{12}\R_{13}\R_{23} = \R_{23}\R_{13}\R_{12}. 
\end{equation}
In the present context this constrains the allowed form of the $R$ matrix. 
\subsection{Symmetry action on scattering states}
The formalism in the previous section gives us a natural description of scattering states and the $S$ matrix. It cannot by itself, however, constrain the $S$ matrix far enough to bootstrap it completely, since the $S$ matrix depends on the symmetry algebra of the theory under consideration. Therefore, to determine the $S$ matrix we first have to think how our symmetry algebra acts on scattering states. We can write the general action of a symmetry generator $\mathbb{J}\in \mathfrak{J}$ on the zero-, one- and two-particle states in an arbitrary linear representation as
\begin{align}
\label{eq:symmetryactiononstates}
\mJ\ket{\mathbf{0}} &= 0, \nn
\mJ A^{\dagger}_k(p)\ket{\mathbf{0}} &= J^{k'}_k(p;C) A^{\dagger}_{k'}(p)\ket{\mathbf{0}}, \nn
\mJ A^{\dagger}_{k_1}(p_1)A^{\dagger}_{k_2}(p_2)\ket{\mathbf{0}} &= J^{k_1',k_2'}_{k_1,k_2}(p_1,p_2;C_1,C_2) A^{\dagger}_{k_1}(p_1)A^{\dagger}_{k_2}(p_2)\ket{\mathbf{0}},
\end{align}
which incorporates the natural assumption that $\mJ$ commutes with the world-sheet momentum and the particle number operator as well as the entire tower of conserved charges present due to integrability. This means that representations of $\mathfrak{J}$ are labelled by momentum as well as possible central charges $\{C\}$ present in the algebra, which leads to the dependence on these quantities of the matrix elements of $\mJ$ in eqn. \eqref{eq:symmetryactiononstates}. Consistent with our previous formalism, let us note that for the action on two-particle states we can write
\begin{equation}
\label{eq:factoringofbJ}
\mJ_{12}(p_1,p_2;C_1,C_2) = \mJ\left(p_1;C_1\right)\otimes \mathbb{I} + \mathbb{I}^g \left( \mathbb{I} \otimes \mJ\left(p_2;C_2\right)\right)\mathbb{I}^g,
\end{equation}
where $\mathbb{I}^g$ is the graded identity operator. 
\\[5mm]
Since by assumption $\mJ$ generates a symmetry of the theory we should have that
\begin{equation}
\mJ \ket{p_1,p_2}^{\text{in}}_{k_1,k_2} = \mathbb{S} \mJ \ket{p_1,p_2}^{\text{out}}_{k_3,k_4},
\end{equation}
which leads to the following relation:
\begin{equation}
\label{eq:Rintertwiner}
\mathbb{R}_{12}(p_1,p_2) \mJ_{12}(p_1,p_2) =\mJ_{21}(p_2,p_1) \mathbb{R}_{12}(p_1,p_2).
\end{equation}
This is a very powerful relation that, combined with the Yang-Baxter equation, in some cases can completely constrain the $R$ matrix. From an algebraic viewpoint this equation looks like an intertwining relation for the algebra and equation \eqref{eq:factoringofbJ} is nothing else but the definition of an abstract coproduct. In the next section we will use these observations to construct a Hopf algebra out of the symmetry algebra $\mathfrak{J}$ and describe how this can be used to bootstrap the $S$ matrix.
\subsection{Hopf algebra}
\label{sec:Hopfalgebra}
Our motivation to study Hopf algebras is twofold: they arise naturally from the scattering theory we have discussed and are also the most natural framework to discuss \emph{quantum groups}, which form the basis for the $\eta$-deformed theory that is central in this thesis. The fact that quantum groups were in fact Hopf algebras was observed by Drinfel'd and Jimbo a while after the first appearance of quantum groups in the physics literature, in particular in the work of the Leningrad school \cite{Drinfeld:1985rx,Jimbo:1985vd}. An excellent introduction to quantum groups is the book \cite{Chari:1995gqg}, a good review of Hopf algebras in the scattering theory of AdS/CFT is \cite{Torrielli:2010kq}. We will now consider the construction of a Hopf algebra out of the symmetry algebra $\mathfrak{J}$. 

\paragraph{Construction of $U\left(\mathfrak{J}\right)$}
 The construction starts with a Lie (super)algebra, in our case the symmetry algebra $\mathfrak{J}$, and we consider its \emph{universal enveloping algebra} $\mathfrak{H} = U\left(\mathfrak{J}\right)$: this is defined as the quotient of the tensor algebra 
\begin{equation}
T\left(\mathfrak{J}\right) = \bigoplus_{n\geq 0}^{\infty} \mathfrak{J}^{\otimes n}
\end{equation}
over the two-sided ideal generated by the elements
\begin{align}
x\otimes y - y\otimes x -& \left[x,y\right] \text{ for all bosonic } x,y\in \mathfrak{J}, \nn 
x\otimes y + y\otimes x -& \left\{x,y\right\} \text{ for all fermionic } x,y\in \mathfrak{J}, 
\end{align}
with $\left[\, , \, \right]$ ($\left\{\, , \, \right\}$) being the usual bosonic (fermionic) Lie bracket of $\mathfrak{J}$. This quotient roughly realises the Lie brackets in $\mathfrak{H}$ as the usual commutator. The tensor structure of $\mathfrak{H}$ is motivated from the fact that we want to consider multi-particle states and the action of $\mathfrak{J}$ on those states. It defines the \emph{multiplication map}\footnote{Despite their roles in the rest of this thesis, we do keep the standard notation $\mu$ and $\Delta$ the multiplication and coproduct maps as we think it will not lead to confusion. }  $\mu : \mathfrak{H} \otimes \mathfrak{H} \rightarrow \mathfrak{H}$ as $\mu(x,y) = x \otimes y$ for any elements $x,y \in \mathfrak{H}$ which therefore introduces a formal product of Lie algebra elements. It comes together with a \emph{unit} which can be embedded into $\mathfrak{H}$ with the \emph{unit map} usually denoted by ``$\eta$", but in light of the importance of $\eta$ as a deformation parameter we will opt for the notation $\zeta : \C \rightarrow \mathfrak{H}$. This structure is enough to consider the action of $\mathfrak{J}$ on one-particle states. To extend this to multiparticle states, we can introduce a \emph{coproduct} $\Delta : \mathfrak{H} \rightarrow \mathfrak{H}  \otimes \mathfrak{H}$ defined as 
\begin{equation}
\Delta(x) = x \otimes 1 + 1 \otimes x,
\end{equation}
reminding us of eqn. \eqref{eq:factoringofbJ}. This choice for $\Delta$ is not unique as we will see, although we do require $\Delta$ to be an algebra homomorphism to ensure a consistent extension of the action of $\mathfrak{J}$ to multiparticle states. Lastly we define a \emph{counit} $\varepsilon: \mathfrak{J} \rightarrow \C$ as $\varepsilon(x) = 0$ for all $x \in \mathfrak{J}$.\footnote{One can extend this uniquely to $\mathfrak{H}$ using the consistency relations of the various maps (see for example \cite{Chari:1995gqg}). For example, $\varepsilon(z) = z$ for all $z \in \C$.} The maps introduced above satisfy a set of consistency axioms that turn $\mathfrak{H}$ into a \emph{bialgebra}. As we will not really need these axioms we refer the interested reader to \cite{Chari:1995gqg}. In particular, the definition of the counit can be derived from the consistency axioms. 

\begin{figure}[!t]
\centering
\includegraphics[width=6cm]{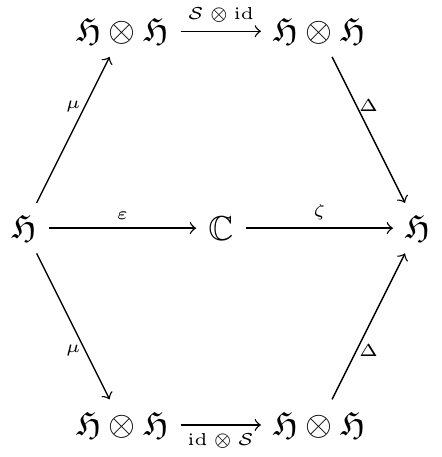}
\caption{The defining diagram for the antipode of the Hopf algebra $\mathfrak{H}$.}
\label{fig:hopfalgebra}
\end{figure}

\paragraph{Into a Hopf algebra.} To turn the bialgebra $\mathfrak{H}$ into a Hopf algebra all we need to do is define one more map $\mathcal{S} : \mathfrak{H}\rightarrow \mathfrak{H}$, called the \emph{antipode}, that should not be confused with the $S$ matrix. It is defined by $\mathcal{S}(x) =-x$ for all $x\in \mathfrak{J}$ and is such that the diagram in fig. \ref{fig:hopfalgebra} commutes.
Although in our current discussion we get this map for free in the current construction, it does have a function in the scattering picture: it helps defining conjugated representations of $\mathfrak{J}$ and therefore defines antiparticles.\footnote{Although relevant for the whole physical picture we will not spend more time on this point, as it is not essential for our discussion.} Moreover, it defines the \emph{crossing equation} which constrains the scalar factor of the $S$ matrix. This completes the construction of the canonical Hopf algebra out of the universal enveloping algebra. 

\paragraph{Quasitriangularity.} In the structure we have defined so far we have not seen a role for either the $S$ or $R$ matrix, confer eqn. \eqref{eq:Rintertwiner}. There is, however, a very natural way for the $R$ matrix to occur: consider the graded permutation operator $\tau$ on $ \mathfrak{H} \otimes  \mathfrak{H}$ defined by $\tau\left( x \otimes y \right) = (-1)^{\e(x)\e(y)}y \otimes x$ for generators $x,y\in \mathfrak{J}$, which we can use to define the opposite coproduct $\Delta^{\text{op}} = \Delta \circ \tau$. A Hopf algebra is called \emph{quasicocommutative} if there exists an invertible element $\R\in \mathfrak{H} \otimes \mathfrak{H}$ such that 
\begin{equation}
\label{eq:cocommutativity}
\Delta^{\text{op}} (x )  = \R \Delta(x) \R^{-1} \text{ for all } x\in \mathfrak{H},
\end{equation}
obviously mimicking eqn. \eqref{eq:Rintertwiner}. Even more, the Hopf algebra is called \emph{quasitriangular} if this element $\R$ additionally satisfies
\begin{align}
\label{eq:quasitriangular}
\left(\Delta\otimes \ \mathbb{I}\right)(\R) &= \R_{13}\R_{23}, \nn
\left( \ \mathbb{I} \otimes \Delta\right)(\R) &= \R_{13}\R_{12},
\end{align}
where $\R_{12} = \R \otimes \mathbb{I}$, $\R_{23} =\mathbb{I} \otimes \R$ and $\R_{13}$ is the generalisation of the preceding definitions to the case where the second tensor leg contains the unit. The way to define this is to write $\R = \sum r_i \otimes r_j$ for some elements $r_i,r_j\in  \mathfrak{J}$ and define $\R_{13} = \sum r_i \otimes \mathbb{I} \otimes r_j$. If the Hopf algebra is quasitriangular, then the element $\R$ is called the \emph{universal $R$ matrix}. The importance of this property is that the universal $R$ matrix satisfies the following two relations:
\begin{align}
\R_{12}\R_{13}\R_{23} &= \R_{23}\R_{13}\R_{12}, \nn
\left( \mathcal{S} \otimes \mathbb{I} \right) \R &= (\left( \mathbb{I} \otimes \mathcal{S}^{-1}\right) \R = \R^{-1},
\end{align}
where we immediately recognise the first relation as the Yang-Baxter equation \eqref{eq:YangBaxterR}. The second relation is called the \emph{crossing equation} and plays an important role in bootstrapping the S matrix\footnote{Just to be explicit: in the present context the S matrix is related to $\R$. The antipode $\mathcal{S}$ plays a different role entirely.}: apart from certain analyticity requirements it is the only constraint on the momentum-dependent normalisation of the $S$ matrix. 
\\[5mm]
Thus, if we can find a universal $R$ matrix that makes $\mathfrak{H}$ into a quasitriangular Hopf algebra we are in principle done: we have defined a way to extend one-particle representations to multiparticle representations and found an operator which can act as an $S$ matrix, relating in states to out states as in the preceding section. Keeping in mind the ZF formalism we conclude that this therefore defines a consistent scattering theory for some integrable field theory. In practice finding the universal $R$ matrix can be very hard and it is easier to consider the problem of cocommutativity for a specific representation. One can then find the $R$ matrix for that particular representation. Since physical particles are usually modelled by a particular representation, this approach can be fruitful, as was the case for the $\ads$ superstring.

\paragraph{Braiding.} As it turns out, the Hopf algebra we just constructed does not accurately describe the scattering theory of the $\ads$ world-sheet QFT, due to our choice of the coproduct $\Delta$ \cite{Plefka:2006ze}. This does not take into account the effect of ``length changing"\footnote{This nomenclature comes from the spin chain picture on the gauge theory side.}: there exists an abelian generator $\mathbf{U}$ in the symmetry algebra that acts as a ladder operator on $P_-$ eigenstates, thereby changing the ``length" of a state. This leads to a mixing of different-length operators in the scattering process. To correct for length changing, we can introduce a different coproduct, thereby changing the action of $\mathfrak{J}$ on multiparticle states. This new coproduct is defined by a \emph{braiding} and amounts to 
\begin{equation}
\Delta_{\text{braid}}(x) = x\otimes 1 + \mathbf{U}^n  \otimes x, 
\end{equation}
where the power $n$ depends on which generator $x$ the coproduct is acting. We will not need its precise form here. The Hopf algebra with this braided coproduct is still quasicocommutative.

\paragraph{The $\ads$ $S$-matrix} The matrix form of the $R$ matrix corresponding to fundamental short representations was determined in \cite{Beisert:2005tm,Arutyunov:2006yd,Arutyunov:2006ak} and its overall scalar factor known as the \emph{dressing factor} \cite{Arutyunov:2006iu,Arutyunov:2004vx,Beisert:2006ez,Janik:2006dc,Beisert:2005tm} was determined by solving the crossing equation. This then defines the $S$ matrix of the $\ads$ world-sheet QFT, thereby defining a candidate for the quantisation of the classical theory. This $S$ matrix is exact, providing an $S$ matrix for every value of the string tension. By expanding it for small values we can compare it with the perturbative $S$-matrix, which turns out to be in perfect agreement \cite{Klose:2006zd, Arutyunov:2006yd}.
\\[5mm]
Our next task is to find the $S$ matrix for the $\eta$-deformed theory, which can be done in a way inspired by the one we just reviewed. We will treat it in detail in the next section. 

\section{$q$ deforming the $\ads$ $S$-matrix}
Quantisation of the classical $\eta$-deformed theory is incredibly complicated due to the complex interactions appearing in the lagrangian, even more so than was the case for the undeformed model. Nevertheless, we can follow the route taken for the undeformed case in bootstrapping an exact $S$ matrix from symmetry. We saw that the undeformed off-shell $S$-matrix was invariant under $\mathfrak{psu}(2|2)_{\text{c.e.}}^{\otimes 2}$ and could be bootstrapped from this knowledge completely. There is a natural construction of the $q$ deformation of centrally extended $\mathfrak{su}(2|2)$ that defines the \emph{quantum group} $U_q\left( \mathfrak{su}(2|2)\right)_{\text{c.e.}}$, which will allow us to bootstrap the $q$-deformed $S$-matrix: indeed, by considering cocommutativity of this Hopf algebra it proved possible to construct an $S$ matrix which is invariant under $U_q\left( \mathfrak{su}(2|2)\right)_{\text{c.e.}}^{\otimes 2}$, thereby providing a natural $q$-deformation of the $\ads$ $S$-matrix.\footnote{\label{footnote:suvspsu} We really do mean $\mathfrak{su}(2|2)$ and $\mathfrak{psu}(2|2)$, although their mixed occurrence might be confusing: note that $\mathfrak{psu}(2|2)$ is really just $\mathfrak{su}(2|2)$ with one generator modded out, as we discussed in section \ref{sec:psu224}, and in that sense after central extension their difference sits in the details of the extension.} This $q$-deformed $S$-matrix was found long before the classical $\eta$-deformed theory was formulated and one could rightfully argue that in this case the search for the classical theory was even motivated by advancements in the quantum model! In the next section we discuss how the $\ads$ $S$-matrix can be $q$ deformed, reviewing the original work by Beisert and Koroteev \cite{Beisert:2008tw}.

\subsection{The universal enveloping algebra U$\left( \mathfrak{su}(2|2)_{\text{c.e.}}\right)$}
\label{sec:uqsu22}
The most convenient point to start is the centrally extended \note{Ideally we connect this more strongly with the original description of PSU$(2,2|4)$ as a matrix quotient group by giving the expressions for these generators for example.}algebra $\mathfrak{g} =\mathfrak{su}(2|2)_{\text{c.e.}}$, which is generated by the $\mathfrak{su}(2)\times \mathfrak{su}(2)$ generators $\mathbf{R}^a_{\,\,\,b}$ and $\mathbf{L}^{\alpha}_{\,\,\,\beta}$, by the supercharges $\mathbf{Q}^{\alpha}_{\,\,\,b}$ and $\mathbf{Q}^{\dagger \alpha}_{\,\,\,\,\,b}$ and the central charges $\mathbf{C},\mathbf{D}$ and $\mathbf{K}$. The algebra relations are given by the following nonvanishing Lie brackets: 
\begin{equation}
\begin{aligned}[c]
\label{eq:su22algebrarelations}
\left[\mathbf{R}^a_{\,\,\,b}, \bR^c_{\,\,\,d} \right] &= \delta^c_b \bR^a_{\,\,\,d}-\delta^a_d \bR^c_{\,\,\,b}, \\
\left[\mathbf{R}^a_{\,\,\,b}, \bQ^{\gamma}_{\,\,\,d} \right] &= -\delta^a_d \bQ^{\gamma}_{\,\,\,b}+\frac{1}{2}\delta^a_b \bQ^{\gamma}_{\,\,\,d}, \\
\left[\mathbf{R}^a_{\,\,\,b}, \bQ^{\dagger c}_{\,\,\,\,\,\delta} \right] &= \delta^c_{b} \bQ^{\dagger a}_{\,\,\,\,\,\delta}-\frac{1}{2}\delta^a_b \bQ^{\dagger c}_{\,\,\,\,\,\delta},
\end{aligned}\quad
\begin{aligned}[c]
\left[\mathbf{L}^{\alpha}_{\,\,\,\beta}, \bL^{\gamma}_{\,\,\,\delta} \right] &= \delta^{\gamma}_{\beta} \bL^{\alpha}_{\,\,\,\delta}-\delta^{\alpha}_{\delta} \bL^{\gamma}_{\,\,\,\beta}, \\
\left[\mathbf{L}^{\alpha}_{\,\,\,\beta}, \bQ^{\gamma}_{\,\,\,d} \right] &=\delta^{\gamma}_{\beta} \bQ^{\alpha}_{\,\,\,d}-\frac{1}{2}\delta^{\alpha}_{\beta} \bQ^{\gamma}_{\,\,\,d},\\
\left[\mathbf{L}^{\alpha}_{\,\,\,\beta}, \bQ^{\dagger c}_{\,\,\,\,\,\delta} \right] &=- \delta^{\alpha}_{\delta} \bQ^{\dagger c}_{\,\,\,\,\,\beta}+\frac{1}{2}\delta^{\alpha}_{\beta} \bQ^{\dagger c}_{\,\,\,\,\,\delta},
\end{aligned} \\
\end{equation}
where we recognise the standard $\mathfrak{su}(2)$ Lie brackets in the first line. We furthermore have   
\begin{equation}
\{\bQ^{\alpha}_{\,\,\, b}, \bQ^{\gamma}_{\,\,\, d}   \} = \e^{\alpha \gamma } \e_{b d} \bC,\quad \{\bQ^{\dagger a}_{\,\,\,\,\,\beta}, \bQ^{\dagger c}_{\,\,\,\,\, \delta}   \} = \e^{a c } \e_{\beta \delta} \mathbf{D},
\end{equation}
generating the central charges and the Lie brackets of the supercharges
\begin{equation}
\{\bQ^{\alpha}_{\,\,\, b} , \bQ^{\dagger c}_{\,\,\,\,\, \delta}  \} = \delta^{\alpha}_{\delta} \bR^{c}_{\,\,\, b}+\delta^c_b \bL^{\alpha}_{\,\,\, \delta} +\delta^c_b \delta^{\alpha}_{\delta} \mathbf{K}.
\end{equation}
Following section \ref{sec:Hopfalgebra} we can straightforwardly define the universal enveloping algebra $U \left( \mathfrak{g}\right)$. Note that $\mathfrak{g}$ as generated by the above can be regarded as $\mathfrak{su}(2|2)_{\text{c.e.}}$ with two central extensions or equivalently as $\mathfrak{psu}(2|2)_{\text{c.e.}}$ with three central extensions, cf. the footnote \footnotemark[11]. 

\paragraph{Chevalley basis.} We do not need all the original generators to describe $U \left( \mathfrak{g}\right)$. We can define a Chevalley basis of $U \left( \mathfrak{g}\right)$ by three Cartan generators $\mathbf{H}_j$, three simple positive roots $\mathbf{E}_j$ and three simple negative roots $\mathbf{F}_j$. Expressed in the Lie algebra generators they read
\begin{equation}
\begin{aligned}[c]
\mathbf{H}_1 &= \mathbf{R}^2_{\,\,\, 2} -\mathbf{R}^1_{\,\,\, 1}, \\
\mathbf{H}_2 &= -\mathbf{K}-\frac{1}{2} \mathbf{H}_1-\frac{1}{2} \mathbf{H}_3,\\
\mathbf{H}_3 &= \mathbf{L}^2_{\,\,\, 2} -\mathbf{L}^1_{\,\,\, 1},
\end{aligned}\quad
\begin{aligned}
\mathbf{E}_1 &= \mathbf{R}^2_{\,\,\, 1}, \\
\mathbf{E}_2 &= \mathbf{Q}^2_{\,\,\, 2}, \\
\mathbf{E}_3 &= \mathbf{L}^1_{\,\,\, 2},
\end{aligned}\quad
\begin{aligned}
\mathbf{F}_1 &= \mathbf{R}^1_{\,\,\, 2}, \\
\mathbf{F}_2 &= \mathbf{Q}^{\dagger 2}_{\,\,\,\,\, 2}, \\
\mathbf{F}_3 &= \mathbf{L}^2_{\,\,\, 1}. 
\end{aligned}
\end{equation}
In this basis the symmetric Cartan matrix has the form
\begin{equation}
A_{jk} = \begin{pmatrix}
2 & -1 & 0 \\
-1 & 0 & 1 \\
0 & 1 & 2 \\
\end{pmatrix},
\end{equation}
which has rank two. 
\paragraph{Commutation relations.} In this basis the commutation relations can be written compactly as follows: for $j,k=1,2,3$
\begin{equation}
\left[ \mathbf{H}_j, \mathbf{H}_k  \right] = 0, \quad 
\left[ \mathbf{H}_j, \mathbf{E}_k  \right] = A_{jk} \mathbf{E}_k, \quad
\left[ \mathbf{H}_j, \mathbf{F}_k  \right] = - A_{jk} \mathbf{F}_k.
\end{equation}
The commutators between positive and negative roots are
\begin{equation}
\left[ \mathbf{E}_1, \mathbf{F}_1 \right] = \mathbf{H}_1, \quad
\left\{ \mathbf{E}_2, \mathbf{F}_2 \right\} = -\mathbf{H}_2, \quad
\left[ \mathbf{E}_3, \mathbf{F}_3 \right] = -\mathbf{H}_3. 
\end{equation}
All other commutators vanish. 
\paragraph{Serre relations.} The Serre relations, which are the usual result of imposing consistency of higher order algebra relations, put additional restrictions on positive and negative simple roots
\begin{align}
\label{eq:Serrerelations}
\left[ \mathbf{E}_1, \mathbf{E}_3  \right] &=  \mathbf{E}_2  \mathbf{E}_2 = \left[  \mathbf{E}_1, \left[  \mathbf{E}_1, \mathbf{E}_2\right] \right] = \left[  \mathbf{E}_3, \left[  \mathbf{E}_3, \mathbf{E}_2\right] \right] = 0, \nn
\left[ \mathbf{F}_1, \mathbf{F}_3  \right] &=  \mathbf{F}_2  \mathbf{F}_2 = \left[  \mathbf{F}_1, \left[  \mathbf{F}_1, \mathbf{F}_2\right] \right] = \left[  \mathbf{F}_3, \left[  \mathbf{F}_3, \mathbf{F}_2\right] \right] = 0. 
\end{align}
Two additional Serre relations, stemming from the fact that $\mathfrak{su}(2|2)$ is a superalgebra, can be simplified to read 
\begin{equation}
\label{eq:Serrerelations2}
\mathbf{C} = \mathbf{D} = 0,
\end{equation}
but we in fact are not forced to impose these relations. Instead of imposing them, we can consistently drop these relations to obtain a centrally extended algebra. 

\paragraph{Central charges.} The central charges can be expressed in terms of the Chevalley basis as
\begin{align}
\mathbf{K} &= - \tfrac{1}{2}\bH_1-\bH_2- \tfrac{1}{2}\bH_3, \nn
\mathbf{C} &= \left\{\left[\mathbf{E}_1,\mathbf{E}_2 \right] , \left[\mathbf{E}_3,\mathbf{E}_2 \right]   \right\}, \nn
\mathbf{D} &= \left\{\left[\mathbf{F}_1,\mathbf{F}_2 \right] ,\left[\mathbf{F}_3,\mathbf{F}_2 \right]   \right\}. 
\end{align} 

\subsection{The quantum group U$_q\left( \mathfrak{psu}(2|2)_{\text{c.e.}}\right)$}
Now we are ready to introduce the $q$ deformation of U$\left( \mathfrak{psu}(2|2)_{\text{c.e.}}\right)$ known as the \emph{quantum group} $U_q\left( \mathfrak{psu}(2|2)\right)_{\text{c.e.}}$, which we will denote as $\mathfrak{psu}_q(2|2)_{\text{c.e.}}$ or simply $\mathfrak{h}$. The deformation will manifest itself as deformations of the right-hand sides of the commutation relations discussed in the previous paragraphs. For this we need to define the \emph{quantum number} $\left[ x \right]_q$ for any complex number $q$ as
\begin{equation}
\label{eq:quantumnumber}
\left[ x \right]_q \defeq \frac{q^x -q^{-x}}{q-q^{-1}},
\end{equation}
where $x \in U\left( \mathfrak{su}(2|2)\right)_{\text{c.e.}}$. There are two standard ways to understand this definition, which coincide when both make sense: either $q^x$ is defined by its power series (convergent or formal)
\begin{equation}
q^x = 1 + x \log q +\bO{\left(\log q\right)^2},
\end{equation}
or we can simply include the element $q^x$ in the Hopf algebra, such that $q^{-x}$ acts as its inverse. Importantly, in the limit $q\rightarrow 1$ we should find $\left[ x\right]_q \rightarrow x$. 

\paragraph{Deformed commutation relations.} Not all the commutation relations get deformed under the $q$ deformation, but we will list all of them for completeness: for $j,k=1,2,3$ we have the undeformed relations
\begin{equation}
\left[ \mathbf{H}_j, \mathbf{H}_k  \right] = 0, \quad 
\left[ \mathbf{H}_j, \mathbf{E}_k  \right] = A_{jk} \mathbf{E}_k, \quad
\left[ \mathbf{H}_j, \mathbf{F}_k  \right] = - A_{jk} \mathbf{F}_k,
\end{equation}
where the last two relations can be  rewritten in exponentials in the sense discussed above as
\begin{equation}
q^{\bH_j} \mathbf{E}_k = q^{A_{jk}}\mathbf{E}_k q^{\bH_j}, \quad 
q^{\bH_j} \mathbf{F}_k = q^{-A_{jk}}\mathbf{F}_k q^{\bH_j}. 
\end{equation}
The commutators between positive and negative roots get deformed into
\begin{equation}
\left[ \mathbf{E}_1, \mathbf{F}_1 \right] = \left[\mathbf{H}_1\right]_q, \quad
\left\{ \mathbf{E}_2, \mathbf{F}_2 \right\} = -\left[\mathbf{H}_2\right]_q, \quad
\left[ \mathbf{E}_3, \mathbf{F}_3 \right] = -\left[\mathbf{H}_3\right]_q,
\end{equation}
whereas all other commutators still vanish. 
\paragraph{Deformed Serre relations.} The Serre relations \eqref{eq:Serrerelations} get deformed as well, but we will not need their explicit form here. By dropping the Serre relations \eqref{eq:Serrerelations2} which do not get $q$ deformed we obtain the centrally extended quantum group $\mathfrak{psu}_q(2|2)_{\text{c.e.}}$ containing three central charges. 

\paragraph{Central charges.} The expression of the central charge $\mathbf{K}$ in terms of Chevalley generators does not get $q$ deformed, but the expressions for $\mathbf{C}$ and $\mathbf{D}$ do:
\begin{align}
\mathbf{C} &= \mathbf{E}_1 \mathbf{E}_2 \mathbf{E}_3 \mathbf{E}_2+
\mathbf{E}_2\mathbf{E}_3\mathbf{E}_2\mathbf{E}_1 +
\mathbf{E}_3 \mathbf{E}_2 \mathbf{E}_1 \mathbf{E}_2+
\mathbf{E}_2\mathbf{E}_1\mathbf{E}_2\mathbf{E}_3 -\left( q +q^{-1}\right) \mathbf{E}_2\mathbf{E}_1\mathbf{E}_3\mathbf{E}_2, \nn
\mathbf{D} &= \mathbf{F}_1 \mathbf{F}_2 \mathbf{F}_3 \mathbf{F}_2+
\mathbf{F}_2\mathbf{F}_3\mathbf{F}_2\mathbf{F}_1 +
\mathbf{F}_3 \mathbf{F}_2 \mathbf{F}_1 \mathbf{F}_2+
\mathbf{F}_2\mathbf{F}_1\mathbf{F}_2\mathbf{F}_3 -\left( q +q^{-1}\right) \mathbf{F}_2\mathbf{F}_1\mathbf{F}_3\mathbf{F}_2. 
\end{align} 

\subsection{$\mathfrak{psu}_q(2|2)_{\text{c.e.}}$ as a Hopf algebra}
As anticipated by our discussion of Hopf algebras, we can now equip $\mathfrak{h} = \mathfrak{psu}_q(2|2)_{\text{c.e.}}$ with a Hopf algebra structure. As in the undeformed case the multiplication $\mu$ and unit $\zeta$ follow from the present tensor structure. We therefore only need to introduce the counit and the $q$-deformed coproduct and antipode. 
\paragraph{Braided coproduct.} Following our discussion in section \ref{sec:Hopfalgebra} we will straightforwardly introduce a non-standard coproduct, based on a new abelian generator $\mathbf{U}$ that associates non-trivial charges $\{2,1,-1,-2\}$ to the generators $\{\mathbf{C},\mathbf{E}_2,\mathbf{F}_2,\mathbf{D}\}$ respectively, whereas other generators remain uncharged. This leads to the following relations for the coproduct
\begin{equation}
\begin{aligned}
\Delta(1) & = 1\otimes 1, \\
\Delta\left(\bH_j\right)  &= \bH_j \otimes 1 + 1 \otimes \bH_j, \\
\Delta\left(\mathbf{E}_{k'}\right)  &= \mathbf{E}_{k'} \otimes 1 + q^{-\bH_{k'}} \otimes \mathbf{E}_{k'}, \\
\Delta\left(\mathbf{F}_{k'}\right)  &= \mathbf{F}_{k'} \otimes q^{\bH_{k'}} + 1 \otimes \mathbf{F}_{k'}, \\
\Delta\left( \mathbf{K} \right) &= \mathbf{K} \otimes 1 + 1 \otimes \mathbf{K},
\end{aligned} \quad
\begin{aligned}
\Delta\left(\mathbf{E}_{2}\right)  &= \mathbf{E}_{2} \otimes 1 + q^{-\bH_{2}} \mathbf{U} \otimes \mathbf{E}_{2}, \\
\Delta\left(\mathbf{F}_{2}\right)  &= \mathbf{F}_{2} \otimes q^{\bH_{2}} + \mathbf{U}^{-1} \otimes \mathbf{F}_{2}, \\
\Delta\left(\mathbf{C}\right)  &= \mathbf{C} \otimes 1 + q^{2\mathbf{K}} \mathbf{U}^2 \otimes \mathbf{C}, \\
\Delta\left(\mathbf{D}\right)  &= \mathbf{D} \otimes q^{-2\mathbf{K}} +  \mathbf{U}^{-2} \otimes \mathbf{D}, \\
\Delta\left( \mathbf{U} \right)& = \mathbf{U} \otimes \mathbf{U}. 
\end{aligned}
\end{equation}
where $j = 1,2,3$ and $k'= 1,3$ and the unbraided relations follow by setting $\mathbf{U}=1$.

\paragraph{Counit.} The counit $\varepsilon: \mathfrak{h} \rightarrow \C$ is defined as
\begin{equation}
\varepsilon(1) = 1,\quad \varepsilon\left( \mathbf{U} \right) = 1, \quad \varepsilon\left(\mathbf{M}\right) = 0 
\quad \text{ for all } \mathbf{M} \in \{\bH_j,\mathbf{E}_j,\mathbf{F}_j\}_{j=1,2,3}. 
\end{equation}
\paragraph{Antipode.} The antipode $\mathcal{S}: \mathfrak{h} \rightarrow \mathfrak{h}$ is uniquely fixed by the preceding definitions and the antipode consistency condition (see fig. \ref{fig:hopfalgebra}) and reads\footnote{We have corrected what we believe is a typo in the $\mathbf{F}_2$ relation, compare with (2.46) in \cite{Beisert:2008tw}.}
\begin{equation}
\begin{aligned}
\mathcal{S}\left( 1 \right) &= 1, \\
\mathcal{S}\left( \mathbf{H}_j \right) &= -\mathbf{H}_j, \\
\mathcal{S}\left( \mathbf{E}_{k'} \right) &= -q^{\mathbf{H}_{k'}}\mathbf{E}_{k'}, \\
\mathcal{S}\left( \mathbf{F}_{k'} \right) &= -\mathbf{F}_{k'} q^{-\mathbf{H}_{k'}}, \\
\mathcal{S}\left( \mathbf{K} \right) &= -\mathbf{K}, \\
\end{aligned} \quad 
\begin{aligned}
\mathcal{S}\left( \mathbf{E}_2 \right) &= -q^{\mathbf{H}_2}\mathbf{U}^{-1} \mathbf{E}_2, \\
\mathcal{S}\left( \mathbf{F}_2 \right) &= -\mathbf{U} \mathbf{F}_2 q^{-\mathbf{H}_2}, \\
\mathcal{S}\left( \mathbf{C} \right) &= - q^{-2\mathbf{K}}\mathbf{U}^{-2} \mathbf{C}, \\
\mathcal{S}\left( \mathbf{D} \right) &= - q^{2\mathbf{K}}\mathbf{U}^{2} \mathbf{D}, \\
\mathcal{S}\left( \mathbf{U} \right) &= \mathbf{U}^{-1}.
\end{aligned}
\end{equation}
\paragraph{Reality.} To restrict to a real algebra, we have to identify what the conjugate expression is for our generators. Hermiticity is consistent with the identification
\begin{equation}
\mathbf{H}_j^{\dagger} = \mathbf{H}_j, \quad \mathbf{E}_j^{\dagger} = q^{-\mathbf{H}_j} \mathbf{F}_j,
\end{equation}
implying that the central charges are related as
\begin{equation}
\mathbf{K}^{\dagger} = \mathbf{K}, \quad \mathbf{C}^{\dagger} = q^{2\mathbf{K}} \mathbf{D}. 
\end{equation}
The braiding element $\mathbf{U}$ satisfies $\mathbf{U}^{\dagger} = \mathbf{U}^{-1}$. Importantly, hermiticity also requires us to set $q\in \R$, which we will do for the rest of this thesis by parametrising 
\begin{equation}
q = e^{-\ad}
\end{equation}
for a real number $\ad \in \R$. To make some formulae look a bit simpler we will continue to write $q$ whenever this is convenient. 
\\[5mm]
In the following we will not discuss the existence of the universal $R$-matrix, whose existence is still under investigation \cite{Beisert:2016qei,Beisert:2017xqx}. We will focus on the $R$ matrix we need and functions as an intertwiner for the fundamental short representations.

\subsection{Fundamental representation} The representation theory of $\mathfrak{h}$ for real $q$ is structurally very similar to the representation theory of $\mathfrak{su}(2|2)$ \cite{Ky:1994cr}: in the following we will only need the structure of short multiplets, which are characterised by the central-charge eigenvalues satisfying the \emph{shortening condition}
\begin{equation}
\label{eq:shorteningcondition}
\left[K \right]_q^2 - C D = \left[ \frac{1}{2} (m+n+1) \right]_q^2,
\end{equation}
where $m,n$ are Dynkin labels corresponding to $\mathfrak{su}(2)$ representations. These labels descend from the existence of an automorphism on $\mathfrak{h}$ that keeps the combination of central-charge eigenvalues on the left-hand side constant while rotating any non-trival triplet $(K,C,D)$ to $(K_0,0,0)$. The algebra with eigenvalues $(K_0,0,0)$ is isomorphic to non-extended $U_q\left( \mathfrak{su}(2|2)\right)$, whose representation theory is analogous to that of $\mathfrak{su}(2|2)$. Such representations are characterised by the two Dynkin labels $m,n$ corresponding to its $\mathfrak{su}(2)\times \mathfrak{su}(2)$ subalgebra. Therefore we can parametrise any representation by the Dynkin labels $m,n$ along with the eigenvalues $(K,C,D)$ of the three central charges $(\mathbf{K},\mathbf{C},\mathbf{D})$. 
\\[5mm]
The fundamental representation is a short multiplet and can be regarded as a $q$ deformation of the fundamental $\mathfrak{su}(2|2)$ representation with Dynkin labels $m=n=0$. We can realise the fundamental representation on a four-dimensional super vector space with two bosonic and two fermionic basis vectors $\{ \ket{\phi^1},\ket{\phi^2}  \}$ and $\{ \ket{\psi^1},\ket{\psi^2}  \}$ respectively. The most general action of the Chevalley generators on this basis is specified by 
\begin{equation}
\begin{aligned}
\mathbf{H}_1 \ket{\phi^1} &= -\ket{\phi^1}, \\
\mathbf{H}_1 \ket{\phi^2} &= +\ket{\phi^2}, \\
\mathbf{H}_3 \ket{\psi^1} &= -\ket{\psi^1}, \\
\mathbf{H}_3 \ket{\psi^2} &= +\ket{\psi^2}, \\
\end{aligned}\quad
\begin{aligned}
\mathbf{H}_2 \ket{\phi^1} &= -\left(K-\tfrac{1}{2}\right) \ket{\phi^1}, \\
\mathbf{H}_2 \ket{\phi^2} &= -\left(K+\tfrac{1}{2}\right)\ket{\phi^2}, \\
\mathbf{H}_2 \ket{\psi^1} &= -\left(K-\tfrac{1}{2}\right)\ket{\psi^1}, \\
\mathbf{H}_2 \ket{\psi^2} &= -\left(K+\tfrac{1}{2}\right)\ket{\psi^2}, \\
\end{aligned}\quad
\begin{aligned}
\mathbf{E}_1 \ket{\phi^1} &= q^{1/2}\ket{\phi^2}, \\
\mathbf{E}_1 \ket{\phi^2} &= a \ket{\psi^2}, \\
\mathbf{E}_3 \ket{\psi^1} &= b\ket{\phi^1}, \\
\mathbf{E}_3 \ket{\psi^2} &= q^{-1/2}\ket{\psi^2}, \\
\end{aligned}\quad
\begin{aligned}
\mathbf{F}_1 \ket{\phi^1} &= c\ket{\psi^1}, \\
\mathbf{F}_1 \ket{\phi^2} &= q^{-1/2} \ket{\phi^1}, \\
\mathbf{F}_3 \ket{\psi^1} &= q^{1/2}\ket{\psi^2}, \\
\mathbf{F}_3 \ket{\psi^2} &= d\ket{\phi^2}, \\
\end{aligned}
\end{equation}
where $a,b,c,d$ are parameters. The other actions follow straightforwardly from others by using the relations between generators. All the basis states are eigenvectors of  $\mathbf{U}$ with eigenvalue $U$. Not all these parameters are independent: the closure of the algebra requires that these parameters are related to the central charges as
\begin{equation}
\label{eq:abcdconstraints}
ad = \left[ K+\tfrac{1}{2}\right]_q, \quad 
bc = \left[ K-\tfrac{1}{2}\right]_q, \quad
ab = C, \quad cd = D.
\end{equation}
Combining the first two relations we find a relation for the four coefficients independent of the central charges
\begin{equation}
\label{eq:abcdquadraticconstraint}
(ad - q bc) ( ad - q^{-1} bc) = 1. 
\end{equation}

\paragraph{Requiring cocommutativity.} Another constraint on the representation ultimately descends from the fact that we are interested in Hopf algebras that allow for a non-trivial $R$ matrix. As we have seen in section \ref{sec:Hopfalgebra}, this at least requires us to consider only cocommutative Hopf algebras for which there exists an $R$ matrix $\R$ that satisfies eqn. \eqref{eq:cocommutativity}. This can only be if for elements $\mathbf{M}$ in the center of $\mathfrak{h}$ we find $\Delta^{\text{op}}\left(\mathbf{M}\right) = \Delta\left(\mathbf{M}\right)$. This is obviously true for the central element $\mathbf{K}$, but not for $\mathbf{C}$ and $\mathbf{D}$. To ensure cocommutativity, these must be related to $\mathbf{V}\defeq q^{\mathbf{K}}$ and the braiding element $\mathbf{U}$ as 
\begin{equation}
\mathbf{C} = \tfrac{h}{2} \alpha \left( 1- \mathbf{V}^2 \mathbf{U}^2\right), \quad 
\mathbf{D} = \tfrac{h}{2} \alpha^{-1} \left( \mathbf{V}^{-2}-  \mathbf{U}^{-2}\right),
\end{equation}
where $h,\alpha$ are free parameters. The parameter $\alpha$ adjusts the normalisation of $\mathbf{E}_2$ compared to $\mathbf{F}_2$ and is thus ultimately a basis choice. We set $\alpha = 1$ as its value is irrelevant. The parameter $h$ on the other hand will play the role of coupling constant in the quantised theory, as we will see shortly. Writing $V = q^K$ for the eigenvalue of $\mathbf{V}$ and $\mathbf{K}$ we find that $U$ and $V$ are related by the shortening condition as
\begin{equation}
\label{eq:UVshortening}
\left(\frac{V-V^{-1}}{q-q^{-1}}\right)^2-\frac{h^2}{4} \left(1-U^2V^2\right)\left(V^{-2}-U^{-2}\right) = 1. 
\end{equation}

\subsection{$x^{\pm}$ parameters.} As in the undeformed case, we can introduce a convenient parametrisation of the four parameters $a,b,c,d$ in new parameters $x^+,x^-,\gamma$ that take the constraints \eqref{eq:abcdconstraints} and \eqref{eq:abcdquadraticconstraint} into account:
\begin{equation}
\begin{aligned}
a &= \sqrt{\frac{h}{2}}\gamma, \\
c &= \ii \sqrt{\frac{h}{2}}\gamma q^{1/2}\frac{V^{-1}}{x^+} ,
\end{aligned}\quad
\begin{aligned}
b &=\sqrt{\frac{h}{2}}  \frac{1}{ \gamma x^-}\left( x^- - q^{-1} V^2 x^+\right), \\
d &=\sqrt{\frac{h}{2}} \frac{\ii}{\gamma} q^{1/2} V\left( x^- - q^{-1} V^{-2} x^+\right).
\end{aligned}
\end{equation}
The parameter $\gamma$ does not play a role in satisfying the constraints and only adjusts the normalisation of the bosonic basis vector relative to the fermionic ones.\footnote{A more detailed discussion of the various normalisations can be found in \cite{Beisert:2011wq}. It also discusses an alternative approach to the analysis of the quantum deformed representation theory using an affinisation based on doubling the fermionic generators.} Its value will not be important in the remainder, but let us mention that to follow the undeformed case it is convenient to set 
\begin{equation}
\gamma = \sqrt{-\ii q^{1/2} VU\left(x^+ - x^-  \right)}.
\end{equation}
The constraints can now be written in terms of the $x^{\pm}$ and $q$ and $h$ only, reading
\begin{equation}
\frac{1}{q} \left( x^+ +\frac{1}{x^+} + \xi + \frac{1}{\xi}\right) = q \left( x^- +\frac{1}{x^-} + \xi + \frac{1}{\xi}\right),
\end{equation}
where we introduce 
\begin{equation}
\label{eq:xirepresentation}
\xi = -\frac{\ii}{2} \frac{ h \left(q-q^{-1}\right)}{\sqrt{1-\tfrac{h^2}{4}\left(q-q^{-1}\right)^2}}.
\end{equation}
In terms of these quantities we can also rewrite the central-charge eigenvalues $V,U$ as
\begin{equation}
\label{eq:UVinx}
V^2 = q \frac{x^+}{x^-}\frac{x^-+\xi}{x^++\xi}, \quad U^2 = \frac{1}{q} \frac{x^++\xi}{x^-+\xi}.
\end{equation}
As in the undeformed case it turns out that all important quantities can be written in terms of $q,h$ and the $x^{\pm}$ parameters. A final reparametrisation of the algebra parameters consists in introducting $\theta$ as 
\begin{equation}
\xi = \ii \tan \frac{\theta}{2},
\end{equation}
which will turn out to be very natural, as we will see shortly. In particular, we will restrict to $\theta \in (-\pi,\pi]$ which will ensure unitarity of the $S$ matrix later.

\begin{figure}[!t]
\centering
\begin{subfigure}{6cm}
\includegraphics[width=6cm]{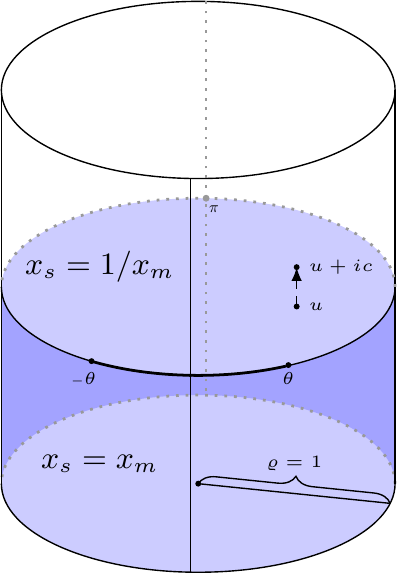}
\subcaption{$x_s$}
\end{subfigure} \quad 
\begin{subfigure}{7cm}
\includegraphics[width=7cm]{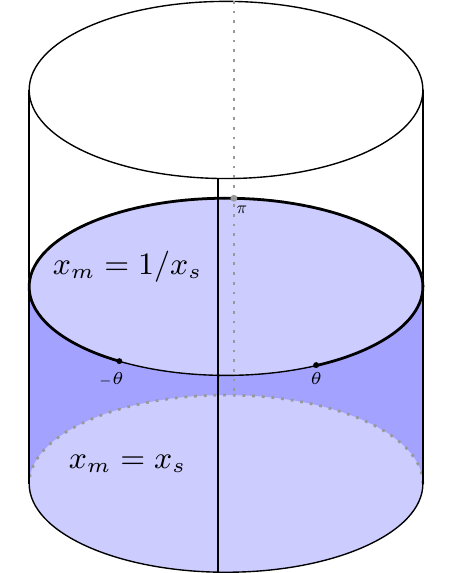}
\subcaption{$x_m$}
\end{subfigure}

\caption{The analytic structure of the deformed $x$-functions: the branch cuts are indicated by the thick line, the domain on which $x_m = x_s$ is indicated in blue. In (a) we have indicated the shift distance $\ad$ and the cylinder radius $\rho$.}
\label{fig:xfunctions}
\end{figure}

\paragraph{Solving the shortening condition.} \label{sec:xfunctions}For a fixed coupling constant $h$ and deformation parameter $q$ a fundamental short representation corresponds to a point on a torus, which uniformises the shortening condition \eqref{eq:shorteningcondition}:
this torus has real period $2\omega_1 ( \kappa )  = 4$K$(m)$ and imaginary period $2\omega_2 ( \kappa )  = 4 \ii $K$(1-m)-4$K$(m)$, where K$(m)$ is the elliptic integral of the first kind depending on the elliptic modulus $m=\kappa^2$. The parameters $(\kappa,z_0)$ where $z_0 \in \C$ parametrise the pair $(q,h)$ through the following equations:
\begin{align}
q = e^{\ii \text{am} (2z_0)} = \frac{\text{cs}(z_0) + \ii \text{dn}(z_0)}{\text{cs}(z_0) - \ii \text{dn}(z_0)}, \qquad
h = - \frac{-\ii \kappa}{2\text{dn}(z_0)} \sqrt{ 1- \kappa^2 \text{sn}(z_0)^2}, 
\end{align}
where $\text{am},\text{cs}$ and $\text{dn}$ are all Jacobi elliptic functions. In principle we could continue in this parametrisation treating all values of $q$ simultaneously a little longer, but this is not the most convenient perspective for the restriction to real $q$, our primary case of interest.\footnote{A convenient description for the root-of-unity case can be found in \cite{Arutyunov:2012zt}.} By introducing a spectral parameter $u$ it is possible to write an arbitrary solution to the shortening condition as $x^{\pm} = x(u\pm \ii \ad)$ for a function $x$ that satisfies  
\begin{equation}
\label{eq:deformedzhukovsky}
e^{\ii u} =  \frac{x(u) + \frac{1}{x(u)} + \xi + \frac{1}{\xi}}{\frac{1}{\xi} - \xi},
\end{equation}
where we should restrict the complex parameter $u$ to a cylinder, i.e. Re$(u) \in (-\pi,\pi]$. To describe the general solution we only need to consider the two special solutions $x_s$ -- called the ``string" $x$-function --
\begin{equation}
\label{eq:xsfunction}
x_s(u) = -\ii \csc \theta \left( e^{\ii u } -\cos \theta -\left( 1-e^{iu}\right) \sqrt{\frac{\cos u - \cos \theta}{\cos u - 1}}\right),
\end{equation}
and $x_m$ -- the ``mirror" $x$-function\footnote{These names will become clear in due course.} --
\begin{equation}
\label{eq:xmfunction}
x_m(u) = -\ii \csc \theta \left( e^{\ii u } -\cos \theta +\left( 1+e^{iu}\right) \sqrt{\frac{\cos u - \cos \theta}{\cos u + 1}}\right). 
\end{equation}
These $2\pi$-periodic functions have branch points at $\pm \theta$. The $x_s$ function has what we will call a \emph{short} cut running on the real line through the origin, whereas $x_m$ has a \emph{long} cut running on the real line through $\pi$. $x_s$ and $x_m$ coincide on the lower half-plane and are each others analytic continuation through their cuts, such that they are inverse on the upper half-plane. This is illustrated in fig. \ref{fig:xfunctions}.
These functions are going to play a central role in this thesis.

\paragraph{Undeformed limit.}\label{par:undlimit} We will often consider the relation between our $\eta$-deformed expressions and the corresponding quantity in the undeformed $\ads$ superstring. The \emph{undeformed limit} that takes us from $\eta$ deformed to undeformed consists of a rescaling and a limit and its exact form depends on which conventions one wishes to follow: if we rescale $u\rightarrow \ad g u$ and then send $\ad \rightarrow 0$ we obtain the $x$ functions from \cite{Arutyunov:2009zu}:
\begin{align}
\label{eq:undeformedx}
x^{\text{und}}_s(u) = \frac{u}{2} \left( 1 + \sqrt{1-\frac{4}{u^2}}\right), \qquad 
x^{\text{und}}_m(u) = \frac{1}{2} \left( u - \ii \sqrt{4-u^2}\right),
\end{align}
which have cuts on $(-2,2)$ and on $(-\infty,-2)\cup (2,\infty)$ respectively. In the conventions of \cite{Gromov:2009bc, Gromov:2014caa} the undeformed limit has to be defined as a rescaling $u \rightarrow 2 \ad u$ and the limit $\ad \rightarrow 0$, which differ from the convention in eqn. \eqref{eq:undeformedx} by agreeing in the upper half-plane instead of on the lower half-plane. It is an unfortunate fact about the literature on this subject that both these conventions are so prolific: to stay in line with the literature we will start using the conventions from \cite{Arutyunov:2009zu,Bombardelli:2009ns} until chapter \ref{chap:QSC}, where we will switch to the 
conventions from \cite{Gromov:2009bc, Gromov:2014caa}. In chapter \ref{chap:solutionstate} we use the first set of conventions once more. The undeformed $x$ functions (also known as \emph{Zhukovsky variables}) \eqref{eq:undeformedx} obtained in the undeformed limit obey the \emph{Zhukovksy identity} 
\begin{equation}
x(u)+\frac{1}{x(u)} = u,
\end{equation}
which is the shortening condition for the undeformed case. In appendix \ref{app:xfunctions} we will collect all the relevant properties of these $x$ functions. 

\subsection{On the $q$-deformed $S$-matrix for $\mathfrak{psu}_q(2|2)$} 
We are now in a good position to understand the construction of the $S$ matrix: we exhibited the Hopf algebra structure present in $\mathfrak{h}$ and restricted the central charges as to no longer obstruct cocommutativity. Moreover we have found an efficient description of the shortening condition in terms of $x^{\pm}$ and $h$. In order to find the $S$ matrix 
\begin{equation}
\mathbb{S} : \mathfrak{h} \times \mathfrak{h} \rightarrow \mathfrak{h} \times \mathfrak{h}
\end{equation}
we consider the cocommutativity condition \eqref{eq:cocommutativity} on the whole algebra to find the $R$ matrix, which differs from the $S$ matrix by a graded permutation. Since their difference is so small, we will directly talk about the result for the $S$ matrix, since this is what we are ultimately after. The $\mathfrak{su}(2) \times \mathfrak{su}(2)$ subalgebra restricts the most general ansatz to ten unrestricted functions $a_k(p_1,p_2)$, where the $p_i$ are chosen to correspond to $U$ as $U^2 = e^{i p}$ and will be interpreted as quasi-momenta: consider elementary matrices $\mathbf{E}_{ij}$ in the basis $\{\phi^1,\phi^2,\psi^1,\psi^2 \}$ with as only nonzero entry the $(i,j)$th one, then we can define the matrices
\begin{equation}
\mathbb{E}_{kilj} = (-1)^{\e(l) \e(k)} \mathbf{E}_{ki} \otimes \mathbf{E}_{lj}
\end{equation}
acting on the (irreducible) tensor product of fundamental short representations. The most general ansatz now is 
\begin{equation}
\label{eq:psuSmatrix}
\mathbb{S}_{12}(p_1,p_2)=\sum_{k=1}^{10}a_k(p_1,p_2)\mathbf{\Lambda}_k\, ,
\end{equation}
where the ten $\mathfrak{su}(2) \times \mathfrak{su}(2)$-invariants are given by\note{Did the page get separated properly?}
\begin{equation}
\begin{aligned}
\mathbf{\Lambda}_1=&\mathbb{E}_{1111}+\frac{q}{2}\mathbb{E}_{1122}+\frac{1}{2}(2-q^2)\mathbb{E}_{1221}+\frac{1}{2}\mathbb{E}_{2112}+\frac{q}{2}\mathbb{E}_{2211}+\mathbb{E}_{2222}\, ,\nonumber\\
\mathbf{\Lambda}_2=&\frac{1}{2}\mathbb{E}_{1122}-\frac{q}{2}\mathbb{E}_{1221}-\frac{1}{2q}\mathbb{E}_{2112}+\frac{1}{2}\mathbb{E}_{2211}\, , \nonumber \\
\mathbf{\Lambda}_3=&\mathbb{E}_{3333}+\frac{q}{2}\mathbb{E}_{3344}+\frac{1}{2}(2-q^2)\mathbb{E}_{3443}+\frac{1}{2}\mathbb{E}_{4334}+\frac{q}{2}\mathbb{E}_{4433}+\mathbb{E}_{4444} \, , \nonumber\\
\mathbf{\Lambda}_4=&\frac{1}{2}\mathbb{E}_{3344}-\frac{q}{2}\mathbb{E}_{3443}-\frac{1}{2q}\mathbb{E}_{4334}+\frac{1}{2}\mathbb{E}_{4433}\, , \nonumber\\
\mathbf{\Lambda}_5=&\mathbb{E}_{1133}+\mathbb{E}_{1144}+\mathbb{E}_{2233}+\mathbb{E}_{2244}\, ,\\
\mathbf{\Lambda}_6=&\mathbb{E}_{3311}+\mathbb{E}_{3322}+\mathbb{E}_{4411}+\mathbb{E}_{4422}\, , \nonumber\\
\mathbf{\Lambda}_7=&\mathbb{E}_{1324}-q\mathbb{E}_{1423}-\frac{1}{q}\mathbb{E}_{2314}+\mathbb{E}_{2413}\, , \nonumber\\
\mathbf{\Lambda}_8=&\mathbb{E}_{3142}-q\mathbb{E}_{3214}-\frac{1}{q}\mathbb{E}_{4132}+\mathbb{E}_{4231}\, , \nonumber\\
\mathbf{\Lambda}_9=&\mathbb{E}_{1331}+\mathbb{E}_{1441}+\mathbb{E}_{2332}+\mathbb{E}_{2442}\, , \nonumber\\
\mathbf{\Lambda}_{10}=&\mathbb{E}_{3113}+\mathbb{E}_{3223}+\mathbb{E}_{4114}+\mathbb{E}_{4224}.
\end{aligned}
\end{equation}
Up to an overall factor the unknown functions $a_j$ are completely fixed by the cocommutativity requirement for the generators $\mathbf{E}_2$ and $\mathbf{F}_2$: we set $a_1 =1$ yielding
\begin{equation}
\begin{aligned}
a_2=&-q+\frac{2}{q}\frac{x^-_1(1-x^-_2x^+_1)(x^+_1-x^+_2)}{x^+_1(1-x^-_1x^-_2)(x^-_1-x^+_2)},\\
a_3=&\frac{U_2V_2}{U_1V_1}\frac{x^+_1-x^-_2}{x^-_1-x^+_2}, \\
a_4=&-q\frac{U_2V_2}{U_1V_1}\frac{x^+_1-x^-_2}{x^-_1-x^+_2}+\frac{2}{q}\frac{U_2V_2}{U_1V_1}\frac{x^-_2(x^+_1-x^+_2)(1-x^-_1x^+_2)}{x^+_2(x^-_1-x^+_2)(1-x^-_1x^-_2)}, \\
a_5=&\frac{x^+_1-x^+_2}{\sqrt{q}\, U_1V_1(x^-_1-x^+_2)},\\
a_6=&\frac{\sqrt{q}\, U_2V_2(x^-_1-x^-_2)}{x^-_1-x^+_2},\\
a_7=&\ii\frac{(x^+_1-x^-_1)(x^+_1-x^+_2)(x^+_2-x^-_2)}{\sqrt{q}\, U_1V_1 (1-x^-_1 x^-_2) (x^-_1-x^+_2)\gamma_1\gamma_2},
\\
a_8=&\ii\frac{U_2V_2\,  x^-_1x^-_2(x^+_1-x^+_2)\gamma_1\gamma_2}{q^{\frac{3}{2}} x^+_1x^+_2(x^-_1-x^+_2)(x^-_1x^-_2-1)},\\
a_9=&\frac{(x^-_1-x^+_1)\gamma_2}{(x^-_1-x^+_2)\gamma_1}, \\
a_{10}=&\frac{U_2V_2 (x^-_2-x^+_2)\gamma_1}{U_1V_1(x^-_1-x^+_2)\gamma_2},
\end{aligned}
\end{equation}
where the subscripts indicate to which subspace the variables belong. As indicated before the parameters $\gamma_i$ play a gauge-like role and do not influence the spectrum of the QFT we are describing by this $S$ matrix. We will therefore set them to one in the remainder. 

\paragraph{Quasitriangularity.} The $S$ matrix in eqn. \eqref{eq:psuSmatrix} functions as the intertwiner of the coproduct and the opposite coproduct for the fundamental representation of $\mathfrak{psu}_q\left(2|2\right)$. It also satisfies the Yang-Baxter equation. These are all necessary elements, but one could hope for more: indeed, in our initial discussion of Hopf algebras the $R$ matrix satisfies these last two conditions only because it satisfies the quasitriangularity condition \eqref{eq:quasitriangular}. This turns out to be a very strong condition: the resulting universal $R$-matrix is representation independent and for any representation  immediately gives the explicit form of the $S$. Since we want to consider the scattering of all possible particles and bound states this seems like a necessity to solve the spectral problem, but we will see that we are lucky: the particular bound-state representations we need can be obtained from the fundamental representation as well as the corresponding $S$ matrix, thereby circumventing the need for a universal $R$-matrix. The question of whether such an $R$ matrix exists is still very interesting from the mathematical point of view and has been studied \cite{Beisert:2016qei,Beisert:2017xqx}, without a definite answer yet.

\paragraph{Yang Baxter equation and crossing symmetry.} As we discussed in section \ref{sec:Hopfalgebra} for the $R$ matrix to give rise to an $S$ matrix that defines a consistent scattering theory it should satisfy both the Yang-Baxter equation and the crossing equation. \note{We might need to expand on this. }By checking every element it was shown that the $S$ matrix indeed satisfies the Yang-Baxter equation as long as the $x^{\pm}$ satisfy the shortening condition \eqref{eq:shorteningcondition}. To ensure that the $R$ matrix has crossing symmetry, i.e. solves the crossing equation, we need to put a restriction on its overall normalisation $1/\sigma$ known as the \emph{dressing phase}: for any $z_1,z_2\in \C$
\begin{equation}
\label{eq:deformedcrossing}
\sigma(z_1,z_2)\sigma(z_1,z_2-\omega_2) = \frac{1}{q} \frac{x_1^- + \xi}{x_1^+ + \xi} \frac{x_1^--x_2^+}{x_1^--x_2^-}\frac{1-\frac{1}{x_1^+x_2^+}}{1-\frac{1}{x_1^+x_2^-}},
\end{equation}
where $x_i = x(z_i)$ and $\omega_2$ is the imaginary period of the uniformisation torus of the shortening condition. Its solutions were studied in \cite{Hoare:2011wr} for the case when $q$ lies on the unit circle. The real-$q$ case was solved in \cite{Arutynov:2014ota}: writing 
\begin{equation}
\label{eq:dressingphase}
\sigma(z_1, z_2) = \exp \ii \left( \chi\left( x_1^+, x_2^+\right) - \chi\left( x_1^-, x_2^+\right)-\chi\left( x_1^+, x_2^-\right) + \chi\left( x_1^-, x_2^-\right)\right),
\end{equation}
we can specify $\sigma$ by the following double integral representation of the function $\chi$:
\begin{equation}
\label{eq:Phi}
\chi(x_1,x_2) = \Phi(x_1, x_2) = \ii \oint_{\mathcal{C}} \frac{dz}{2\pi \ii} \frac{1}{z-x_1} \oint_{\mathcal{C}} \frac{dw}{2\pi \ii} \frac{1}{w-x_2} \log \frac{\Gamma_{q^2}\left( 1+ \tfrac{\ii}{2\ad} \left( u(z) - u(w)\right)\right)}{
\Gamma_{q^2}\left( 1- \tfrac{\ii}{2\ad} \left( u(z) - u(w)\right)\right)},
\end{equation}
where $\Gamma_q$ is the $q$-gamma function and the contour $\mathcal{C}$ is specified by
\begin{equation}
\label{eq:contourC}
\mathcal{C} : |y|^2 -1 + (y^* - y) \xi = 0. 
\end{equation}
This is just a shift and rescaling of the unit circle. The notation $u$ simply represents the inverse of $x$, i.e. $u(z) = x^{-1}(z)$. We consider other representations of the dressing phase in appendices \ref{app:Definitions} and \ref{app:dressing}. This solution is not the unique solution of the deformed crossing equation \eqref{eq:deformedcrossing}, but at present it is the only explicitly-known solution. It was constructed using knowledge of the dressing phase for the undeformed case, where the ``physical" solution was found by comparing with other data, coming from both the string (strong coupling) and the gauge theory (weak coupling) side. Indeed, taking the undeformed limit after rescaling the rapidity one finds the undeformed $S$-matrix \cite{Beisert:2005tm,Janik:2006dc,Beisert:2006ez}. 

\subsection{The $\eta$-deformed $S$ matrix}
As in the undeformed case we can use the $S$ matrix for $\mathfrak{h}$ to construct the $S$ matrix that is invariant under two copies of that algebra. This is the basis for the $q$ deformation of the light-cone gauge-fixed $\ads$ world-sheet theory.\footnote{An extensive introduction of this model can be found in \cite{Arutynov:2014ota}.} The full $S$ matrix is defined as
\begin{equation}
\label{eq:etaSmatrix}
S^{\text{full}}(p_1,p_2) = S_{\mathfrak{su}(2)} \mathbb{S}\, \check{\otimes}\, \mathbb{S},
\end{equation}
where $\mathbb{S}$ is the $\mathfrak{h}$-invariant $S$-matrix that we just introduced in eqn. \eqref{eq:psuSmatrix} and 
\begin{equation}
\label{eq:Ssu2def}
S_{\mathfrak{su}(2)}(p_1,p_2)=\frac{1}{\sigma^2(p_1,p_2)}\frac{x_1^+ + \xi}{x_1^- + \xi}\frac{x_2^- + \xi}{x_2^+ + \xi}\cdot
\frac{x_1^- - x_2^+}{x_1^+ - x_2^-}\frac{1-\frac{1}{x_1^-x_2^+}}{1-\frac{1}{x_1^+x_2^-}},
\end{equation}
is the overall normalisation of the $S$ matrix, where the $\eta$-deformed dressing phase was defined in eqn. \eqref{eq:dressingphase}. The graded tensor product $\check{\otimes}$ absorbs various minus signs, as follows: written out in matrix components eqn. \eqref{eq:etaSmatrix} reads
\begin{equation}
S_{M\dot{M},N\dot{N}}^{P\dot{P} Q \dot{Q}}(p_1,p_2) = S_{\mathfrak{su}(2)}(-1)^{\e\left(\dot{M}\right)\e\left(N\right)+ \e\left(P\right) \e\left(\dot{Q}\right)} \mathbb{S}_{MN}^{PQ}(p_1,p_2) \dot{\mathbb{S}}_{\dot{M}\dot{N}}^{\dot{P}\dot{Q}},
\end{equation}
where the dots indicate that the corresponding object belongs to the second space in the tensor product and the indices represent bosonic ($1,2$) or fermionic ($3,4$) basis elements.

\subsection{Matching with the classical theory}
In the previous section we have reviewed the bootstrapping of the $\eta$-deformed $S$ matrix, which defines the (factorised) scattering theory of some integrable QFT with $\mathfrak{psu}(2|2)^{\otimes 2}$ off-shell symmetry. Indeed, a priori this is the only thing we can be certain about and relating it to the classical $\eta$-deformed theory is based on our intuition about the scarcity of integrable field theories and the uniqueness of the $q$ deformed $S$ matrix as it follows from its construction: a slightly different Hopf algebra could lead to a different $S$ matrix and thus to a different QFT and the crossing equation generically has more than one physically allowed solution leading to different $S$ matrices. Also, it is not clear whether the entire classical symmetry group of the $\eta$-deformed theory should survive quantisation. However, there are some checks one can perform to see whether the $S$ matrix could correspond to the quantised $\eta$-deformed model. In the undeformed case the AdS/CFT correspondence allowed for a plethora of tests, see the review \cite{Arutyunov:2009ga} and the paper \cite{Arutyunov:2007tc} and references therein. In the deformed case, however, the absence of a dual gauge theory and the immense complexity of the classical deformed string theory makes any test very difficult to execute. Nevertheless some tests have been done, showing clear agreement between the classical and quantised theory. 
\\[5mm]
In \cite{Arutyunov:2013ega} the tremendous task of perturbatively quantising the $\eta$-deformed theory was tackled starting from the Polyakov action \eqref{eq:etadeformedbosonicPolyakov}. Following the methods explained in section \ref{subsec:perturbativeSmatrix} the tree-level bosonic $S$-matrix was constructed and successfully compared to the exact $\eta$-deformed $S$-matrix by expanding the latter in the large-tension limit, showing agreement of the two if $q$ is related to $\eta$ and the string tension $g$ as 
\begin{equation}
\label{eq:semiclassicalq}
q= e^{-\frac{2\eta}{g\left(1+\eta^2\right)}}. 
\end{equation}
In \cite{Engelund:2014pla} the integrability of this perturbative $S$-matrix was further investigated by considering particle production and factorisation in the bosonic sector:  it was shown that, after a unitary basis transformation on the two-particle states, the $2\rightarrow 4$ and $4\rightarrow 6$ scattering processes are absent and that the $3 \rightarrow 3$ scattering factorises, i.e. satisfies the Yang-Baxter equation. This puts the integrability of the theory described by the perturbative $S$ -matrix on a firmer footing. 

\paragraph{Unitarity.} One subtle issue of the $S$ matrix is that of unitarity. We saw that the ZF algebra directly implies that $\mathbb{R}$ satisfies the braiding unitarity relation \eqref{eq:braidingunitarity}. However, for a physical system the property we really should impose is matrix unitarity of the $R$ matrix, i.e. $\mathbb{R}_{12}\mathbb{R}_{12}^{\dagger} = \mathbb{I}$ as a matrix, because it implies unitary time evolution of the theory. Braiding unitarity and matrix unitarity are equivalent when the $R$ matrix satisfies the axiom of \emph{hermitian} analyticity, i.e. $\mathbb{R}_{12}^{\dagger} = \mathbb{R}_{21}$, which is usually the case for relativistic field theories. It was observed that a necessary condition for these properties to exist is for $q$ to be real, since hermitian analyticity is lost for other $q$, in particular for the interesting case of $q$ on the unit circle \cite{Beisert:2008tw,Hoare:2011wr}. We refer to the discussion in section 7.2 of \cite{vanTongeren:2013gva} for more information. Also, reality of $q$ is not a sufficient condition for unitarity: the $S$ matrix is only unitary for 
\begin{equation}
\label{eq:unitaritybound}
0\leq h^2 \sinh^2 \ad \leq 1,
\end{equation}
which in terms of $\xi$ means that the $S$ matrix is unitary exactly when $\xi$ is imaginary.

\paragraph{Renormalisation of $q$ and $h$.} The identification of the deformation parameter $q$ with the deformation parameter $\eta$ and the string tension $g$ as in eqn. \eqref{eq:semiclassicalq} and the unitarity bound \eqref{eq:unitaritybound} together pose an interesting question, as satisfying both simultaneously is not straightforward: the exact quantum theory described by the $S$ matrix is parametrised by $q$ and the ``algebraic" coupling constant $h$, but the relation of these parameters to their classical counterparts $\eta$ and $g$ might depend on renormalisation. On the other hand, unitarity and the undeformed limit also pose constraints on the dependence of $q$ and $h$ on $\eta$ and $g$. In particular, $q(\eta,g)$ and $h(\eta,g)$ should be such that
\begin{itemize}
\item the $S$ matrix is unitary for all values of $\eta$ and $g$, i.e. eqn. \eqref{eq:unitaritybound} is satisfied,
\item $q$ satisfies eqn. \eqref{eq:semiclassicalq} and $h=g$ in the semiclassical regime,
\item the effect disappears as $\eta \rightarrow 0$, since the undeformed theory has a vanishing $\beta$ function and $h$ does not renormalise.
\end{itemize}
The only proposal \cite{Arutynov:2014ota} that satisfies all these constraints has $h=g$ and 
\begin{equation}
\ad = \arcsinh\left( \frac{2\eta}{g\left(1+\eta^2\right)} \right). 
\end{equation}
\subsection{Labelling of states}
Thus far our discussion of the quantum theory has been mostly algebraic: we only used the classical action once to find the perturbative $S$-matrix to check the exact $S$-matrix. Moreover, the exact $S$-matrix was bootstrapped in such a way as to ensure that the quantum theory has $\mathfrak{psu}_q(2|2)^{\otimes 2}$ on-shell symmetry. 
\\[5mm]
To complete the description of the quantum model we need to include a description of states: they can be organised in supermultiplets and labelled by their eigenvalue  under the Cartan charges of $U_q\left( \mathfrak{psu}(2,2|4)\right)$, since it is possible to diagonalise all of them simultaneously due to their mutual commutativity. This labelling is not unique, however, since different bases for the Cartan subalgebra are possible. A crucial observation is that the $q$ deformation did not touch the algebra relations for the Cartan subalgebra and therefore we can use the same description for the multiplets as was used in the undeformed case. Since we are working towards the $\eta$-deformed QSC we will follow the conventions of \cite{Gromov:2014caa}: an undeformed multiplet is characterised by the sextet
\begin{equation}
\label{eq:quantumnumbers}
\{J_1,J_2,J_3, E, S_1, S_2 \},
\end{equation}
which are the three angular momenta on $S^5$, the string energy and the two Lorenz spins respectively. It might at first be strange to use the Cartan charges of $U_q\left( \mathfrak{psu}(2,2|4)\right)$, as the quantum theory defined by the $S$ matrix only has $\mathfrak{psu}_q(2|2)^{\otimes 2}$ symmetry. At the hamiltonian level these labels are indeed overcomplete: the charge $J_1=J$ does not commute with the hamiltonian and it hence seems an inconvenient label in the hamiltonian description. This, however is an artefact of the light-cone gauge, which demoted $J$ from a quantum number into the circumference of the circle on which our hamiltonian theory is defined. We will see shortly that in the thermodynamic Bethe ansatz $J$ gets restored and the quantum numbers \eqref{eq:quantumnumbers} become an appropriate set of labels.

\paragraph{Energy and dispersion.}
Before we can start discussing bound states one particularly important identification that we have to make is which (combination) of the central charges in the fundamental representation corresponds to the world-sheet excitation energy and momentum. As in the undeformed case, we associate the world-sheet excitation energy $E$ to the eigenvalue of the central charge $\mathbf{K}$ ($\mathbf{V}$) and its momentum $p$ to the eigenvalue $U$ of the braiding $\mathbf{U}$ as follows:
\begin{equation}
\label{eq:energyascharges}
V^2 = q^K = q^E, \quad U^2 = e^{\ii p},
\end{equation}
which relates energy and momentum to the $x$ functions through eqn. \eqref{eq:UVinx} and has the expected undeformed limit . Using the shortening condition \eqref{eq:UVshortening} we find the dispersion relation
\begin{equation}
\label{eq:dispersion}
\cos^2 \frac{\theta}{2} \sinh^2 \frac{ \ad E}{2} - \sin^2 \frac{\theta}{2} \sin^2 \frac{p}{2} = \sinh^2 \frac{\ad}{2},
\end{equation}
which when solved for positive energies yields
\begin{equation}
\label{eq:positiveenergies}
E(p) = \frac{2}{\ad} \arcsinh \sqrt{\sec^2 \frac{\theta}{2} \sinh^2 \frac{\ad}{2} +\tan^2 \frac{\theta}{2} \sin^2 \frac{p}{2}}. 
\end{equation}
A noteworthy property of this dispersion relation is that it is non-relativistic, implying it is not invariant under the mirror transformation
\begin{equation}
\label{eq:mirrortransformation}
E\rightarrow \ii \tilde{p}, \quad p \rightarrow \ii \tilde{E}
\end{equation}
induced by a double Wick-rotation, where $\tilde{P}$ and $\tilde{E}$ are the mirror momentum and energy. However, the result of a double Wick-rotation is not completely unfamiliar: performing a double Wick-rotation on eqn. \eqref{eq:dispersion} with $\theta=\theta_0$ and rescaling the energy and momentum as $E\rightarrow \pm E/\ad$ and $p \rightarrow \pm p/\ad$ we obtain the exact same dispersion relation but with $\theta = \theta_0 + \pi$.\footnote{All signs in the rescaling are allowed.} This duality was dubbed \emph{mirror duality} in \cite{Arutynov:2014ota} and is worth reviewing in more detail. 

\subsection{Mirror duality}
The relation between dispersion relations at different values of $\theta$ was found to extend to other aspects of the theory. The mirror transformation \eqref{eq:mirrortransformation} can be recast in the language of the $x$ functions: all real momenta and energies satisfying eqn. \eqref{eq:positiveenergies} can be parametrised by real rapidities through the $x_s$ function and eqn. \eqref{eq:UVinx} and eqn. \eqref{eq:energyascharges}. Considering the mirror transformation on the level of $x$ functions one finds that, requiring the mirror energy to be positive and the mirror momentum to be real, the transformation is equivalent to replacing $x_s \rightarrow x_m$. Since the difference between the two $x$ functions is determined by the choice of the branch cuts this effectively means that the mirror transformation amounts to changing long cuts into short ones and vice versa. Moreover, the $x$ functions satisfy
\begin{equation}
\label{eq:xshifts}
x_s(u+\pi)\big|_{\theta=\theta_0 + \pi} = x_m(u)\big|_{\theta=\theta_0},
\end{equation}
for all $u$ on the cylinder with cuts. Comparing this with the implementation of the mirror transformation we see that shifting $u\rightarrow u+\pi$ and $\theta\rightarrow \theta_0 +\pi$ is an equivalent transformation. 

As the $S$ matrix is built from the $x$ functions it is perhaps not surprising that mirror duality can be lifted to the theory as a whole: for example one finds that for the energy and momentum
\begin{equation}
\label{eq:mirrorenergyandmomenta}
\ad E( u + \pi) \big|_{\theta=\theta_0 + \pi} = \tilde{E}(u)\big|_{\theta=\theta_0}, \quad 
p( u + \pi) \big|_{\theta=\theta_0 + \pi} = -\ad \tilde{p}(u)\big|_{\theta=\theta_0}.
\end{equation}
An analysis of the $S$ matrix shows that the theories at $\theta= \theta_0+ \pi$ not only have identical dispersion relations as their mirror cousins at $\theta= \theta_0$, also their scattering properties coincide, making them effectively the same theory. Apart from a technicality that yields a minus sign\footnote{Shifting $u$ and $\theta_0$ as in eqn. \eqref{eq:xshifts} takes $\xi$ outside the positive imaginary axis on which our initial family of theories were defined. Including a sign change $\xi \rightarrow -\xi$  in the mirror duality transformation ensures that $\xi$ stays in its original domain.} this shows mirror duality on the level of the sigma model: it takes a member out of the $\eta$-deformed family with $\theta=\theta_0$ and relates it to the double Wick-rotated member at $\theta = \pi - \theta_0$. This was checked semiclassically as well as on a giant-magnon solution \cite{Arutynov:2014ota}, inspired by the solutions found in the undeformed case \cite{Hofman:2006xt,Arutyunov:2006gs}. 

\paragraph{Spectrum and thermodynamics.}\label{sec:spectrumandthermodynamics} One important aspect of mirror duality that might not be immediately obvious is that for the models under consideration it provides a relation between thermodynamics and the spectral problem: we will use the thermodynamic Bethe ansatz on the mirror theory to obtain the spectrum, but the primary result of this method is of course the thermodynamics of the mirror theory itself. The identification in the previous paragraph therefore shows that by shifting $u$ and $\theta$ we can potentially interpolate between these two regimes. One application of this idea is the computation of the Hagedorn temperature\footnote{See \cite{Harmark:2017yrv}. Note that their approach does not rely on the $\eta$ deformation.}: one can either consider the thermodynamics of the $\ads$ superstring or the spectral problem of its mirror. Due to mirror duality we can reach the spectral problem for the mirror theory by taking the ``undeformed" mirror limit $\theta\rightarrow \pi$\footnote{To avoid singular behaviour this limit has to be accompanied by a shift in $u$ by $\pi$, rescale $u\rightarrow 2 \ad u$ and take $\ad \rightarrow 0^+$ with $\theta = 2\arccos\left(2 g \sinh \ad\right)$.} of the $\eta$-deformed theory. In this thesis we will show that this spectral problem takes the form of a QSC. 

\paragraph{Undeformed $\ads$ superstring.} The mirror duality of the the $\eta$-deformed theory does not have an undeformed analogue. Although the mirror transformation in that case can also be achieved by replacing $x_s\rightarrow x_m$, one cannot achieve this result by shifting the parameters a finite amount: the undeformed $x$ functions have fundamentally different cut structures, the short cut being of finite length and the long cut being a union of two (infinite) half-lines (see eqn. \eqref{eq:undeformedx}). 

\section{TBA for the $\eta$-deformed model}
We have now collected all the basic ingredients to consider the central question in this thesis:
\\[5mm]
\textbf{What is the simplest set of equations we can find that describe the spectrum of the $\eta$-deformation of the $\ads$ superstring?}
\\[5mm]
At this point in the derivation we have found the fundamental two-body $S$-matrix and the accompanying dispersion relation. In addition, we have a description of multi-particle states through the Zamolodchikov-Faddeev algebra. The particle spectrum will consist both of fundamental particles as well as bound states. Generically, one would have to reconsider the details of the scattering theory to determine the $S$-matrix elements for these bound states, but integrability helps us to circumvent this: using the concept of factorised scattering we can determine these bound-state $S$-matrix elements directly using a procedure usually called \emph{fusion}. Therefore our present knowledge allows us to completely solve the two-dimensional QFT on a flat space. This, however, is not the whole story: we are interested in the spectrum of this QFT when considered on a cylinder and only obtained a flat space description by taking the decompactification limit in section \ref{sec:lightconegauge}. To honestly consider the theory on a cylinder -- with a compact space-direction -- we will have to take into account wrapping corrections to the $S$ matrix coming from particles travelling around the compact circle. This is very messy business, but can be resolved in a much cleaner way using a method put forward by Zamolodchikov in \cite{Zamolodchikov:1989cf}. In the following we will describe this method, also used for the undeformed case, and the way it can be used to find the spectrum of the $\eta$-deformed model.\footnote{Our description is somewhat brief and we refer to the reviews \cite{vanTongeren:2013gva,vanTongeren:2016hhc} for a much more detailed description.}

\subsection{Finding the spectrum from the mirror model}
A first step to finding the spectrum of the $\eta$-deformed model is to find the ground-state energy $E_0$. One way to characterise $E_0$ is as the leading term in the low-temperature expansion of the (Euclidean) partition function
\begin{equation}
\label{eq:partitionfunction}
Z(\beta, L) = \sum_{j\geq 0} e^{-\beta E_j},
\end{equation}
where $\beta= 1/T$ and $L$ is the volume our theory lives in. The standard way to obtain $Z$ is in the imaginary-time formalism, by first performing a \emph{Wick rotation} $\tau \rightarrow \tilde{\sigma} = \ii \tau$ and then computing the path integral over all the fields periodic in $\tilde{\sigma}$ with period $\beta$. Thus, in addition to a compact space-direction after the Wick rotation also the time direction becomes periodic, effectively turning the world-sheet into a torus with periods $L$ and $\beta$. An interesting observation is that in this setting the roles of time and space have been put on completely equal footing. This implies that the partition function not only describes the thermodynamics of our original theory with hamiltonian $H$  generating $\tau$ translations but also of the so-called \emph{mirror} theory with hamiltonian $\tilde{H}$ generating $\tilde{\sigma}$ translations. In particular, $Z$ describes the thermodynamics of the mirror theory in the volume $\beta$ at temperature $1/L$. The mirror theory can be obtained by analytically continuing the Euclidean theory $\sigma \rightarrow \tilde{\tau} = -\ii \sigma$. The combination of these two Wick rotations is known as the \emph{mirror transformation} and acts on the hamiltonian and momentum as 
\begin{equation}
H\rightarrow \ii \tilde{p}, \quad p \rightarrow \ii \tilde{H},
\end{equation}
where $\tilde{p}$ is the mirror momentum. We can turn the idea that the partition function describes both the original and the mirror theory around and say that we can \emph{determine} the partition function from either of these two theories: we have the original description in eqn. \eqref{eq:partitionfunction} from the original theory, but could also characterise it through the Helmholtz free energy $\tilde{F}$ of the mirror theory as
\begin{equation}
Z(\beta,L)= e^{-L \tilde{F}(\beta)},
\end{equation}
where we indicated that $\tilde{F}$ generically depends on the mirror volume $\beta$. In the low-temperature limit $\beta \rightarrow \infty$ these two perspectives now lead to the identification 
\begin{equation}
L \frac{\tilde{F}(\beta)}{\beta} = L \tilde{f}(\beta) \rightarrow E_0,
\end{equation}
where $\tilde{f}$ is the mirror free-energy density. In other words, by computing the mirror free energy in the infinite volume at finite temperature we find an expression for the ground-state energy of the original model in finite volume! This might not sound like such an improvement, but remember that all the methods we have discussed so far only worked in the infinite-volume limit: only in the infinite-volume does a scattering theory based on asymptotic states make sense, so the $S$ matrix we have introduced can only aptly be used in that setting. It turns out that it is much more tractable to consider the problem of finite temperature than that of finite volume. In the next section we will describe how the methods known as asymptotic Bethe ansatz and Thermodynamic Bethe ansatz can be used to find the ground-state energy. Let us finish this section with two important points:
\begin{itemize}
\item In the analysis above we have for simplicity assumed that all fields were bosonic, which are periodic in the imaginary-time direction as discussed. Fermionic fields, however, are anti-periodic, which means that in the mirror theory we should not compute the partition function $Z$ but rather \emph{Witten's index}
\begin{equation}
Z_W = \text{Tr}\left( (-1)^F e^{-\beta \tilde{H}}\right),
\end{equation}
where $F$ is the fermion number operator. 
\item What should worry us much more is that generically we have no guarantee that the mirror theory described by the hamiltonian $\tilde{H}$ is integrable, which is paramount for the application of the ideas that we have discussed in the previous sections. For relativistic theories it turns out that their mirror is equal to the original theory and hence integrable, circumventing possible problems here, but the $\eta$-deformed theory is not relativistic as follows from the dispersion relation \eqref{eq:dispersion}.\footnote{Usually it is possible to Wick rotate enough of the charges to ensure integrability of the mirror model \cite{vanTongeren:2013gva}.} Nevertheless, we are in luck, since for the $\eta$-deformed model we know that its mirror is also integrable due to mirror duality: its mirror is just another $\eta$-deformed model at a different value of $\theta$, hence we immediately have a description of the mirror model to work with. 
\end{itemize}
\subsection{Asymptotic and thermodynamic Bethe ansatz}
\begin{figure}[!t]
\centering
\includegraphics[width=6cm]{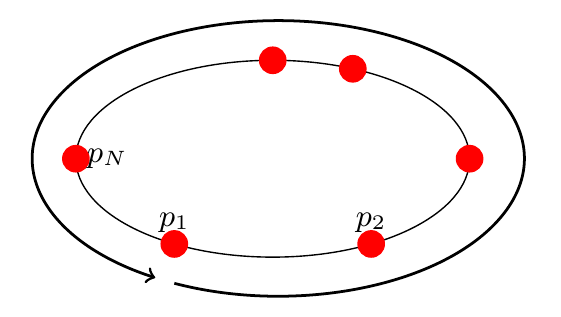}
\caption{Moving a particle with momentum $p_1$ around the circle forces it to interact with every other particle. }
\label{fig:scattering}
\end{figure}

With Zamolodchikov's mirror trick we have turned the problem of finding the ground-state energy of a QFT in finite volume into finding the Helmholtz free energy density of the mirror theory in infinite volume but at finite temperature. To find out the quantisation conditions on the momenta of the scattering we employ the \emph{asymptotic Bethe ansatz}. 
\\[5mm]
Remember that we ended up in the infinite-volume in the mirror theory because we are considering the low-temperature limit of the original theory. For finite volume the mirror theory is defined on a circle with circumference $R=\beta$. Periodicity of the wave function for free particles then leads to quantisation conditions as it must be that 
\begin{equation}
e^{\ii p_j R}  = 1.
\end{equation}
Now consider $N$ particles with momenta $p_1 > p_2 >\ldots > p_N$ on a circle with very large circumference that allows for the particles to be well-separated at positions 
\begin{equation}
\label{eq:xpositions}
x_1 \ll x_2 \ll \ldots \ll x_N \text{ at } \tau = -\infty.
\end{equation}
The fact that the particles are well-separated implies that the $N$-particle wave-function is that of $N$ free particles. As time goes on these particles will scatter, and following our discussion in section \ref{sec:integrablefieldtheories} we can describe the scattering process using the infinite-volume two-body $S$-matrix and factorisation. After the scattering process the strict ordering of the momenta causes the particles to become well-separated and free again at $\tau =\infty$.  The fact that these particles live on a circle means we should be careful in interpreting the position ordering \eqref{eq:xpositions}: there is no absolute ordering on a circle. In particular the ordering $x_2 \ll \ldots x_N \ll (x_1+R)$ leads to a different wave function, but since this ordering and the ordering in eqn. \eqref{eq:xpositions} are equivalent on the circle so should the wave functions. Moving the particle at $x_1$ to the last position to obtain the second ordering from the first requires this particle to scatter of off every other particle, as illustrated in fig. \ref{fig:scattering}, such that we find a restriction on the wave function known as the Bethe-Yang equations:
\begin{equation}
\label{eq:BetheYang}
e^{\ii p_j R} \prod_{\substack{k=1 \\k\neq j}}^N S_{jk}(p_j,p_k) = 1,
\end{equation}
where $S_{jk}$ here is the relevant $S$-matrix coefficient for the scattering of the particles with momentum $p_j$ and $p_k$. 
In theories with more than one particle species these equations become more complicated and can contain several $S$-matrices describing the scattering of different particle species. We sketch the general approach here but refer to the details of this procedure to the review \cite{vanTongeren:2013gva} and the book \cite{Korepin}. 
\\[5mm]
The general form of the Bethe-Yang equations can be quite complicated if not written down in a convenient basis. The \emph{nested Bethe ansatz} \cite{Yang:1967bm} describes how to find a convenient basis and write the equations in an efficient form: we first introduce a \emph{transfer matrix} $T$ as the (super) trace over the product of $S$ matrices
\begin{equation}
T\left(p'| \{p_j\}\right) = \text{tr}_a \prod_{j=1}^N S_{a j}\left( p', p_j\right),
\end{equation}
where the label $a$ refers to the auxiliary space an auxiliary particle with momentum $p'$ lives. The trace is taken in the auxiliary space and the product of matrices is ordered left to right. Plugging in one of the momenta of the particles $p'=p_k$ the transfer matrix reproduces the right-hand side of the Bethe-Yang equations \eqref{eq:BetheYang} with a minus sign. Moreover, by virtue of the Yang-Baxter equation the transfer matrix has the crucial property that it commutes with itself for different values of the auxiliary momentum
\begin{equation}
\left [ T\left( p | \{p_j\}\right), T\left( p' | \{p_j\}\right)\right] = 0,
\end{equation}
allowing us to diagonalise the transfer matrix for all values of $p$ simultaneously! After diagonalisation the Bethe-Yang equations read
\begin{equation}
e^{\ii p_j R} \Lambda\left( p_j | \{p_k\}\right) = -1,
\end{equation}
where $\Lambda\left( p_j | \{p_k\}\right)$ is any eigenvalue of $T\left( p_j | \{p_k\}\right)$. Using the Bethe ansatz one can explicitly perform this diagonalisation, although the fact that our $S$ matrix is of higher rank -- higher than two, that is -- we have to perform this diagonalisation in stages known as the nested Bethe ansatz. Alternatively one can derive commutation relations for the $S$ matrix and the transfer matrix and employ the algebraic Bethe ansatz \cite{Faddeev:1979gh}. Extensive accounts of the application of the nested Bethe ansatz for the undeformed case can be found in \cite{Beisert:2005tm,deLeeuw:2010nd,Arutyunov:2007tc}. The Bethe-Yang equations for the $q$-deformed case were obtained in \cite{Arutyunov:2012zt} using the algebraic Bethe ansatz. 

\paragraph{$\eta$-deformed Bethe-Yang equations.} For the $\eta$-deformed mirror model the resulting equations were found in \cite{Arutyunov:2013ega} and are
\begin{align}
\label{eq:BYmirror}
1 = e^{\ii \tilde{p}_j R} \prod_{\substack{k=1 \\ k\neq j}}^{\tilde{K}^{\mathrm{I}}} S_{\mathfrak{sl}(2)}(\tilde{p}_j,\tilde{p}_k)\prod_{\alpha=l,r}\prod_{k=1}^{\tilde{K}^{\mathrm{II}}_{\alpha}} \sqrt{q}\frac{y_j^{(\alpha)} - x_k^-}{y_j^{(\alpha)} - x_k^+}\sqrt{\frac{x_k^+}{x_k^-}}\, ,
\end{align}
which come accompanied with a set of auxiliary Bethe equations for each $\alpha = l,r$ (and every $1 \leq j \leq N$)
\begin{align}
\label{eq:auxBEy}
1&= \prod_{k=1}^{\tilde{K}^{\mathrm{I}}}\sqrt{q}\frac{y_j - x^-_k}{y_j - x^+_k}\sqrt{\frac{x^+_k}{x^-_k}}
\prod_{k=1}^{\tilde{K}^{\mathrm{III}}}\frac{\sin{\frac{1}{2}(\nu_j - w_k-\ii \ad)}}{\sin{\frac{1}{2}(\nu_j - w_k+\ii \ad)}}\, ,\\
\label{eq:auxBEw}
-1&= \prod_{k=1}^{\tilde{K}^{\mathrm{II}}} \frac{\sin{\frac{1}{2}(w_j - \nu_k + \ii \ad)}}{\sin{\frac{1}{2}(w_j - \nu_k - \ii \ad)}}\prod_{l=1}^{\tilde{K}^{\mathrm{III}}}\frac{\sin{\frac{1}{2}(w_j - w_l - 2 \ii \ad)}}{\sin{\frac{1}{2}(w_j - w_l + 2 \ii \ad)}}\, ,
\end{align}
where the tilde indicates mirror quantities, $x_k^{\pm} = x\left(p_k\pm \ii \ad\right)$ and the $\tilde{K}$ count the different types of excitations (world-sheet or auxiliary) in the nested Bethe ansatz and the $y$ and $w$ denote the rapidities of auxiliary excitations. The $\nu_j$ are to $y_j$ what $u$ is to $x$ and thus are defined through 
\begin{equation}
\label{eq:nu}
 y = x_s(\nu)^{\pm},
\end{equation}
where both options are possible. The dressing phase $S_{\mathfrak{sl}(2)}$ can be obtained from the $S_{\mathfrak{sl}(2)}$ dressing phase defined in eqn. \eqref{eq:Ssu2def}:
\begin{equation}
\label{eq:sl2dressingphase}
S_{\mathfrak{sl}(2)}\left( - \tilde{p}_1 , - \tilde{p}_2 \right)\big|_{\theta=\theta_0} = 
S_{\mathfrak{su}(2)}\left( p_1 , p_2 \right)\big|_{\theta = \theta_0+\pi},
\end{equation}
using the identification of momenta as in eqn. \eqref{eq:mirrorenergyandmomenta}. The undeformed Bethe-Yang equations follow by plugging in the undeformed $x$-function and sending $q\rightarrow 1$. 

\paragraph{Matching of the excitation parameters.} The $K$ counting excitation numbers can be matched with four out of the six quantum numbers specifying a multiplet in the $\eta$-deformed theory. The precise relation with the quantum numbers \eqref{eq:quantumnumbers} is \note{Check this!}
\begin{equation}
\label{eq:labelsinKs}
q_{\alpha} = \tilde{K}^{II}_{\alpha}-2\tilde{K}^{III}_{\alpha}, \quad 
s_{\alpha} =  \tilde{K}^{I} - \tilde{K}^{II}_{\alpha},
\end{equation}
with
\begin{equation}
\begin{aligned}
J_2 & = \frac{q_l + q_r}{2}, \\
J_3 & = \frac{q_l - q_r}{2},
\end{aligned}\quad 
\begin{aligned}
S_1 & = \frac{s_l + s_r}{2}, \\
S_2 & = \frac{s_l - s_r}{2}.
\end{aligned}
\end{equation}
The remaining two quantum numbers $J_1 = L$ and $E$ are the inverse temperature of the mirror model and the energy respectively and we will encounter them below in the TBA description.

\subsection{Including bound states}
In order to find the infinite-volume thermodynamics of the mirror model we wish to compute the mirror free energy $f = e- T s$, where $e$ is the energy density and $s$ is the entropy density. Considering the infinite-volume limit $R\rightarrow \infty$ we note that the thermodynamic behaviour of the system will be dominated by states with a finite particle density $N/R$. In our case we should also consider a finite density for the auxiliary particles to allow for the most general contribution of a state to the thermodynamics. Concretely this means we need to analyse the Bethe-Yang equations \eqref{eq:BYmirror}-\eqref{eq:auxBEw} for ever-increasing mirror length $R$ and excitation numbers $K$, while keeping their ratios fixed. The infinite-length solutions of the Bethe-Yang equations are known as \emph{strings} since they often form a string pattern in the complex plane. A first step towards knowing the full thermodynamics is then to analyse the possible solutions of the Bethe-Yang equations. 
\\[5mm]
To understand how this analysis works, consider the simple Bethe-Yang equations \eqref{eq:BetheYang}: as long as all the momenta are real increasing $N$ will simply lead to more solutions, but these solutions are not the strings we are after. Consider instead for the case $N=2$ that $p_1$ has positive imaginary part. As $R\rightarrow \infty$ this causes the leading exponential to go to zero and for the left-hand side to yield $1$ in the limit it must be that the product of $S$ matrices diverges. This means that the momentum $p_2$ must be such that $S_{12}(p_1,p_2)$ diverges as $R\rightarrow \infty$. Vice versa, if $p_1$ has negative imaginary part $p_2$ must be such that in the limit $S_{12}(p_1,p_2) = 0$. The precise form of the solutions of course depends on the $S$ matrix. However, since in the generic situations the zeroes and poles occur for momenta related as $p_j = p_{j+1}+ a  \ii $ for some $a\in \R$ these solutions tend to form vertical patterns on the complex plane leading to the name ``string solutions". Additionally, for physical solutions we must demand that they have real total momentum and energy, after which we can interpret them as bound states of particles moving with the average momentum of its constituent particles. In particular, we can consider the scattering of these bound states, yielding a new $S$ matrix containing these bound states. The energy, momentum and $S$ matrix elements for the bound states are obtained by a procedure called \emph{fusion}. 
\\[5mm]
Analysing the zeroes and poles of the $\eta$-deformed $S$ matrix\footnote{Again we refrain from giving all the details, as they have been discussed in great detail elsewhere \cite{vanTongeren:2013gva,Arutyunov:2012zt}. A general account of how to construct string solutions can also be found in \cite{Klabbers:2016cdd}.} yields different types of solutions -- different bound states -- which come in the same types as the analysis of the undeformed $\ads$ Bethe-Yang equations \cite{Arutyunov:2007tc}: \begin{itemize}
\item \emph{$Q$ particles} are bound states of fundamental particles and the only bound states carrying a non zero amount of energy,
\item \emph{$vw$ strings} are complexes of $y$ and $w$ roots,
\item \emph{$w$ strings} are complexes of $w$ roots,
\item \emph{$y$ particles} are single $y$-roots.
\end{itemize}
With these bound states in hand we can consider what happens in the thermodynamic limit. 
\subsection{From the Bethe-Yang equations to TBA}
The true thermodynamics of our model we find only when taking the thermodynamic limit, i.e. taking the length to infinity ($R\rightarrow \infty$) as well as the number of particles. We argued that we can use the knowledge of string solutions to take this limit properly. Note, however, that the analysis we followed in the previous section to construct string solutions is by no means a rigorous way to determine the solutions which are relevant in the thermodynamic limit: indeed, instead of a pole in one of the $S$ matrices we could imagine a situation where it is the ever-growing product of finite-valued $S$-matrices that yields a pole-like contribution to the Bethe-Yang equation in the thermodynamic limit. Nevertheless, these type of solutions seem relatively uncommon and we continue assuming that the bound states we just singled out represent the solutions that contribute a measurable amount to the free energy in the thermodynamic limit.\footnote{There are examples of solutions which are not of ``string form" in the thermodynamic limit, see for example \cite{Woynarovich:1982} for an example in the Heisenberg \textsc{xxx} spin chain.} In the thermodynamic Bethe ansatz this assumption goes under the name of \emph{string hypothesis} \cite{Korepin}.\footnote{One could argue that the string hypothesis also includes the exact form of the entropy term that we put into the analysis as well and in some cases that the counting functions, which we will not introduce here, are monotonously increasing functions.} 
\\[5mm]
Assuming the string hypothesis it is straightforward to derive the thermodynamic Bethe ansatz (TBA) equations which relate the particle densities of all the present bound-states through coupled non-linear integral equations, see \cite{vanTongeren:2013gva,Klabbers:2016cdd} for examples how to derive them from the knowledge of bound states and the Bethe-Yang equations. In order to introduce the TBA equations for the $\eta$-deformed mirror model we need to introduce some notation. 

\paragraph{Branch cuts and convolutions.} Branch cuts will play a prominent role in our further analysis. They will almost exclusively be of square-root type\footnote{The only exception being the logarithmic branch cut responsible for the reconstruction of the driving term in section \ref{sec:discdelta}.} and for the $\eta$-deformed case are located on one of the following line segments (for $N\in \Z$)
\begin{equation}
\label{eq:Zs}
\begin{aligned}
Z_N &= \left\{ u \in \C \, | \, u=v+\ii\ad N, \, v\in (-\pi,\pi]\right\},\\
\hat{Z}_N &= \left\{ u \in \C \, | \, u=v+\ii\ad N, \, v\in [-\theta,\theta]\right\},\\ \check{Z}_N &= \left\{ u \in \C \, | \, u=v+\ii\ad N, \, v\in [-\pi,-\theta]\cup [\theta,\pi]\right\}.
\end{aligned}
\end{equation}
Branch cuts located on $\hat{Z}_N$ we will refer to as short and those located on $\check{Z}_N$ as long. We define three convolutions based on these definitions: for two functions $f$ and $h$
\begin{align}
\label{eq:convolutionsdef}
f\star h(u,v)=\,  \int_{Z_0}\, dt\, f(u,t)h(t,v)  &= \int_{-\pi}^{\pi}\, dt\, f(u,t)h(t,v) \, , \nonumber \\
f\, \hat{\star}\,  h(u,v)=\, \int_{\hat{Z}_0}\, dt\, f(u,t)h(t,v)  &=  \int_{-\theta}^{\theta}\, dt\, f(u,t)h(t,v)\, , \\
f\, \check{\star}\, h(u,v)=\, \int_{\check{Z}_0}\, dt\, f(u,t)h(t,v)  &= \int_{-\pi}^{-\theta}\, dt\, f(u,t)h(t,v)+\int_{\theta}^{\pi}\, dt\, f(u,t)h(t,v)  \, \nonumber.
\end{align}

\begin{figure}[!t]
\centering
\begin{subfigure}{7.5cm}
\includegraphics[width=7.5cm]{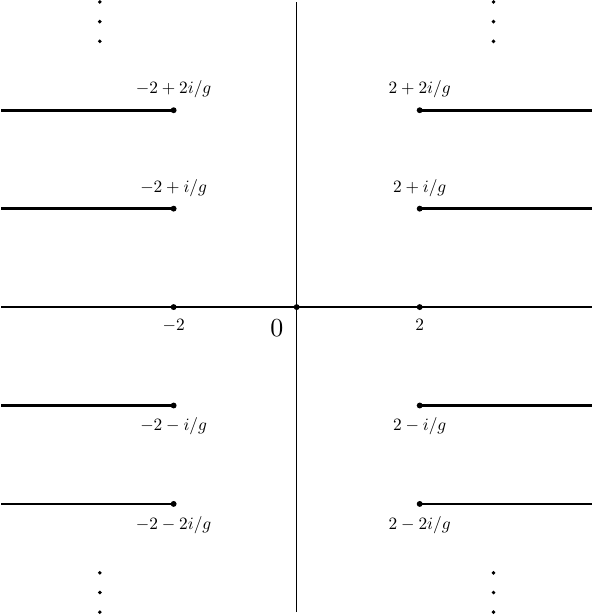}
\caption{}
\end{subfigure}
\qquad \qquad
\begin{subfigure}{6cm}
\includegraphics[width=6cm]{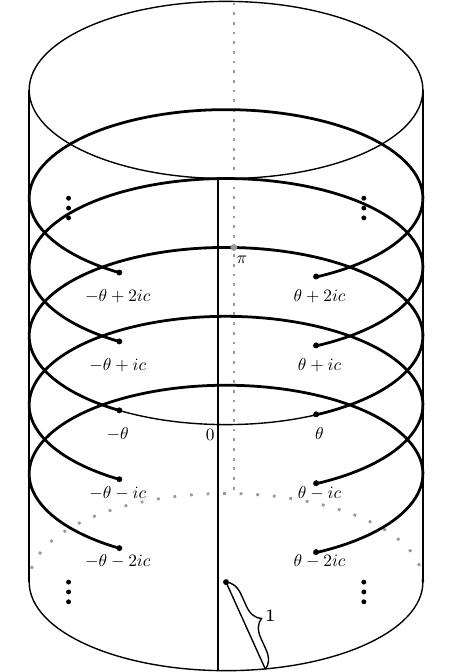}
\caption{}
\end{subfigure}
\caption{The cut structures of the undeformed (a) and deformed (b) models. Possible branch cuts are indicated by thick lines.}
\label{fig:cutstructures}
\end{figure}

\begin{figure}[t]
\centering
\includegraphics[width=14cm]{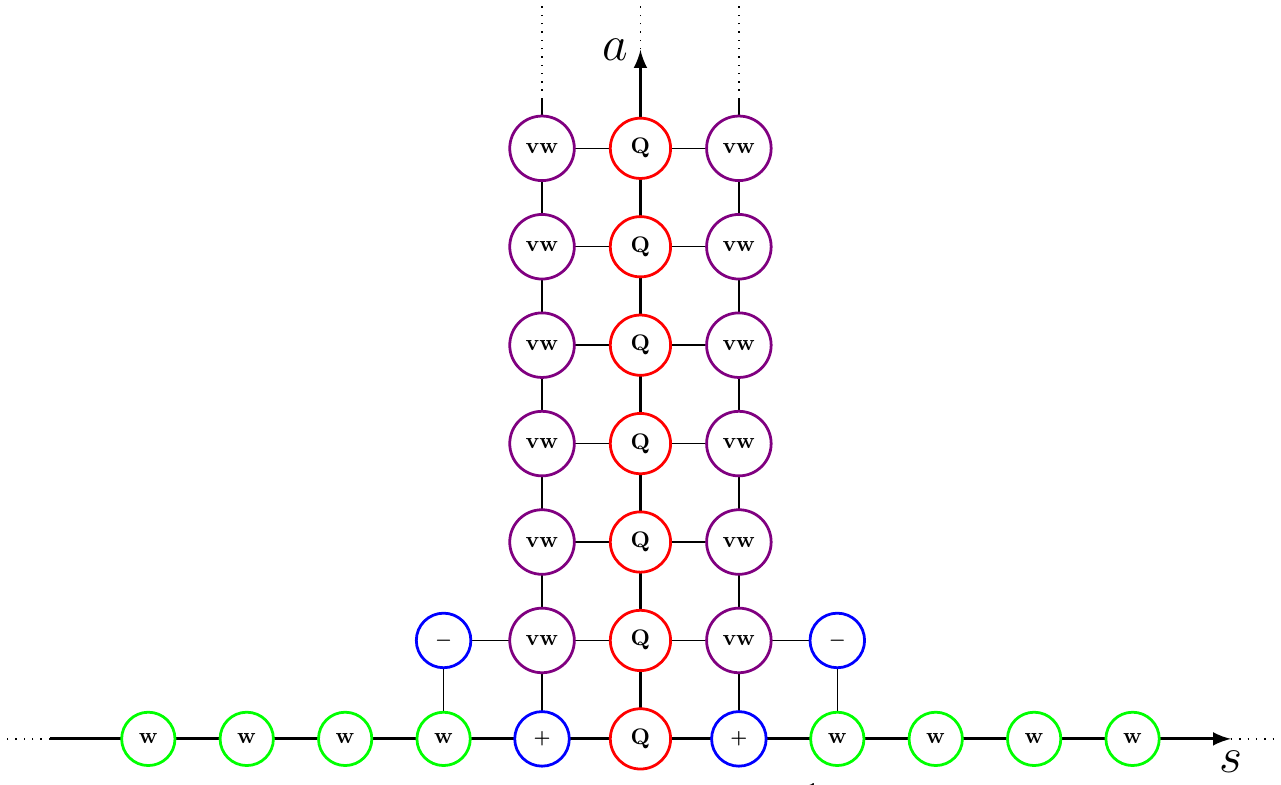}
\caption{The $Y$ hook organising all the $Y$ functions, cf. eqn. \eqref{YsinTs}. The central red nodes are the $Y_Q$, the violet nodes the $Y_{vw}^{(\alpha)}$, the green nodes the $Y_{w}^{(\alpha)}$ and the blue nodes the $Y_{\pm}^{(\alpha)}$ functions.}
\label{fig:Yhook}
\end{figure}

\paragraph{TBA equations.} With these definitions we can introduce the TBA equations for the $\eta$-deformed model:
\begin{equation}
\label{eq:TBAeqns}
\begin{aligned}
\log Y_{Q} =&\, -J \tilde{E}_Q +\Lambda_{P} \star K^{PQ}_{\mathfrak{sl}(2)} + \sum_\alpha \Lambda^{(\alpha)}_{M|vw}\star K^{MQ}_{vwx} + \sum_\alpha L^{(\alpha)}_{\beta}\, \hat{\star} \,K_{\beta}^{yQ},  \\
\log Y^{(\alpha)}_{\beta} =& \,- \Lambda_P \star K^{Py}_{\beta} + \left(L^{(\alpha)}_{M|vw}-L^{(\alpha)}_{M|w}\right)\star K_M,\\
\log Y^{(\alpha)}_{M|w} =&\, L^{(\alpha)}_{N|w} \star K_{NM}  + \left(L^{(\alpha)}_{-}-L^{(\alpha)}_{+}\right)\,\hat{\star}\, K_M, \\
\log Y^{(\alpha)}_{M|vw} =&\, L^{(\alpha)}_{N|vw} \star K_{NM}  +  \left(L^{(\alpha)}_{-}-L^{(\alpha)}_{+}\right)\,\hat{\star}\, K_M  -\Lambda_{Q} \star K_{xv}^{QM},
\end{aligned}
\end{equation}
where the $Y$ functions are the unknown functions and represent the ratio of hole-to-particle density for all the four types of bound-states: $Y_Q$ for $Q$-particles, $Y_{\beta}$ for $y$ particles\footnote{The notation $\beta=\pm$ for $y$ particles derives from the fact that they come in two flavours depending on the sign of the  imaginary part of the $y$ rapidity.} and  $Y_{M|(v)w}^{\alpha}$ for $M|(v)w$ strings. For generic $Y$ functions we have defined
\begin{equation}
L_\chi = \,\log(1+1/Y_\chi), \qquad \Lambda_\chi =  \,\log(1+Y_\chi),
\end{equation}
with the exception of $L_{\pm}$ and $\Lambda_{\pm}$, defined as
\begin{equation}
L_\pm = \,\log(1-1/Y_\pm), \qquad \Lambda_\pm = \,\log(1-Y_\pm).
\end{equation}
The functions $K$ are the \emph{TBA kernels} and their explicit expressions can be found in appendix \ref{app:Definitions}. The $Q$-particle bound-state mirror energy $\tilde{E}_Q$ -- sometimes known as \emph{driving term} -- is given by
\begin{equation}
\tilde{E}_Q(u) = \log q^Q \frac{ x\left( u- \ii Q \ad \right) + \xi}{ x\left( u +\ii Q \ad \right) + \xi},
\end{equation}
where we use $\xi$ as defined in eqn. \eqref{eq:xirepresentation}. Repeated indices are summed over, $M,N,\ldots \in \mathbb{N}$, $\beta=\pm$, and $\alpha = l,r$ distinguishes a so-called left and right set of $Y$ functions. All $Y$ functions are periodic with period $2\pi$, and have branch cuts of square-root type on some $\check{Z}_N$s as illustrated in fig. \ref{fig:cutstructures}. 
We can organise the $Y$ functions on the \emph{$Y$ hook} (see fig. \ref{fig:Yhook}). 
We will discuss this convenient representation in more detail in chapter \ref{chap:Tsystem}. 
\\[5mm]
Since bound states can occur with any possible length there are infinitely many TBA equations, one for each of the bound states. Only the $y$ particles can occur in two forms only. This makes the TBA equations an infinite set of coupled non-linear integral equations. 
\\[5mm]
Finally, note the re-occurrence of the quantum number $J$ here: it features in the TBA equations since we are investigating the ground-state of the mirror theory at temperature $1/L = 1/J$ (consistent with our light-cone-gauge choice below eqn. \eqref{eq:lccquantities}).

\paragraph{Ground-state energy.} A solution of the TBA equations is a set of functions\begin{displaymath}
\{ Y_Q, Y_{M|w}^{(\alpha)}, Y_{M|vw}^{(\alpha)}, Y_{\beta}^{(\alpha)}\},
\end{displaymath} 
which implicitly depend on the length $J$ present in front of the driving term in the $Y_Q$ TBA-equation. The energy corresponding to a solution of these equations is given by
\bea
\label{eq:Energy} E(J) &=&-\int_{Z_0} du\, \sum_Q \frac{1}{2\pi}\frac{d\tilde{p}^Q}{du}\log\left(1+Y_Q\right),
\eea
where the $Q$-particle mirror momentum is given by
\begin{equation}
\label{eq:Qmirrormomentum}
\tilde{p}^Q(u) = \frac{\ii}{\ad}\log \left( q^Q \frac{x^+}{x^-}\frac{x^-+\xi}{x^++\xi}\right).
\end{equation}
One thing to note is that the energy only depends on the form of the $Y_Q$ functions, all the other bound states play only an auxiliary role. Although we have not really pressed this point before, it is important to note that the energy here is in fact the ground-state energy of the $\eta$-deformed model in a volume $L$. This seems underwhelming: did we go through all these steps from the Bethe-Yang equations to end up with an equation for the ground-state only? Remember that the Bethe-Yang equations describe the entire spectrum. We will see soon though, that we can use the TBA equations to access the energies of excited states as well.

\paragraph{Undeformed TBA-equations.} Since we will want to compare our own analysis with that done for the undeformed case it will be useful to have access to the undeformed TBA-equations, i.e. the TBA equations for the undeformed $\ads$ superstring as well. It is both lucky and unlucky that the undeformed TBA equations can be written in exactly the same form \eqref{eq:TBAeqns} by reinterpreting the symbols in all the equations: first off, the undeformed counterpart of the $Z$ line segments are\footnote{For easy comparison with the literature we stick to the conventions of \cite{Arutyunov:2007tc,Cavaglia:2010nm}.} 
\begin{equation}
\begin{aligned}
\label{eq:undZs}
Z_N^{\text{und}} &= \left\{ u \in \C \, | \, u=v+\ii N/g, \, v\in (-\infty,\infty)\right\},\\
\hat{Z}_N^{\text{und}} &= \left\{ u \in \C \, | \, u=v+\ii N/g, \, v\in [-2,2]\right\},\\ \check{Z}_N^{\text{und}} &= \left\{ u \in \C \, | \, u=v+\ii N/g, \, v\in [-\infty,-2]\cup [2,\infty]\right\},
\end{aligned}
\end{equation}
which straightforwardly define the undeformed convolutions as we did for the deformed case in eqn. \eqref{eq:convolutionsdef}. After defining the undeformed kernels (see appendix \ref{app:Definitions}) and the driving term and $Q$-particle mirror momentum as
\begin{equation}
\label{eq:undeformeddrivingtermandmomentum}
\tilde{E}_Q^{\text{und}}(u) = \log \frac{x\left( u - \frac{\ii Q}{g}\right) }{x\left( u + \frac{\ii Q}{g}\right)},  \quad \tilde{p}^Q_{\text{und}} = g\, x\left( u - \frac{\ii Q}{g}\right)- g \,x\left( u + \frac{\ii Q}{g}\right) +\ii Q
\end{equation}
with $x$ the undeformed $x_s$ function \eqref{eq:undeformedx}, we can write the undeformed TBA equations exactly as in eqn. \eqref{eq:TBAeqns}. The $Y$ functions are not periodic and have branch cuts on the undeformed $Z$ segments. Also the ground-state energy can be computed in the same way as in the deformed case using eqn. \eqref{eq:Energy} using undeformed quantities. 

\subsection{Using the TBA equations}
The TBA equations are derived originally to compute the ground-state energy for the original model in finite volume. As these equations are extremely complicated, in most cases numerical approaches are the only viable way forward: forgetting about the current context of the mirror model, this usually yields the free energy as a function of the temperature, which can in turn be used to compute all kinds of thermodynamic quantities, see for example \cite{Klabbers:2016cdd} where the numerics are explained for the Heisenberg \textsc{xxx} and long-range Inozemtsev spin chains. 

\paragraph{Analytic continuation.}\label{sec:analyticalcontinuation} It turns out, however, that in practice this is just the beginning: using other techniques it is possible to get access to the energies of other states as well. The most important mathematical tool for this purpose is \emph{analytic continuation}. Since this will also prove to be a very important tool for us, let us spend some time on how these techniques work: consider as a toy model a simple eigenvalue problem for a finite-dimensional matrix $H(z)$ depending polynomially on a parameter $z$: in turn the eigenvalues and eigenvectors will be $z$-dependent as well:
\begin{equation}
H(z) \mathbf{v}(z) = E(z) \mathbf{v}(z). 
\end{equation}
Now suppose that we have found an eigenvalue-eigenvector pair $(E(z), \mathbf{v}(z))$ that solves the eigenvalue equation for all $z \in \R$. By the simple dependence of $H$ on $z$ we know that $E$ is a meromorphic function on some Riemann surface, but it will generically have branch points. If we now consider a path from some $z_0$ on the real axis around one such branch point that returns to $z_0$ we will find that the resulting value for $E$ does not coincide with the starting value $E(z_0)$, but it should in fact again be an eigenvalue! Indeed, since the eigenvalue problem has not changed the new value of $E$ should still solve the eigenvalue problem for $H(z_0)$. If the original $E(z_0)$ was the ground-state energy, what we found must be the energy of one of the excited states. Note that we do not need to bother with the analytic continuation of the eigenvector $\mathbf{v}$: it might change as well under the continuation, but if all we are interested in is finding out the energy spectrum we do not need to consider this change. 

\begin{figure}
\centering
\begin{subfigure}{4cm}
\includegraphics[width=4cm]{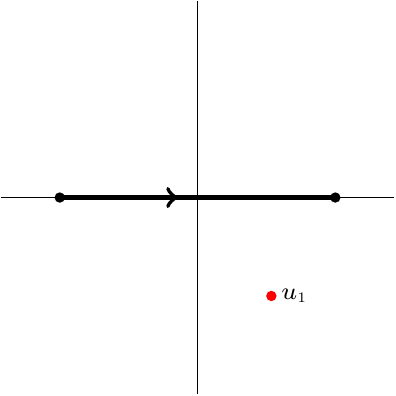}
\subcaption{}
\end{subfigure} \quad 
\begin{subfigure}{4cm}
\includegraphics[width=4cm]{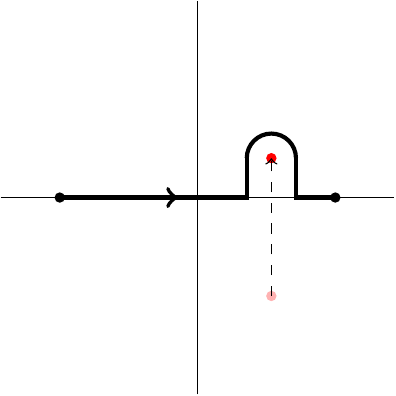}
\subcaption{}
\end{subfigure} \quad 
\begin{subfigure}{4cm}
\includegraphics[width=4cm]{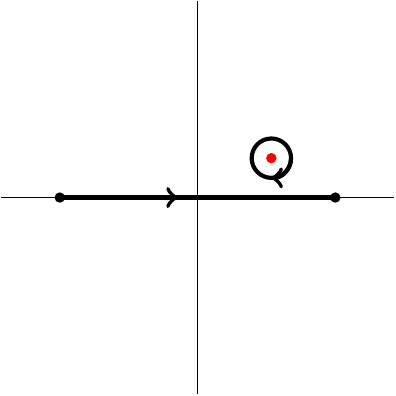}
\subcaption{}
\end{subfigure}
\caption{We start out with some integral over the thick line (see (a)) with an integrand with a pole at $u_1$. By moving the pole across the integration contour we are forced to move the contour to ensure continuity (see (b)). Finally, by deforming further one observes that the resulting integral is given in (c): it is the old integration contour plus a residue integral around the shifted pole.}
\label{fig:continuation}
\end{figure}

\paragraph{Analytic continuation in the TBA.}\label{sec:analyticcontinuationTBA} This technique can be applied in more complicated settings as well. In particular, one can use analytic continuation to find excited-state energies from the ground-state TBA-equations \cite{Dorey:1996re,Arutyunov:2009ax} although mathematically the problem looks a little different. Consider a non-linear integral equation resembling the TBA equations for two functions $Y_1,Y_2$: 
\begin{equation}
Y_{1}(u) = \log \left(1+Y_2(u)\right) \star K (u),
\end{equation}
with $K$ some known kernel of the form 
\begin{equation}
K (v,u) = \frac{1}{2 \pi \ii} \frac{d}{dv} \log S(v, u),
\end{equation}
for some $S$ matrix $S$. Suppose $Y_2$ has a pole at $u_1$ and by analytic continuation in some other variable, such as the coupling constant, we can move this pole around in the complex plane. If during the continuation $u_1$ collides with the integration contour of the convolution we can start deforming the contour as long as we meet no other singularities along the way. This will ensure continuity of the expression. Instead of deforming, we can also pick up the residue of the integral at $u_1$, such that after continuation the TBA equation becomes
\begin{equation}
Y_{1}(u) = \log \left(1+Y_2(u)\right) \star K (u) \pm \log S(u_1, u),
\end{equation}
where the sign depends on whether the pole hits the contour from below or from above. 
This idea is illustrated in fig. \ref{fig:continuation}. 
We see that the result of continuation is a TBA equation with an additional term also called \emph{driving term}.\footnote{There does not seem to be complete consensus about which terms can be called ``driving terms": in this thesis we will call all terms that lead to non-trivial asymptotics for the $Y$ functions driving terms, whereas some authors reserve this term for terms resulting from analytic continuation in the sense discussed here.} Subsequently solving these equations numerically then yields the energy related to some excited state. This approach and the very related contour deformation trick \cite{Arutyunov:2009ax}, which gives a presciption how to write down the TBA equations for any excited state, has been beneficial for the undeformed TBA-equations, for example for the entire $\mathfrak{sl}(2)$-sector of the theory the relevant TBA-equations were found in \cite{Balog:2011nm,Arutyunov:2009ax}. 
\\[5mm]
We will consider analytic continuation of the TBA equations extensively: in chapter \ref{chap:Ysystem} we will consider what happens when we continue the TBA equations in the free variable $u$, to investigate the analytic structure of the $Y$ functions.

\section{Beyond the TBA}
This completes our review of the derivation of the ground-state TBA equations all the way from the first definitions of the ($\eta$-deformed) model. These equations and all the related machinery we have introduced will form the basis of our analysis in the rest of this thesis. We will venture down one of the possible paths that might lead to a significant simplification of the TBA equations. Many such paths have been found over the years for a variety of models, in most cases yielding simpler sets of non-linear integral equations (NLIE), see for example \cite{Klumper:1992vt,bazhanov1997quantum,Fioravanti:1996rz, Juttner:1997tc,Destri:1992qk,Destri:1994bv,Takahashixxz,Tsuboi:2004ni}. The fact that it is possible to simplify the TBA equations for these models suggests that there is a hidden structure present that we are not fully exposing at the level of the TBA equations. We will ultimately derive a set of equations -- the $\eta$-deformed QSC -- which are no longer integral equations and in that sense do not fit in the simplifications cited above. However, this derivation is inspired by many of these previous works: the first step towards the QSC is to derive the analytic $Y$-system, which is also a step along the way to derive the NLIE equations of \cite{Klumper:1992vt,bazhanov1997quantum,Fioravanti:1996rz}. We will consider this in detail in the next chapter. 


%% file: etaQSC.tex
In the previous chapter we have discussed the early steps taken for the spectral problem of the $\eta$-deformed $\ads$ superstring, indicating how these steps were related to the ones taken for the undeformed spectral problem. This culminated in a set of TBA equations \eqref{eq:TBAeqns} describing the ground state of the theory in finite volume. Although this is already a huge achievement for such a complicated theory, it turns out that the $\eta$-deformed theory admits a further generalisation and even simplification. In this part we will discuss how one can derive the $\eta$-deformed QSC -- a finite set of functional equations on a Riemann surface -- from the TBA equations.\footnote{This part is based on the publication \cite{Klabbers:2017vtw}.}  This derivation in many respects stays close to the derivation of the undeformed QSC, although we will find that many of the original results are ``trigonometrised": where in the undeformed problem we were dealing with rational functions, these become trigonometric functions in the $\eta$-deformed case. The trigonometrisation is consistent with the fact that all the $\eta$-deformed functions have to be defined on a cylinder. A useful definition for these functions is the following: a function $f:\C \rightarrow \C$ is \emph{real periodic} if there exists some $\omega\in \R$ such that $f(z+\omega) = f(z)$ for all $z\in \C$. In particular, if we do not specify $\omega$ it is understood that the function is defined on the standard cylinder with circumference $\omega= 2\pi$. The undeformed case has no real periodicity.
\\[5mm]
Because of the similarity in the derivations of the $\eta$-deformed and undeformed case we will take the opportunity to compare the two such that this chapter can be read as a detailed review of the historical derivation of the QSC for the $\ads$ superstring. Our derivation will be quite explicit, showing all the necessary calculations. 
\\[5mm]
One can divide the derivation of the QSC (in both the deformed and undeformed cases) from the TBA equations in three steps:
\begin{align}
\text{TBA equations} \rightarrow \text{Analytic $Y$ system}\rightarrow \text{Analytic $T$ system}\rightarrow \text{Quantum Spectral Curve.} \nonumber
\end{align}
Alternatively we could divide the derivation also into an algebraic an an analytic part, where the algebraic part is by far the easiest: deriving the equations of the $T$ system from the TBA equations is completely straightforward, it is only at the final step that analytic and algebraic considerations have to be combined to end up with the $\Pf\mu$ system, one of the incarnations of the QSC. We will see this in due course. 
\\[5mm]
Let us summarise the derivation ahead of us: the first step is to derive the analytic $Y$-system: it consists of a set of equations for the $Y$ functions appearing in the TBA equations as well as a set of analytic constraints that we should impose on a solution of these equations. Alternatively we could interpret these constraints as a definition of the function space on which the $Y$-system equations are defined. Deriving the $Y$-system equations follows the canonical steps of applying first an inverse integral kernel $K^{-1}$ to the TBA equations followed by the operator $s$ and was already done partially in \cite{Arutynov:2014ota}. We then derive a set of discontinuity equations from the TBA equations that we should impose on the solution of the $Y$-system equations in the spirit of \cite{Cavaglia:2010nm}. We call the discontinuity equations, $Y$-system equations and the prescribed analyticity strips together the \emph{analytic $Y$-system}. By specifying a few more properties we can isolate the solution of the analytic $Y$-system that also satisfies the ground-state TBA-equations, thereby proving equivalence of the two sets of equations. 
\\[5mm]
The next step is to derive the analytic $T$-system: obtaining the $T$-system equations -- more commonly known as the Hirota equation -- from the $Y$-system equations is straightforward by a reparametrisation of the $Y$ functions. Transferring the discontinuity relations to the analytic $T$ system in a convenient way is a bit more difficult. First of all the reparametrisation of $Y$s into $T$s introduces a redundancy, because this reparametrisation contains a gauge freedom. However, despite the gauge symmetry it is not possible to find a set of $T$s such that all of them have nice properties. We therefore settle for two sets of $T$s that both partially have nice properties such that for every $Y$ function we have at least one nice set of $T$-functions and we know how to transform $T$s in the first set to the second set and vice versa. This derivation is inspired by \cite{Gromov:2011cx}. As it turns out this solution can be further simplified by noticing that it can be recast into a beautifully-simple-looking form as was derived for the undeformed case in \cite{Gromov:2010km}.
\\[5mm]
This allows us to perform the next step: the basic building blocks for one of the $T$ gauges can be interpreted as the basic $\Pf$-functions featuring in the QSC. Since the $T$s expressed in these $\Pf$ functions automatically satisfy the Hirota equations we seem to have no further constraints on them. However, the analytic constraints of the $T$ functions can be captured by introducing new functions and subsequently impose certain analyticity properties on these functions. Also, introducing the gluing object $\mu_{12}$ we find that the second $T$-gauge also imposes some constraints on the $\Pf$ functions. This ultimately yields the $\Pf\mu$ equations, which forms the basis of the QSC. 
\\[5mm]
The final step is to derive the boundary conditions that we should impose on the QSC equations to find the energy of a particular deformed-string state. By tracing back our parametrisation to the TBA equations we can derive that certain asymptotics of $\mu_{12}$ contain the energy of the state under consideration. We then argue that the other five quantum number should be present in similar asymptotics of some of the other functions, which is then fully fixed by consistency of the QSC and an appeal to the weak-coupling regime.

%% file: Ysystem.tex
Our goal in this chapter is to derive the analytic $Y$-system, a set of functional equations for the $Y$ functions featuring in the TBA equations \eqref{eq:TBAeqns} along with an additional set of analytic constraints. We will treat the derivation of the undeformed and deformed case simultaneously. To avoid needless repetition and unnatural notation our derivations will be given from the perspective of the deformed case when it comes to notation, trusting that the undeformed result can be read off easily using the identifications that we will make along the way. For example, the undeformed shift distance $1/g$ follows by the replacement $c \rightarrow 1/g$. When the two cases differ we will discuss the differences. 

\section{Setup}
A lot of the model's properties is encoded in the analytic properties of the $Y$ functions, so our first task is to derive some of them from the TBA equations. We will see that these properties, when interpreted correctly, are very similar for the undeformed and deformed $Y$-functions. The fundamental difference between the undeformed and deformed case is that the undeformed functions are defined on the complex plane with branch cuts whereas the deformed functions are defined on a cylinder with radius $1$, also with branch cuts. All branch cuts are located on either $\hat{Z}_N$ (short branch cuts) or $\check{Z}_N$ (long branch cuts) and are of square-root type, meaning that continuing a function two full circles around any branch point gives the original function back as a result. The nomenclature ``short" and ``long" has been inherited from the undeformed case, although geometrically it does not make sense in the deformed case: for $\theta>\pi/2$ short cuts are in fact longer than the long cuts. The fact that we cannot clearly distinguish long and short cuts is related to the mirror duality of the $\eta$-deformed model: shifting the spectral parameter and the branch-point location simultaneously changes short cuts into long ones and vice versa. We will return to this point in section \ref{sec:mirrordualityandlimit}. 

The compactification of the deformed rapidity plane can be interpreted as a geometric consequence of the deformation: the undeformed superstring has only one parameter, the string tension $g$. Geometrically it determines the ratio between the location of the branch points on the real line and the fundamental shift distance (the distance between two successive $Z_i$). Since a full rescaling of the rapidity variable is immaterial in our models it follows that one parameter is exactly enough to parametrise the inequivalent models.   The deformation parameter adds a second parameter to the model, but in order for this parameter to give a meaningful deformation we need to introduce an extra scale to the system. By compactifying the complex plane to a cylinder we end up with three scales: the branch-point location on the real line, the shift distance and the cylinder's radius. Using the rescaling invariance we can get rid of one of the three scales, resulting in two parameters $\theta$ (for the branch-point location) and $\ad$ (for the shift distance), see fig. \ref{fig:cutstructures}. Confer our discussion in the previous chapter \note{Reference?} these parameters are constrained as $\ad \geq 0$ and $\theta \in [0,\pi]$, which fits our geometric description well. We will often encounter ``shifted" functions for which we introduce the notation
\begin{align}
\label{eq:shift}
f^{[n]}(u) \defeq f(u+i\ad n), \quad f^{\pm} = f^{[\pm 1]},\quad f^{\pm\pm}= f^{[\pm2]}.
\end{align}

\section{Analytic properties of the $Y$ functions}
\subsection{Analyticity strips}
The region around the real axis plays a special role in our analysis. In particular, the region with $|\text{Im}(u)| < \ad$ we will call the \emph{physical strip}. Generically, the $Y$ functions are especially nicely behaved on a strip with $|\text{Im}(u)| < \ad M $ for some $M\in \N$, which can be derived directly from the TBA equations \eqref{eq:TBAeqns} using only knowledge of the integration kernels $K$. 
\\[5mm]
As an example, consider the basic undeformed kernel 
\begin{equation}
K_M^{\text{und}} ( u) = \frac{1}{\pi} \frac {g M }{M^2 + g^2 u^2},
\end{equation}
which has poles at $\pm  \ii M/g$. The deformed kernel 
\begin{equation}
K_M (u) = \frac{1}{2\pi} \frac{ \sinh M\ad}{\cosh M\ad- \cos u}
\end{equation}
has poles at $\pm  M \ii \ad $ and many more on the complex plane due to its $2\pi$ periodicity. In the following we will consider the deformed quantities always on the cylinder $-\pi < \text{Re}(u) \leq \pi$, where it only has the mentioned poles. 

The locations of these poles are typical for our models: they are in fact located on the strips $Z_M$ that we introduced in eqn. \eqref{eq:Zs} (and in eqn. \eqref{eq:undZs} for the undeformed case). 

\begin{figure}[!t]
\centering
\begin{subfigure}{7.5cm}
\includegraphics[width=7.5cm]{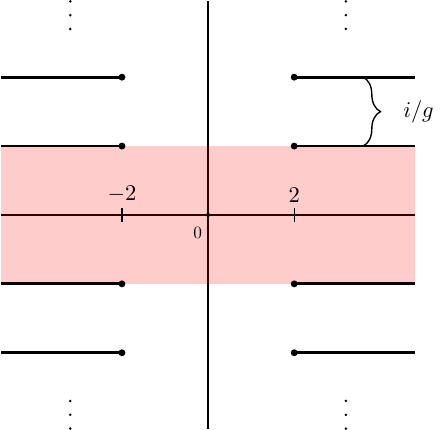}
\caption{}
\end{subfigure}
\quad
\begin{subfigure}{7cm}
\includegraphics[width=7cm]{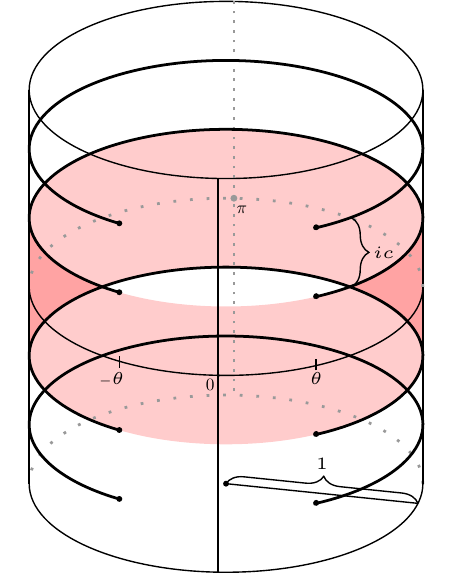}
\caption{}
\end{subfigure}
\caption{The analytic structure of the undeformed $Y$ functions on the plane (a) and the deformed $Y$ functions on a cylinder with radius one. The red-filled band indicates the analyticity strip associated to $\mA_1$.}
\label{fig:Yanalytics}
\end{figure}

\paragraph{Analyticity strips.} The solutions of the TBA equations we are interested in are $Y$ functions which are analytic in a wider strip around the real axis. To write this succinctly we introduce some notation: the functions which are analytic in the strip
\be
\label{eq:Astrips}
\left\{ u \in \C\, | \, |\text{Im}(u)| < M\ad\right\},
\ee
possibly with the exception of a finite number of poles form the set $\mathcal{A}_M$. If we want $f\in \mA_M$ to also be pole free, i.e. be analytic, on the strip, we write $f \in \mAp_M$. 
\\[5mm]
The behaviour of the $Y$ functions around the real axis can be characterised as follows: $Y_Q, Y_{Q|(v)w}^{(\alpha)} \in \mA_Q$. This is illustrated in fig. \ref{fig:Yanalytics}. When considering the ground state we have the stronger property $Y_Q, Y_{Q|(v)w}^{(\alpha)} \in \mAp_Q$. 

\subsection{Branch points and discontinuities}
All the $Y$ functions have branch cuts: the behaviour of these $Y$s around the branch cuts can be captured in \emph{discontinuities}. We can therefore use these discontinuities to pick up exactly the solution to the $Y$-system equations that we are interested in. In order to understand how they arise from the TBA equations, we will have to dig a little deeper: first notice that the $Y$ functions are always integrated over in the convolutions on the right-hand side of the TBA equations. This implies that their branch cut structure is not visible there. However, there are two sources of branch cuts on the right-hand side: 
\begin{itemize}
\item Branch cuts of the integration kernels $K$.
\item Convolutions over an ``incomplete" interval: in the undeformed case this means a convolution over a finite interval, but for the deformed case this means an interval for which at least one of the endpoints does not lie at $\pm \pi$. 
\end{itemize}

\paragraph{$y$-particle $Y$-functions.} The easiest source of branch cuts is branch cuts in the integration kernels. For example, consider the kernel $K_{\beta}^{Qy}$ (see \eqref{eq:TBAkernels2}). When continuing it around $u=\theta$ ($u=2$ for the undeformed case) we find that
\begin{equation}	
K^{Qy}_-(z,u^*) = \frac{1}{2\pi i} \frac{d}{dz} \log \left(q^{Q/2} \frac{x^-(z) -1/x(u)}{x^+(z) -1/x(u)}\sqrt{\frac{x^+(z)}{x^-(z)}} \right) = K^{Qy}_+(z,u),
\end{equation}
where we use the $*$ to indicate clockwise continuation around its right branch point if Im$(u)>0$ and counter clockwise continuation if Im$(u)<0$. We will argue below that the other terms in the TBA equation for $Y_{\pm}$\footnote{when the derivation is equal for both values of $\alpha$ we often suppress this index.} do not contribute any branch cuts and hence we find that $\log Y_{-}(u^*) = \log Y_{+}(u)$. Moreover, continuing again will return us to $ K^{Qy}_-(z,u)$ and hence the branch point at $u=\theta$ is of square-root type. Therefore we must have $Y_{-}(u^*)=Y_{+}(u)$, i.e. $Y_-$ and $Y_+$ are two different-sheet evaluations of one two-sheeted function. 

\subsubsection{Discontinuity from an incomplete interval}
\label{sec:discontinuityfiniteinterval}
\begin{figure}[!t]
\centering
\begin{subfigure}{4.5cm}
\includegraphics[width=4.5cm]{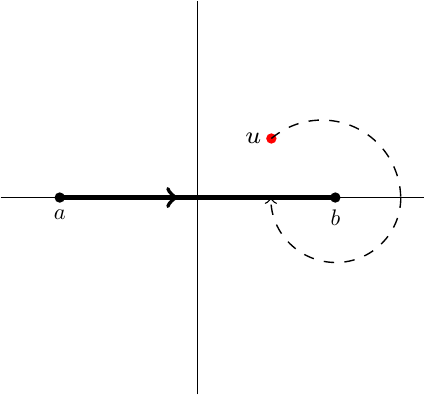}
\caption{}
\end{subfigure}
\quad
\begin{subfigure}{4.5cm}
\includegraphics[width=4.5cm]{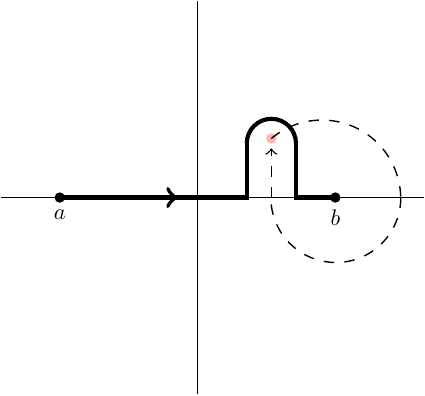}
\caption{}
\end{subfigure}
\quad
\begin{subfigure}{4.5cm}
\includegraphics[width=4.5cm]{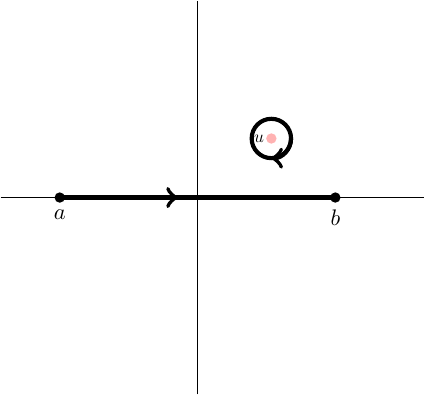}
\caption{}
\end{subfigure}
\caption{We continue $f$ in $u$ clockwise around $b$: in (a) the continuation hits the integration contour represented by the thick line. By deforming the contour as in (b) we avoid the collision, but the resulting function is no longer the same as our original $f$. It is equivalent to the original function $f$ plus the clockwise contour integral around the point $u$.}
\label{fig:continuations}
\end{figure}

\begin{figure}[!t]
\centering
\begin{subfigure}{4.5cm}
\fbox{\includegraphics[width=4.5cm]{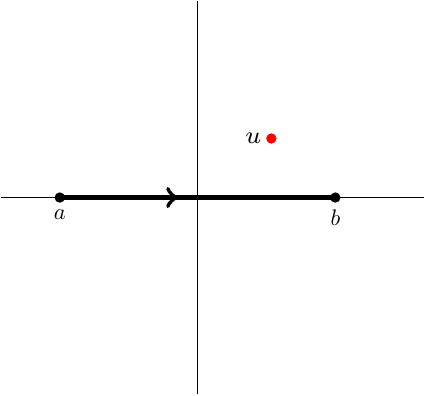}}
\caption{}
\end{subfigure}
\quad
\begin{subfigure}{10.3cm}
\fbox{\includegraphics[width=10.3cm]{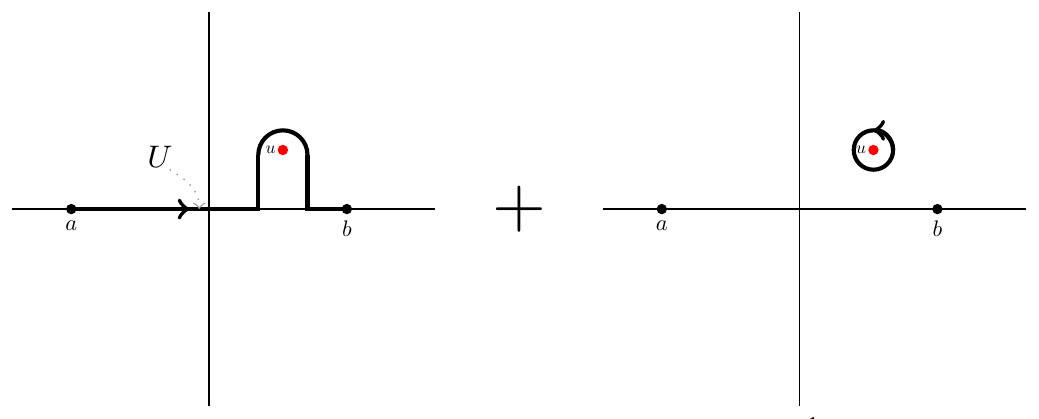}}
\caption{}
\end{subfigure}
\quad
\begin{subfigure}{10.3cm}
\fbox{\includegraphics[width=10.3cm]{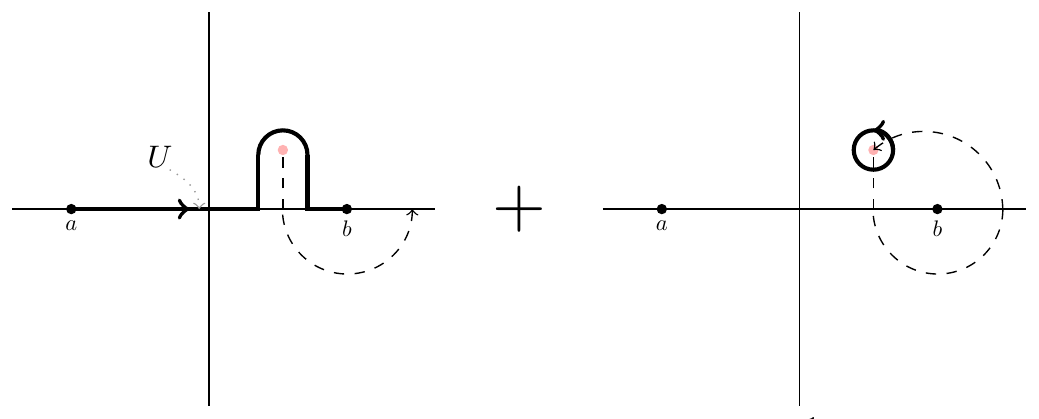}}
\caption{}
\end{subfigure}
\quad
\begin{subfigure}{10.3cm}
\fbox{\includegraphics[width=10.3cm]{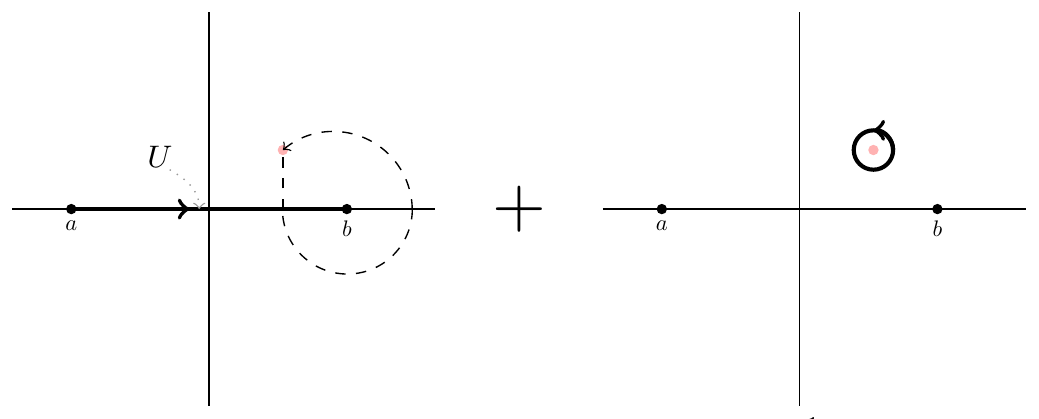}}
\caption{}
\end{subfigure}
\quad 
\begin{subfigure}{4.5cm}
\fbox{\includegraphics[width=4.5cm]{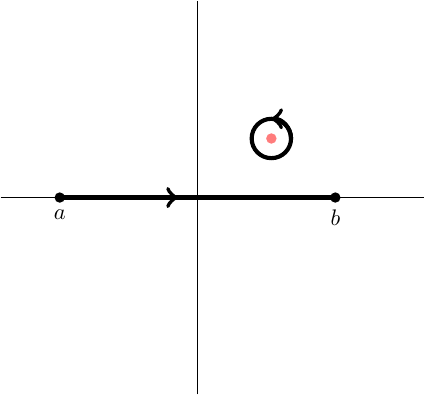}}
\caption{}
\end{subfigure}
\caption{The counter clockwise continuation around $b$: we start with the definition of $f$ in (a). This is equivalent to the sum of integrals in (b), where the contour has been shifted to allow for the continuation. In (c) we continue the first term half way and the second term completely. This allows for the deformation of $U$ back to the standard interval $[a,b]$, as we see in (d), as well as completing the continuation. The result is the sum of integrals in (e): the sign of the contour integral is opposite that of the result in \ref{fig:continuations}. Additionally, we note that the contour integral has been continued around $b$ as well, thus branch points in the integrand here will be felt by the result.}
\label{fig:continuations2}
\end{figure}

To understand the origin of branch points from an incomplete-interval convolution let us discuss some general theory. We will focus on the non-periodic case, it is quite straightforward to extend these results to the periodic case. 
\\[5mm]
Consider a function $f : \C \rightarrow \C$ defined as 
\begin{equation}
f(u) = \int_{[a,b]} dz g(z) K(z-u),
\end{equation}
for $u$ in the upper half-plane (Im$(u)>0$) and where $a<b$ are real and $g$ is analytic. Moreover, let the integration kernel $K$ have one unique first-order pole at the origin. When continuing $f$ around either of the end points $a$ or $b$, at some point the pole will have to cross the contour as illustrated fig. \ref{fig:continuations}, similar to our discussion in section \ref{sec:analyticcontinuationTBA}. By deforming the contour continuously, this can be avoided (since we assumed there are no other singular points), but the integral over the resulting contour is usually not equivalent to the original one and a Cauchy residue has to be picked up. Moreover, in general the direction of continuation ((counter) clockwise) results in a sign in front of the residue. 
For Im$(u) >0$ the continuations  around $b$ are
\bea
f\Big(
\cwanalyticcontinuation{u}{b}
\Big) = f(u)-2\pi i g(u)\mbox{Res}(K)(u), \nonumber \\
f\Big(
\ccwanalyticcontinuation{u}{b}
\Big) = f(u)+2\pi i g(u)\mbox{Res}(K)(u),
\eea
where the subscripts indicate in which direction (see fig. \ref{fig:continuations} for clockwise or see fig. \ref{fig:continuations2} for counter clockwise) and around which point. At the left end point $a$ of the integration contour the continuations are given by
\bea
f\Big(
\cwanalyticcontinuation{u}{a}
\Big) = f(u)+2\pi i g(u)\mbox{Res}(K)(u), \nonumber \\
f\Big(
\ccwanalyticcontinuation{u}{a}
\Big) = f(u)-2\pi i g(u)\mbox{Res}(K)(u),
\eea
where each time we continue exactly the same contribution is picked up. This might suggest that these types of continuation are always of $\log$ type, but if $g$ has non-trivial behaviour around these branch points this does not have to be the case. Note that having square-root branch-points is very convenient, since this implies that does not matter in which direction we are continuing. 
\\[5mm]
This leads us to consider a slightly more complicated case: the case where $g$ has a branch point at $a$ and at $b$. Consider therefore the exact same case as we did above (so still Im$(u)>0$), but this time, let $g$ have square-root branch points at both $a$ and $b$. There are again four possible continuations as we treated in the first case. The easiest cases remain to be the clockwise continuation around $b$ and the counter clockwise continuation around $a$; the disturbance of the contour is minimal. The two other cases can be seen best by letting the pole cross the contour before continuing, as we do in fig. \ref{fig:continuations2}. This leads to extra Cauchy residues which change when we are continuing. Let us give formulas for these cases: we start with the expression 
\be
f(u) = \int_{[a,b]} dz g(z) K(z,u),
\ee
for $u$ in the upper half-plane and such that the pole of $K(z,u)$ lies just above the integration contour. Deform the contour as in fig. \ref{fig:continuations2} to $U$, picking up the residue to obtain
\be
f(u) =  \int_{U} dz g(z) K(z,u)+2\pi i \mbox{Res}(K) g(u).
\ee
Now we can continue $f$ and end up with
\be
f\Big(
\ccwanalyticcontinuation{u}{b}
\Big) = \int_{[a,b]} dz g(z) K\Big(z,\ccwanalyticcontinuation{u}{b}\Big)+2\pi i \mbox{Res}(K) g\Big(\ccwanalyticcontinuation{u}{b}\Big),
\ee
where we deformed the contour back to the original interval somewhere during the continuation and can actually forget about the continuation of the integral since the kernel has no branch points. 
Note that the end result depends on the branch point behaviour of $g$. Similarly, for the point $a$ we get
\be
f(u) =   \int_{U} dz g(z) K(z,u)+2\pi i \mbox{Res}(K) g(u),
\ee
which after continuation becomes 
\be
f\Big(\cwanalyticcontinuation{u}{a}\Big) = \int_{[a,b]} dz g(z) K\Big(z,\cwanalyticcontinuation{u}{a}\Big)+2\pi i \mbox{Res}(K) g\Big(\cwanalyticcontinuation{u}{a}\Big).
\ee
To summarise, the results for these cases are
\bea
\label{disc1}
f\Big(\cwanalyticcontinuation{u}{b}\Big) &=& f(u)-(2\pi i) \mbox{Res}(K) g(u), \nonumber \\
f\Big(\ccwanalyticcontinuation{u}{b}\Big) &=& f(u)+(2\pi i) \mbox{Res}(K) g\Big(\ccwanalyticcontinuation{u}{b}\Big),\nonumber \\
f\Big(\cwanalyticcontinuation{u}{a}\Big) &=& f(u)+(2\pi i) \mbox{Res}(K) g\Big(\cwanalyticcontinuation{u}{a}\Big), \nonumber \\
f\Big(\ccwanalyticcontinuation{u}{a}\Big) &=& f(u)-(2\pi i)\mbox{Res}(K) g(u).
\eea
Continuing $f$ twice in the same direction leads to
\begin{equation}
f\Big(u_{
\begin{tikzpicture}
\draw[<-] (0.15,-0.03) arc (350:0:0.15cm);
\node at (0,0) {\scalebox{0.75}{b}};
\end{tikzpicture}
\begin{tikzpicture}
\draw[<-] (0.15,-0.03) arc (350:0:0.15cm);
\node at (0,0) {\scalebox{0.75}{b}};
\end{tikzpicture}
}
\Big) = f(u) + (2\pi i) \text{Res}(K) \left(g\Big(\ccwanalyticcontinuation{u}{b}\Big)-g\Big(u_{
\begin{tikzpicture}
\draw[<-] (0.15,-0.03) arc (350:0:0.15cm);
\node at (0,0) {\scalebox{0.75}{b}};
\end{tikzpicture}
\begin{tikzpicture}
\draw[<-] (0.15,-0.03) arc (350:0:0.15cm);
\node at (0,0) {\scalebox{0.75}{b}};
\end{tikzpicture}
}
\Big)  \right),
\end{equation}
such that for functions $g$ satisfying 
\begin{equation}
g\Big(\ccwanalyticcontinuation{u}{b}\Big)-g\Big(u_{
\begin{tikzpicture}
\draw[<-] (0.15,-0.03) arc (350:0:0.15cm);
\node at (0,0) {\scalebox{0.75}{b}};
\end{tikzpicture}
\begin{tikzpicture}
\draw[<-] (0.15,-0.03) arc (350:0:0.15cm);
\node at (0,0) {\scalebox{0.75}{b}};
\end{tikzpicture}
}
\Big)  =0
\end{equation}
-- such as a pure square-root -- the double continuation returns the original function $f$. We will see that these results are going to be central in the analysis of TBA equations.  
\\[5mm]
It is clear from this analysis that only the points $a$ and $b$ are possible branch points of $f$. Continuing around any other point will return the function to the same value as one can see as follows: take any point to continue around and start continuing. As the pole gets close to the contour we shift the contour far enough to allow for the pole to complete its circle around our point under investigation. We relax the contour back to its original form after the circle has been completed. Thus only $a$ and $b$ are special points for $f$. Moreover, if the integration contour is not ``incomplete" there are no special points and therefore $f$ does not inherit any branch cuts. 
\\[5mm]
The periodic case might deserve special attention here: the ``full" interval in that case is simply $(-\pi,\pi]$, which seems to define what we have previously called incomplete-interval convolution. The kernel $K$, however, is periodic and therefore cannot have a unique pole on the complex plane, invalidating the results we have obtained. We can instead consider $f$ as a function on the complex cylinder $-\pi < \text{Re}(u) \leq \pi$, such that $K$ does have a unique pole. In turn, the integration domain becomes closed on the cylinder and the points $-\pi$ and $\pi$, which are identified on the cylinder, are not special: we can follow the deformation from the previous paragraph to show that a continuation around those points is trivial. The other results from this section do carry over to the periodic case, as they do not depend on any properties away from the branch cut.
 
\subsection{Discontinuities, continuations and short- and long-cutted functions}  The analytic continuation of a function can be encoded in a lot of ways and can be subtle to define. In this thesis we will often talk about both analytic continuation and \emph{discontinuities} or \emph{jump} of functions, so we should make sure that our definitions are clear and consistent. Although the notation introduced in the previous section is very explicit, it will be cumbersome to use in the rest of this thesis. Instead we will use the following:

\paragraph{Discontinuity.}\label{sec:discontinuity} For a function $f: \C \rightarrow \C$ we define its discontinuity across its cut on $Z_N$ as 
\be
\label{eq:discontinuity}
\left[ f \right]_N := \lim_{\e \rightarrow 0^+} f(u+ i N \ad +i \e) - f(u+ i N \ad -i \e),
\ee
where $u$ lies on the cut.\footnote{The undeformed versions of all these definitions follow directly by replacing $\ad \rightarrow 1/g$.} In practice we will almost exclusively use this notation for functions with only $\check{Z}$ cuts, such that $u \in \check{Z}_N$. Therefore this function is initially only defined on this interval. It can be continued to the whole complex plane, where it coincides with $f(u+i N \ad) - f(u^*+i N \ad)$, where here the continuation is performed after the shift. Thus we can regard discontinuities as functions on the whole complex domain as 
\be
\left[ f \right]_N (u) = f(u+i N \ad) - f(u^*+i N \ad).
\ee
Another way to denote this is as follows, using our shift notation \eqref{eq:shift}
\begin{equation}
\left[ f \right]_N (u) = f^{[N]} - \widetilde{f^{[N]}}.
\end{equation}
Here the tilde on top of the function indicates we are considering its second-sheet evaluation, i.e. 
\begin{equation}
\tilde{f}(u) = f\left(u^*\right). 
\end{equation}
Both notations have their merits and we will use both accordingly. The function $\tilde{f}$ can not be regarded as an analytic continuation of $f$ in the strict sense, since by definition $\tilde{f}$ does not coincide with $f$ anywhere on the first sheet. Instead, $\tilde{f}(u)$ for $u$ in the upper half-plane is part of the analytic continuation of $f(u)$ with $u$ in the lower half-plane and vice versa, consistent with the definition of the $*$. 

\paragraph{Short- and long-cutted functions.}\label{sec:shortcuttedfunctions} Basically all the branch points in the spectral problem are of square-root type and they come in pairs. This creates the freedom how to connect these branch points with a branch cut, which we have resolved by choosing them to be ``long" or ``short".  Whenever we want to emphasise the choice of branch cuts we will write $\hat{f}$ if $f$ has short cuts and $\check{f}$ when it has long cuts.\footnote{This goes against a convention in part of the literature which uses the hat and the check in exactly the opposite situations to match the definition of the convolutions present in the TBA equations. It is in line with more modern literature on the quantum spectral curve, which is why we follow it here.} The freedom of connecting branch points allows us in some cases to consider functions defined on some open domain and consider either the \emph{long-cutted} or the \emph{short-cutted} extension of this function. 
\\[5mm]
The most convenient choice here depends on the definition of the basic functions in the model: this poses an inconvenience for us, since these basic functions -- in our case the $x$ functions -- have been defined differently throughout the literature (see the appendix G of \cite{Cavaglia:2010nm} for the ``Rosetta stone" for AdS/CFT). The literature we follow here are the TBA papers \cite{Arutyunov:2009ur, Bombardelli:2009ns, Cavaglia:2010nm,Balog:2011nm,Arutynov:2014ota} and the later quantum spectral curve papers \cite{Gromov:2011cx,Gromov:2013pga,Gromov:2014caa,Marboe:2014gma, Kazakov:2015efa}, where the former use what we will call \emph{lower-half-plane conventions} and the latter \emph{upper-half-plane conventions}. In the lower- (upper)-half-plane conventions the $x_s$ and $x_m$ functions are chosen to coincide in the lower (upper) half-plane. This means that if a function is specified by its dependence on the $x$ function we can easily change its cuts from short to long or vice versa by changing the $x$ function, but its precise form will depend on the choice of conventions. To stay close to the literature we will use the lower-half-plane conventions for most of the derivation and switch to the upper-half-plane conventions when we have extracted all the information from the TBA equations as we need. 
\\[5mm]
This has an effect on the precise definition of $\hat{f}$ and $\check{f}$, which we have not been very elaborate about so far. Functions with branch cuts do not have to be either short- or long-cutted, since in principle we could choose to connect different pairs of branch points with different cuts, leading to mixed structures and a seemingly endless stream of possible extensions of a function defined only on the physical strip. We will restrict in this thesis (almost)\footnote{The only exceptions are the two functions $\dot{\mbT}_{1,\pm 1}$ in the analsis of the $\mbT$ gauge (see \ref{sec:step41}).} exclusively to short- and long-cutted functions. In the first part of this thesis we will let $\hat{f}$ ($\check{f}$) denote the short-cutted (long-cutted) extension of a function defined on the \emph{lower} half of the physical strip, in line with the lower-half-plane conventions. In the second part we will switch to the upper-half-plane conventions, such that $\hat{f}$ ($\check{f}$) denotes the short-cutted (long-cutted) extension of a function defined on the \emph{upper} half of the physical strip. 
\section{Simplified TBA-equations}
\label{sec:simpTBA}
The (canonical) TBA-equations \eqref{eq:TBAeqns} can be simplified to the appropriately named \emph{simplified} TBA-equations, by acting on both sides of the TBA equations with the operator
\begin{equation}
\label{eq:K+1invdef}
(K+1)^{-1}_{PQ} \defeq \delta_{P,Q} - (\delta_{P,Q+1}+ \delta_{P,Q-1}) \s,
\end{equation}
where $\s$ is the doubly-periodic analogue of the undeformed kernel $s(u) = (4 \cosh{\frac{\pi u}{2}})^{-1}$,
\begin{equation}
\label{eq:sigmakerneldef}
\mathbf{s}(u)  \defeq \sum_{n \in \mathbb{Z}} \tfrac{1}{\ad} s\left(\tfrac{u + 2\pi n}{\ad}\right) =  \sum_{n\in \mathbb{Z}} \frac{1}{4 \ad \cosh \frac{\pi (u+ 2\pi n)}{2 \ad}} = \tfrac{K(m^\prime)}{2\pi \ad} \mbox{dn}(u)\, ,
\end{equation}
where dn is the corresponding Jacobi elliptic function with real period $2\pi$ and imaginary period $4 \ad$, and $K(m^\prime)=\sum_{l \in \mathbb{Z}} \frac{\pi}{2\cosh\pi^2 l/\ad}$ is the elliptic integral of the complementary elliptic modulus. This gives the set of simplified TBA equations presented in \cite{Arutynov:2014ota}
\begin{equation}
\begin{aligned}
\label{eq:simpTBA}
\log Y_1 = \, & \sum_\alpha L_{-}^{\alpha} \, \hat{\star}\, \s - L_2 \star \s -\ewD\,\check{\star}\, \s\, , \\
\log Y_Q = \, &  -(L_{Q-1} + L_{Q+1})\star \s +  \sum_\alpha L_{Q-1|vw}^{(\alpha)} \star \s, \qquad \qquad Q>1,\\
\log Y_{+}^{(\alpha)}/Y_{-}^{(\alpha)} = & \,\Lambda_Q \star K_{Qy}\,,\\
\log Y_{-}^{(\alpha)}Y_{+}^{(\alpha)}  = & \,\Lambda_Q \star (2K_{xv}^{Q1}\star \s - K_{Q})+
 2(\Lambda_{1|vw}^{(\alpha)}-\Lambda_{1|w}^{(\alpha)}) \star \s\,,\\
\log{Y_{M|vw}^{(\alpha)}} = & \,(\Lambda_{M+1|vw} + \Lambda_{M-1|vw}) \star  \s - \Lambda_{M+1} \star  \s + \delta_{M,1}(\Lambda_{-}^{(\alpha)}-\Lambda_{+}^{(\alpha)})  \, \hat{\star} \, \s\, ,\\
\log{Y_{M|w}^{(\alpha)}} = & \, (\Lambda_{M+1|w} + \Lambda_{M-1|w})\star  \s + \delta_{M,1} (L_{-}^{(\alpha)}-L_{+}^{(\alpha)}) \, \hat{\star} \, \s\, ,
\end{aligned}
\end{equation}
where $Y_{0|(v)w}^{(\alpha)}=0$ and $\ewD$ is defined as\footnote{Our usage of the hat on $\ewD$ is different from previous literature on the TBA equations, see footnote  \footnotemark[3]}
\begin{align}
\ewD = J \hat{\mathcal{E}} + \sum_{\alpha} \left( L^{(\alpha)}_{-} +  L^{(\alpha)}_{+} \right) \star \hat{K}  + 2 \Lambda_Q\star \hat{K}_Q^{\Sigma} + \sum_{\alpha} L_{M|vw}^{(\alpha)}\star \hat{K}_M,
\end{align}
with $\hat{\mathcal{E}}$ defined as
\be
\label{eq:energyterm}
\hat{\mathcal{E}}(u) = \log \frac{x(u-i\e)+\xi}{x(u+i\e)+\xi}, \quad \hat{\mathcal{E}}^{\text{und}}(u) = \log \frac{x(u-i\e)}{x(u+i\e)}.
\ee
To work out the action of $(K+1)^{-1}_{PQ}$ on the TBA equations we can make use of various identities between the integration kernels $K$, which are valid in both the deformed and undeformed cases some of which can be found in appendix \ref{app:Definitions}. We refer to \cite{vanTongeren:2013gva} for more details on the simplified TBA-equations. 

\section{$Y$-system equations}
A remarkable property of the TBA equations \eqref{eq:TBAeqns} is that they can be considered as a set of equations constraining solutions with particular boundary conditions of a much more general system of equations known as the \emph{$Y$ system}. The $Y$ system is a set of functional finite-difference equations whose form seems to only depend on the representation-theoretical structure of the model under consideration. In particular, since the representation theory of the $\eta$-deformed and undeformed model is essentially the same the associated $Y$ system equations are identical for the two cases!\footnote{As a side note, let us mention that the root-of-unity $Y$-system is very different and has only a finite number of $Y$ functions \cite{Arutyunov:2012zt}.} The deformed and undeformed case are distinguished only by which solutions of the $Y$-system equations we are considering, or differently put on which function space we try to solve these equations. We specify the wanted class of solutions by imposing additional analytical data on the $Y$ system. This data comes in the form of analyticity strips and discontinuity equations, both of which are encoded in the TBA equations. It is in this sense that the TBA equations are a refinement of the $Y$ system to a specific model and can be used in practice to find numerical solutions and consider uniqueness and existence of solutions with certain boundary conditions. THe relation between the TBA and $Y$ system has been analysed in great detail in the recent work \cite{Hilfiker:2017jqg}. 

\subsection{Deriving the $Y$-system equations}
To get to the $Y$ system we define the operator $\s^{-1}$ by its action on a function $f$ as
\begin{equation}
\label{eq:sinvdef}
 f \circ \s^{-1}(u) = \lim_{\epsilon\rightarrow 0} f(u + i \ad - i\epsilon) +  f(u - i \ad + i\epsilon)\, ,
\end{equation}
so that it acts as a right inverse of $f\mapsto f\star \s$:
\begin{equation}
(f \star \s) \circ \s^{-1}(u) = f(u)\, , \, \, \, \, \mbox{for} \, \, u \in Z_0 \, .
\end{equation}
The fact that $\s^{-1}$ has a non-trivial kernel -- note that it kills all $2i\ad$ functions -- implies that analytical information is lost when acting with $\s^{-1}$ on the TBA equations\footnote{This also implies $\s^{-1}$ is not a full inverse of $\s$.}, consistent with the fact that the deformed and undeformed TBA-equations both lead to the same $Y$ system (up to a rescaling of the shift parameter), which for $u\in Z_0$ reads as follows:
\paragraph{Q particles}
\begin{align}
\label{Ysys:Q}
\frac{Y_1^+ Y_1^-}{Y_2} & = \frac{\prod_{\alpha}
\left(1-\frac{1}{Y^{(\alpha)}_{-}}\right)}{1+Y_2}\, , \, \, \, \, \, \mbox{for} \, \,
u\in \hat{Z}_0\, ,\\
\frac{Y_Q^+ Y_Q^-}{Y_{Q+1}Y_{Q-1}} & = \frac{\prod_{\alpha} \Bigg(
1+\frac{1}{Y_{Q-1|vw}^{(\alpha)}}\Bigg)}{(1+Y_{Q-1})(1+Y_{Q+1})} \, .
\end{align}
\paragraph{$w$ strings}
\begin{align}
\label{Ysys:w}
Y_{1|w}^+ Y_{1|w}^- & =(1+Y_{2|w})\left(\frac{1-Y_-^{-1}}{1-Y_+^{-1}}\right)^{\vartheta(\theta-|u|)},  \\
Y_{M|w}^+ Y_{M|w}^- & =(1+Y_{M-1|w})(1+Y_{M+1|w})\,,
\end{align}
\paragraph{$vw$ strings}
\begin{align}
\label{Ysys:vw}
Y_{1|vw}^+ Y_{1|vw}^- & =\frac{1+Y_{2|vw}}{1+Y_2}\left(\frac{1-Y_-}{1-Y_+}\right)^{\vartheta(\theta-|u|)},  \\
Y_{M|vw}^+ Y_{M|vw}^- & =\frac{(1+Y_{M-1|vw})(1+Y_{M+1|vw})}{1+Y_{M+1}} ,
\end{align}
\paragraph{$y$ particles}
\begin{equation}
\label{Ysys:y}
Y_{-}^{+}Y_{-}^{-} = \frac{1+Y_{1|vw}}{1+Y_{1|w}}\frac{1}{1+Y_1}\, .
\end{equation}
As for the undeformed string, there is no $Y$ system equation for $Y_+$, and to get this equation for $Y_-$ we used the identities $K^{Qy}_- \circ s^{-1} =  K_{xv}^{Q1} + \delta_{Q,1}$ and $K_M \circ s^{-1} = K_{M1} + \delta_{M,1}$. In the equations for $(v)w$ strings and $y$ particles we have suppressed the $\alpha$ index. $\vartheta$ is the usual Heaviside function. It is noteworthy that the $Y$-system equations for a particular $Y$ function only feature $Y$ functions which are adjacent to it on the $Y$ hook (see fig. \ref{fig:Yhook}). 
\\[5mm]
To find a system that is truly equivalent to the original TBA-equations our next task is to find the extra analyticity data in the form of discontinuity equations. 

\section{Discontinuity equations}

Our goal is to find enough equations such that when imposed on the $Y$-system equations we have enough information to rederive the TBA equations from them, thereby showing equivalence of the two systems. Naturally, the discontinuities of the $Y$ functions are related through the $Y$-system equations. This allows us to only impose a very small set of discontinuities and therefore we only need to analyse the TBA equations for a small number of cases. We will go through all of them here. The undeformed case was analysed in \cite{Cavaglia:2010nm}.

\subsection{$Y_{w}$}
\label{sec:discYw}
The first discontinuity for the $Y_{1|w}$ (for both values of $\alpha$) lies on $Z_{\pm 1}$. As argued in section \ref{sec:discontinuityfiniteinterval} the only term generating a discontinuity there is the one with the $\hat{Z}$ convolution. For $Y_{1|w}$ this term reads
\be
\left(L_{-}-L_{+}\right)\hat{\star} K_1,
\ee
where the kernel $K_1$ has poles at $\pm i\ad$ in the physical strip. Continuing the function $H_{\pm}$ defined by
\be
H_{\pm}(u) \defeq \int_{\hat{Z}_0} dv \left(L_{-}-L_{+}\right)(v) K_1(v-u\mp \ad i),
\ee
around the branch point at $\theta$ we find that
\be
\left[H \right]_{\pm1} (u) = L_{-}(u)-L_{+}(u),
\ee
using eqn. \eqref{disc1}. Since $Y_+$ and $Y_-$ are related by analytic continuation, the discontinuity of $\log Y_{1|w}^{(\alpha)}$ can be written as
\be
\left[\log Y_{1|w}\right]_{\pm1}(u) =\left[L_{-}\right]_0(u).
\ee
\subsection{$Y_{vw}$}
The derivation for $Y_{1|vw}^{(\alpha)}$ is very similar, but relies on kernel identities to first rewrite the TBA equation for $Y_{1|vw}^{(\alpha)}$ in a convenient form:
\be
\label{eq:rewriteYvw1TBA}
\log Y_{M|vw}^{(\alpha)} = L_{N|vw}^{(\alpha)} \star K_{NM} + \left(\Lambda_{-}^{(\alpha)}-\Lambda_{+}^{(\alpha)}\right)\hat{\star} K_M -\sum_{Q=2}^{\infty}\Lambda_Q \star K_{Q-1}.
\ee
In this form we can continue by analogy with the previous case and conclude that
\be
\label{eq:discYvw}
\left[\log Y_{1|vw}^{(\alpha)}\right]_{\pm1}(u) =  \left(\Lambda_{-}^{(\alpha)}-\Lambda_{+}^{(\alpha)}\right)(u) =\left[\Lambda_{-}^{(\alpha)}\right]_0(u).
\ee
In order to arrive at the form \eqref{eq:rewriteYvw1TBA} of the TBA equation we use the kernel identity 
\begin{equation}
K_{xv}^{Q1}(u,v) = K_{Q-1}(u-v) +\int_{\hat{Z}_0} dx K_{Qy} (u,x)K_1(x-v),
\end{equation}
which holds on the physical strip as can be checked numerically. The last two terms in the TBA-equation for $Y_{1|vw}$ \eqref{eq:TBAeqns} can now first be rewritten as
\bea
\left(L_{-}^{(\alpha)}-L_{+}^{(\alpha)}\right)\hat{\star} K_1 -\Lambda_Q \star K_{xv}^{Q1} = \left(\Lambda_{-}^{(\alpha)}-\Lambda_{+}^{(\alpha)}\right)\hat{\star} K_1 -\log Y_-/Y_+ \hat{\star} K_1 -\Lambda_Q \star K_{xv}^{Q1}, \quad \,\,
\eea
where the first term is of the form we wanted. Continuing with the other terms and using the equation \eqref{eq:Y-nonlocal} derived below we find
\bea
\left(\Lambda_Q \star K_{Qy}\right)\hat{\star} K_1 - \Lambda_Q \star K_{xv}^{Q1} &=&\Lambda_Q \star \left(K_{Qy}\hat{\star} K_1\right) - \Lambda_Q \star K_{xv}^{Q1} \nonumber\\
= \sum_{Q=1}^k \Lambda_Q \star \left( K_{Qy}\hat{\star} K_1 - K_{xv}^{Q1}\right) 
&=&  -\sum_{Q=2}^k \Lambda_Q \star K_{Q-1},
\eea
where we used that $K_0=0$. This shows that the TBA equation can be rewritten in the form \eqref{eq:rewriteYvw1TBA} and the discontinuity result \eqref{eq:discYvw} follows. 

\subsection{$Y_-$}
Since the $Y$ functions $Y_{\pm}$ are each others continuation finding the $Z_0$ discontinuity of $Y_-$ is very easy: it follows directly that
\begin{equation}
\label{eq:Y-nonlocal}
\left[  \log Y_- \right]_0 = \log Y_- -\log Y_+ =  -\Lambda_P \star K_{Py},
\end{equation}
using the TBA equations \eqref{eq:TBAeqns}. This equation is not very convenient though, since it still contains a convolution, i.e. it is \emph{non-local}. In order to continue our simplification it is necessary to derive an equation for the $Y_{\pm}$ discontinuities that only contains the discontinuities of other $Y$ functions.
\subsection{A local equation for $Y_-$}
\label{sec:derivelocalequation}
Finding a local equation can be done by analysing the non-local one we just derived: we simply compute the $Z_{2N}$ discontinuity of the second and third expression in the non-local equation \eqref{eq:Y-nonlocal}: for $N\in \N$ we find
\begin{equation}
\label{eq:Y-local}
\left[   \log Y_-/Y_+ \right]_{\pm 2N} = \left[   -\Lambda_P \star K_{Py} \right]_{\pm 2N} = - \sum_{P=1}^N \left[ \Lambda_P\right]_{\pm(2N-P)},
\end{equation}
which follows from the pole structure of $K_{Qy}$. We refer to appendix \ref{app:Ysystemcomputations} for the explicit computation. 
\subsubsection{Showing it is equivalent to the non-local equation}
\label{sec:discYmin}
To make sure that the local discontinuity equation \eqref{eq:Y-local} for $Y_-$ can be imposed on the $Y$-system equations instead of the non-local equation \eqref{eq:Y-nonlocal} we consider whether they are equivalent while assuming the $Y$-system equations. We have already shown that the local equation can be derived from the non-local one in the previous section, so to prove equivalence we now consider how to derive the non-local equation from the local one.\footnote{See \cite{Cavaglia:2010nm} for this derivation in the undeformed case.} 

\begin{figure}[!t]
\centering
\begin{subfigure}{7cm}
\includegraphics[width=7cm]{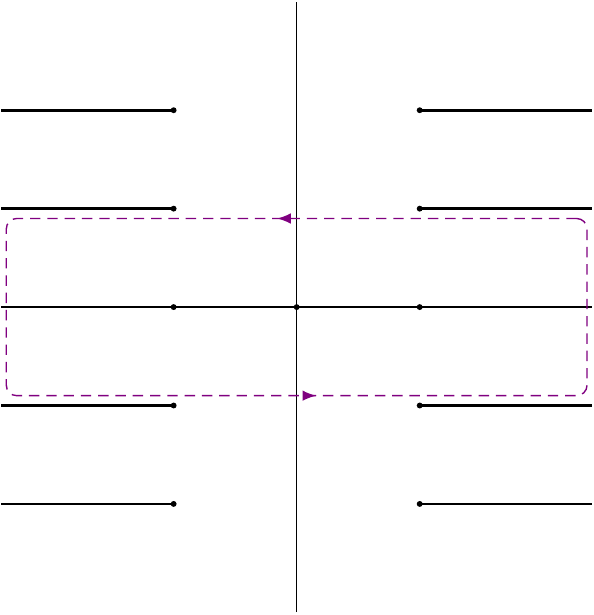}
\caption{}
\end{subfigure}
\quad \,\,
\begin{subfigure}{7cm}
\includegraphics[width=7cm]{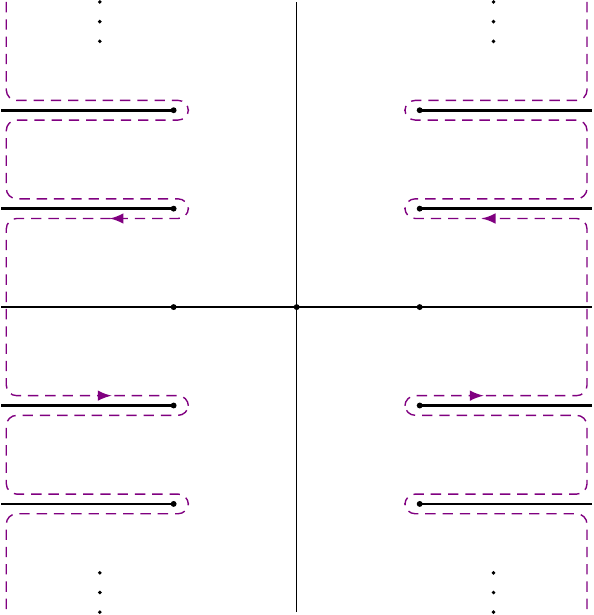}
\caption{}
\end{subfigure}
\caption{Undeformed case: the integration contours $\gamma$ (in (a)) and $\Gamma$ (in (b)) are indicated by the dashed lines, which extend to $\pm \infty$. The dots in (b) indicate that the contour continues to $\pm \i \infty$. The solid lines are the branch cuts connecting $2+ i N/g $ to $-2+ i N/g$ for $N\in \Z$.}
\label{fig:undeformedgamma}
\end{figure}

\begin{figure}[!t]
\centering
\begin{subfigure}{5cm}
\includegraphics[width=5cm]{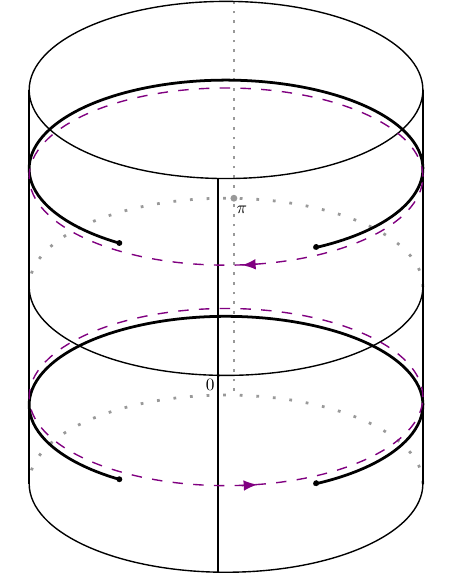}
\caption{}
\end{subfigure}
\quad
\begin{subfigure}{7cm}
\includegraphics[width=7cm]{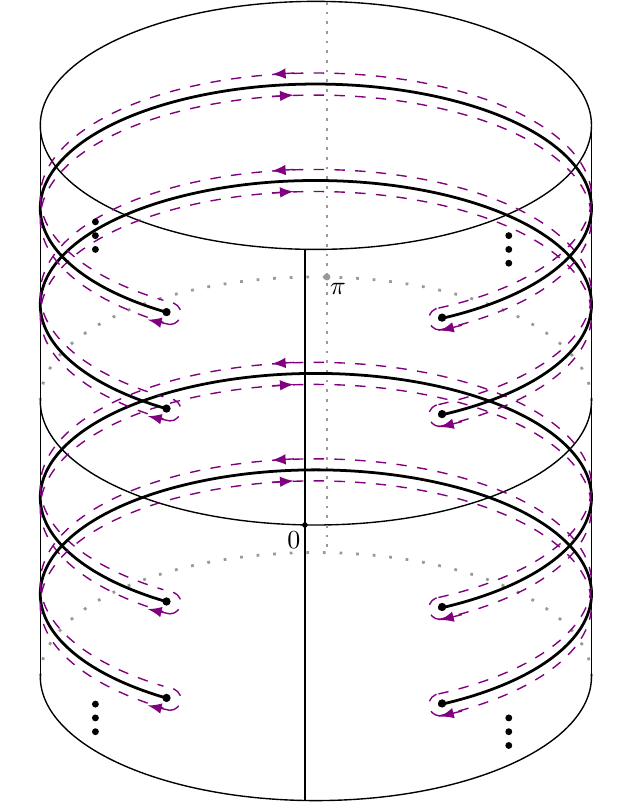}
\caption{}
\end{subfigure}
\caption{Deformed case: the integration contours $\gamma$ (in (a)) and $\Gamma$ (in (b)) are indicated by the dashed lines. The dots in (b) indicate that the contour continues to $\pm i\infty$ along the cylinder in both directions. The solid lines are the branch cuts connecting $\theta+ i\ad N $ to $-\theta+ i\ad N$ for $N\in \Z$.}
\label{fig:deformedgamma}
\end{figure}

\paragraph{Two approaches.} We can treat both the deformed and undeformed case simultaneously. The algebraic details of the deformed and undeformed derivations are almost identical, the crucial differences sit in the analytical properties of the functions involved: the basic idea of this derivation is to consider the function $F(u)=\log \frac{Y_-}{Y_+}(u)$ as some contour integral of the form
\begin{equation}
F(u) = \oint_{\gamma} \frac{dz}{2\pi i} P(z-u) F(z),
\end{equation}
with $\gamma$ some contour and $P$ some function with exactly one first-order pole with residue $1$ inside $\gamma$. We can then deform the contour and obtain integrals over the local discontinuity equations to end up with the non-local ones. Since $F$ has a square-root branch cut on the real line we cannot write it as a contour integral directly. We instead try to find a function $G$ such that the product $G\cdot F$ is branch-cut free on the real line and we can proceed. It is tempting to choose the simplest function $G$ that has a square-root branch cut, and in the undeformed case this is indeed enough as we will see. However, the fact that $G$ should satisfy a further constraint makes its construction a bit harder in the deformed case. We ultimately want to find the non-local discontinuity which contains the kernel $K_{Py}$, which has to be built up from two ingredients: the functions $G$ and $P$. In the undeformed case the simplest function $P^{\text{und}}$ is the canonical $1/u$. A natural periodic analogue of this is 
\begin{equation}
P(u) = \frac{1}{2 \sin \frac{u}{2}},
\end{equation}
which will suffice for this derivation.\footnote{Later in this thesis we will need a function with the same properties, but with \emph{all} its poles having residue $1$. To achieve this we turn to the $\cot$ version of the kernel considered here.} The condition to ultimately find $K_{Py}$ later in the derivation is given by
\begin{equation}
K(z,u) = \frac{1}{2\pi i} P(z-u) \frac{G(z)}{G(u)},
\end{equation}
which we can solve for both cases. In the undeformed case it follows directly from the definition of $K(z,u)$ that we can use 
\begin{equation}
G^{\text{und}}(u) =\frac{1}{\sqrt{4-u^2}},
\end{equation}
which is the simplest function with a square-root branch cut that falls off as $u \rightarrow \pm \infty$. In the  deformed case we find the solution more implicitly as
\begin{equation}
\label{eq:squarerootG}
G(u) = \frac{1}{4\pi \ii} \frac{1}{\sin\tfrac{1}{2}(1-u)K(1,u)},
\end{equation}
where the choice to evaluate in $z=1$ is arbitrary. Finally, to make all of this fit together properly we need to define a proper contour $\gamma$. The undeformed case is illustrated in fig. \ref{fig:undeformedgamma}, the deformed case in fig.  \ref{fig:deformedgamma}. Note that both $\gamma$ contours enclose exactly one pole of the pole function $P$ to make the rewriting possible and that the undeformed $\gamma$ only yields the correct result because we assume that $F$ falls off sufficiently fast as $u\rightarrow \pm \infty$. In the deformed case we do not need this assumption, since the $2\pi$ periodicity allows for exact cancellation of the vertical contour parts. The same is true for the contour deformation into $\Gamma$, see fig. \ref{fig:undeformedgamma} and \ref{fig:deformedgamma}.
\paragraph{Rewriting the result.} With these definitions we can proceed. First, we define the function
\be
B(z,u) =P(z-u)G(z) = G(u) 2\pi i K(z,u),
\ee
where $B$ is $2\pi$-periodic for the deformed case. A simple rewriting shows that 
\bea
\label{eq:Tproof}
&-&\sum_{Q=1}^{\infty} \int_{Z_0} \frac{dz}{2\pi i} \Lambda_Q(z)\left(B(z-iQ \ad,u) -B(z+iQ\ad,u) \right) \nonumber \\
&=& \sum_{Q=1}^{\infty} \int_{Z_0} dz G(u) \Lambda_Q(z)\left(K(z+iQ\ad,u) -K(z-iQ\ad,u) \right) \nonumber \\
&=& G(u)\sum_{Q=1}^{\infty} \int_{Z_0} dz \Lambda_Q(z) K_{Qy}(z,u) = G(u)\left(\Lambda_Q\star K_{Qy}(u)\right).
\eea
With this result and the non-local discontinuity equation \eqref{eq:Y-nonlocal} we see that the equation
\be
\label{eq:Grel}
G(u)\log \frac{Y_-}{Y_+}(u) = \sum_{Q=1}^{\infty} \int_{Z_0} \frac{dz}{2\pi i} \Lambda_Q(z)\left(B(z-iQ\ad,u) -B(z+iQ\ad,u) \right)
\ee
is equivalent to the equation
\be
\log \frac{Y_-}{Y_+}(u) =  -\Lambda_P \star K_{Py}.
\ee
So in order to prove equivalence of the two sets of discontinuities it is enough to derive eqn. \eqref{eq:Grel} from the local discontinuities \eqref{eq:Y-local} as we will do below.
\paragraph{Rewriting a contour integral.} We can write the left-hand side of eqn. \eqref{eq:Grel} as
\be
G(u) \log \frac{Y_-}{Y_+}(u) = \oint_{\gamma} \frac{dz}{2\pi i}\log \frac{Y_-}{Y_+}(z)B(z,u),
\ee
since $\log \frac{Y_-}{Y_+}(z)G(z)$ is analytic on the inside of $\gamma$ and the integrand has a first-order pole with residue $1$. We can deform $\gamma$ into $\Gamma$ (see fig. \ref{fig:undeformedgamma} and fig. \ref{fig:deformedgamma}) using that there are no other poles present. We can rewrite the contour integral using the local discontinuities:
\bea
&-&\sum_{N,\tau}\left(\int_{Z_{-2N\tau}+i\e}- \int_{Z_{-2N\tau} -i\e} \right)\frac{dz}{2\pi i} \sum_{P=1}^{N}   \Lambda_P (z+i \tau P \ad)B(z,u) \nonumber \\
&=&-\sum_{N,\tau} \int_{Z_0}\frac{dz}{2\pi i} \sum_{P=1}^{N} \left[\Lambda_P\right]_{-\tau (2N-P)}(z)  B(z-2i\tau N \ad,u) \nn
&=& \sum_{N,\tau}\int_{Z_0}\frac{dz}{2\pi i} \left[ \log \left(Y_-/Y_+\right) \right]_{-\tau 2 N}(z) B(z-2i\tau N \ad,u)\nonumber \\
&=& \sum_{N,\tau}\int_{Z_0}\left(  \log \left(Y_-/Y_+\right)(z-i2N\tau\ad+i\e)-\log Y_-/Y_+(z-i2N\tau \ad-i\e) \right)B(z-2i\tau N \ad,u) \nonumber \\
&=&\oint_{\Gamma}\frac{dz}{2\pi i}\log \left(Y_-/Y_+\right)(z) B(z,u) =\oint_{\gamma}\frac{dz}{2\pi i} \log \left(Y_-/Y_+\right)(z) B(z,u) ),
\eea
where $\tau$ sums over $\pm 1$ and $N\in \N$ and the $\e$ is infinitesimal.\footnote{Concretely this means that one should read expressions containing $\e$ as in the limit $\e\rightarrow 0^+$. } This shows that
\bea
\label{eq:discYy1}
& &\oint_{\gamma}\frac{dz}{2\pi i}\log \left(Y_-/Y_+\right)(z) B(z,u) \nn
&=& -\sum_{N,\tau}\left(\int_{Z_{-2N\tau}+i\e}- \int_{Z_{-2N\tau} -i\e} \right)\frac{dz}{2\pi i} \sum_{P=1}^{N}   \Lambda_P (z+i \tau P \ad)B(z,u).
\eea
Using the fact that $\Lambda_P$ and $B$ do not have any poles except at the branch points allows us to cancel integrals on the right-hand side by deforming the relevant contours:
\bea
&\phantom{=}& -\sum_{\tau} \sum_{N} \sum_{P=1}^{N} \left(\int_{Z_{-2N\tau }+i\e}- \int_{Z_{-2N\tau } -i\e} \right)\frac{dz}{2\pi i}  \Lambda_P (z+i \tau P \ad)B(z,u)\nonumber \\
&=& -\sum_{\tau} \sum_{P=1}^{\infty} \sum_{N=P}^{\infty}\left(\int_{Z_{-2N\tau }+i\e}- \int_{Z_{-2N\tau } -i\e} \right)\frac{dz}{2\pi i}  \Lambda_P (z+i \tau P \ad)B(z,u)\nonumber \\
&=& \sum_{\tau} \tau \sum_{P=1}^{\infty} \int_{Z_{-2N\tau }+\tau i\e} \frac{dz}{2\pi i}  \Lambda_P (z+i \tau P \ad)B(z,u) \nn
&=& \sum_{\tau} \tau \sum_{P=1}^{\infty} \int_{Z_0} \frac{dz}{2\pi i}  \Lambda_P (z)B(z-i \tau P\ad,u).
\eea
We can rewrite eqn. \eqref{eq:discYy1} as follows:
\bea
\label{eq:finaldiscYv2}
& &\oint_{\gamma}\frac{dz}{2\pi i}\log \left(Y_-/Y_+\right)(z) B(z,u) =
\sum_{\tau} \tau \sum_{P=1}^{\infty} \int \frac{dz}{2\pi i}  \Lambda_P (z)B(z-i \tau P\ad,u) \nonumber \\
&=& \sum_{P=1}^{\infty} \int \frac{dz}{2\pi i} \Lambda_P (z) \left(  B(z-i P\ad,u)-B(z+i P\ad,u) \right) = - G(u)\left(\Lambda_P\star K_{Py}(u)\right),\qquad \quad 
\eea
using the equation \eqref{eq:Tproof}. After recognising the left-hand side as $G(u)\log \left(Y_-/Y_+\right)(u)$ we see that equation \eqref{eq:finaldiscYv2} is equivalent to
\be
\log \frac{Y_-}{Y_+}(u) =  -\Lambda_P \star K_{Py},
\ee
which is what we were after. In conclusion, this proves that the local discontinuities \eqref{eq:Y-local} for $Y_-$ contain exactly the same amount of information as the non-local version \eqref{eq:Y-nonlocal} and we can impose the local discontinuity relations on the $Y$ system.

\subsection{$Y_Q$}
The $Y_Q$ functions are the most intricate functions in the TBA equations and the equations necessary to rederive the TBA equations from the $Y$ system are of a more complicated nature than those we found for the other functions. We will derive a set of discontinuities for the function $\Delta$ which is the long-cutted version of the discontinuity of $\log Y_1$ on $Z_1$. We therefore need to go through two steps: first we understand the $\log Y_1$ discontinuity to find an expression for $\Delta$ and then we derive $\Delta$'s discontinuities. A particular ingredient needed to finish this analysis is finding a simpler expression for the dressing-phase kernel, which in the undeformed case was derived in \cite{Cavaglia:2010nm} based on kernel identities derived in \cite{Arutyunov:2009ux}. To not distract from our main derivation we treat this simplification for the deformed case in appendix \ref{app:dressing}.
\subsubsection{Deriving the $\log Y_1$ discontinuity}
We will first concentrate on the discontinuity of the $Y_1$ function. Analysing the simplified TBA-equation \eqref{eq:simpTBA} for $\log Y_1$ we know immediately from our previous discussions that two of the terms are potential sources of discontinuities and as we will see both in fact are. 
\\[5mm]
Acting with $\s^{-1}$ on the simplified TBA-equation yields the $Y$-system equation \eqref{Ysys:Q} for $\log Y_1$, which we only wrote for $u\in \hat{Z}_0$. The complete expression contains $\ewD$ and reads
\be
\label{line1}
\log Y_1(u +i\ad-i\e) Y_1(u-i\ad+i\e) = \sum_{\alpha} L_-^{(\alpha)}(u) \vartheta(\theta-|u|)-L_2(u) -\ewD(u) \vartheta(|u|-\theta).
\ee
An alternative way to write the discontinuity of $\log Y_1$ is, for a $u$ with $|$Re$(u)|>\theta$ 
\begin{align}
\left[ \log Y_1 \right]_{1} (u) = \log Y_1(u+i\ad+i\e)-\log Y_1(u+i\ad-i\e).
\end{align}
From eqn. \eqref{line1} we already have an expression for $\log Y_1(u+i\ad-i\e)$. We can use the $Y$ system to find the other expression: We have for $|$Re$(u)|<\theta$ that
\be
\log Y_1(u +i\ad) Y_1(u-i\ad) = \sum_{\alpha}L_-^{(\alpha)}(u)-L_2(u).
\ee
We can analytically continue this equation to approach the branch cut on $\check{Z}_1$ from above:
\be
\log Y_1(u +i\ad +i\e) Y_1(u-i\ad +i\e) = \sum_{\alpha}L_-^{(\alpha)}(u+i\e)-L_2(u) \quad \mbox{ with } |\mbox{Re}(u)|>\theta.
\ee
Plugging in this expression into the discontinuity gives us (now for $|$Re$(u)|>\theta$) 
\begin{align}
\label{eq:Delta1}
\hat{\Delta} &\defeq \left[ \log Y_1 \right]_{1} (u) = \log Y_1(u+i\ad+i\e)-\log Y_1(u+i\ad-i\e) \nn
&=  \sum_{\alpha}L_-^{(\alpha)}(u+i\e)+\ewD(u).
\end{align}
The full definition of $\hat{\Delta}$ is the continuation of the functions in the last expression to the entire complex plane, where we should be careful in the interpretation of $L_-$: the $Y$ functions have long cuts, but as a discontinuity $\hat{\Delta}$ has short cuts. So we need to continue the expression \eqref{eq:Delta1} to the lower half-plane such that $\hat{\Delta}$ has a short cut on the real axis. To define this we introduce $\hat{L}_y^{(\alpha)}$, the double-valued continuation of $L_-^{(\alpha)}$ from the upper half-plane on the first sheet with a short cut. On the first sheet it can be characterised as
\be
\hat{L}_y^{(\alpha)}(u) =
\left\{
	\begin{array}{ll}
		L_-^{(\alpha)}(u)  & \mbox{if Im}(u)>0  \\
		L_+^{(\alpha)}(u)  & \mbox{if Im}(u)<0
	\end{array}
\right..
\ee
Explicitly plugging in the definition of $\ewD$ we find that $\hat{\Delta}$ is given by
\be
\label{deftD}
\hat{\Delta} =  J \hat{\mathcal{E}}+ \sum_{\alpha}\hat{L}_y^{(\alpha)}  + \sum_{\alpha} \left( L^{(\alpha)}_{-} +  L^{(\alpha)}_{+} \right) \star \hat{K}  + 2 \Lambda_Q\star \hat{K}_Q^{\Sigma} + \sum_{\alpha} L_{M|vw}^{(\alpha)}\star \hat{K}_M,
\ee
where we defined $\hat{\mathcal{E}}$ in eqn. \eqref{eq:energyterm}. The definitions of the other kernels can be found in appendix \ref{App:kernelsandenergies} and in eqn. \eqref{eq:Ksigmadef}. This defines $\hat{\Delta}$ as a function with short cuts, which is consistent with its identification as the discontinuity of $\log Y_1$. However, since our tools are more suited to analyse long-cutted functions we consider a new function that coincides with $\hat{\Delta}$ on the upper half-plane\footnote{This is indeed an unfortunate choice to work in the upper-half-plane conventions, although in line with the work \cite{Cavaglia:2010nm} for the undeformed case. It is immaterial for the results and we stick to this choice to stay close to \cite{Cavaglia:2010nm}.}:
\be
\label{delta2}
\Delta = -J \log \frac{x(u)+\xi}{1/x(u)+\xi} + \sum_{\alpha}L_-^{(\alpha)}  - \sum_{\alpha} \left( L^{(\alpha)}_{-} +  L^{(\alpha)}_{+} \right) \star K  - 2 \Lambda_Q\star K_Q^{\Sigma} - \sum_{\alpha} L_{M|vw}^{(\alpha)}\star K_M,
\ee
where 
\be
\widecheck{\mathcal{E}} = \log \frac{x(u)+\xi}{1/x(u)+\xi},\qquad 
\widecheck{\mathcal{E}}_{\text{und}} = \log\left( x(u)^2\right)
\ee
are long-cutted versions of $\hat{\mathcal{E}}$. This concludes the first step in our derivation. 
\subsection{Deriving discontinuities for $\Delta$}
\label{sec:discdelta}
Next we find an expression for the discontinuities of $\Delta$ in terms of discontinuities of $Y_{\pm}$ and $Y_{M|vw}$. The derivation of the term containing $L_{M|vw}$ functions follows closely the analysis we followed in section \ref{sec:discYw} and we will not repeat it here. The $Y_{\pm}$-contribution is a bit more involved:
the second term in $\Delta$ has discontinuities
\be
\left[\sum_{\alpha}L_{-}^{(\alpha)}(u)\right]_{\pm2N}. 
\ee
The contribution coming from the term $\left(L^{(\alpha)}_{-} +L^{(\alpha)}_{+} \right) \star K$ generates discontinuities only for positive $N$, which read
\be
-\left[L^{(\alpha)}_{-} +L^{(\alpha)}_{+}\right]_{-2N},
\ee
whereas the contributions for negative $N$ vanish. Adding both contributions together we see that the total contributions from $Y_{\pm}$ functions sum up to be
\be
\pm \left[\sum_{\alpha}[L^{(\alpha)}_{\mp}\right]_{\pm 2N}.
\ee
Finally the discontinuity of the dressing phase kernel follows directly from its simplified expression\footnote{See appendix \ref{app:dressing} discussing this simplification. The result can be found in eqn. \eqref{eq:simplifieddressing}.}
\begin{align}
2 L_Q \star K_Q^{\Sigma}  = \sum_\alpha \oint_{\gamma_x} \log Y^{(\alpha)}_-(z) \left(\sum_{N=1}^{\infty} K(z + 2 i N \ad, u) + K(z - 2 i N \ad, u)\right).
\end{align}
discussed in appendix \ref{app:dressing}. Putting together these ingredients we find the following discontinuity for $\Delta$:
\be
\left[\Delta\right]_{\pm 2N} (u) = \pm \sum_{\alpha} \left( \left[L^{(\alpha)}_{\mp}\right]_{\pm 2N} + \sum_{M=1}^N \left[   L_{M|vw}^{(\alpha)}\right]_{\pm(2N-M)} + \left[ \log Y_-^{(\alpha)}\right]_{0} \right).
\ee
This is one of the two discontinuities that we have to impose on the $Y$-system equations to be able to rederive the TBA equations. Looking closely at the expression \eqref{delta2} for $\Delta $ we see that it has one additional branch cut: the energy gives rise to a logarithmic branch cut on the imaginary axis
\be
\label{eq:logarithmicjump}
\Delta(iv +\e) - \Delta(iv -\e) = 2\pi i J \quad \text{ for } v \in \R,
\ee
using the principal branch description of the logarithm. 
\section{The analytic $Y$-system}
\label{sec:analyticYsystem}
For the reader's convenience we summarise the results from the previous sections: we have found a set of equations known as the $Y$ system which the $Y$ functions for both the undeformed and $\eta$-deformed models should satisfy. These equations are\footnote{It is possible to write these equations in a more compact form by redefining some of the functions (see \cite{Gromov:2011cx}). Since it does not help us in our current analysis we refrain from this rewriting.}
\begin{equation}
\label{eq:Ysystem}
\begin{aligned}
\frac{Y_1^+ Y_1^-}{Y_2} & = \frac{\prod_{\alpha}
\left(1-\frac{1}{Y^{(\alpha)}_{-}}\right)}{1+Y_2}\, , \, \, \, \, \, \mbox{for} \, \,
u\in \hat{Z}_0\, ,\\
\frac{Y_Q^+ Y_Q^-}{Y_{Q+1}Y_{Q-1}} & = \frac{\prod_{\alpha} \Bigg(
1+\frac{1}{Y_{Q-1|vw}^{(\alpha)}}\Bigg)}{(1+Y_{Q-1})(1+Y_{Q+1})} \,,\\
Y_{1|w}^+ Y_{1|w}^- & =(1+Y_{2|w})\left(\frac{1-Y_-^{-1}}{1-Y_+^{-1}}\right)^{\vartheta(\theta-|u|)},
\end{aligned}
\begin{aligned}
Y_{M|w}^+ Y_{M|w}^- & =(1+Y_{M-1|w})(1+Y_{M+1|w})\,,\\
Y_{1|vw}^+ Y_{1|vw}^- & =\frac{1+Y_{2|vw}}{1+Y_2}\left(\frac{1-Y_-}{1-Y_+}\right)^{\vartheta(\theta-|u|)}, \\
Y_{M|vw}^+ Y_{M|vw}^- & =\frac{(1+Y_{M-1|vw})(1+Y_{M+1|vw})}{1+Y_{M+1}} ,\\
Y_{-}^{+}Y_{-}^{-} &= \frac{1+Y_{1|vw}}{1+Y_{1|w}}\frac{1}{1+Y_1}\, .
\end{aligned}
\end{equation}
We have derived additional analytical data that the $Y$ functions should satisfy. They come in the form of discontinuity equations: 
\begin{equation}
\label{discs1}
\begin{aligned}
\left[\log Y_{1|w}^{(\alpha)}\right]_{\pm1}(u) &= \left[L_{-}^{(\alpha)}\right]_0(u), \\
\left[\log Y_{1|vw}^{(\alpha)}\right]_{\pm1}(u) &= \left[\Lambda_{-}^{(\alpha)}\right]_0(u),\\
\left[\log \frac{Y_-}{Y_+} \right]_{\pm2N}(u) &= -\sum_{P=1}^{N} \left[\Lambda_P\right]_{\pm(2N-P)}(u) \text{ for } N\geq 1,\\
\left[\Delta\right]_{\pm 2N} (u) &= \pm \sum_{\alpha} \left( \left[L^{(\alpha)}_{\mp}\right]_{\pm 2N}(u) + \sum_{M=1}^N \left[   L_{M|vw}^{(\alpha)}\right]_{\pm(2N-M)}(u) + \left[ \log Y_-^{(\alpha)}\right]_{0}(u) \right),
\end{aligned}
\end{equation}
where
\begin{equation}
\begin{aligned}
\Delta(u) &=
\left\{
	\begin{array}{ll}
		\cD(u)  & \mbox{if Im}(u)>0  \\
		\cD(u_*)  & \mbox{if Im}(u)<0
	\end{array}\right. ,\\
\cD (u) &= \left[\log Y_1 \right](u), \\
\Delta(i u +\e) - \Delta(i u -\e) &= 2\pi J i, \text{ for } u\in \R.
\end{aligned}
\end{equation}
These two sets of equations \eqref{eq:Ysystem} and \eqref{discs1} together with the analyticity strips for the $Y$ functions we call the \emph{analytic $Y$-system} and will form the basis for our further reduction of the spectral problem. Note that the functional form of the analytic $Y$-system is absolutely identical for the undeformed and deformed model. The only difference at this point sits in the fact that the undeformed $Y$-functions vanish as $u \rightarrow \pm \infty$ whereas the deformed $Y$-functions are $2\pi$-periodic. 
\\[5mm]
Before we continue our reduction, we take a step back: we have derived the analytic $Y$-system from the ground-state TBA-equations, i.e. the TBA equations which only describe the $Y$ functions for the ground-state of the model(s). The resulting equations in the analytic $Y$-system make no reference to a specific state (apart from the $J$ appearing in the logarithmic branch cut), however, which are characterised by zeroes, poles and asymptotic behaviour of the $Y$ functions. We will continue our reduction of the TBA equations keeping this in mind, but at the same time anticipating that because of the general form of the equations they might be applicable to a wider range of states.

\section{Rederiving the TBA equations}
It is prudent to at least make sure that we have extracted the correct information from the TBA equations. We do this by rederiving the TBA equations from the analytic $Y$-system assuming that the $Y$ functions have no isolated öpoles. From numerical analysis this is the expected behaviour for the ground-state $Y$-functions. In addition we have to require some asymptotic behaviour for the $Y$ functions. Before we start rederiving the TBA equations let us first derive these asymptotic conditions.

\subsection{Asymptotics} 
\label{sec:TBAasymptotics}
The undeformed and deformed case are different when it comes to the final pieces of analyticity data. The natural direction for asymptotics in the undeformed case is to consider $u\rightarrow \infty$, usually just above the real axis.\footnote{The direction $u\rightarrow -\infty$ would be natural as well, but lead to additional minus signs that might cloud the presentation.} Since the $Y$ functions have long cuts it generically is not possible to connect asymptotics found here to other asymptotic regions of the complex plane. By considering the $u\rightarrow \infty$ limit of the undeformed TBA-equations directly we find that all $Y$ functions vanish there. It is clear that this can no longer hold true in the deformed case: in fact, the $u\rightarrow \infty$ limit does not make sense on the cylinder. The only asymptotic regions present on the cylinder are those for which $u\rightarrow \pm i \infty$. As we will see shortly this is the right idea: in order to correctly reproduce the TBA equations we need to impose that 
\begin{equation}
\label{eq:YQasymptotics}
\log Y_Q \rightarrow \mp \ad Q J \text{ as } u\rightarrow \pm i \infty,
\end{equation}
whereas all other $\log Y_{\circ}$ go to zero in these limits. This is consistent with a direct analysis of the TBA equations: all the integration kernels vanish in these regimes, only the driving term $\tilde{E}_Q$ has non-vanishing asymptotics leading to 
the non-vanishing asymptotics \eqref{eq:YQasymptotics}. 
\subsubsection{Other assumptions}
To make the entire rederivation as clear as possible we list the other assumptions we make: we assume the $Y$ functions have 
\begin{itemize}
\item no poles other than possibly at branch points. 
\item the analyticity strips as we derived at the beginning of this chapter: $Y_Q, Y_{Q|(v)w}^{(\alpha)} \in \mAp_Q$ and $Y_-^{(\alpha)}$ and $Y_+^{(\alpha)}$ are each others analytic continuation through their cut on the real axis. 
\end{itemize}
\subsection{General strategy}
To rederive the TBA equations \eqref{eq:TBAeqns} we will follow (almost) the same strategy for all four sets of equations: 
\begin{enumerate}
\item We write the left-hand side of the TBA equations as a contour integral,
\item we deform the contour using our knowledge of the branch cut structure of the integrand to yield a sum over discontinuities,
\item We use the $Y$ system to derive expressions for these discontinuities in terms of simpler ones and the ones we imposed in eqn. \eqref{discs1},
\item We plug these simpler expressions into the sum of discontinuities and use a telescoping of integrals to reduce it to a few convolutions, producing the right-hand side of the TBA equations.
\end{enumerate}
In the $Y_{-}$ case we use a slightly different approach, since we can use the non-local discontinuity equation to produce part of the TBA equation already. The $Y_Q$ case is by far the most complicated because we encoded its information in discontinuities of $\Delta$. To practice in a simple case, let us start with the $Y_-$ case.  
\subsection{$Y_-$}
We would like to to use the analytic $Y$-system to derive the TBA equations
\begin{equation}
\label{eq:repeatedY-TBA}
\log Y^{(\alpha)}_{\beta} = \,- \Lambda_P \star K^{Py}_{\beta} + \left(L^{(\alpha)}_{M|vw}-L^{(\alpha)}_{M|w}\right)\star K_M. 
\end{equation}
One piece of information we use is that $Y_{\pm}^{(\alpha)}$ are each others analytic continuation, which implies that we can derive the $\beta= +$ case from the $\beta=-$ case, allowing us to restrict ourselves to the latter. We then note that in section \ref{sec:discYmin} we proved that the local discontinuity equation we imposed for $Y_{\pm}$ is equivalent to the non-local one in eqn. \eqref{eq:Y-nonlocal}. So without computation we can use the non-local equation 
\begin{equation}
\log \frac{Y_-}{Y_+}(u) =  -\Lambda_P \star K_{Py}.
\end{equation}
Now we see that 
\begin{equation}
\label{eq:rederiveY-1}
\log Y_- = 1/2 \left( \log Y_-Y_+ + \log Y_-/Y_+ \right) = 1/2 \left( \log Y_-Y_+ -\Lambda_Q\star \left(K_-^{Qy} -K_+^{Qy} \right)\right). 
\end{equation}
This already yields part of the TBA equation for $Y_-^{(\alpha)}$. To complete this derivation we need to find an expression for $\log Y_-Y_+$, which we will do with the method sketched in the previous section. First we write $\log Y_-Y_+$ as a contour integral
\be
\log Y_-Y_+(u) = \oint_{\gamma} \frac{dz}{2\pi i}\log Y_-Y_+(z)H(z-u),
\ee
which is valid for $u$ in the physical strip and where $\gamma$ is illustrated in fig. \ref{fig:deformedgamma} and fig. \ref{fig:undeformedgamma} and we introduced the function $H$ which is defined as
\be
H(u) = \frac{1}{2}\cot \frac{u}{2}, \quad H^{\text{und}}(u) = \frac{1}{u}.
\ee
We can deform the contour $\gamma$ immediately to $\Gamma$, see the figures \ref{fig:deformedgamma} and \ref{fig:undeformedgamma}, using the asymptotic boundary conditions that we introduced in the section \ref{sec:TBAasymptotics}. For both the undeformed as well as the deformed case $\log Y_-Y_+$ vanishes in the asymptotic regime ($u\rightarrow \infty$ for the undeformed and  $u\rightarrow \pm i \infty$ for the deformed case). This gives
\bea
\label{intrep1}
& &\oint_{\Gamma} \frac{dz}{2\pi i} \log Y_-Y_+(z)H(z-u) \nn
&=& \sum_{N=1}^{\infty}\sum_{\tau} \left(\int_{Z_0 +2i \tau N \ad +i\e}- \int_{Z_0 +2i \tau N \ad -i\e} \right)\frac{dz}{2\pi i}H(z-u) \log Y_-Y_+(z) \nonumber \\
&=& \sum_{N=1}^{\infty}\sum_{\tau}  \int \frac{dz}{2\pi i}H(z+2i \tau N \ad-u) \left[\log Y_-Y_+ \right]_{2\tau N}(z).
\eea
So to express $\log Y_-Y_+$ we need expressions for the discontinuities 
\begin{equation}
\label{eq:Y-2ndiscs}
\left[\log Y_-Y_+ \right]_{2\tau N} \text{ for } N\in \N,
\end{equation} 
which we will derive from the $Y$-system equations alone. The result is 
\be
\left[\log Y_- Y_+  \right]_{\pm 2N} = 2\sum_{J=1}^N \left[L_{J|vw} - L_{J|w}\right]_{\pm(2N-J)}-\sum_{Q=1}^{N}\left[\Lambda_Q \right]_{\pm(2N-Q)},
\ee
as we will derive in the next section. 
\subsubsection{Deriving discontinuities from the $Y$-system}
Seeing the details of a computation at least once can help in understanding the concepts involved, which is why we will show a detailed derivation of the discontinuities \eqref{eq:Y-2ndiscs}. Those who are more interested in the general approach can skip this section and continue with section \ref{sec:derivingY-TBA}. 
\\[5mm]
The $Y$-system equation for $Y_{\pm}$ is 
\be
Y_-^+Y_-^- = \frac{1+Y_{1|vw}}{1+Y_{1|w}} \frac{1}{1+Y_1}
\ee
and taking logs gives us
\be
\log Y_-^+Y_-^- = \Lambda_{1|vw} - \Lambda_{1|w}-\Lambda_1. 
\ee
Shifting upwards $u\rightarrow u+(2N+1)\ad \ii$ and looking at the discontinuities by taking brackets gives
\be
\left[\log Y_- \right]_{2N} +\left[\log Y_- \right]_{2N-2}  = \left[ \Lambda_{1|vw} - \Lambda_{1|w}-\Lambda_1\right]_{2N-1}
\ee
leading to
\bea
\label{rec1}
& &\left[\log Y_- Y_+ + \log Y_-/Y_+ \right]_{2N} +\left[\log Y_- Y_+ + \log Y_-/Y_+ \right]_{2N-2}  \nn
&=& 2\left(\left[ \Lambda_{1|vw} - \Lambda_{1|w}-\Lambda_1\right]_{2N-1}\right) = 2\left(\left[ L_{1|vw} - L_{1|w}+\log Y_{1|vw}/Y_{1|w}-\Lambda_1\right]_{2N-1}\right).\nn
\eea
We need an expression for $\left[\log Y_{1|vw}/Y_{1|w}\right]_{2N-1}$. Using the $Y$-system equations
\bea
\label{Yvw}
Y_{1|vw}^+Y_{1|vw}^- = \frac{1+Y_{2|vw}}{1+Y_2}\frac{1-Y_-}{1-Y_+}, \quad
Y_{1|w}^+Y_{1|w}^- = (1+Y_{2|w})\frac{1-Y^{-1}_-}{1-Y^{-1}_+}
\eea
we immediately see that
\bea
\label{start1}
\left[\log Y_{1|vw}/Y_{1|w}\right]_{\pm(2N-1)} &=& -\left[\log Y_{1|vw}/Y_{1|w}\right]_{\pm(2N-3)}- \left[\Lambda_2+L_{2|vw} - L_{2|w}\right]_{\pm(2N-2)} \nonumber\\ 
&+& \left[\log Y_-/Y_+\right]_{\pm(2N-2)} +\left[\log Y_{2|vw}/Y_{2|w} \right]_{\pm(2N-2)}.
\eea
We can now continue to rewrite the last term using the $Y$-system equations, which for $N>0$ give rise to  
\bea
\label{rule1}
& & \left[\log Y_{M|vw}/Y_{M|w} \right]_{\pm N} = -\left[\log Y_{M|vw}/Y_{M|w} \right]_{\pm(N-2)} +\sum_{J=1}^{\infty} A_{MJ}\left[L_{J|vw} - L_{J|w}\right]_{\pm(N-1)} \nonumber \\
&-&\left[\Lambda_{M+1} \right]_{\pm(N-1)} + \left[\log Y_{M+1|vw}/Y_{M+1|w} \right]_{\pm(N-1)}+ \left[\log Y_{M-1|vw}/Y_{M-1|w} \right]_{\pm(N-1)}, \qquad
\eea
where 
\begin{equation}
\label{eq:AMJ}
A_{MJ} = \delta_{J,M+1} + \delta_{J,M-1}  \text{ for } M>1.
\end{equation}
By repeated application of this rule to eqn. \eqref{start1} we will be able to find the expression we are looking for. Note that we introduced the infinite sums here only to have convenient notation, all of these sums have a finite number of non-zero terms. 
\\[5mm]
If we define $A_{1J} = \delta_{2,J}$ we can rewrite equation \eqref{start1} as 
\bea
& &\left[\log Y_{1|vw}/Y_{1|w}\right]_{\pm(2N-1)}= -\left[\log Y_{1|vw}/Y_{1|w}\right]_{\pm(2N-3)}- \left[\Lambda_2 \right]_{\pm(2N-2)} \nonumber\\ 
&+& \left[\log Y_-/Y_+\right]_{\pm(2N-2)}  +\sum_{J=1}^{\infty} A_{1J}\left[L_{J|vw} - L_{J|w}\right]_{\pm(2N-2)}+\left[\log Y_{2|vw}/Y_{2|w} \right]_{\pm(2N-2)},\nn
\eea
and applying our rule \eqref{rule1} to the last term leads to 
\bea
& &\left[\log Y_{1|vw}/Y_{1|w}\right]_{\pm(2N-1)}\nn
&=& \left[\log Y_-/Y_+\right]_{\pm(2N-2)} -\sum_{Q=2}^3 \left[\Lambda_Q \right]_{\pm(2N-Q)} + \sum_{Q=1}^2\sum_{J=1}^{\infty} A_{QJ}\left[L_{J|vw} - L_{J|w}\right]_{\pm(2N-1-Q)} \nonumber \\
&-&\left[\log Y_{2|vw}/Y_{2|w} \right]_{\pm(2N-4)} + \left[\log Y_{3|vw}/Y_{3|w} \right]_{\pm(2N-3)}.
\eea
Repeated application of eqn. \eqref{rule1} to the last term now leads to
\bea
& &\left[\log Y_{1|vw}/Y_{1|w}\right]_{\pm(2N-1)} \nonumber \\
&=& \left[\log Y_-/Y_+\right]_{\pm(2N-2)} -\sum_{Q=2}^{N} \left[\Lambda_Q \right]_{\pm(2N-Q)} + \sum_{Q=1}^{N-1}\sum_{J=1}^{\infty} A_{QJ}\left[L_{J|vw} - L_{J|w}\right]_{\pm(2N-1-Q)} \nonumber \\
&-&\left[\log Y_{N-1|vw}/Y_{N-1|w} \right]_{\pm(N-1)} + \left[\log Y_{N|vw}/Y_{N|w} \right]_{\pm N}.
\eea
Applying eqn. \eqref{rule1} one last time to the final term here gives
\bea
& &\left[\log Y_{N|vw}/Y_{N|w} \right]_{\pm N} =-\left[\log Y_{N|vw}/Y_{N|w} \right]_{\pm(N-2)} +\sum_{J=1}^{\infty} A_{NJ}\left[L_{J|vw} - L_{J|w}\right]_{\pm(N-1)} \nonumber \\
&-&\left[\Lambda_{N+1} \right]_{\pm(N-1)} + \left[\log Y_{N+1|vw}/Y_{N+1|w} \right]_{\pm(N-1)}+ \left[\log Y_{N-1|vw}/Y_{N-1|w} \right]_{\pm(N-1)}  \nonumber \\
&=&\left[L_{N-1|vw} - L_{N-1|w}\right]_{\pm(N-1)}+ \left[\log Y_{N-1|vw}/Y_{N-1|w} \right]_{\pm(N-1)}.
\eea
The vanishing discontinuities here disappear due to our assumption on the analyticity strips of the $Y$ functions. So the result is
\bea
\label{res1}
& &\left[\log Y_{1|vw}/Y_{1|w}\right]_{\pm(2N-1)} = \left[\log Y_-/Y_+\right]_{\pm(2N-2)} -\sum_{Q=2}^{N}\left[\Lambda_Q \right]_{\pm(2N-Q)}  \nonumber \\
&+& \sum_{Q=1}^{N-1}\sum_{J=1}^{\infty} A_{QJ}\left[L_{J|vw} - L_{J|w}\right]_{\pm(2N-1-Q)}+\left[L_{N-1|vw} - L_{N-1|w}\right]_{\pm(N-1)}. \nonumber \\
\eea
Plugging in the assumed discontinuity relations we get for the original relations \eqref{rec1}
\bea
\label{eq:resultY-Ysystem}
& &\left[\log Y_- Y_+  \right]_{\pm 2N} +\left[\log Y_- Y_+\right]_{\pm(2N-2)} \nonumber \\
&=& 2\left(\left[ L_{1|vw} - L_{1|w}-\Lambda_1\right]_{\pm(2N-1)}\right) - \left[\log Y_-/Y_+\right]_{\pm(2N-2)} -\sum_{Q=2}^{N}\left[\Lambda_Q \right]_{\pm(2N-Q)}  \nonumber \\
&+& 2\sum_{Q=1}^{N-1}\sum_{J=1}^{\infty} A_{QJ}\left[L_{J|vw} - L_{J|w}\right]_{\pm(2N-1-Q)}+2\left[L_{N-1|vw} -  L_{N-1|w}\right]_{\pm(N-1)} \nonumber \\
&=& -\sum_{Q=1}^{N-1}\left[\Lambda_Q \right]_{\pm(2N-2-Q)} -\sum_{Q=2}^{N}\left[\Lambda_Q \right]_{\pm(2N-Q)} \nonumber \\
&+&2\sum_{J=1}^N \left[L_{J|vw} - L_{J|w}\right]_{\pm(2N-J)} + 2\sum_{J=1}^{N-1} \left[L_{J|vw} - L_{J|w}\right]_{\pm(2N-2-J)}.
\eea
Starting with $N=1$, we see that 
\be
\left[\log Y_- Y_+  \right]_{2} = 2\sum_{J=1}^1 \left[L_{J|vw} - L_{J|w}\right]_{\pm(2-J)}-\sum_{Q=2}^{2}\left[\Lambda_Q \right]_{\pm(2-Q)}
\ee
and invoking mathematical induction using the eqs. \eqref{eq:resultY-Ysystem} we end up with
\be
\label{res2}
\left[\log Y_- Y_+  \right]_{\pm 2N} = 2\sum_{J=1}^N \left[L_{J|vw} - L_{J|w}\right]_{\pm(2N-J)}-\sum_{Q=1}^{N}\left[\Lambda_Q \right]_{\pm(2N-Q)}.
\ee
It is interesting to see how much information is stored in the $Y$-system equations: the previous derivation is based almost exclusively on the $Y$-system equations.
\subsubsection{Rederiving the $Y_-$ TBA-equation}
\label{sec:derivingY-TBA}
With the result for the discontinuities \eqref{eq:Y-2ndiscs} in hand we can finally rederive the $Y_-$ TBA-equation. Plugging in the discontinuities \eqref{res2} in the integral expression \eqref{intrep1} yields
\bea
\log Y_- Y_+(u) = \sum_{N=1}^{\infty}\sum_{\tau}  \int \frac{dz}{2\pi i}H(z+2i \tau N \ad-u)  \sum_{J=1}^N \left[ \left( 2L_{J|vw} - 2L_{J|w} -\Lambda_J \right)\right]_{\tau(2N-J)}.\nn
\eea
Defining $F_J =  2L_{J|vw} - 2L_{J|w} -\Lambda_J$ we can simplify this expression by noticing that the sum contains pairs of integrals which when paired vanish exactly because their respective contours can be deformed to overlap. To see this we relabel the sums:
\bea
\log Y_- Y_+(u) &=&  \sum_{\tau} \sum_{N=1}^{\infty} \sum_{J=1}^N   \int \frac{dz}{2\pi i}H(z+2i \tau N \ad-u)  \left[F_J\right]_{\tau(2N-J)} \nonumber \\
&=&  \sum_{\tau}  \sum_{J=1}^{\infty} \sum_{N=J}^{\infty}  \int \frac{dz}{2\pi i}H(z+2i \tau N \ad-u)  \left[F_J\right]_{\tau(2N-J)} \nonumber \\
&=&  \sum_{\tau}  \sum_{J=1}^{\infty} \sum_{N=J}^{\infty}  \int \frac{dz}{2\pi i}H(z+2i \tau N \ad-u)\cdot \nonumber \\
& &\left( F_J(z+\tau(2N-J)i\ad+i\e) - F_J(z+\tau(2N-J)i\ad-i\e) \right)\nonumber \\
&=&  -\sum_{\tau} \tau \sum_{J=1}^{\infty}  \int \frac{dz}{2\pi i}H(z+2i \tau J \ad-u) F_J(z-\tau (Ji-i\e)).
\eea
Now we see that almost all the integrals cancel, leaving us only with the innermost integrals with $N=J$ in the last line. Shifting $z\rightarrow z-i \tau J\ad$ and using analyticity of $F_J$ we find
\bea
 &-&\sum_{\tau} \tau \sum_{J=1}^{\infty}  \int \frac{dz}{2\pi i}H(z+i \tau J \ad-u) F_J(z) \nonumber \\
&=& - \sum_{J=1}^{\infty}  \int dz  F_J(z)\frac{1}{2\pi i}\left(H(z+i J \ad -u)-H(z-i J \ad -u) \right)\nonumber \\
&=&   \sum_{J=1}^{\infty}  (2L_{J|vw} - 2L_{J|w} -\Lambda_J)\star K_J.
\eea
This is the result we were after. Plugging this into our partial result \eqref{eq:rederiveY-1} we find, employing the kernel identity \eqref{kernelids} that
\bea
\log Y_- &=&  (L_{J|vw} - L_{J|w})\star K_J  -1/2\sum_{J=1}^{\infty} \Lambda_J\star K_J -1/2\Lambda_Q\star \left(K_-^{Qy} -K_+^{Qy} \right) \nonumber \\
&=& (L_{J|vw} - L_{J|w})\star K_J  -\Lambda_J\star K_-^{Qy},
\eea
which is precisely the TBA-equation as written in eqn. \eqref{eq:TBAeqns} and in the beginning of this section in eqn. \eqref{eq:repeatedY-TBA}. 
\subsection{$Y_{(v)w}$}
\label{sec:rederiveYvwTBA}
Our next task is to prove that we can rederive the TBA equations of the $Y_{w}$ and $Y_{vw}$ functions. Since their TBA equations are very similar we will be able to combine their derivation into one. The starting point is a rewriting of the left-hand side of their TBA equation as a contour integral: for $u$ in the physical strip we find
\be
\log Y_{M|(v)w}(u) = \oint_{\gamma} \frac{dz}{2\pi i} \log Y_{M|(v)w}(z) H(z-u).
\ee
Deforming in the usual way to $\Gamma$, we get 
\bea
\label{eq:contourintegralYvw}
& &\oint_{\gamma} \frac{dz}{2\pi i} \log Y_{M|(v)w}(z) H(z-u) = \oint_{\Gamma} \frac{dz}{2\pi i} \log Y_{M|(v)w}(z) H(z-u)\nonumber \\
  &=& \sum_{\tau}\sum_{l=0}^{\infty} \left( \int_{Z_0 +\tau(M+2l)i\ad +i\e} -\int_{Z_0+\tau(M+2l)i\ad -i\e} \right)  \log Y_{M|(v)w}(z) H(z-u) \nn
  &=& \sum_{\tau}\sum_{l=0}^{\infty}  \int_{Z_0}  \frac{dz}{2\pi i} H(z+2\ii \tau N \ad -u) \left[ \log Y_{M|(v)w} \right]_{\pm 2M+l},
\eea
so we will need discontinuities of the form
$$
\left[ \log Y_{M|(v)w} \right]_{\pm 2M+l},
$$
where $M,l\in \mathbb{N}$. We use the $Y$ system to derive their expressions. The relevant $Y$-system equations are
\bea
Y_{1|w}^+Y_{1|w}^- &=& (1+Y_{2|w}) \left(L_--L_+\right) \nonumber \\
Y_{M|w}^+Y_{M|w}^- &=& (1+Y_{M+1|w})(1+Y_{M-1|w}),\nn
Y_{1|vw}^+Y_{1|vw}^- &=& (\frac{1+Y_{2|vw}}{1+Y_2}) \left(\Lambda_--\Lambda_+\right) \nonumber \\
Y_{M|vw}^+Y_{M|vw}^- &=& (1+Y_{M+1|vw})(1+Y_{M-1|vw})(1+Y_{M+1}). 
\eea
To combine the derivations of the discontinuities we note that these $Y$-system equations are very similar: we can obtain those for $Y_{M|w}$ functions by the replacements $\Lambda_{\beta} \rightarrow L_{\beta}$ and $Y_{M|vw}\rightarrow Y_{M|w}$ combined with setting $Y_Q\rightarrow 0$. The discontinuity equations that we have imposed on the $Y$ system have the same similarity, thus we can truly perform both computations simultaneously: we will do the derivation using $Y_{M|vw}$ functions and use the above mentioned rules to find the $Y_{M|w}$ result. We will this time present only the main points of the derivation, referring the reader to the appendix \ref{app:Yvwdiscs} for more details.
\\[5mm]
We use the $Y$-system equations to find the necessary discontinuity equations. Taking logs of the $Y$-system equations \eqref{eq:Ysystem} for $Y_{M|vw}$ we find for $M\geq 2$
\be
\log Y_{M|vw}^+ +\log Y_{M|vw}^- =\sum_{N=1}^{\infty} A_{MN} \Lambda_{N|vw}-\Lambda_{M+1},
\ee
with $A_{MN}$ as in eqn. \eqref{eq:AMJ}. Taking discontinuity brackets at $M+2l-1$ yields
\bea
& &\left[ \log Y_{M|vw} \right]_{(M+2l)\tau} =  \sum_{N=1}^{\infty} A_{MN} \left[L_{N|vw}\right]_{(M+2l -1)\tau} +  \left[\log Y_{M-1|vw}\right]_{(M+2l-1)\tau} \nn 
&+& \left[\log Y_{M+1|vw}\right]_{(M+2l-1)\tau} -\left[\Lambda_{M+1}\right]_{(M+2l-1)\tau} - \left[\log Y_{M|vw}\right]_{(M+2(l-1))\tau}. 
\eea
This is the rule that we can use repeatedly to bring the discontinuity containing $\log Y_{Q|vw}$ with $Q<M$ closer and closer to the real line. Repeated application, plugging in the $Y_{1|vw}$ discontinuity equation and some rewriting ultimately yield 
\bea
\left[ \log Y_{M|vw} \right]_{(M+2l)\tau} = \left[D^{M|vw}_{(M+2l)\tau} -\delta_{l,0}\Lambda_-\right]_0,
\eea
where we have defined a set of $D$ functions as
\bea
& &D^{M|vw}_{(M+2l)\tau}(u) =\left(\Lambda_--\Lambda_+ \right)(u+2l\ad\tau i) +\sum_{Q=1}^{M-1} L_{Q|vw}(u+i(Q+2l)\ad \tau) \nn
&+& \sum_{J=1}^l \left( 2 \sum_{Q=1}^{M}  L_{Q+J|vw}(u+i(Q+2l-J)\ad \tau) + L_{M+J|vw}(u+i(M+2l-J)\ad \tau)\right)\nn
&+&  \sum_{J=1}^l   L_{J|vw}(u+i(2l-J)\ad \tau)-\sum_{J=1}^{l}\sum_{Q=1+J}^{M+J} \Lambda_Q(u+i(Q+2l-2J)\ad \tau).\nn
\eea
This is not just a packaging of a sum of complicated objects: the $D$ functions satisfy a recursive identity that will simplify their contribution significantly:
\bea
D^{M|vw}_{(M+2l)\tau }(u) &-& D^{M|vw}_{(M+2l-2)\tau}(u+2\tau i \ad ) = L_{l|vw}(u+\tau li \ad)+L_{M+l|vw}(u+\tau (M+l)i \ad ) \nn
&+& 2 \sum_{Q=l+1}^{M-1+l}  L_{Q|vw}(u+\tau Q\ad i) -\sum_{Q=1+l}^{M+l} \Lambda_Q(u+\tau Q \ad i).
\eea
We can now plug the discontinuities into our expression \eqref{eq:contourintegralYvw} for $Y_{M|vw}$ and obtain
\bea
\log Y_{M|vw}(u) &=&\sum_{\tau}\sum_{l=0}^{\infty} \int_{Z_0}  \left[\log Y_{M|vw}\right]_{\tau (M+2l)}(z) H(z+\tau (M+2l)-u) \nn
&=& - \left(\int_{Z_0+i\e}-\int_{Z_0-i\e}\right)\left(\Lambda_-(z) H(z+M \ad i -u) + \Lambda_-(z) H(z-M \ad i -u)\right) \nn
&+& \sum_{\tau}\sum_{l=0}^{\infty} \int \left(\left[D_{(M+2l)\tau}^{M|vw}\right]_{0}(z) \right) H(z+\tau (M+2l)\ad i-u).
\eea
After separating the contributions of the different types of $Y$ functions we use different identities for the integral kernels and a telescoping of integrals as we had in the $Y_{\beta}$ case to recognise that this expression is the right-hand side of the $Y_{M|vw}$ TBA-equation, such that we ultimately derive that
\bea
\log Y_{M|vw}^{(\alpha)} = L_{N|vw}^{(\alpha)} \star K_{NM} +\left(L_-^{(\alpha)}-L_+^{(\alpha)}\right) \hat{\star} K_M-\Lambda_Q \star K_{xv}^{QM}.
\eea
Using the rules that we discussed above we can also obtain the $Y_w$ TBA-equation. To do this most transparently we first rewrite the term 
\bea
\left(L_-^{(\alpha)}-L_+^{(\alpha)}\right) \hat{\star} K_M &=& \left(\Lambda_-^{(\alpha)}-\Lambda_+^{(\alpha)}\right) \hat{\star} K_M -\log Y_-^{(\alpha)}/Y_+^{(\alpha)} \hat{\star} K_M \nn
&=&  \left(\Lambda_-^{(\alpha)}-\Lambda_+^{(\alpha)}\right) \hat{\star} K_M +\left(\Lambda_Q \star K_{Qy} \right)\hat{\star} K_M,
\eea
which shows that replacing $\Lambda_{\beta} \rightarrow L_{\beta}$ yields an expression which is equivalent to the previous one modulo terms proportional to $\Lambda_Q$. Since these vanish in the transformation from $Y_{M|vw}$ to $Y_{M|w}$ we can safely proceed and obtain the $Y_w$ TBA-equation:
\bea
\log Y_{M|w}^{(\alpha)} = L_{N|w}^{(\alpha)} \star K_{NM} +\left(L_-^{(\alpha)}-L_+^{(\alpha)}\right) \hat{\star} K_M. 
\eea

\subsection{$Y_Q$}
\label{sec:rederiveYQ}
The final TBA equation we want to extract from the analytic $Y$-system is the one for $Y_Q$ functions. As we already saw in the derivation of suitable discontinuity equations its analysis is a lot more intricate than for other $Y$ functions. In order to follow the method we used so far we need to go through a few extra steps: since we only know the discontinuities of $\Delta$ and not $\Delta$ itself we will first derive an expression for $\Delta$ from its discontinuities, in the same spirit as we have done so far. We can then use the fact that $\Delta$ is closely related to the $\log Y_Q$ discontinuities to derive the TBA equations for $Y_Q$ from a contour integral expression. 

\subsubsection{Retrieving $\Delta$}
\begin{figure}[!t]
\centering
\begin{subfigure}{7.5cm}
\includegraphics[width=7.5cm]{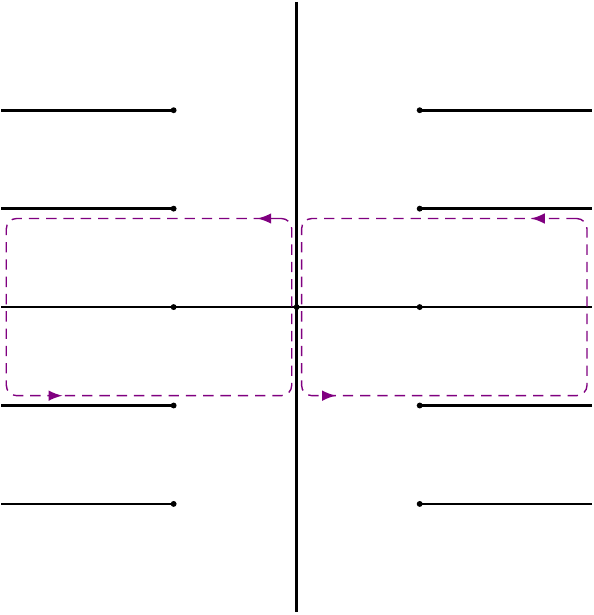}
\caption{}
\end{subfigure}
\qquad \qquad
\begin{subfigure}{6cm}
\includegraphics[width=6cm]{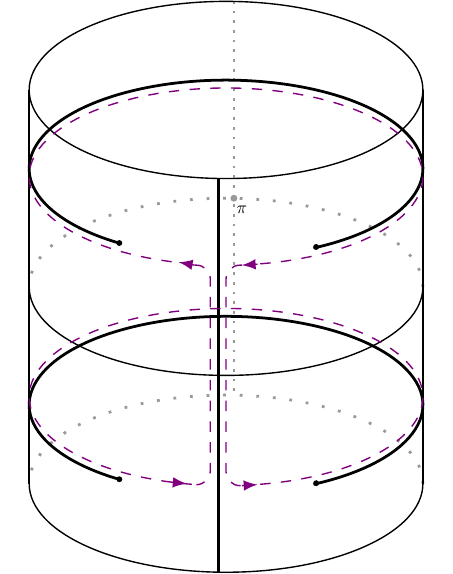}
\caption{}
\end{subfigure}
\caption{The contour $C_1$ for both the undeformed (a) and deformed (b) case. Note that in (a) the outer vertical parts of the contour are pushed to $\pm \infty$.}
\label{fig:gammaxC1}
\end{figure}

\begin{figure}[t]
\centering
\begin{subfigure}{7.5cm}
\includegraphics[width=7.5cm]{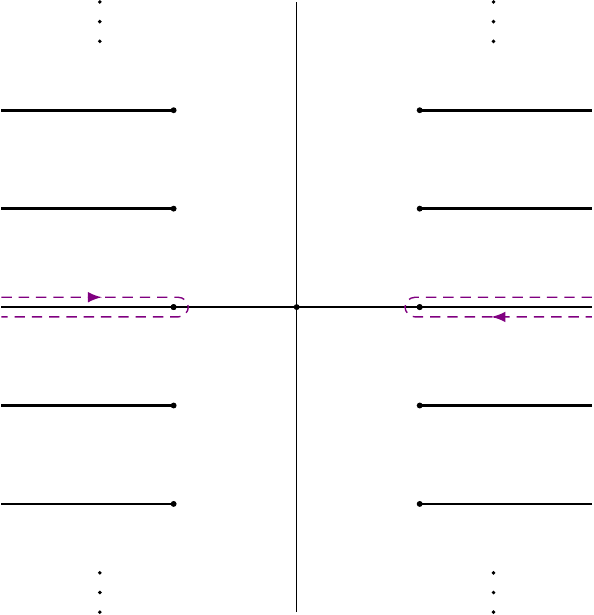}
\caption{}
\end{subfigure}
\qquad \qquad
\begin{subfigure}{6cm}
\includegraphics[width=6cm]{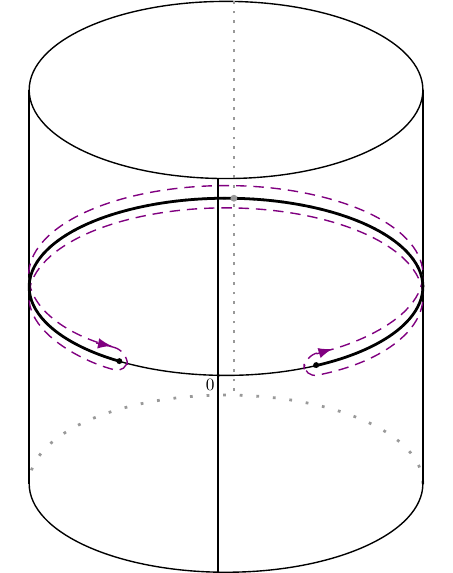}
\caption{}
\end{subfigure}
\caption{The contour $\gamma_x$ for both the undeformed (a) and deformed (b) case, wrapping around the long $\check{Z}_0$ cut. Note that in (a) the contour extends all the way to $\pm \infty$.}
\label{fig:gammax}
\end{figure}

First we will find $\Delta$ from the local discontinuity relations given above. In the physical strip we can write $\Delta$ as
\begin{align}
G(u) \Delta(u) = \oint_{C_1} \frac{dz}{2\pi i}  \Delta(u)B(z,u),
\end{align}
where the contour $C_1$ is depicted in fig. \ref{fig:gammaxC1}. 
Note that we cannot use our previous contour $\gamma$ due to $\Delta$'s extra branch cut on the imaginary axis. Deforming the contour now towards imaginary infinity to give it a $\Gamma$-like shape we reach the contour that encircles all horizontal branch cuts separately in the clock-wise direction (we will call this part $C_2$) and runs along the vertical branch cut on $i \R$ towards $+i\infty$ on the left and $-i\infty$ on the right. This shows that we can write our original integral expression as
\begin{align}
\label{der1}
G(u) \Delta(u) &= \oint_{C_2} \frac{dz}{2\pi i}  \Delta(z)B(z,u) + \left(\int_{i\R -\e}-\int_{i\R +\e}\right) \frac{dz}{2\pi i}  \Delta(z)B(z,u) \nn
&= I_{C_2} + J G(u)\widecheck{\mathcal{E}},
\end{align}
where we use the logarithmic jump of $\Delta$ given in eqn. \eqref{eq:logarithmicjump} and used the relation between $G$ and $K(z,u)$, as well as the definition of the kernel $K(z,u)$, to find the energy term \eqref{eq:energyterm} hidden in $\Delta$. Now we can work on massaging the integral $I_{C_2}$, defined as
\begin{align}
I_{C_2} &= \oint_{C_2} \frac{dz}{2\pi i}  \Delta(z)B(z,u)\nn
&= \sum_{N=1}^{\infty} \sum_{\tau} \oint_{\gamma_x} \frac{dz}{2\pi i}  \Delta(z+\tau i 2N \ad)B(z+\tau i 2N \ad,u),
\end{align}
with $\gamma_x$ as in fig. \ref{fig:gammax}. 
Now we can recognise the discontinuity of $\Delta$ and using its definition in eqn. \eqref{discs1} we find
\begin{align}
I_{C_2} &= \sum_{N=1}^{\infty} \sum_{\tau} \oint_{\gamma_x} \frac{dz}{2\pi i}  \Delta(z+\tau i 2N \ad)B(z+\tau i 2N \ad,u) \nn
&= \sum_{N=1}^{\infty} \sum_{\tau}\int_{\hat{Z}_{0}}\frac{dz}{2\pi i}  \left[ \Delta(z)\right]_{\tau 2N} B(z+\tau i 2N \ad,u)\nn
&= G(u)\sum_{N,\tau,\alpha}\tau \oint_{\gamma_x}dz
\left( \factornumbering{L^{(\alpha)}_{-\tau}(z+i\tau 2N\ad)}{$1$}
 + \factornumbering{\sum_{M=1}^N L_{M|vw}^{(\alpha)}(z+\tau(2N-M)i\ad)}{$2$}
 + \factornumbering{\log Y_-^{(\alpha)}(z)}{$3$} \right)\cdot\nn & \phantom{=} K(z+\tau i 2N \ad,u).
\end{align}
We have already factored out $G$ from the expression, but to continue we will split the contributions of the numbered factors. The first term can be treated as follows: we deform $\gamma_x$ using the analyticity of the integrand into $\gamma$. Now we can rewrite the first term using the analyticity of $L_{\pm}$ in the $\substack{\mbox{\small lower}\\ \mbox{\small
upper}}$ half-plane and using that $K(z_*,u) =-K(z,u)$:
\begin{align}
& G(u) \sum_{N=1}^{\infty} \sum_{\tau,\alpha}\tau\oint_{\gamma_x}dz L^{(\alpha)}_{-\tau}(z+i\tau 2N\ad)K(z+\tau i 2N \ad,u)\nn
&= -G(u)\sum_{\alpha} \left(\int_{Z_{-2}+i\e}dz L^{(\alpha)}_{+}K(z,u)+\int_{Z_{2} -i\e}dz L^{(\alpha)}_{-}(z)K(z,u)\right),\nn
&= G(u)\sum_{\alpha} \left( L^{(\alpha)}_{-} -  \left(L_-^{(\alpha)}+L_+^{(\alpha)}\right)\star K (u) \right) ,
\end{align}
exactly giving us the relevant terms in the definition of $\Delta$ in \ref{delta2}.
\\[5mm]
The second term follows the standard procedure:
\begin{align}
& G(u) \sum_{N=1}^{\infty} \sum_{\tau,\alpha}\tau\oint_{\gamma_x}dz \sum_{M=1}^N L_{M|vw}^{(\alpha)}(z+\tau(2N-M)i\ad) K(z+\tau i 2N \ad,u) \nn
&= G(u) \sum_{M=1}^{\infty}\sum_{N=M}^{\infty} \sum_{\tau,\alpha}\tau\left(\int_{Z_{2N\tau}+i\e}-\int_{Z_{2N\tau} -i\e}\right)dz  L_{M|vw}^{(\alpha)}(z-\tau M i\ad) K(z,u)\nn
&=- G(u) \sum_{M=1}^{\infty} \sum_{\alpha}\int_{Z_{0}} dz L_{M|vw}^{(\alpha)}(z) \left( K_M(z,u)\right)  = -G(u) \sum_{M=1}^{\infty} \sum_{\alpha} L_{M|vw}^{(\alpha)}\star  K_M,
\end{align}
which also exactly matches the definition given in \ref{delta2}.
\\[5mm]
The third term can immediately be seen to give
\be
 G(u) \sum_{\alpha}\oint_{\gamma_x}dz \log Y_-^{(\alpha)}(z)\sum_{N=1}^{\infty} \left( K(z+ i 2N \ad,u)-K(z- i 2N \ad,u)\right),
\ee
which matches the expression \eqref{eq:simplifieddressing} for the dressing phase factor. So we see that we can indeed reconstruct $\Delta$ from its discontinuities, which means we can use it in the derivation of the $Y_Q$ TBA-equation.
\subsection{Reconstructing the $Y_Q$ TBA-equation}
\label{sec:rederivingYQ}
The final thing we need to do to derive the ground-state TBA-system from the analytic $Y$-system is to rederive the TBA equation for $Y_Q$-particles. First we write $\log Y_Q$ as a contour integral on the physical strip
\begin{align}
\log Y_Q (u) = \oint_{\gamma} \frac{dz}{2\pi i} \log Y_{Q}(z) H(z-u).
\end{align}
Following the analysis as in the previous section the following step is to deform the contour $\gamma$ into $\Gamma$. It is important to note that $\Gamma$ is the union of infinitely many contours and therefore we should worry about the convergence of this transition. The limit describing the $\gamma\rightarrow \Gamma$ transition depends on the behaviour of $\log Y_Q$ as $u\rightarrow \pm  i \infty$. As discussed in section \ref{sec:TBAasymptotics} the deformed case requires us to be a little more careful here: the undeformed $Y_Q$ functions are such that $\log Y_Q$ vanishes as $u\rightarrow i \infty$ and therefore we do not need to alter our usual analysis. As we saw the deformed $Y_Q$ functions have a non-vanishing limit
\be
\log Y_Q \rightarrow \mp \ad QJ \text{ as } u \rightarrow \pm i \infty
\ee
requiring us to be more careful. Writing $\gamma^{\pm}$ for the parts of $\gamma$ in the upper and lower half-plane respectively we now find
\begin{align}
\log Y_Q &= \oint_{\gamma} \frac{dz}{2\pi i} \log Y_{Q}(z) H(z-u) \nn
&= \int_{\gamma^+} \frac{dz}{2\pi i} \left( \log Y_{Q}(z) +\ad QJ - \ad Q J \right)H(z-u)  \nn
&+ \int_{\gamma^-} \frac{dz}{2\pi i} \left( \log Y_{Q}(z) +\ad QJ - \ad Q J \right)H(z-u) \nn
&= \ad Q J \oint_{z=u}\frac{dz}{2\pi i} H(z-u) +    \int_{\gamma^+} \frac{dz}{2\pi i} \left( \log Y_{Q}(z) +\ad QL  \right)H(z-u)  \nn
&+ \int_{\gamma^-} \frac{dz}{2\pi i} \left( \log Y_{Q}(z) - \ad Q J \right)H(z-u).
\end{align}
The contour integral can be done easily to give $\ad Q J$ and deforming $\gamma^{\pm}$ in the usual way to $\Gamma$, we get
{
\bea
\label{eq:LogYQcontour}
\oint_{\gamma} \frac{dz}{2\pi i} \log Y_{Q}(z) H(z-u) &=& \ad Q J + \sum_{\tau}\sum_{l=0}^{\infty} \int_{Z_0} \left[ \log Y_{Q}(z) \right]_{\tau(Q+2l)} H(z-u).\qquad 
\eea
}
The undeformed version of this formula can be obtained by omitting the first factor on the right-hand side. 
\\[5mm]
To continue we will need discontinuities of the form
$$
\left[ \log Y_{Q} \right]_{\pm (Q+2l)},
$$
where $Q,l\in \mathbb{N}$, which we will derive from the $Y$ system. Its derivation is very similar to the one for the $Y_{(v)w}$ functions (see \ref{app:Yvwdiscs}) and we discuss it in detail in appendix \ref{app:YQdiscs}. It uses the knowledge of the analyticity strips and branch cut locations of the $Y$ functions to combine discontinuities of the $Y$ system. The result is given in terms of $D$ functions, which are defined for $l\geq 0$ as
\begin{align}
D_{\tau (Q+2l)}^Q(u) &= \sum_{J=1}^{l+1}\sum_{M=J}^{Q+J-2} L_{vw|M}^{(\alpha)}(u+\tau(M+2l-2J+2)i\ad)  + L_-^{(\alpha)}(u+2\tau l i \ad)  \nn
&-\sum_{J=1}^l \left( 2 \sum_{M=1}^{Q-1}  \Lambda_{M+J}(u+\tau(M+2l-J)i\ad) + \Lambda_{Q+J}(u+\tau(Q+2l-J)i\ad)\right)\nn
&-\sum_{M=1}^{Q-1} \Lambda_{Q}(u+\tau(M+2l)i\ad) -  \sum_{J=1}^l  \Lambda_{J}(u+\tau(2l-J)i\ad) ,
\end{align}
and obey the following recursion relation\footnote{This is similar to the corresponding expression in \cite{Cavaglia:2010nm} except for the lower bound on $M$ for the $vw$ terms: it starts at $l+1$ instead of $l$.}:
\begin{align}
\label{recYQD}
D_{\tau (Q+2l)}^Q(u) - D_{\tau (Q+2l-2)}^Q(u +\tau 2 i\ad ) &=\sum_{M=l+1}^{Q+l-1} L_{vw|M}^{(\alpha)}(u+\tau M i\ad)-   \Lambda_{l}(u+\tau li\ad)  \nn
&- 2 \sum_{M=1+l}^{Q-1+l}  \Lambda_{M}(u+\tau Mi\ad) -\Lambda_{Q+l}(u+\tau (Q+l)i\ad).
\end{align}
The discontinuity of $Y_Q$ can now be written as follows for all $l\geq0$:
\begin{align}
\left[ \log Y_{Q} \right]_{\tau(Q+2l)}(u) &= \left[D_{\tau (Q+2l)}^Q\right]_0 (u) -\delta_{l,0}\left[\log Y_{1}\right]_{-\tau 1}(u).
\end{align}
Plugging this into our expression \eqref{eq:LogYQcontour} for $\log Y_Q$ leaves 
\begin{align}
&\oint_{\gamma} \frac{dz}{2\pi i} \log Y_{Q}(z) H(z-u)= \ad Q L
\nn &+ \sum_{\tau}\sum_{l=0}^{\infty} \int_{Z_0}\frac{dz}{2\pi i} \left(\left[D_{\tau (Q+2l)}^Q\right]_0 (u) -\delta_{l,0}\left[\log Y_{1}\right]_{-\tau 1}(u) \right) H(z+\tau(Q+2l)\ad i-u).
\end{align}
Let us massage these integrals in steps and consider the terms associated to different $Y$ functions separately. As a first step we simplify the contributions coming from the $D$ functions and $\log Y_1$ separately. The $D$-function contribution can be written as
\begin{align}
\label{eq:Dcontribution}
&\sum_{\tau}\sum_{l=0}^{\infty} \int_{Z_0}\frac{dz}{2\pi i} \left[D_{\tau (Q+2l)}^Q\right]_0 (u) H(z+\tau(Q+2l)\ad i-u) \nn
&=- \sum_{\tau}\sum_{l=1}^{\infty}\tau  \int_{Z_0 -i\tau \e } \left(D_{(Q+2l)\tau}^{Q}(z) -D_{(Q+2l-2)\tau}^{Q}(z+2\tau \ad)\right) H(z+\tau (Q+2l)\ad i-u) \nn
&-\sum_{\tau} \tau \int_{Z_0 -i\tau \e }\left(L_- + \sum_{M=1}^{Q-1} \Lambda_{M}(z+\tau i \ad M )+\sum_{M=2}^{Q} L_{vw|M-1}^{(\alpha)}(z+\tau (M-1) i\ad ) \right)\cdot \nn &H(z+\tau Q\ad i-u),
\end{align}
whereas the $\log Y_1$ contribution can be rewritten as
\begin{align}
\label{eq:Y1contribution}
&-\sum_{\tau}\int_{Z_0}\left[\log Y_{1}\right]_{-\tau 1}(u) H(z+\tau Q \ad i-u)\nn
&=-\int_{Z_0}\left[L_-^{(\alpha)} \right]_0H(z+ Q\ad i-u) -\oint_{\gamma_x}\log Y_{1}(u+i \ad) K_Q(z-u),
\end{align}
using the $Y$-system property
\be
\label{eq:Y1property}
\left[\log Y_1\right]_1+\left[\log Y_1\right]_{-1} = \left[L_-^{(\alpha)}\right]_0. 
\ee
Simplifying the contour integral over $\gamma_x$ containing $\log Y_1$ will be quite complicated, so we will first treat the $Y_-$-terms, collecting all contributions in both the rewritten expressions in eqn. \eqref{eq:Dcontribution} and eqn. \eqref{eq:Y1contribution}:
\begin{align}
&\sum_{\tau}\sum_{l=0}^{\infty} \int_{Z_0}\frac{dz}{2\pi i} \left(\left[L_-^{(\alpha)}\right]_{\tau 2l}(u) -\delta_{l,0}\delta_{\tau,+1}\left[L_-^{(\alpha)} \right]_0(u) \right) H(z+\tau(Q+2l)\ad i-u)\nn
&=-\oint_{\gamma_x}\frac{dz}{2\pi i} L_-^{(\alpha)}(u)H(z+Q\ad i-u) + \sum_{\tau}\tau \int_{Z_0+i\tau \e }\frac{dz}{2\pi i} L_-^{(\alpha)}(u) H(z-\tau Q\ad i-u)\nn
&= \int_{Z_0+i \e }  L_-^{(\alpha)}(u) K_Q(z-u).
\end{align}
Now let us treat the remaining terms in eqn. \eqref{eq:Dcontribution}, including the bulk terms from the $D$s, applying the recursion relation \eqref{recYQD}:
\begin{align}
\label{derivYQ1}
&- \sum_{\tau}\sum_{l=1}^{\infty}\tau  \int_{Z_0 -i\tau \e } \left(D_{(Q+2l)\tau}^{Q}(z) -D_{(Q+2l-2)\tau}^{Q}(z+2\tau \ad)\right) H(z+\tau (Q+2l)\ad i-u) \nn
&-\sum_{\tau} \tau \int_{Z_0 -i\tau \e } \sum_{M=1}^{Q-1} \Lambda_{M}(z+\tau i \ad M) H(z+\tau Q\ad i-u)\nn
&=
- \sum_{\tau}\sum_{l=1}^{\infty}\tau  \int_{Z_0 -i\tau \e } \Bigg(-   \Lambda_{l}(z+\tau li\ad ) -\Lambda_{Q+l}(z+\tau (Q+l)) \nn
&\left.- 2 \sum_{M=1+l}^{Q-1+l}  \Lambda_{M}(z+\tau Mi\ad)+\sum_{M=l+1}^{Q+l-1} L_{vw|M}^{(\alpha)}(z+\tau M i\ad) \right) H(z+\tau (Q+2l)\ad i-u) \nn
&-\sum_{M=1}^{Q-1} \left(\Lambda_{M}-L_{vw|M}\right) \star K_{Q-M}.
\end{align}
The terms in the previous expression containing $Y_Q$ functions can be simplified to\\ $-\sum_{M=1}^{\infty} \Lambda_M \star K_{M Q}$, whereas the $Y_{vw}$ functions can be simplified to the term
\begin{align}
&- \sum_{\tau}\sum_{l=1}^{\infty}\tau  \int_{Z_0 -i\tau \e } \sum_{M=1+l}^{Q+l-1} L_{vw|M}^{(\alpha)}(z+\tau M i\ad) H(z+\tau (Q+2l)\ad i-u)+\sum_{M=1}^{Q-1} L_{vw|M}\star K_{Q-M}\nn
&=\sum_{l=1}^{\infty}\sum_{M=1+l}^{Q+l-1} L_{vw|M}^{(\alpha)}(z)\star K_{Q+2l-M} +\sum_{M=1}^{Q-1} L_{vw|M}\star K_{Q-M} \nn
&=\sum_{l=1}^{\infty}\sum_{M=l}^{Q+l-2} L_{vw|M}^{(\alpha)}(z)\star K_{Q+2l-M-2}
=\sum_{M=1}^{\infty}\sum_{l=\max(0,M-Q+1)}^{M} L_{vw|M}^{(\alpha)}(z)\star K_{Q+2l-M} \nn
&=\sum_{M=1}^{\infty}\sum_{l=0}^{M} L_{vw|M}^{(\alpha)}(z)\star K_{Q+2l-M},
\end{align}
using that the terms we introduce in the last line sum up to zero due to the antisymmetry of $K_M$ in $M$. Let us summarise what our results thus far are, by writing the simplest expression we have for the right-hand side of the $Y_Q$ TBA equation.
\begin{align}
\label{prelres}
\log Y_Q(u) &= -\oint_{\gamma_x}\log Y_{1}(u+i\ad) K_Q(z-u)-\sum_{M=1}^{\infty} \Lambda_M \star K_{MQ}(u)\nn 
&+\int_{Z_0+i \e }  L_-^{(\alpha)}(u) K_Q(z-u)+\sum_{M=1}^{\infty} \sum_{l=0}^{M} L_{M|vw}\star K_{Q+2l-M}(u)+ \ad Q L.
\end{align}
The next step is using our knowledge of $\Delta$ to rewrite the first contour integral:
\begin{align}
\label{eq:y1expl}
&\oint_{\gamma_x}\log Y_{1}(u+i\ad) K_Q(z-u) = \int_{\check{Z}_0}\cD(u) K_Q(z-u)
\nn
 &= \frac{1}{2}\oint_{\gamma_x} \left(\Delta-\sum_{\alpha}L_-^{(\alpha)}\right)(u) K_Q(z-u)
+ \int_{\check{Z}_0}\left( \sum_{\alpha}\check{L}_-^{(\alpha)}(u)\right) K_Q(z-u)
\nn &= \frac{1}{2}\oint_{\gamma_x} \left(-J \widecheck{\mathcal{E}} - \sum_{\alpha} \left( L^{(\alpha)}_{-} +  L^{(\alpha)}_{+} \right) \hat{\star} K  - 2 \Lambda_M\star K_M^{\Sigma} - \sum_{\alpha} L_{M|vw}^{(\alpha)}\star K_M\right)K_Q(z-u)
\nn &+ \int_{\check{Z}_0}\left( \sum_{\alpha}\check{L}_-^{(\alpha)}(z)\right) K_Q(z-u)
\end{align}
We can now look at all the separate terms to recognise the TBA-contributions.
\subsubsection{Driving term $\tilde{E}_Q$}
First we treat the energy contribution: it is retrieved from
\begin{align}
- \frac{J}{2}\oint_{\gamma_x} \widecheck{\mathcal{E}} K_Q(z-u).
\end{align}
We first consider the slightly more general integral
\begin{align}
\oint_{\gamma_x}dz f(z) K_Q(z-u),
\end{align}
for both the deformed and undeformed cases.\footnote{This means in the deformed case $f$ is $2\pi$ periodic, whereas in the undeformed case it vanishes as $u\rightarrow \pm \infty$.} We can directly apply Cauchy's theorem to the domain on the outside of this contour, picking up residues at all the poles. $K_M$ only has two poles, so we get for a function $f$ which is analytic and regular on this domain the result
\begin{align}
\label{fsimple}
\oint_{\gamma_x}dz f(z) K_Q(z-u) =  f(u+i Q \ad) - f(u-i Q \ad)
\end{align}
We cannot directly apply this formula to the energy term, since that has a branch cut at the imaginary axis. However, incorporating the branch cuts is quite straightforward:
\begin{align}
\label{eq:rederiveenergy}
\oint_{\gamma_x}dz \widecheck{\mathcal{E}} K_Q(z-u)&= \widecheck{\mathcal{E}}(u+i Q\ad)- \widecheck{\mathcal{E}}(u-i Q\ad)+ \left( \int_{i \R-\epsilon}-\int_{i \R+\epsilon}\right) dz \widecheck{\mathcal{E}}(u)K_Q(z-u) \nn
&= -2\tilde{E}_Q
+\int_{i \R}2\pi i K_Q(z-u)dz = -2\tilde{E}_Q- 2 \ad Q,
\end{align}
where the second term in the last line is due to the branch cut along the imaginary axis and we used information about the jump of $\widecheck{\mathcal{E}}$ over the imaginary axis. The undeformed kernel $K_Q$ actually integrates to zero over $i\R$, such that in the undeformed case the second factor is absent altogether. The discrepancy between the two derivations here is due to different asymptotics of the $\log Y_Q$ functions in the two cases. After correctly accounting for prefactors this term exactly cancels the extra term $\ad Q J$ we picked up from the boundary condition on $\log Y_Q$, which is absent in the undeformed case. 
\subsubsection{The $L_{M|vw}$ terms}
For rewriting the contribution of the $L_{M|vw}$ we can explicitly write it as
\be
-\frac{1}{2}\oint_{\gamma_x}dz \sum_{\alpha} L_{M|vw}^{(\alpha)}\star\left(K(t+i M\ad,z)+K(t-i M \ad,z) \right)K_Q(z-u).
\ee
This can further be simplified by contour deformation of $\gamma_x$ for some of the terms involved, which governs the integration over $z$:
by moving both halves of the contour onto the second sheet we have formed a contour that looks exactly like $\gamma_x$, but it runs in the other direction and on the second sheet.\footnote{Note the subtlety involved here: when deforming we are first left with a closed contour on the second sheet plus two contours enclosing the
branch points on the first sheet. We can take these last two contours
to the second sheet by reversing the direction of integration, leaving us with $\gamma_x$ on the second sheet.}
This leads to
\begin{align}
&-\frac{1}{2}\oint_{\gamma_x}dz \sum_{\alpha} L_{M|vw}^{(\alpha)}\star\left(K(t+i M\ad,z)+K(t-i M \ad,z) \right)K_Q(z-u)
\nn
&= \frac{1}{2\pi i}\oint_{\gamma_x}dz \sum_{\alpha} L_{M|vw}^{(\alpha)}\star_t
\left(\frac{d}{dt}  \log \left(x(t+i M \ad) -x(z)\right)\left(x(t-i M \ad) -x(z)\right)  \right)K_Q(z-u),
\end{align}
which can be simplified even further by deforming $\gamma_x$ again using an extension of eqn. \eqref{fsimple}: if $f$ has poles itself, the right-hand side of eqn. \eqref{fsimple} will also feature the poles of $f$. In this case
\be
f(z) = \frac{d}{dt}  \log \left(x(t+i M \ad) -x(z)\right)\left(x(t-i M \ad) -x(z)\right),
\ee
which gives for the $L_{M|vw}$ contribution
\begin{align}
& \frac{1}{2\pi i}\oint_{\gamma_x}dz \sum_{\alpha} L_{M|vw}^{(\alpha)}\star_t
\left(\frac{d}{dt}  \log \left(x(t+i M \ad) -x(z)\right)\left(x(t-i M \ad) -x(z)\right)  \right)K_Q(z-u)
\nn &= \sum_{\alpha} L_{M|vw}^{(\alpha)}\star \frac{1}{2\pi i} \frac{d}{dt}\Bigg( \log \left(
\frac{S( t+i M \ad-u+i Q \ad)}{S( t+i M \ad-u-i Q \ad)}
\frac{S( t-i M \ad-u+i Q \ad)}{S( t-i M \ad-u-i Q \ad)} \right)
\nn &+ \log \left(\frac{x(t+i M \ad) -x(u+iQ\ad)}{x(t+i M\ad) -x(u-iQ\ad)}\frac{x(t-i M\ad) -x(u+iQ\ad)}{x(t-i M\ad) -x(u-iQ\ad)}
\right) \Bigg)\nn
&=  -\sum_{\alpha} L_{M|vw}^{(\alpha)}\star \left( K^{MQ}_{vwx}(u) -  \sum_{i=0}^{M} K_{Q-M+2i} \right),
\end{align}
where we introduced 
\begin{equation}
\label{eq:Sfunction}
S(u) = \sin u/2, \quad S^{\text{und}}(u) = u.
\end{equation}
If we now combine all the $Y_{vw}$ contributions on the right-hand side of the $Y_Q$ TBA equation (see also eqn. \eqref{prelres}) we get
\begin{align}
\sum_{l=0}^{M} L_{M|vw}\star K_{Q+2l-M}(u) -\frac{1}{2}\oint_{\gamma_x}dz \sum_{\alpha} L_{M|vw}^{(\alpha)}\star K_M(z)K_Q(z-u) \nn
= \sum_{\alpha} \sum_{M=1}^{\infty} L_{M|vw}^{(\alpha)}\star K^{MQ}_{vwx}(u)
\end{align}
which is the correct TBA contribution due to the $vw$ functions.
\subsubsection{The $L_-$ terms}
The contributions of $L_-$ in eqn. \eqref{eq:y1expl} to the TBA equation can be summed up as
\begin{align}
& \frac{1}{2}\oint_{\gamma_x}\sum_{\alpha} \left( L^{(\alpha)}_{-} +  L^{(\alpha)}_{+} \right) \hat{\star} K(z) K_Q(z-u) \nn
&- \int_{\check{Z}_0} \left( \sum_{\alpha}\check{L}_-^{(\alpha)}(z)\right) K_Q(z-u)  + \int_{Z_0+i \e }  L_-^{(\alpha)}(u) K_Q(z-u).
\end{align}
We can rewrite the first integral by contracting the integral contour. For this, we need to continue the integral through the $\hat{Z}_0$ cut, resulting in a residue being picked up due to the convolution on this interval. Therefore, the first integral reads
\begin{align}
&\frac{1}{2}\oint_{\gamma_x}\sum_{\alpha} \left( L^{(\alpha)}_{-} +  L^{(\alpha)}_{+} \right) \hat{\star} K(z) K_Q(z-u)  \nn
&= -\left( L^{(\alpha)}_{-} \hat{\star}K_-^{yQ} - L^{(\alpha)}_{+} \hat{\star}K_+^{yQ}\right)
-\int_{Z_0+i \e }  L_-^{(\alpha)}(u) K_Q(z-u)\nn
&+ \int_{\check{Z}_0}\left( \sum_{\alpha}\check{L}_-^{(\alpha)}(z)\right) K_Q(z-u)
\end{align}
such that after combining them we find as the total $L_{\pm}$ contribution
\be
L_{\beta} \hat{\star} K_{\beta}^{yQ}.
\ee
\subsubsection{Restoring the dressing phase contribution}
\label{sec:restoringdressing}
The dressing phase kernel in the TBA equations gets restored by the term
\be
\oint_{\gamma_x} \Lambda_P\star K_P^{\Sigma}(z) K_Q(z-u).
\ee
We can first rewrite the product $\Delta^{\Sigma}= \Lambda_P\star K_P^{\Sigma}$ using the result \eqref{eq:simplifieddressing} from appendix \ref{app:dressing}. Using the kernel $K_{q\Gamma}^{[Q]}$ and the identity \eqref{eq:Kqgammaidentity}
 we replace the infinite sum containing $K$ kernels in eqn. \eqref{eq:simplifieddressing} by the terms containing $K_{q\Gamma}^{[N]}$. Using the TBA equation for $Y_-$ we then obtain
\be
\Delta^{\Sigma}(u) = 2 \Lambda_P \star \oint_{\gamma_x} ds K^{Py}_-(s) \left( \oint_{\gamma_x} dt K_{q\Gamma}^{[2]}(s-t)K(t,u)-K_{q\Gamma}^{[2]}(s-u) \right) \text{ for } u \not\in \check{Z}_0.
\ee
We can recognise the right-hand side of this equation as the discontinuity of the function
\be
2 \Lambda_P \star \oint_{\gamma_x} ds K^{Py}_-(s) \oint_{\gamma_x} dt K_{q\Gamma}^{[2]}(s-t)\frac{1}{2\pi i} \frac{d}{dt} \log \left( x(t) -x(u) \right),
\ee
such that we can immediately rewrite
\be
\oint_{\gamma_x}dz \, \Delta^{\Sigma}(z) K_Q(z-u) =
- 2 \Lambda_{P} \star \oint_{\gamma_x} ds \, K^{Py}_-(s) \oint_{\gamma_x} dt \, K_{q\Gamma}^{[2]}(s-t)K^{yQ}_-(t,u).
\ee
Following \cite{Cavaglia:2010nm} and in particular using equation (4.17) in \cite{Arutynov:2014ota} we find that indeed
\be
\oint_{\gamma_x}dz \, \Delta^{\Sigma}(z) K_Q(z-u) =
- 2 \Lambda_{P} \star K_{P  Q}^{\Sigma}.
\ee
Together with the already present $\Lambda_P \star K_{PQ}$ term this forms the dressing phase term in the TBA equations.

\subsection*{Combining all partial results}
Combining all the partial results in the previous subsections and plugging them into eqn. \eqref{prelres} we get
\be
\log Y_Q(u) = -J \tilde{E}_Q + \sum_{\alpha}\left( \sum_{M=1}^{\infty} L_{M|vw}^{(\alpha)}\star K^{MQ}_{vwx}(u)+ L_{\beta}^{\alpha} \hat{\star} K_{\beta}^{yQ}\right)   + \Lambda_P \star K_{\mathfrak{sl}(2)}^{PQ},
\ee
which indeed coincides with the TBA equation \eqref{eq:TBAeqns} for $Y_Q$.

\section{Conclusions}
The main result of this chapter is the analytic $Y$-system \eqref{eq:Ysystem} and \eqref{discs1} as follows from the TBA equations we started with. Particular solutions to the analytic $Y$-system allow us to compute energies of string states of the $\ads$ superstring and its $\eta$ deformation. The analytic $Y$-system consists of a general set of finite-difference equations for the $Y$ functions already featuring in the TBA equations and has to be supplemented with additional analyticity data. Deriving a complete yet minimal set of analyticity data is not straightforward though.
\\[5mm]
In the second part of this chapter we have checked the completeness of the found analyticity data: we showed that we can reproduce the ground-state TBA equations from the analytic $Y$-system assuming only some properties which are specific to the ground-state $Y$-functions: in that case the $Y$ functions are pole-free except possibly at branch points and vanish at $\pm i \infty$ except for the deformed $Y_Q$ functions. These have non-vanishing asymptotics reflected by the driving term in its TBA equation.

%% file: Tsystem.tex
\section{Introduction}
In the previous chapter we have reformulated the TBA equations as an analytic $Y$-system, thereby turning a set of coupled integral equations into a set of coupled finite-difference equations. This comes at the price of losing information that was built-in in the TBA equations about the analytic properties of the solution that one has to supplement in a different form in order to find the same solution. Therefore one could wonder whether this reformulation is actually worthwhile. 
\\[5mm]
Before explaining why we (of course) think this transition is important, let us note that from a computational point of view the analyic $Y$-system does not necessarily simplify the spectral problem. It is not easy to impose the discontinuity relations in practice and hence in many cases one actually progresses in the other direction: a given $Y$ system is transformed into a set of TBA-like equations, which allow for direct numerical computations. Moreover, addressing questions concerning the existence and uniqueness of solutions historically has been easier for integral equations than for finite-difference equations. The standard argument uses the Banach fixed-point theorem on the integral operator that relates the left- and right-hand sides of the TBA equation as follows: we can schematically write the TBA equation as 
\begin{equation}
\label{eq:BanachTBA}
\mathbf{Y} = L\left( \mathbf{Y} \right),
\end{equation}
where $L$ is some integral operator and $\mathbf{Y}$ is the vector containing all the $Y$ functions. Under certain conditions one can prove that $L$ has a unique fixed point, which then must solve the TBA equations as is immediate from eqn. \eqref{eq:BanachTBA}. Curiously, many of the analytical questions concerned with such a proof were left for a long time: the recent paper \cite{Hilfiker:2017jqg} treats these questions systematically for a wide range of TBA equations and $Y$ systems. 
\\[5mm]
One of the main reasons why the $Y$ system provides such an appealing description of the spectral problem is its generality: its functional form is the same for all excited states and can be written down based on general information about the model alone.\footnote{Roughly put one only requires knowledge of the locations of the non-zero entries of the $S$ matrix and the shift distance.} The $Y$ system for certain integral models are even identical, a somewhat trivial example being the $Y$ system related to the $\ads$ string and mirror model. Much more interestingly the $Y$-system equations for the undeformed and deformed case are also identical up to a trivial reparametrisation of the shift parameter. The additional periodicity requirements determine the difference between the two cases. A similar observation can be made for many other models, such as the $Y$ system of the Lee Yang minimal models, where the equations look the same for all geometries and excitations \cite{Bajnok:2014fca}. We will not pursue the question of how this uniformity arises -- which is a deep question in integrability -- in this thesis, but invite the readers to look at the review \cite{Gromov:2010kf}, the PhD thesis \cite{Leurent:2012xc} and the classic paper \cite{bazhanov1997quantum}. \note{Am I missing an important reference here?}
\\[5mm]
The next step to take is to reformulate the analytic $Y$-system as an \emph{analytic $T$-system}. The new $T$ functions are introduced as a decomposition of the $Y$ functions and hence it is not straightforward that the $T$ functions are ``nice" in any way. However, their explicit relation to the $Y$ functions is also deeply rooted in integrability, as they homogenise the $Y$-system equations further: we will see that the $T$ functions satisfy the Hirota equation, a ubiquitous equation in the study of quantum integrable models. Where the $Y$-system equations still depended on the model under investigation, the Hirota equation is universal for many integrable models. Since for us the $T$ system is just a stop on our way to the quantum spectral curve we will not venture into the discussion of this topic. The references mentioned above together with \cite{Krichever:1996qd,Gromov:2010km} are great reading material for the interested reader.  
\\[5mm]
In this chapter we will discuss how to perform the reformulation of the analytic $Y$-system as an analytic $T$-system. This discussion will be almost completely about analytic questions, since on an algebraic level the transition is straightforward. It is much more involved to prove that the decomposition into $T$ functions leads to a consistent set of $T$ functions with good analytic properties. The parametrisation of $Y$ functions in terms of $T$ functions has a gauge freedom that will allow us to tune these properties to some extent, but will make the discussion technically involved. It turns out that it seems impossible to define a set of $T$ functions such that all its $T$ functions have nice analytic properties. Instead, we will resort to two sets of $T$ functions, each of which has a subset of nicely-looking $T$s, such that for every $T$ function there is at least one nice description of it. These different sets are related by a gauge transformation usually called \emph{gluing condition}.

\paragraph{Strategy.} The strategy to derive the $T$ functions is very much inspired by the strategy used in \cite{Gromov:2011cx} for the undeformed case: 
\begin{itemize}
\item we will start by introducing the parametrisation and the corresponding $T$ hook and discuss the gauge freedom it introduces and how it simplifies the $Y$ system. 
\item We then get into the construction of different $T$ gauges, i.e. different sets of $T$ functions, all of which we will give their own distinctive version of the letter $T$. We first construct the $\JT$ gauge from the $Y$ system. Using the gauge freedom we can improve this gauge into a nice gauge $\bfT$ for the \emph{upper band} (see fig. \ref{fig:thook}).
\item Next we introduce a construction for $T$ gauges which are nice for the left and right band called $\mbT_L$ and $\mbT_R$. Collectively we denote these gauges with $\mbT$.
\item To check that $\mbT$ has nice properties we introduce yet another gauge $\mcT$. This gauge can be used to prove that $\mbT$ has good analyticity properties and has $\Z_4$ symmetry. 
\item Finally we prove that $\mbT$ can be decomposed again using the \emph{Wronskian parametrisation}, which is the last step before introducing the quantum-spectral-curve parametrisation.
\end{itemize}
The analyticity properties of the resulting gauges $\bfT$ and $\mbT$ are summarised in table \ref{tab:Tsanalyticity} using the notation introduced around eqn. \eqref{eq:Astrips}. We will consider this derivation for both the undeformed and the deformed case simultaneously, discussing differences along the way.
\begin{table}
\begin{center}
\begin{tabular}[t]{ l | l}
For $a\geq |s|$ & For $s\geq a$ \\
\hline
$\bfT_{a,0} \in \mathcal{A}_{a+1}$ & $\mathbb{T}_{0, \pm s } = 1$ \\
$\bfT_{a,\pm 1} \in \mathcal{A}_{a}$ & $\mathbb{T}_{1, \pm s } \in \mathcal{A}_s$ \\
$\bfT_{a,\pm 2} \in \mathcal{A}_{a-1}$ & $\mathbb{T}_{2, \pm s } \in \mathcal{A}_{s-1}$
\end{tabular}
\end{center}
\caption{Analyticity strips of the two $T$ gauges.}
\label{tab:Tsanalyticity}
\end{table}
\section{The $T$ parametrisation}
\begin{figure}[t]
\centering
\includegraphics[width=10cm]{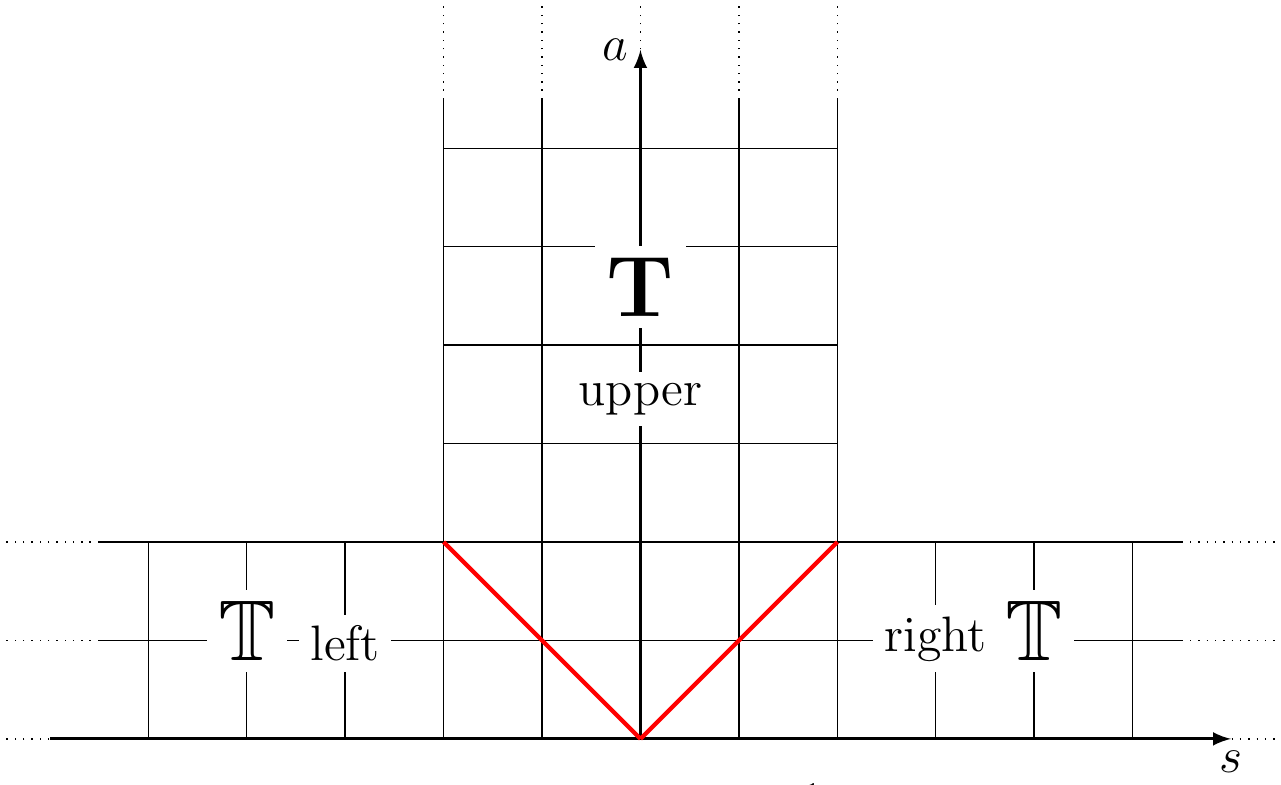}
\caption{The $T$ hook organising all the $T$ functions. The  \emph{upper}, \emph{left} and \emph{right} band are separated at the red lines. The $T$ functions live on the nodes.}
\label{fig:thook}
\end{figure}

On an algebraic level the transition to the $T$ functions works the same for both cases, matching the underlying representation theory.\footnote{For contrast, when the deformation parameter $q$ is a root of unity the TBA equations have a more intricate, truncated, structure \cite{Arutyunov:2012zt}, reflecting itself in the parametrisation in terms of $T$ functions \cite{Arutyunov:2012ai}.} Indeed, we have seen in the previous section that the main difference between the analytic $Y$-system for the $\eta$-deformed and the undeformed case is the periodicity of the $Y$ functions. With the standard Hirota map we decompose the $Y$ functions as
\begin{align}
\label{YsinTs}
Y^{(\pm)}_{M|w} &\defeq Y_{1,\pm(M+1)} = \frac{ T_{1,\pm(M+2)} T_{1,\pm M} }{T_{0,\pm(M+1)} T_{2,\pm (M+1)}}, \quad &Y^{(\pm)}_{M|vw}&\defeq Y_{(M+1),\pm 1}^{-1} = \frac{ T_{M+2,\pm 1} T_{M,\pm 1} }{  T_{M+1,0} T_{M+1,\pm 2} },\nn
Y_{+}^{(\pm)}&\defeq Y_{2,\pm 2} = -\frac{ T_{2,\pm1} T_{2,\pm 3} }{T_{1,\pm 2} T_{3,\pm2}}, \quad &Y_{-}^{(\pm)}&\defeq Y_{1,\pm 1}^{-1} = -\frac{ T_{0,\pm1} T_{2,\pm 1} }{T_{1,0} T_{1,\pm2}},\nn
Y_{M}&\defeq Y_{M,0} = \frac{ T_{M,1} T_{M,-1} }{T_{M-1,0} T_{M+1,0}},
\end{align}
where the $T$ functions $T_{a,s}$ live on a \emph{$T$ hook}, see fig. \ref{fig:thook}, and we have also introduced another notation for the $Y$ functions in liedx3ne with the notation used in \cite{Gromov:2011cx,Gromov:2009tv}. This puts the $Y$ functions on the \emph{$Y$ hook}, illustrated in fig. \ref{fig:Yhook}. 
$T$ functions with indices that fall outside of this hook are zero by definition. The $T$ hook is naturally partitioned into \emph{upper} ($a\geq s$), \emph{left} ($-s\geq a$) and \emph{right} ($s\geq a$) band. The $T$ functions on the $T$ hook homogenise the $Y$-system equations (\ref{Ysys:Q}-\ref{Ysys:y}) in the sense that if the $T$s satisfy the \emph{Hirota equation}
\be
\label{hirota}
T_{a,s}^+T_{a,s}^- = T_{a-1,s}T_{a+1,s}+ T_{a,s-1}T_{a,s+1},
\ee
then the $Y$s parametrised by these $T$s satisfy the $Y$ system equations (\ref{Ysys:Q}-\ref{Ysys:y}). This can be checked directly by plugging in this parametrisation into the $Y$-system equations. Using the $(a,s)$ notation for the $Y$ functions we can compactly write \eqref{YsinTs} as 
\begin{equation}
Y_{a,s}  = \frac{ T_{a,s+1} T_{a,s-1} }{T_{a-1,s} T_{a+1,s}},
\end{equation}
which, using the Hirota equation \eqref{hirota}, admits other representations such as
\begin{equation}
\label{eq:otherYinT}
1+Y_{a,s} = \frac{T_{a,s}^+T_{a,s}^-}{T_{a+1,s}T_{a-1,s}}, \quad 1+\frac{1}{Y_{a,s}} =  \frac{T_{a,s}^+T_{a,s}^-}{T_{a,s+1}T_{a,s-1}}.
\end{equation}
The real question here is whether there exists a set of $T$s that parametrise the $Y$ functions in such a way that those satisfy the additional analyticity properties we derived in the previous chapter. To find out which solutions of the Hirota equation we are interested in we should transfer the analyticity properties from the $Y$ functions to the $T$ functions. Since more than one $T$ function features in the parametrisation of each of the $Y$ functions there is a choice how to distribute the analytic properties of the $Y$ functions over the $T$s. In addition to this there is a gauge freedom present.

\paragraph{Gauge freedom.} The parametrisation of $Y$s in terms of $T$s is left invariant under redefining
\be
\label{eq:Tgaugetransformation}
T_{a,s} \rightarrow g_1^{[a+s]}g_2^{[a-s]}g_3^{[-a+s]}g_4^{[-a-s]}T_{a,s},
\ee
where the $g_i$ are arbitrary functions. What is more, this gauge freedom is also present in the Hirota equation: if $\{T_{a,s}\}$ is a solution to the Hirota equation, so is
\begin{displaymath}
\left\{g_1^{[a+s]}g_2^{[a-s]}g_3^{[-a+s]}g_4^{[-a-s]}T_{a,s}\right\}.
\end{displaymath}
Therefore this gauge freedom is a genuine gauge freedom of the $T$ system, which we can use to improve any given set of $T$s. In the undeformed case we do not need to impose any further restrictions on the $g_i$, but we should carefully consider the deformed case: since the $g_i$ drop out of the parametrisation of the $Y$s we are in principle free to consider any $g_i$ we want, even non-periodic ones, thereby leaving us with $T$ gauges which have no periodicity properties. Looking at this from the perspective of the parametrisation \eqref{YsinTs} where the $T$ features quadratically, one would argue that we do not have to insist on $2\pi$ periodicity of the $T$ functions, although $2\pi$ (anti)-periodicity, being
\begin{equation}
T_{a,s}(u+2\pi) = - T_{a,s}(u),
\end{equation}
does seem sensible. The gauge freedom \eqref{eq:Tgaugetransformation} shows that we can further allow for a non-periodic piece that can be gauged away. In the following we will only define $2\pi$-periodic $T$-gauges, but one should keep in mind that at this point this is motivated by naturalness more than by mathematical necessity. We will see later, however, that to allow for proper asymptotics it is essential that the $T$ functions are periodic.
\section{Constructing $T$ gauges}
We are now ready to construct explicit $T$-gauges. We will contruct these gauges for both the undeformed and deformed cases simultaneously, commenting on the differences along the way. 
\subsection{Constructing $\JT$}
We start building the gauge $\JT$ for the upper band, adapting the method outlined in \cite{Balog:2011nm}. First we set $\JT_{0,0}=1=\JT_{a,\pm2}$ for $a\geq 2$. From eqn. \eqref{eq:otherYinT} we find that
\be
\label{YsinJTs}
\lambda_a = 1+Y_{a,0} = \frac{\JT_{a,0}^+\JT_{a,0}^-}{\JT_{a+1,0}\JT_{a-1,0}} \text{ for } a>0.
\ee
We can solve this equation for $\JT_{a,0}$s using \emph{the chain lemma}.

\paragraph{The chain lemma.} The chain lemma, proposed in \cite{Balog:2011nm} only for the undeformed case, solves equations of the form

\be
\frac{\sigma_j^+ \sigma_j^-}{\sigma_{j+1} \sigma_{j-1}} = \xi_j \text{ for } j\in \N \text{ with } \sigma_0 = 1,
\ee
where the unknown functions $\sigma_j$ do not have poles or zeroes near the physical strip. Also, $\xi_j$ do not have poles and zeroes near the real axis or take negative real values and only $\xi_1$ is allowed to have discontinuities on the real axis with bounded branch points there. In those cases we assume that only $\sigma_1$ has discontinuities on the first lines $Z_{\pm1}$. The wanted solution is given by
\be
\sigma_a = \JT_{a,0} = \exp \left(\ssum{j} \Lambda_j \star l_a^j \right),
\ee
with
\be
l_a^j = \sum_{n=0}^{j-1} K_{a+1-j+2n}.
\ee
The deformation of this lemma to the $\eta$-deformed case is done easily by interpreting the convolution and the kernel $K$ as the deformed versions. The proof that the deformed lemma also holds follows by an adaptation of the undeformed proof in \cite{Balog:2011nm}, which we will go over soon. 

\paragraph{Checking the assumptions.} First we note that for the application of this lemma we check that there are
\begin{itemize}
\item no poles and zeroes in the strip with $|$Im$(u)<(a-1)\ad$, and $1+Y_{a,0} \in \mA_a$, for $a\geq 2$, where we use the notation introduced just below \eqref{eq:Astrips}.
\item no poles and zeroes in the physical strip for $1+Y_1$.
\item that $1+Y_a$ does not attain negative real values to avoid unwanted branch cut behaviour.
\end{itemize}
These properties can be analysed in the asymptotic large-volume solution of the $Y$ system and appear to be respected.\footnote{We performed this check numerically for the deformed case. The authors of \cite{Balog:2011nm} did the same for the undeformed case.} We will assume these properties remain satisfied at finite volume. The cut structure of all the $Y$s of course follows rigorously from the TBA equations as we discussed before. In addition to the list above, we find that $Y_1$ does not have a $Z_0$ cut and that all $Y$ functions are real on the real line. This will imply that also the $\JT$ are real. 

\paragraph{Applying the chain lemma.} Assuming the above requirements are met, we can follow the chain lemma and write down the suspected solution
\be
\label{rand1}
\sigma_a = \JT_{a,0} = \exp \left(\ssum{j} \Lambda_j \star l_a^j \right),
\ee
with 
\be
l_a^j = \sum_{n=0}^{j-1} K_{a+1-j+2n},
\ee
using the convention that we expand $K_N$ to non-positive integers by demanding $K_{-N} = -K_N$ for all $N\in \Z$.\footnote{We will assume convergence of the infinite sum.} Note that for this solution we find
\begin{align}
\log \frac{ \sigma_a^+ \sigma_a^-}{\sigma_{a+1}\sigma_{a-1}} &= \ssum{j} \left( \left(\left(\Lambda_j \star l_a^j\right)^+ +\left(\Lambda_j \star l_a^j\right)^- -\Lambda_j \star l_{a+1}^j-\Lambda_j \star l_{a-1}^j \right)   \right) \nn
&=
\ssum{j}\sum_{n=0}^{j-1} \left(\left(\left(\Lambda_j \star K_{a+2n+1-j}\right)^+ + \left(\Lambda_j \star K_{a+2n+1-j}\right)^- \right.\right. \nn
  & \left.\left.- \Lambda_j \star K_{a+2n-j} -\Lambda_j \star K_{a+2n+2-j}
 \right)   \right).
\end{align}
The crucial point now is that not in all cases 
\be
\label{cp1}
\left(\Lambda_j \star K_{a+2n+1-j}\right)^{\pm}=\Lambda_j \star K_{a+2n+1-j}^{\pm}.
\ee
Indeed, if the two cases coincide we find that the summand of kernels vanishes: 
\begin{align}
\label{Ksvanish}
K_M^+ + K_M^- - K_{M-1}-K_{M+1} &\sim \frac{d}{du} \log\left( \frac{S^{[-(M-1)]}}{S^{[M+1]}}\frac{S^{[-(M+1)]}}{S^{[M-1]}}\frac{S^{[(M-1)]}}{S^{[-(M-1)]}}\frac{S^{[(M+1)]}}{S^{[-(M+1)]}} \right) =0,
\end{align}
where $S$ was defined in eqn. \eqref{eq:sinkernel} for the deformed case and in eqn.  \eqref{eq:undeformedsinkernel} for the undeformed case. However, \eqref{cp1} does not hold if the integration contour hits a pole during the shift. The only relevant pole is that of $K_1$. Now, for $j < a$ $K_1$ is not part of the summand, so it vanishes by the above computation. If $j > a$ it can contain $K_1$ , but if it does it also contains $K_{-1}= -K_1$ so that their contributions cancel by construction. Only if $j=a$ do we get a contribution from the pole of $K_1$: in order to make use of the identity \eqref{Ksvanish} we must push the contours past the pole, leading to a Cauchy integral. For small positive imaginary part the pole with negative residue is picked up using a clockwise contour, whereas for small negative imaginary part exactly the opposite happens. Therefore in both cases the contribution is exactly $\Lambda_a$, so we find 
\be
\log \frac{ \JT_{a,0}^+ \JT_{a,0}^-}{\JT_{a+1,0}\JT_{a-1,0}} = \Lambda_a,
\ee 
as we wanted. This therefore defines $\JT_{a,0}$.

\paragraph{Extending the solution.} The solution \eqref{rand1} solves eqn. \eqref{YsinJTs} at least in a neighbourhood of the physical strip, but we can in most cases extend this: the solution already implies that the $\JT_{a,0}$ do not have a cut at $Z_{\pm1}$. For the rest we use induction: suppose we have shown that for all $a\leq k$ $\JT_{a,0}$ has no cuts until possibly $Z_{\pm (a+1)}$. Using the defining relation \eqref{YsinJTs} and the knowledge that $Y_a \in \mathcal{A}_a$ we can now derive that $\JT_{k+1,0}$ has no cuts until possibly $Z_{\pm (k+2)}$: from the defining relation for $a=k$ we immediately see that $\JT_{k+1,0}$ has no cuts until possibly $Z_{\pm k}$. If it has a cut at $Z_{\pm k}$, the $a=k+1$ equation tells us that $\JT_{k+2,0}$ has to have a cut at $Z_{\pm(k-1)}$. Continuing up the chain of equations we find now that there is a $\JT_{m,0}$ with a cut at $Z_{\pm 1}$, which we know is not possible. So $\JT_{k+1,0}$ has no cut at $Z_{\pm k}$. Completely analogously one proves that also the cuts at $Z_{\pm (k+1)}$ are not possible.
\\[5mm]
Secondly, we look at poles. We know that $1+Y_a \in \mAp_{a-1}$. Moreover, $1+Y_1,\JT_{a,0} \in \mAp_1$.  Using the same arguments as for the cut structure we find that $\JT_1 \in \mAp_2$ and $\JT_a \in \mAp_{a-1}$ for $a\geq 2$.
\\[5mm]
Now we have completely defined and analysed the central $\JT$s. We can find the remaining $\JT$s in the upper band using the Hirota equation and the second equation in \eqref{eq:otherYinT}. First we fill the band for $a\geq 2$ solving the finite-difference equation
\be
\JT_{a,1}^+\JT_{a,1}^- = \JT_{a,0}(1+Y_{a,1}^{-1}),
\ee
which also follows from the $T$ parametrisation of the $Y$ functions. We can solve this finite-difference equation by a variation of the solution in the next section. Moreover, from the analyticity of the right-hand side we immediately see that  $\JT_{a,1} \in \mathcal{A}_a$. To complete the upper band we only need to define $\JT_{1,\pm 1}$, which we do using the Hirota equation. It follows immediately that $\JT_{1, \pm 1} \in \mathcal{A}_{1}$. This finishes the construction of the analytic gauge $\JT$, satisfying the analyticity strips in table \ref{tab:Tsanalyticity}. Note that from the construction it follows that the $\JT$ gauge is real.
\subsection{Use $Y_+Y_-$ to construct $\bfT$}
From the $\JT$ gauge we can construct a gauge with even better properties, known as the $\bfT$ gauge which will be our final gauge for $T$ functions in the upper band. In addition to the analyticity properties listed in table \ref{tab:Tsanalyticity} the $\bfT$s can be made to also satisfy the following identities, which were dubbed ``group-theoretical'' in \cite{Gromov:2011cx}:
\be
\label{eq:grouptheoretical}
\bfT_{3,\pm 2} =\bfT_{2,\pm 3}, \quad \bfT_{0,0}^+=\bfT_{0,0}^-, \quad \bfT_{0,s} = \bfT_{0,0}^{[+s]}.
\ee
We start from the discontinuity relation for $Y_{\pm}^{(\alpha)}$, which on the $Y$ hook reads
\be
\left[ \log Y_{1,\pm 1} Y_{2 \pm,2} \right]_{2n} = - \sum_{a=1}^n \left[ \log 1+Y_{a,0} \right]_{2n-a} \text{ for } n>0.
\ee
Notice that the right-hand side of this expression is independent of $\pm$ ($\alpha=l,r$ in our TBA notation). This is due to the fact that the product $Y_{1,\pm 1} Y_{2,\pm2}$ is the same on both sides of the $Y$ hook, so
\be
\label{qmodlity}
Y_{1, 1} Y_{2 ,2} = Y_{1, -1} Y_{2 ,-2}
\ee
as follows from the TBA equations \eqref{eq:TBAeqns}. Replacing the $Y$s by their parametrisation in $\JT$s leads to a telescoping cancellation of terms and we are left with
\be
\left[ \log \frac{1}{Y_{1,\pm 1} Y_{2,\pm 2}} \frac{\JT_{1,0}}{\JT_{0,0}^-} \right]_{2n} = 0 \text{ for } n>0.
\ee
Thus it follows that for a solution of the $Y$ system
the function $\bfB$ defined by
\be
\textbf{B} = \frac{1}{Y_{1,\pm 1} Y_{2,\pm 2}} \frac{\JT_{1,0}}{\JT_{0,0}^-}
\ee
is analytic in the upper half-plane.
\\[5mm]
Now we are ready to define the $\bfT$ gauge: to modify the $\JT$ gauge we consider the gauge transformation
\be
\label{bfT}
\textbf{T}_{a,s} = f_1^{[a+s]}f_2^{[a-s]}\bar{f_1}^{[-a-s]}\bar{f_2}^{[-a+s]} \JT_{a,s}
\ee
for two unknown functions $f_1,f_2$ such that $f_1^-,f_2^-$ are analytic in the upper half-plane. The function $\bar{f}$ is the complex conjugate of the function $f$, \emph{not} of the images it produces, i.e. somewhat awkwardly we have $\bar{f}(u) = \overline{f\left(u^*\right)}$. This gauge transformation is the most general one that does not destroy the reality nor the analyticity of the $\bfT$s. On $f_1,f_2$ we will impose that
\be
\mathbf{B} = \frac{\left(f_1 f_2\right)^-}{\left(f_1 f_2\right)^+}.
\ee
If we can find $f_1,f_2$ satisfying this constraint, then the $\bfT$s satisfy the identities
\bea
\label{eq:bft1}
\frac{\bfT_{3,\pm 2}\bfT_{0,\pm 1}}{\bfT_{2,\pm 3}\bfT_{0,0}^-} = 1 =
\frac{\bfT_{3,\pm 2}\bfT_{0,\pm 1}}{\bfT_{2,\pm 3}\bfT_{0,0}^+},
\eea
where the second equality is just complex conjugation of the first equality. We have more freedom in choosing $f_1,f_2$ still. Imposing
\be
\bfT_{0,1}=\bfT_{0,-1}
\ee
leads to the equation
\be
\label{realdiffff}
\frac{\JT_{0,-1}}{\JT_{0,1}}= \frac{\left(f_1/f_2\right)^+}{\left(f_1/f_2\right)^-}\frac{\left(\bar{f}_1/\bar{f}_2\right)^-}{\left(\bar{f}_1/\bar{f}_2\right)^+},
\ee
where we notice that the two fractions containing the $f_i$ are complex conjugate functions. The left hand side is real and can be decomposed\footnote{The fact that this is possible is non-trivial, but we will not provide a proof here.} as
\be
\frac{\JT_{0,-1}}{\JT_{0,1}} = \bfH \overline{\bfH},
\ee
with $\bfH$ analytic in the upper half-plane, implying we should solve
\be
\mathbf{H} = \frac{\left(f_1/f_2\right)^+}{\left(f_1/f_2\right)^-}.
\ee
To find $f_1,f_2$ we now have to solve
\be
\mathbf{B} = \frac{\left(f_1 f_2\right)^-}{\left(f_1 f_2\right)^+}, \quad \mathbf{H} = \frac{\left(f_1/f_2\right)^+}{\left(f_1/f_2\right)^-},
\ee
which are two finite-difference equations of the exact same form. Their solution can be found as we will discuss in the next subsection, solving the logarithmic version of this equations. With these solutions we can find $f_1,f_2$ and define the $\bfT$ gauge. The group-theoretical properties in eqn. \eqref{eq:grouptheoretical} follow from eqn. \eqref{eq:bft1}.
\subsubsection{Solving periodic difference equations}
To find the gauge transformation in the previous section we are supposed to solve the finite-difference equations
\begin{align}
\label{ffromB}
\log \textbf{B} &= \log \left(F^-\right)-\log \left(F^+\right), \nn
- \log \textbf{H} &= \log \left(G^-\right)-\log \left(G^+\right),
\end{align}
where $F=f_1 f_2$ and $G= f_1/f_2$.
\\[5mm]
Let us consider the general finite-difference equation
\begin{equation}
\label{eq:generalfinitedifference}
\Omega = \zeta^--\zeta^+.
\end{equation}
A formal solution of this equation can be written down immediately as the infinite sum
\be
\label{formalsol1}
\zeta = \ssum{n} \Omega^{[2n-1]},
\ee
but this will in general not be convergent. We can regularise this sum by considering a spectral representation of $\Omega$ in the upper half-plane. This spectral representation is different for the non-periodic and periodic cases. 

\paragraph{Non-periodic case.} In the undeformed case we are looking for non-periodic solutions of eqn. \eqref{eq:generalfinitedifference}. If $\lim_{u\rightarrow \infty} \Omega = 0$ we can write the solution $\zeta$ in the form
\begin{equation}
\zeta(u) = -\frac{1}{2\pi i} \int_{Z_0} dv \frac{\rho_{\Omega}(v) }{u-v},
\end{equation}
for Im$(u)>0$. The spectral density $\rho_{\Omega}$ is defined as
\begin{equation}
\rho_{\Omega}(u) = 2 \lim_{\e \rightarrow 0} \text{Re}\left( \Omega(u+i\e)\right). 
\end{equation}
By summing the series \note{Check this formula}
\begin{equation}
\Psi(u)  \defeq \frac{\psi\left( -\frac{i v g}{2}\right)}{2\pi} = \frac{\gamma}{2\pi}+\sum_{n=0}^{\infty} \left(  \frac{1}{v+2i n/g } - \frac{g}{2i (n+1) }\right)  ,
\end{equation}
with $\psi$ the polygamma function and $\gamma$ the Euler-Mascheroni constant, we find a finite expression for $\zeta$: for Im$(u)>0$
\begin{equation}
\zeta(u) = \Psi^+ \star \rho_{\Omega}(u),
\end{equation}
as directly plugging this into the finite-difference equation \eqref{eq:generalfinitedifference} will show.

\paragraph{Periodic case.} We provide the proof of the periodic case here in details, by proving the following lemma. 
\\[5mm]
\textbf{Lemma.} Let $\Omega:\C\rightarrow \C$ be a $2\pi$-periodic function regular in the upper half-plane obeying
\be
\lim_{u\rightarrow i \infty} \Omega(u) \in \R,
\ee
and which converges uniformly to an $L^1$ function on $[-\pi,\pi]$. Then it admits a spectral representation
\be
\Omega(u) = \int_{Z_0} \mathcal{K}(u-v)\rho_{\small \Omega}(v)dv \text{ for Im}(u)>0,
\ee
where $\mathcal{K}(u) = -\frac{1}{2 \pi i} \cot(u)$ is the natural periodic version of the $1/u$ kernel and where for $u\in Z_0$
\be
\rho_{\small \Omega}(u) = 2 \lim_{\e \rightarrow 0} \text{Re}\left(\Omega(u+i\e)\right) = \lim_{\e \rightarrow 0} \Omega(u+i\e)+\overline{\Omega}(u-i\e)	.
\ee
\\[5mm]
\textbf{Proof.} Consider for Im$(u)>0$ the integral
\be
\int_{-\pi}^{\pi} \mathcal{K}(u-v)\rho_{\small \Omega}(v)dv.
\ee
We can immediately split this integral as
\be
-\frac{1}{2 \pi i} \left(\int_{[-\pi+\pi]+i\e}  \frac{\Omega(v)}{\tan \left( u-v+i \e\right)}dv+
\int_{[-\pi+\pi]-i\e} \frac{\bar{\Omega}(v)}{\tan \left( u-v-i \e\right)}dv \right).
\ee
Pushing the integration contour of the first integral through the pole at $u=v$ to the line Im$(u)i$ and the second integral to $-\text{Im}(u)i$ gives that these two integrals equal
\be
\Omega(u) -\frac{1}{2 \pi i}\left( \int_{[-\pi+\pi]+\text{Im}(u)i +i\e} \, \frac{\Omega(v)}{\tan \left( u-v+i \e\right)}dv +\text{c.c.}\right).
\ee
The integrals in the brackets actually do not depend on the height of the contour in the upper half-plane and we should be able to find the same result as long as the contour is higher up than $\text{Im}(u)i$. Around $u\rightarrow i\infty$ we find
\be
\int_{[-\pi+\pi]+\text{Im}(u)i+i\e}  \frac{\Omega(v)}{\tan \left( u-v+i \e\right)}dv \sim 
i \int_{[-\pi+\pi]+\text{Im}(u)i+i\e} dv\Omega(v)\text{Im}\left(\cot \left( u-v\right)\right),
\ee
which is purely imaginary as long as the contribution of $\Omega$ is real. Using the limit condition on $\Omega$ we therefore find that this integral and its complex conjugate vanish and we are left with
\be
\Omega(u) = -\frac{1}{2 \pi i}\int_{-\pi}^{\pi} dv \frac{\rho_{\small \Omega}(v)}{\tan \left( u-v\right)} \text{ for Im}(u)>0,
\ee
i.e. we have found a spectral representation for $\Omega$. \ensuremath{\blacksquare}
\\[5mm]
We can now use this result: we define the kernel
\be
\psi_{\ad}(u) = \mathcal{K}(u) + \frac{1}{2\pi i} \cot(2i\ad ) + \sum_{n=1}^{\infty} \left( \mathcal{K}^{[2n]}(u) + \frac{1}{2\pi i} \cot(2(n+1) i \ad )\right),
\ee
where the sum gives a combination of $q$-polygamma functions.\note{Make this precise.} Now it is easy to check that
\bea
\zeta &=& \psi_{\ad}^+ \star \rho_{\Omega}
\eea
solves the finite-difference equation \eqref{eq:generalfinitedifference} we started with.

\paragraph{Finding solutions.} It is easy to check numerically in the asymptotic finite-volume solution that $\log \bfB$ and $\log \bfH$ satisfy the requirements mentioned above in both the undeformed and deformed case. We assume this extends beyond the asymptotic solution, yielding a solution of the finite-difference equation \eqref{eq:generalfinitedifference}. In particular, we can find the gauge transformations $f_1,f_2$ and hence define the $\bfT$ gauge. 

\subsection{Constructing $\mbT$}
Now we have constructed an upper-band gauge with good properties we can continue building good gauges for the other two wings of the $T$ hook. We will focus on the construction of a good gauge $\mbT_R$ (simply written as $\mbT$ to avoid heavy notation) for the right hook, since the construction for the left and right band is completely analogous. The specific properties we are after for $\mbT$ are:
\begin{itemize}
\item the $\mbT$ functions should be analytic in a region of the real axis as specified in table \ref{tab:Tsanalyticity}
\item the gauge should be $\Z_4$-symmetric, which means that
\begin{equation}
\label{eq:Z4symmetry}
\hbT_{a,s} = (-1)^a \hbT_{a,-s},
\end{equation}
where the hat indicates that we consider the short-cutted version of the usually long-cutted $\mbT$ functions, as explained in section \ref{sec:shortcuttedfunctions}. \end{itemize}
The construction of the right gauge $\mbT$ will follow these steps:
\begin{enumerate}
\item Define the global $\mbT$ gauge from the $\bfT$ gauge,
\item Define a $\Z_4$-symmetric analytic gauge $\mcT$ in the right band,
\item Use the fact that $\mbT$ and $\mcT$ are related by a gauge transformation together with the discontinuity equations for the $Y$ functions to conclude that $\mbT$ must have the same analyticity strips as $\mcT$,
\item Show that a small redefinition of the $\bfT$ forces $\hat{\mbT}_{1,\pm 1}$ to have only two cuts,
\item Use this fact to prove that also $\Z_4$ symmetry carries over from the $\mcT$ gauge to the $\mbT$ gauge,
\item Conclude that this is exactly the gauge we wanted.
\end{enumerate}
Note that this construction does not follow the same steps as that of the $\bfT$ gauge, even though the form of the $T$ hook suggests we could. The reason for this is the $\Z_4$-symmetry property that we have not considered so far: we will see that a gauge with this property is so constrained that it can be parametrised by only two independent functions, which is essential for the introduction of the quantum spectral curve parametrisation in terms of $\Pf$ functions. Therefore we would like to ensure that our starting gauge already is $\Z_4$-symmetric, which leads us to consider a different approach. 

\subsection{Defining $\mbT$ from $\bfT$} We directly define the $\mbT$ gauge from the $\bfT$ gauge through the transformation
\be
\label{mbt}
\mbT_{a,s} = (-1)^{as}  \bfT_{a,s} \left(\bfT_{0,1}^{[a+s-1]}\right)^{\frac{a-2}{2}}.
\ee
One can check directly that this gauge satisfies the Hirota equation on the $T$ hook using the fact that $\bfT$ does. However, whether it has the properties listed above is far from obvious from this definition. In order to find out, we will show that it is related to another gauge with these properties and that the relation is such that the $\mbT$ gauge must have the same properties as well. 

\subsection{Defining $\mcT$} We define a gauge $\mcT$ which behaves nicely in the right band from a solution of the analytic $Y$ system. This construction follows \cite{Gromov:2011cx} closely: 
\\[5mm]
Consider a set of $\mcT$ functions in the right band ($s\geq a$) parametrised by resolvents $G$ and $\bar{G}$ as follows:
\begin{align}
\mcT_{0,s} &= 1, \nn
\mcT_{1,s} &= s + G^{[+s]}+\bar{G}^{[-s]}, \nn
\mcT_{2,s} &= \left(1 + G^{[s+1]}-G^{[s-1]}\right)\left(1 + \bar{G}^{[-s-1]}-\bar{G}^{[-s+1]}\right),
\end{align}
The resolvents $G,\bar{G}$ are parametrised by a density function $\rho$ through the definitions
\begin{align}
G(u) &= \int_{Z_0} dv\, \mK(u-v) \rho(v) \text{ for Im}(u)>0,\nn
\bar{G}(u) &= \int_{Z_0} dv\, \bar{\mK}(u-v) \rho(v) \text{ for Im}(u)<0,
\end{align}
such that we have $\rho(u) = G(u+i\e)+\bar{G}(u-i\e)$. The function $\mK$ is defined as
\begin{equation}
\label{eq:Gkernel}
\mK(u) = -\frac{1}{2\pi i}\cot(u), \quad  \mK^{\text{und}}(u) = -\frac{1}{2\pi i}\frac{1}{u}.
\end{equation}
This gauge has the required analyticity properties and solves the Hirota equation on the right band, so we only need to check the presence of $\Z_4$ symmetry. Ultimately, this follows from the fact that
\be
\label{YplusYmin}
Y_{1,1}(u+i\e)=1/Y_{2,2}(u-i\e) \text{ for } u \in \check{Z}_{0},
\ee
which is true for both the undeformed and deformed $Y$-functions. Indeed, the ratio
\be
\label{rY1Y2}
r= \frac{1+1/Y_{2,2}}{1+Y_{1,1}} = \frac{\mcT_{2,2}^+\mcT_{2,2}^-\mcT_{0,1}}{\mcT_{1,1}^+\mcT_{1,1}^-\mcT_{2,3}}
\ee
is expressed in right-band $\mcT$ functions only and satisfies $r(u+i\e)=1/r(u-i\e)$ for $ u \in \check{Z}_{0}$ due to eqn. \eqref{YplusYmin}. Using our parametrisation this implies $G(u+i\e)=-\bar{G}(u-i\e)$ for $u\in \check{Z}_0$, which implies that the function $\hat{G}$ defined by
\be
\hat{G}(u) = \mK \star \rho
\ee
has only one short $\hat{Z}_0$ cut on the whole complex plane and coincides with $G$ ($\bar{G}$) on the upper (lower) half-plane. This property is crucial to the further simplification of spectral problem: the long-cutted $G$ and $\bar{G}$ generically are cut-free only on half of the complex plane, making our analysis complicated if not impossible. Rewriting our equations using $\hat{G}$ will make the analysis a lot more straightforward. 

\paragraph{Short-cutted $\hmcT$.} \label{sec:shortcuttedhmcT} First of all, we can consider short-cutted versions of the $\mcT$ gauge. These functions can be found by continuation and are given by
\begin{align}
\label{eq:hmcT}
\hmcT_{0,s} &= 1, \nn
\hmcT_{1,s} &= s + \hat{G}^{[+s]}-\hat{G}^{[-s]}, \nn
\hmcT_{2,s} &= \left(1 + \hat{G}^{[s+1]}-\hat{G}^{[s-1]}\right)\left(1 + \hat{G}^{[-s-1]}-\hat{G}^{[-s+1]}\right) = \hmcT_{1,1}^+\hmcT_{1,1}^-.
\end{align}
The $\hmcT$s not only coincide with the $\mcT$s just above the real axis, we can extend this identification to the whole analyticity strips of $\mcT$s. A much more remarkable property is that in this parametrisation the dependence on the index $s$ 
is analytic. This allows for the following consideration that will prove crucial for the introduction of the quantum spectral curve. 
\noindent The $\hmcT$ satisfy the Hirota equation on the right band of the $T$ hook
\begin{figure}[H]
\centering
\includegraphics[width=10cm]{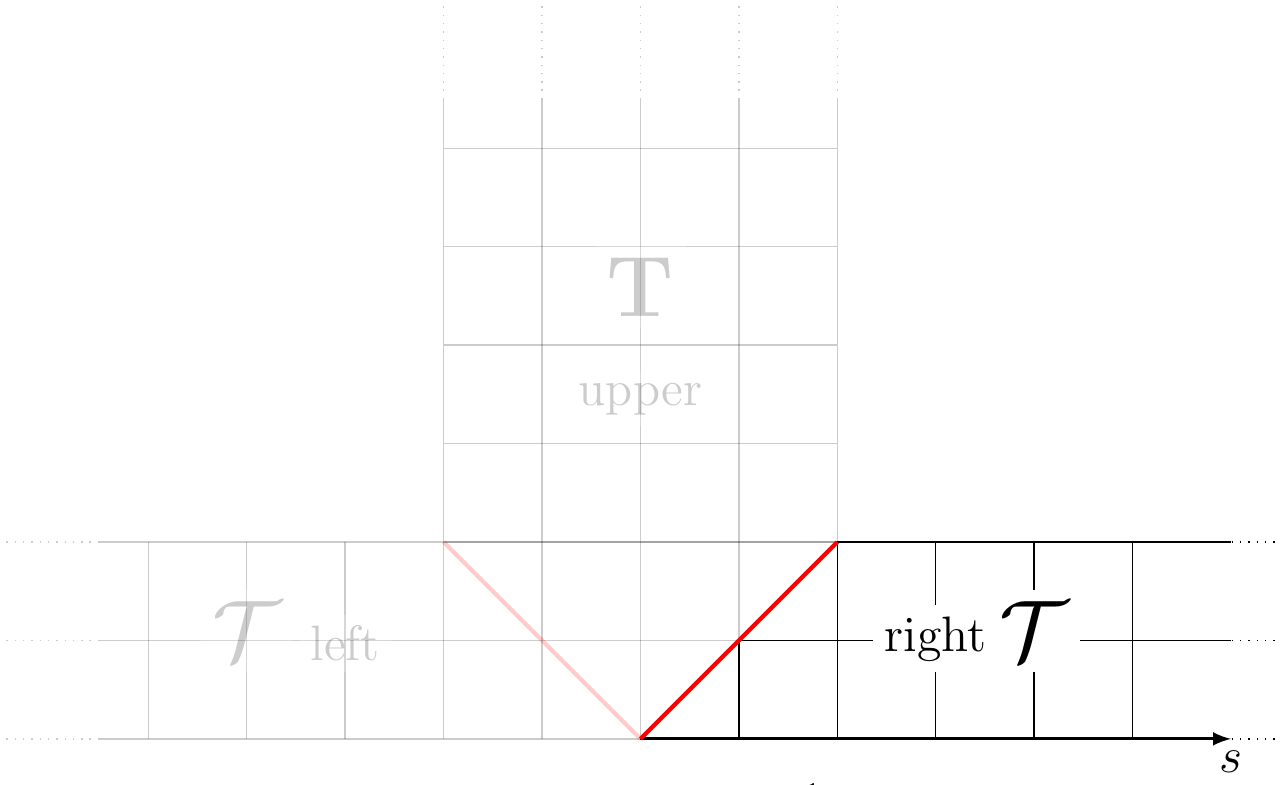}
\label{fig:extendright2}
\end{figure}
\noindent
We can analytically continue the definitions in eqn. \eqref{eq:hmcT}
\begin{figure}[H]
\centering
\includegraphics[width=10cm]{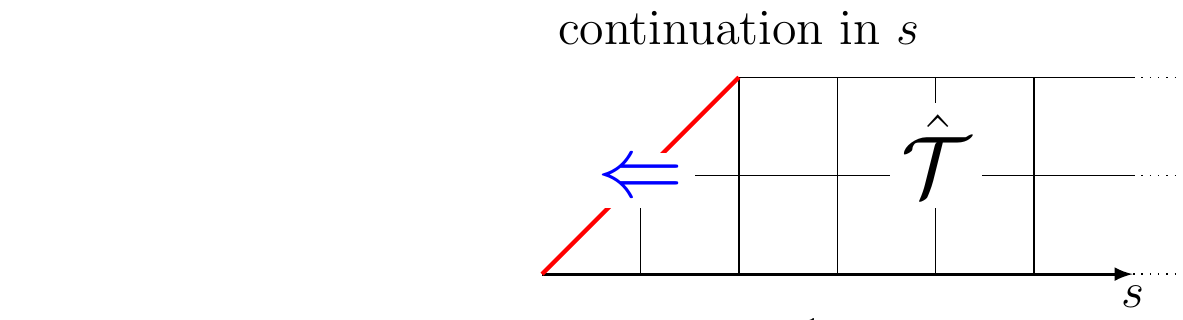}
\label{fig:extendright3}
\end{figure}
\noindent such that now $s\in \Z$ to obtain a set of $\hmcT$ functions defined on an infinite horizontal band
\begin{figure}[H]
\centering
\includegraphics[width=10cm]{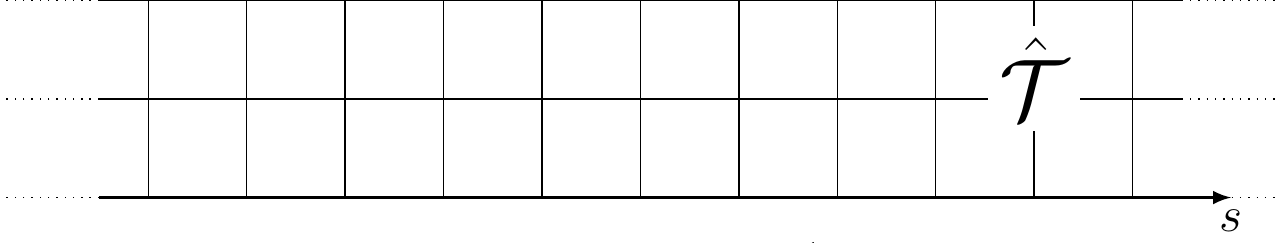}
\label{fig:extendright4}
\end{figure}
\noindent By plugging in these functions into the Hirota equation one finds that the analytically continued $\hmcT$ (which of course coincide with the original $\hmcT$ on the right band) satisfy the Hirota equation everywhere on this infinite band. Moreover, from the definitions of the $\hmcT$s it is straightforward to check that they satisfy
\be
\label{z4hmct}
\hmcT_{a,s} = (-1)^a \hmcT_{a,-s},
\ee
in other words the $\hmcT$s are $\Z_4$-symmetric. This discovery in \cite{Gromov:2011cx} made the further simplification of the analytic $T$-system possible, as we will discuss in more detail in section \ref{sec:Wronskian}. 

\paragraph{$\mcT$ as parametrisation of the $Y$ system.} So far we have not put any restrictions on the density function $\rho$. Indeed, we have only used very mild input from the TBA equations (in eqn. \eqref{YplusYmin}) to constrain the resolvents, but have not made sure that the resulting $\mcT$ gauge actually parametrises a solution to the $Y$ system. To ensure that we should require that the second equality in eqn. \eqref{rY1Y2} holds, leading to the following non-linear integral equation for the density $\rho$:
\be
\label{nonlinearr}
\frac{1+1/Y_{2,2}}{1+Y_{1,1}}= \frac{\left(1+\mK_1^+\slashconv \rho -\rho/2 \right)\left(1+\mK_1^-\slashconv \rho -\rho/2 \right)}{\left(1+\mK_1^+\slashconv \rho +\rho/2 \right)\left(1+\mK_1^-\slashconv \rho +\rho/2 \right)} \text{ for } u\in \hat{Z}_0,
\ee
where
\begin{align}
\mK_s(u) &= \mK^{[s]}(u)-\mK^{[-s]}(u), \nn
K\slashconv f (u) &= \text{P.V.} \int_{\hat{Z}_0} dv\, K(u-v)f(v),
\end{align}
where this last definition involves the principal value integral. In the undeformed case this equation was numerically shown to have a solution \cite{Gromov:2011cx}. For the deformed case this was not pursued, as it would distract from our ultimate goal, the quantum spectral curve. Nevertheless, we are formally required to assume that the equations \eqref{nonlinearr} can be solved so we will do so. 
\subsubsection{Step 3: the $\mbT$ gauge is analytic}
Having defined the $\mcT$ gauge we can see how it relates to the $\mbT$ gauge defined before, since two gauges parametrising the same set of $Y$ functions are related by a gauge transformation, as in eqn. \eqref{eq:Tgaugetransformation}. This will show all the wanted properties of the $\mbT$ gauge.
\\[5mm]
The first thing to check is that the $\mbT$ gauge has the right analyticity properties (see table \eqref{tab:Tsanalyticity}). In order for both gauges to be real we have to restrict the gauge transformation \eqref{eq:Tgaugetransformation} to
\be
\mbT_{a,s} = f_1^{[a+s]}f_2^{[a-s]}\bar{f_1}^{[-a-s]}\bar{f_2}^{[-a+s]} \mcT_{a,s}.
\ee
From $\mcT_{0,s}= \mbT_{0,s} =1 $ we get $f_1f_2^{++}=1$. From the definition of $\mbT_{2,s}$ and the analyticity of the $\bfT$ gauge it follows directly that $\mbT_{2,s}\in \mA_{s-1}$ for $s\geq 2$. Since also $\mcT_{2,s}\in \mA_{s-1}$ for $s\geq 2$ we find that $f_1^{++}/f_1^{--}$ is analytic in the upper half-plane. The last thing to check is the analyticity of the $\mbT_{1,s}$ functions. For this we use the discontinuity relation \eqref{res2}. Parametrising it using the $\bfT$ and $\mcT$ gauge and using that $Y_{2,2} = \frac{\bfT_{2,1}}{\bfT_{1,2}}$ and that
\be
\label{Y11Y22}
Y_{1,\pm 1}Y_{2,\pm 2} = \frac{\bfT_{0,0}^-}{\bfT_{1,0}}
\ee
we find that
\be
\frac{\bfT_{1,2} \mcT_{1,1}^-}{\bfT_{1,1}^- \mcT_{1,2}} =  -\frac{\mbT_{1,2} \mcT_{1,1}^-}{\mbT_{1,1}^- \mcT_{1,2}}
\ee
is also analytic in the upper half-plane. Plugging in our gauge transformation this implies that $f_1^{++}f_1^{--}/f_1^2$ is analytic in the upper half-plane. Together these constraints now imply that $f_1/f_1^{--}$ is analytic in the upper half-plane. For the remaining $\mbT_{1,s}$ the gauge transformation now takes the form\footnote{The conjugate bar on $f_1$ does not affect the shifts. If we want to apply the conjugate bar to the shifts we write $\overline{f^+}$.}
\be
\label{gaugetrsfmmcTR}
\mbT_{1,s} = \left(  \left(\frac{f_1}{f_1^{--}}\right)^{[s]} \times \left(\frac{\overline{f_1}}{\overline{f_1}^{++}}\right)^{[-s]}   \right) \mcT_{1,s} \text{ for } s\geq 1,	
\ee
which shows that $\mbT_{1,s} \in \mA_s$, as desired. So we find that the $\mbT$ gauge indeed has the right analyticity properties. In the next section we will show that $\mbT$ is actually $\Z_4$-symmetric.
\subsubsection{Step 4.1: is $\mbT$ really $\Z_4$-symmetric?}
\label{sec:step41}
The real difficulty is to prove that the $\mbT$ gauge is actually $\Z_4$-symmetric. We will go through the following steps: first we will prove that the product $\hat{\mbT}_{1,1}\hat{\mbT}_{1,-1}$ has only two cuts at $Z_{\pm1}$. Second we will show that we can force each of these functions to have only two cuts. We can then use this to put more constraints on the gauge transformation in eqn. \eqref{gaugetrsfmmcTR} and conclude $\Z_4$ symmetry from there.
\\[5mm]
The derivation starts with the discontinuity relation belonging to the $Y_Q$ functions (see eqn. \eqref{discs1}) and is a generalisation of appendix C.3 of \cite{Gromov:2011cx} to non-symmetric states. On the $Y$ hook it reads for $N\geq 1$
\bea
& &\left[ \left[\log Y_{1,0}^+  \right]_0^{[2N]} \right]_0 = \nn
& &\sum_{\pm} \left( \left[\log\left(1+Y_{1,\pm 1}^{[2N]} \right) + \sum_{m=1}^N \log\left(1+Y_{m+1,\pm 1}  \right)^{[2N-m]}  \right]_0 - \log Y_{1,\pm 1}Y_{2,\pm 2} \right).\qquad
\eea
Using the $\bfT$ gauge we can compute $\left[\log Y_{1,0}^+  \right]_0$:
\be
Y_{1,0} = \frac{\bfT_{1,1}\bfT_{1,-1}}{\bfT_{0,0}\bfT_{2,0}} = \frac{\mbT_{1,1}\mbT_{1,-1}}{\bfT_{2,0}}
\ee
and using that $\bfT_{2,0} \in \mA_{3}$ we find
\be
\left[\log Y_{1,0}^+  \right]_0 = \left[\log \mbT_{1,1}^+\mbT_{1,-1}^+  \right]_0 = \log \mbT_{1,1}^+/\dot{\mbT}_{1,1}^+\log \mbT_{1,-1}^+/\dot{\mbT}_{1,-1}^+,
\ee	
where the $\dot{\mbT}_{1,\pm 1}$ are defined to coincide with the $\mbT_{1,\pm 1}$ below $Z_1$ and have a short cut $\hat{Z_1}$, whereas all its other cuts in the upper half-plane are long. We use the analyticity of the $\bfT$ and $\mbT$ gauges to telescope the sums in the expression above:
\be
\label{telescoped1}
\left[\log \left(\frac{\mbT_{1,1}^+ \mbT_{1,-1}^+ }{\dot{\mbT}_{1,1}^+ \dot{\mbT}_{1,-1}^+}  \frac{\bfT_{0,1}\bfT_{0,-1}}{\bfT_{1,1}^+\bfT_{1,-1}^+}\right)  \right]_{2N} = -2 \log Y_{1,1}Y_{2,2},
\ee
using that $Y_{1,1}Y_{2,2}= Y_{1,-1}Y_{2,-2}$. We know that $\mbT_{1,\pm1} = -\bfT_{1,\pm1} \bfT_{0,0}^{-1/2}$ by definition, so the argument of the log on the left-hand side becomes
\be
\frac{\bfT_{0,0}^+}{\dot{\mbT}_{1,1}^+ \dot{\mbT}_{1,-1}^+}	.
\ee
Now we can use the property that
\be
\left[ \log \bfT_{0,0} \right]_{2N+1} = - 2 \log Y_{1,\pm 1}Y_{2, \pm 2} \text{ for } N\in \Z,
\ee
which follows directly from $Y_{1,\pm 1}Y_{2,\pm 2} = \bfT_{1,0}/\bfT_{0,0}^+$, the periodicity of $\bfT_{0,0}$ and the fact that $\bfT_{1,0} \in \mA_1$. With the previous discontinuity relation fitting perfectly, we find that what remains in eqn. \eqref{telescoped1} is just
\be
\left[\log \dot{\mbT}_{1,1} \dot{\mbT}_{1,-1} \right]_{2N+1} = 0 \text{ for } |N| \geq 1,
\ee
after repeating the argument for negative $N$ as well. Using that the potential cuts are located at $Z_{2N+1}$ with $N\in \Z$ we find that $\dot{\mbT}_{1,1} \dot{\mbT}_{1,-1}$ has just $Z_{\pm 1}$ cuts, finishing step one of our derivation as this directly implies that the product $\hat{\mbT}_{1,1}\hat{\mbT}_{1,-1}$ has only two short $Z_{\pm1}$ cuts.
\\[5mm]
Next we want to prove that we can force each of the functions $\hat{\mbT}_{1,1}$ and $\hat{\mbT}_{1,-1}$ to have only two cuts. To do this we use a gauge freedom of $\bfT$ that we have not used yet to alter the $\mbT$ since this last gauge is based on the former. Going through the analytic and group-theoretical properties of $\bfT$ one finds that nothing changes after transforming them as
\be
\bfT_{a,s} \rightarrow \left(\prod_{k=-(|s|-1)/2}^{(|s|-1)/2} \frac{e^{i \phi^{[a+2k]}}}{e^{i \phi^{[-a+2k]}}}  \right)^{\text{sgn}(s)} \bfT_{a,s},
\ee
as long as $\phi$ is real with one long $\check{Z}_0$ cut. To change the function $\mbT_{1,1}$ we see how this gauge transformation changes the definition of $\mbT$: as a function with short cuts we find
\be
\hbT_{1,1}^{\text{new}} = e^{i \hat{\phi}^+-\bar{\hat{\phi}}^-}\hbT_{1,1}.
\ee
To clarify, the conjugate on the second $\phi$ comes about as follows: by assumption $\phi$ is real analytic when viewed with long cuts, therefore we can write $\phi^-(u) = \overline{\phi^+}(u)$ for $u$ in the physical strip. Now we should take the hatted version of this function, which is easily done: since $\left(\phi^+\right)^{\hat{}} = \hat{\phi}^+$ we find 
\be
\left(\phi^{-} \right)^{\hat{}}(u) = \overline{\hat{\phi}^+}(u) = \overline{\hat{\phi}}^-(u),
\ee
as we used above.
Now, since we want $\hbT_{1,1}^{\text{new}}$ to have only two $Z_{\pm1}$-cuts we should cancel any extra cuts in $\hbT_{1,1}$ through $\phi$:
\be
i \left[\,\hat{\phi} \, \right]_{-2N}= - \left[ \log \hbT_{1,1}\right]_{-1-2N},\quad
i \left[\,\bar{\hat{\phi}}  \, \right]_{2N}= \left[ \log \hbT_{1,1}\right]_{1+2N} \text{ for } N\in \N.
\ee
Constructing a $\phi$ that obeys these rules can be done as follows: we can firstly find a short-cutted function $\psi$ that obeys
\be
i \left[\,\hat{\psi} \, \right]_{-2N}= - \left[ \log \hbT_{1,1}\right]_{-1-2N},\quad
i \left[\,\hat{\psi}  \, \right]_{2N}= \left[ \log \hbT_{1,1}\right]_{1+2N} \text{ for } N\in \N,
\ee
and define $\phi$ as solution to the Riemann-Hilbert problem
\be
\label{eq:RH1}
\phi(u+i\e)+\phi(u-i\e)=  \psi \text{ for } u \in \check{Z}_0.
\ee
One can solve this equation in both the undeformed and deformed cases as we will see in section \ref{sec:solvingRHproblem}, after multiplying the unknown function $\phi$ with a pure square root\footnote{The pure square root on in the undeformed case is just $\sqrt{4-u^2}$, in the deformed case it takes a more complicated form, but we can use $G^{-1}$ as defined in eqn. \eqref{eq:squarerootG}.} and dividing $\psi$ by the same function, turning the relative sign between the shifts of $\phi$ in eqn. \eqref{eq:RH1} into a minus. Any solution is real analytic on the mirror sheet with a long $\check{Z}_0$ cut, so we can use any to force $\hbT_{1,1}$ to have only two cuts. It follows immediately from this and the fact that the product $ \hbT_{1,1} \hbT_{1,-1}$ has only two cuts that now both functions have exactly two cuts at $\hat{Z}_{\pm1}$.
\subsubsection{Step 4.2: actually proving $\hbT$ is $\Z_4$-symmetric}
The gauge transformation \eqref{gaugetrsfmmcTR} can be written like
\be
\mbT_{1,s} = h^{[+s]}\bar{h}^{[-s]}\mcT_{1,s}
\ee
where $h$ is a function analytic in the upper half-plane. Translating this equation for $s=1$ to the short-cutted sheet reads
\be
\hbT_{1,1} = \hat{h}^{+}\bar{\hat{h}}^{-}\hmcT_{1,1}.
\ee
Knowing that the $T$ functions on both sides of this equation have only two short cuts at $\hat{Z}_{\pm1}$ we find that $\hat{h}$ has only one $\hat{Z}_0$ cut. The only remaining thing to prove is that $\overline{\hat{h}}/\hat{h}$ is a constant to conclude that the $\hbT$ gauge is also $\Z_4$-symmetric. Indeed, in that case we find
\be
\label{hbtsimhmct}
\hbT_{0,s}=1,\quad \hbT_{1,s}\sim \hat{h}^{[+s]}\hat{h}^{[-s]}\hmcT_{1,s}, \quad \hbT_{2,s}\sim \hat{h}^{[s+1]}\hat{h}^{[s-1]}\hat{h}^{[-s+1]}\hat{h}^{[-s-1]}\hmcT_{2,s},
\ee
which directly implies $\Z_4$ symmetry using eqn. \eqref{z4hmct}.

The argument starts off with the Hirota equation for $\mbT_{1,1}$ and $\mbT_{2,2}$:
\begin{align}
\label{Hirota1122}
\mbT_{1,1}^+\mbT_{1,1}^- &= \mbT_{1,0}\mbT_{1,2}+\mbT_{0,1}\mbT_{2,1} \nn
\mbT_{2,2}^+\mbT_{2,2}^- &= \mbT_{2,1}\mbT_{2,3}+\mbT_{2,1}\mbT_{2,3}.
\end{align}
Now, defining $\mF = \sqrt{\bfT_{0,0}}$ we can use the properties
\begin{equation}
\mbT_{0,1} =1, \quad \mbT_{1,1}, \mbT_{3,2} = - \mF^+ \mbT_{2,3},\quad \mbT_{1,0} = -\mF^+Y_{1,1}Y_{2,2},
\end{equation}
where the last two properties follow from the $\bfT$ gauge and the gauge transformation \eqref{mbt}. In particular we have used eqn. \eqref{Y11Y22} to obtain the last equation. Consider $u\in \hat{Z}_0$, then we have
\be
\mbT_{2,3}=\hbT_{2,3},\quad \mbT_{2,2}(u\pm i \ad )=\hbT_{2,2}(u\pm i \ad \mp i\e ),\quad \mbT_{1,1}(u\pm i \ad )=\hbT_{1,1}(u\pm i \ad \mp i\e ).
\ee
Additionally we find from the gauge transformation that
\be
\hbT_{2,s} = \frac{\hat{h}^{[s-1]}\bar{\hat{h}}^{[-s+1]}}{\hat{h}^{[-s+1]}\bar{\hat{h}}^{[s-1]}} \hbT_{1,1}^{[+s]} \hbT_{1,1}^{[-s]},
\ee
which for real $h$ simplifies to $\hbT_{2,s} =\hbT_{1,1}^{[+s]} \hbT_{1,1}^{[-s]}$. Excluding $\hbT_{2,1}$ from eqn. \eqref{Hirota1122} and then using these properties we get successively
\be
\mbT_{1,1}^+\mbT_{1,1}^--\mbT_{1,0}\mbT_{1,2} = \frac{\mbT_{2,2}^+\mbT_{2,2}^--\mbT_{3,2}\mbT_{1,2}}{\mbT_{2,3}},
\ee
\be
\hbT_{1,1}^{[+1-\e]}\hbT_{1,1}^{[-1+\e]}+\left(Y_{1,1}Y_{2,2}-1\right)\mF^+\hbT_{1,2} = \frac{\hbT_{2,2}^{[+1-\e]}\hbT_{2,2}^{[-1+\e]}}{\hbT_{2,3}}
\ee
\be
\hbT_{1,1}^{[+1-\e]}\hbT_{1,1}^{[-1+\e]}+\left(Y_{1,1}Y_{2,2}-1\right)\mF^+\hbT_{1,2} = \frac{\hat{h}^{[\e]}\bar{\hat{h}}^{[-\e]}}{\hat{h}^{[-\e]}\bar{\hat{h}}^{[\e]}}  \hbT_{1,1}^{[-1-\e]}\hbT_{1,1}^{[-1-\e]}.
\ee
Now we can consider this equation transformed to the $\mcT$-gauge, using the parametrisation in terms of $\hat{G}$:
\bea
&\hat{h}^{[\e]}\bar{\hat{h}}^{[-\e]}\left(1+\hat{G}^{++}-\hG^{[-\e]}  \right)  \left(1+\hat{G}^{[+\e]}-\hG^{--}   \right)\nn
&+\left(Y_{1,1}Y_{2,2}-1\right)\mF^+\left(2+\hat{G}^{++}-\hG^{--}  \right) = \nn
&\frac{\hat{h}^{[\e]}\bar{\hat{h}}^{[-\e]}}{\hat{h}^{[-\e]}\bar{\hat{h}}^{[\e]}} 
\hat{h}^{[-\e]}\bar{\hat{h}}^{[\e]}\left(1+\hat{G}^{++}-\hG^{[\e]}  \right)  \left(1+\hat{G}^{-\e}-\hG^{--}   \right). 
\eea
We see that the extra $\hat{h}$s conveniently cancel and using that $\rho = \hG^{[\e]}-\hG^{[-\e]}$ we find on $\hat{Z}_0$ that
\be
\hat{h}^{[\e]}\bar{\hat{h}}^{[-\e]} = \frac{\left(1-Y_{1,1}Y_{2,2}\right) \mF^+}{\rho}.
\ee
The right-hand side is analytic on a neighbourhood of the real line, which implies the left hand side is that too, leading to
\be
\hat{h}^{[\e]}\bar{\hat{h}}^{[-\e]} = \hat{h}^{[-\e]}\bar{\hat{h}}^{[\e]},
\ee
which in turn tells us that $\bar{\hat{h}}/\hat{h}$ is cut free on the complex plane. Using the regularity requirement that our $\hbT$s do not have any poles except possibly at the branch points we want $\hat{h}$ to be pole and zero free. For the undeformed case numerical analysis showed in \cite{Gromov:2011cx} that $\hat{h}$ could grow at most polynomially as $u \rightarrow \infty$, implying that $\bar{\hat{h}}/\hat{h}$ must be a constant applying the lemma we prove in appendix \ref{app:constrainingtrigonometricpolynomials}. For the deformed case the analogue is that the Fourier series of $\hat{h}$ is a finite sum, from which it follows directly that $\bar{\hat{h}}/\hat{h}$ can only be a constant, hence leading to eqn. \eqref{hbtsimhmct} and therefore implying the $\Z_4$ symmetry of the $\hbT$ gauge.
\\[5mm]
So now we have proven the existence of a $\Z_4$-symmetric gauge $\mbT$ related to the $\bfT$ gauge by eqn. \eqref{bfT}. It has the analyticity domains in table \ref{tab:Tsanalyticity} we hoped for and we were able to choose $\hbT_{1,1}$ to have exactly two cuts, at $Z_{\pm1}$. This gauge is an excellent starting point to continue our simplification.

\section{A step back}
Let's take a step back and see what we have achieved so far: we started out with the analytic $Y$-system and constructed four gauges $\JT$, $\bfT$, $\mcT$ and $\mbT$. The first two gauges have good analyticity properties (see table \ref{tab:Tsanalyticity}) in the upper band of the $T$ hook \ref{fig:thook}, and the $\bfT$ gauge has particularly nice group-theoretical properties \eqref{eq:grouptheoretical}. We then proceeded to construct a good gauge $\mbT$ in the left and right band by parametrising it directly in terms of the $\bfT$ functions from the upper band. That this gauge is nicely behaved is far from obvious, which is why we introduced the $\mcT$ gauge: by construction it has broad analyticity bands as indicated in table \ref{tab:Tsanalyticity}. In addition, this gauge has $\Z_4$ symmetry (see eqn.  \eqref{eq:Z4symmetry}): from the $\mcT$ in the right band we can construct their short-cutted counterparts $\hmcT$, which can then be analytically continued to form a 
solution of the Hirota equation on an infinite horizontal band. The $\Z_4$ symmetry relates two $T$ functions from this $\hmcT$ gauge. Note that we have only explicitly talked about extending the right band, but one should perform the same operation to the gauge in the left band: the resulting gauge need not coincide with the continuation from the right band, but both are $\Z_4$-symmetric. We could furthermore show that the $\mbT$-gauge was related to the $\mcT$ gauge by a gauge transformation that is so well-behaved that the analyticity properties and $\Z_4$ symmetry carry over from $\hmcT$ to the $\hbT$s, the short-cutted version of the $\mbT$. 
\\[5mm]
This finishes of our construction of the analytic $T$-system. Before delving into the details of the quantum spectral curve we first review how the general theory of solving the Hirota equation can be used to further simplify the parametrisation.

\section{Wronskian parametrisation}
\label{sec:Wronskian}
As anticipated in the previous section we can use further simplify the $\hbT$ gauge by noticing that the analytical continuation discussed in section \ref{sec:shortcuttedhmcT} can also be applied to the $\hbT$ gauge, resulting in a solution to the Hirota equation on a horizontal infinite band of width two. The general solution to the Hirota equation on such bands was analysed in \cite{Gromov:2011cx,Krichever:1996qd,Kazakov:2010kf}, which we can use here. We start by reviewing this solution:
\\[5mm]
For a band of width $n$ consider $2n+2$ functions $q_{\emptyset}, q_i$ and $p_{\emptyset}, p_i$ with $i=1,\cdots,n$. A quick analysis of the structure of the Hirota equation on the band should convince the reader that a full parametrisation requires at most $2n+2$ functions: they saturate already two adjacent columns on the band and using the Hirota equation this allows us to express all other functions in terms of those initial $2n+2$. We can keep track of these functions by packing them into vectors using some auxiliary basis $\{e_i\}$:
\be
p = p_i e^i, \qquad q = q_i e^i.
\ee
Now we can use the language of exterior forms to consider combinations of these functions. We define
\be
 p_{(k)} = \frac{p^{[k-1]} \wedge p^{[k-3]}\cdots p^{[1-k]}}{p_{\emptyset}^{[k-2]} p_{\emptyset}^{[k-4]}\cdots p_{\emptyset}^{[2-k]}}, \quad p_{\emptyset} = p_{(0)}
\ee
and similar definitions for $q_{(k)}$. In particular we set the volume form $e^1\wedge \cdots e^n=1$. The general Wronskian solution for a band of width $n$ is given by
\be
T_{a,s} = q_{(a)}^{[+s]} \wedge p_{(n-a)}^{[-s]}
\ee 
and one can check that these $T$s satisfy the Hirota equations on the band and all $T$s outside the band vanish. For our case, $n=2$ and we obtain
\begin{align}
T_{0,s} = q_{\emptyset}^{[+s]}p_{(2)}^{[-s]}, \quad T_{1,s} = q^{[+s]}\wedge p^{[-s]}, \quad T_{2,s} = p_{\emptyset}^{[-s]}  q_{(2)}^{[+s]}, 
\end{align}
where 
\be
 q_{(2)} = \frac{q^+ \wedge q^-}{q_{\emptyset}}, \quad  p_{(2)} = \frac{p^+ \wedge p^-}{p_{\emptyset}}. 
 \ee
This parametrisation possesses a symmetry -- $H$ symmetry -- rotating the $p$'s and $q$'s using a $u$-dependent matrix $H$ taking values in $\mathfrak{sl}(2)$. However, to be consistent with the Hirota equations $H$ should be $2i\ad$-periodic. Interestingly, in the deformed case the fact that naturally $H$ should also be $2\pi$-periodic ensures that a single-valued $H$ is elliptic, which severely limits the possibilities for $H$ transformations. One can check that there also exists a scalar symmetry 
\begin{align}
p_i \rightarrow A p_i,\quad  q_i \rightarrow A^{-1} q_i, 
\end{align}
where $A$ has the same periodicity properties as the components of $H$. The more general $n$-band solution has another scalar symmetry, but that is trivial for $n=2$. 
\\[5mm]
To find an explicit solution for our case, we will have to impose more constraints, most importantly the $\Z_4$ symmetry
\be
\label{eq:z4sym}
\hbT_{a,-s} = (-1)^a \hbT_{a,s}.
\ee
Indeed,
\begin{itemize}
\item $\Z_4$ symmetry immediately implies $\hat{\mathbb{T}}_{1,0} = 0$, which implies that $\hat{p}= \alpha \hat{q}$, with $\alpha$ a $2\pi$-periodic function of $u$ for the deformed case. 
\item We find that $\alpha^+ = \alpha^-$ from $\hat{\mathbb{T}}_{1,1} = -\hat{\mathbb{T}}_{1,-1}$, another consequence of the $\Z_4$-symmetry. We can now use the scalar symmetry with $A= \sqrt{\alpha}$ to absorb $\alpha$ in $p$ and $q$. 
\item Our gauge choice $\hat{\mathbb{T}}_{0,s} =1$ implies that $\hat{q}_{\emptyset}$ is $i \ad $-periodic and $\hat{p}_{\emptyset} = \hat{q}_{\emptyset} \hat{\mathbb{T}}_{1,1}$.
\end{itemize}
This leaves us with the parametrisation
\begin{align}
\hat{\mathbb{T}}_{0,s} = 1, \qquad \hat{\mathbb{T}}_{1,s} = \hat{q}_1^{[+s]}\hat{q}_2^{[-s]}-\hat{q}_1^{[-s]}\hat{q}_2^{[+s]},\qquad \hat{\mathbb{T}}_{2,s} = \hat{\mathbb{T}}_{1,1}^{[+s]}\hat{\mathbb{T}}_{1,1}^{[-s]}.
\end{align}
for two unknown functions $q_1,q_2$. 
\\[5mm]
We can use the $H$ symmetry to make this parametrisation even nicer: by solving an appropriate Riemann-Hilbert problem we can ensure that the $q_i$ each only have one cut. From the parametrisation we see that the $q_i$ should satisfy a system of Baxter equations
\begin{align}
\label{Be1}
\hat{q}^{[2r-1]}\hat{\mathbb{T}}_{1,1} & = \hat{q}^+\hat{\mathbb{T}}_{1,r}^{[r-1]}   - \hat{q}^-\hat{\mathbb{T}}_{1,r-1}^{[r]} \nn
\hat{q}^{[-2r+1]}\hat{\mathbb{T}}_{1,1} & = \hat{q}^-\hat{\mathbb{T}}_{1,r}^{[-r+1]}   - \hat{q}^+\hat{\mathbb{T}}_{1,r-1}^{[-r]},
\end{align}
which can be used to analyse the solutions: using that $\hat{\mathbb{T}}_{1,1}$ is regular at $3i\ad$ we find that
\be
\label{49}
\hat{\mathbb{T}}_{1,1}^{[-3]} [\hat{q}]_{2r-4} = \hat{\mathbb{T}}_{1,r}^{[r-4]}[\hat{q}]_{-2}-\hat{\mathbb{T}}_{1,r-1}^{[r-4]}[\hat{q}]_{-4}, \quad \text{for } r>2,
\ee
from which one can derive the Riemann-Hilbert problem
\be
\label{RH1}
[H_i^j]_0 \hat{q}_j^{[-2n]}+H_i^j \left[\hat{q}_j^{[-2n]}\right]_0 = 0 \quad \text{ for } j,n=1,2
\ee
for a symmetry transformation $H$ that will regularise $\hat{q}$ to have only one short cut on the real axis. This follows from the fact that eqn. \eqref{49} relates the cuts of $\hat{q}$ in the upper half-plane to the first two cuts above the real line. Taking the conjugate of this equation does the analogous thing in the lower half-plane. Therefore, by making sure that $\hat{q}$ is regular at those two cuts ensures that it has only one cut on the real line. This is achieved by finding a symmetry transformation $H$ that satisfies eqn. \eqref{RH1}. Whether this equation has a solution can be seen by analysing the related integral equation
\be
\label{integralH}
H_i^j = P_i^j + K_{\ad} \, \hat{\star}\left( H_i^k [ A_k^n]_0 (A^{-1})_n^j \right),
\ee
with $A_i^n = \hat{q}_i^{[-2n]}$ an invertible $2\times2$ matrix (the $q_i$ are independent solutions of a Baxter equation). In the undeformed case the $P_i^j$ are branch-cut-free functions with period $i/(2g)$, whereas in the deformed case they are elliptic with half periods $(2\pi, i\ad)$. The integration kernel $K_{1/g}$ for the undeformed formula is 
\be
\label{eq:RHundkernel}
K_{1/g}(u) = g\pi \coth (g \pi u),
\ee
whereas the deformed $K_{\ad}$ is an elliptic integration kernel since it should reflect the real periodicity of $\hbT$s. The simplest choice for this kernel is
\be
\label{eq:RHdefkernel}
K_\ad (u) = \zeta(u) -\zeta(u+2\pi) +\zeta(2\pi),
\ee
with $\zeta$ the quasi-elliptic Weierstra\ss\, function with quasi-halfperiods $(2\pi, i\ad)$. This kernel is elliptic and has exactly two simple poles in its fundamental parallelogram with residues $1$ and $-1$. If we take discontinuities on both sides of eqn. \eqref{integralH} we find for Im$(u)>0$
\begin{align}
\left[ H_i^j \right]_0(u) &= H_i^k [ A_k^n]_0 (A^{-1})_n^j,
\end{align}
showing that a solution to eqn. \eqref{integralH} also solves the Riemann-Hilbert problem \eqref{RH1}. One can view the integral equation \eqref{integralH} as an eigenvalue equation for the integral operator $L$ defined by
\be
L(T) = P^j + K_{\ad} \,\hat{\star} \left( T^k [ A_k^n]_0 (A^{-1})_n^j \right).
\ee
Whether an appropriate eigenfunction with unit eigenvalue exists can in principle be determined using Fredholm theory. As it is not our main aim to have explicit solutions of the $T$ system, we will assume this is the case to be able to continue these properties. Note that for the undeformed case the numerical solutions do exhibit this behaviour. Forgetting for now the condition that det$(H) = 1$ we find a linear combination $\hat{q}_i'$ of the old $\hat{q}_i$ that has only one cut on the real axis. At this stage one can use the fact that the $\hat{\mathbb{T}}_{1,s}$ have only $2$ cuts at $\pm i\ad s$ to conclude that det$H$ is cut free and can be absorbed in $\hat{q}_i'$ without spoiling their cut structure. This shows that $H$ is indeed a symmetry transformation such that the new $\hat{q}_i'$ have exactly one short cut, as we wanted. This reduces the parametrisation of the $\hbT$ gauge to just two functions with a very simple cut-structure. 

\section{Conclusions}
In this chapter we have found basic building blocks for the TBA $Y$-functions in the form of four $T$-gauges. In particular, the $\mbT$ and $\bfT$ gauges have good properties and parametrise the entire $T$-hook in an efficient way. Moreover, after discovering that the short-cutted version of the left and right $\mbT$ gauges have $\Z_4$ symmetry we could map these gauges to solutions on infinite horizontal bands, for which the solutions have been studied extensively and have an elegant representation. The result is that the entire right $\mbT_L$ gauge is parametrised by two functions, let's call them $\Pf_1$ and $\Pf_2$, each with only one $\hat{Z}_0$ cut.\footnote{This notation will become clear in the next chapter.} The left gauge, as promised, follows completely analogously and is therefore parametrised by two functions $\Pf^3$ and $\Pf^4$ with one $\hat{Z}_0$ cut. Using the global gauge transformation \eqref{mbt} this allows us to find an expression for any $T$ function on the $T$ hook using just these four functions and $\bfT_{0,1}$. We will analyse the consequences of this simplicity in the next chapter. 
\\[5mm]
As a final note: the analytic $T$-system we just derived can in principle be used to find (part of) the spectrum of the undeformed and deformed case and in fact in \cite{Gromov:2011cx} an algorithm is presented that can be used to find a numerical solution for the undeformed case. They present results for the Konishi operator. In \cite{Leurent:2013mr} this computation was pushed even further to yield an eight-loop result thereby including double wrapping. Knowing, however, that much more powerful techniques are waiting at the end of the next chapter we do not get sidetracked here and find the analogous algorithm for the deformed case, but push for the final simplification. 

%% file: QSC.tex
\section{Introduction}
The efforts to simplify the spectral problem of the $\ads$ string theory and its dual gauge theory $\mN =4 $ SYM span more than a decade and involve work of dozens of people. This has created a vast landscape of literature and a plethora of roads that lead you through it. This thesis presents one of the shortest routes from the formulation of the model to the simplest known form of the spectral problem, without too much branching off along the way. So far we have 
\begin{itemize}
\item taken the ground-state TBA-equations,
\item extracted a set of $Y$-system equations, 
\item extracted a accompanying set of analyticity requirements to formulate the analytic $Y$-system,
\item and proved that the simplest solution of the analytic $Y$-system is also solution of the ground-state TBA-equations,
\item then introduced the more basic $T$ functions that parametrise the $Y$ functions,
\item and found two particularly nice gauges of these $T$ functions. 
\end{itemize}
At first glance it might seem like we are completely done here, but a closer inspection reveals something strange: our present parametrisation of the $T$ hook in terms of four $\Pf$ functions and $\bfT_{0,1}$ is completely unconstrained. This implies that, suppose we are given the boundary conditions for a particular state, any solution satisfying those boundary conditions should form a solution of the spectral problem. This can mean one of three things: (1) either the boundary conditions are so constraining that they uniquely specify the solution or (2) the spectral problem does not have a unique solution or (3) we have failed to incorporate all of the analytic data available in the analytic $T$-system. Obviously the second option is ludicrous and as we will see the boundary conditions do not uniquely specify the solution, rendering option one false. 
\\[5mm]
It turns out that there are still some constraints coming from the analytic properties of the $T$ system that we have not revisited in the $\Pf$ language: even though the $\Pf$ take into account the branch-cut structure of the $\mbT$ gauge, we have not made sure they do so for the upper-band $\bfT$-functions! Doing this will introduce equations that the $\Pf$ functions should satisfy and are called \emph{$\Pf \mu$ system}, being one incarnation of the QSC. We will do this in the next section. Using the $\Pf \mu$ system we can derive the ``dual" $\Qf \omega$ system that allows for the formulation of the $QQ$ system. The rest of the chapter is devoted to the properties of these systems. We reserve the presentation of the final ingredients for the next chapter. There we will encounter that we in fact left some boundary data back in the TBA equations that we can use to derive the boundary conditions, but let us not skip ahead. 

\section{Deriving the $\Pf \mu$ system}
We left the last chapter with a convenient parametrisation of the left and right band of the $\mbT$ gauge:
\begin{align}
\label{eq:TsinPs}
\hbT_{0,s} &= 1 &\text{for $s\in \Z$}, \nn
\hbT_{1, s} &= \Pf_1^{\phantom{4}[+s]}\Pf_2^{\phantom{4}[-s]}-\Pf_1^{\phantom{4}[-s]}\Pf_2^{\phantom{4}[+s]} &\text{for $s>0$}, \nn
\hbT_{1, s} &= \Pf^{4[+s]}\Pf^{3[-s]}-\Pf^{4[-s]}\Pf^{3[+s]} &\text{for $s<0$}, \nn
\hbT_{2,\pm s} &= \hbT_{1,\pm 1}^{[+s]}\hbT_{1,\pm 1}^{[-s]} &\,\,\text{for $|s|>1$},
\end{align}
where the four $\Pf$-functions all have one $\hat{Z}_0$ cut. This result holds for both the undeformed and deformed case, but in the last case the $\Pf$ functions are at least $2\pi$ (anti-)periodic. To find the $\Pf \mu$ system we need to find expressions for the upper-band $\bfT$-functions in terms of these $\Pf$ functions, for which we use the gauge transformation \eqref{mbt} that we will repeat here for the readers convenience: 
\be
\mbT_{a,s} = (-1)^{as}  \bfT_{a,s} \left(\bfT_{0,1}^{[a+s-1]}\right)^{\frac{a-2}{2}}.
\ee
We see that the upper-band $\bfT$s will not only contain $\Pf$ functions, but also a function that takes care of the dependence on $\bfT_{0,1}$. The process of finding these expressions can be easily programmed, see appendix A of \cite{Gromov:2014caa}. 

\begin{figure}[!t]
\centering
\begin{subfigure}{7.5cm}
\includegraphics[width=7.5cm]{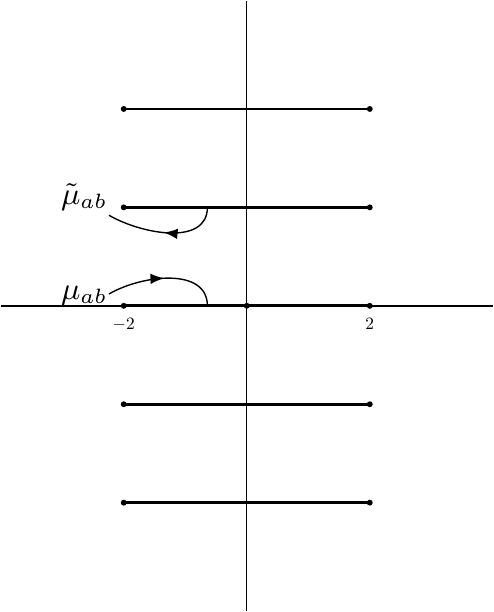}
\caption{}
\end{subfigure}\quad 
\begin{subfigure}{6cm}
\includegraphics[width=6cm]{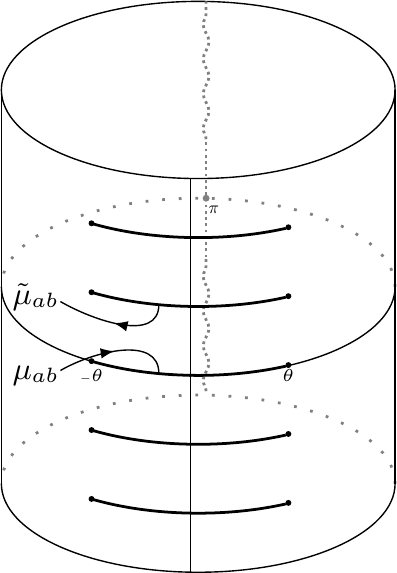}
\caption{}
\end{subfigure}\\[2mm]
\caption{The analytic structure of the $\mu_{ab}$ on the undeformed plane (a) and the deformed cylinder (b). The thick lines indicate branch cuts on the first sheet. The continuation of short-cutted $\mu_{ab}$ can be expressed on the first sheet using its $2i\ad$ periodicity. The squiggly line in (b) indicates that for generic $\theta$ outside the physical strip the $\mu_{ab}$ cannot be put on (a finite cover of) the cylinder.}
\label{fig:mu}
\end{figure}

\paragraph{The middle nodes and $\mu_{12}$.} Following the group-theoretical properties we see that the middle nodes $(0,s)$ of the $\bfT$ gauge are all generated from a single function $\bfT_{0,0}$, which by itself is also $2i\ad$-periodic, such that there are only two unique entries on the middle line of the $\bfT$ hook: $\bfT_{0,0} = \bfT_{0,2} = \ldots$ and $\bfT_{0,1} = \bfT_{0,3} = \ldots$ which are related as $\bfT_{0,s}= \bfT_{0,s+1}^{\pm}$. To simplify notation later we now introduce the first $\mu$ function $\mu_{12}$ just as
\be
\label{defmu}
\check{\mu}_{12} = \left(\bfT_{0,1}\right)^{1/2},
\ee
which most naturally has long cuts as accentuated here and as a long-cutted function is $2i\ad$-periodic (see fig. \ref{fig:mu}).

For our construction the short-cutted version $\hat{\mu}_{12}$ will be more convenient, but its relation to the just-defined $\check{\mu}_{12}$ depends on the choice of upper- or lower-half-plane conventions (see section \ref{sec:shortcuttedfunctions}), since $\bfT_{0,1}$ has a cut on the real line and therefore splits the upper and lower half of the physical strip. In order to stay in line with the literature we will therefore at this point in our construction switch conventions from the lower-half-plane conventions to the upper-half-plane conventions. We will have to return to the lower-half-plane conventions once more in the next chapter, but that should not be problematic. Before formalising our switch, let us mention that the notation $\mu_{12}$ of course suggests that there are other $\mu_{ab}$, which we will indeed introduce shortly.

\begin{figure}[!t]
\centering
\begin{subfigure}{7.5cm}
\includegraphics[width=7.5cm]{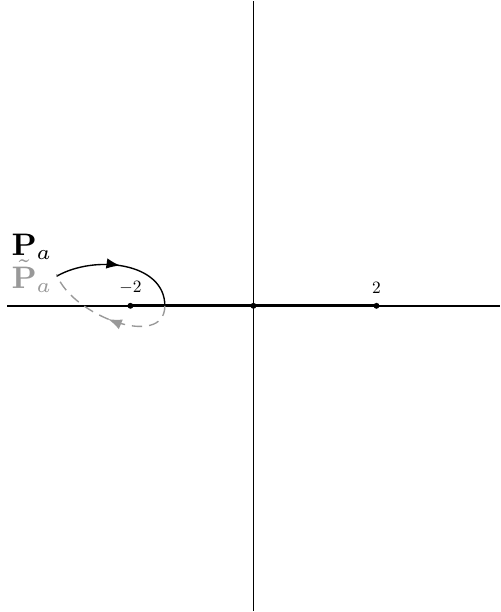}
\end{subfigure}\quad 
\begin{subfigure}{6cm}
\includegraphics[width=6cm]{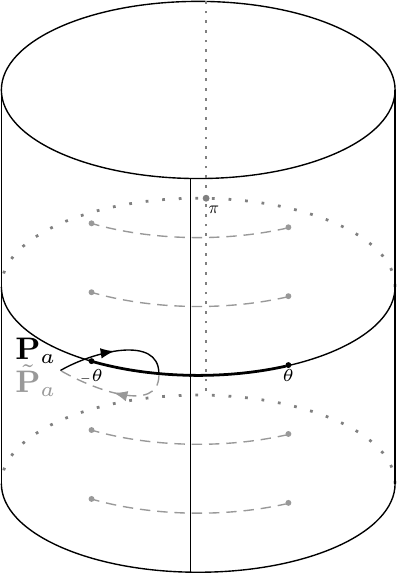}
\end{subfigure}\\[2mm]
\caption{The analytic structure of the $\Pf_a$ in the undeformed (a) and the deformed (b) case. The thick line indicates the only branch cut on the first sheet. The continuation of $\Pf_a$ from the first sheet to the second sheet is indicated by a solid path on the first and a (gray) dashed path on the second sheet. The dotted line through $\pi$ in (b) indicates the possible discontinuity of the $\Pf_a$ due to anti-periodicity.}
\label{fig:P}
\end{figure}

\paragraph{Switching conventions.}\label{sec:switchingconventions} We would like to switch from the lower-half-plane conventions to the upper-half-plane conventions: this can be easily achieved by flipping the sign of $u$, $u\rightarrow -u$. This changes only:
\begin{align}
\label{newconv}
\check{f} = \hat{f} \text{ in the strip } 0> \text{Im}(u)>-\ad \quad &\Longrightarrow \quad \check{f} = \hat{f} \text{ in the strip } 0< \text{Im}(u) < \ad, \nn
x(u) \rightarrow \infty \text{ as } u \rightarrow - i \infty &\Longrightarrow  x(u) \rightarrow \infty \text{ as } u\rightarrow i \infty, \nn
x^{\text{und}}(u-i \e) \rightarrow \infty \text{ as } u \rightarrow  \infty &\Longrightarrow  x^{\text{und}}(u+i\e) \rightarrow \infty \text{ as } u\rightarrow  \infty,
\end{align}
In particular, the $x$ functions $x_s$ and $x_m$ no longer coincide on the lower-half plane but instead on the upper-half plane, in line with the central paper \cite{Gromov:2014caa}. We will use these new conventions in the remaining derivation of the quantum spectral curve. 
\paragraph{Introducing extra $\Pf$ and $\mu$ functions.} As mentioned before, to constrain the $\Pf$ functions and $\mu_{12}$ we need to extract information from the analyticity of the upper band \cite{Gromov:2014caa}. Using the gauge transformation \eqref{mbt} we find for example that 
\begin{equation}
\bfT_{2,1} = \left(\tilde{\Pf}_1 \Pf_{2}^{[-2]} - \tilde{\Pf}_2 \Pf_{1}^{[-2]}  \right) \left(\Pf_1 \Pf_{2}^{[+2]} -\Pf_2 \Pf_{1}^{[+2]}  \right) + \mu_{12}\left(\Pf_1^{[-2]} \Pf_{2}^{[+2]} - \Pf_2^{[-2]} \Pf_{1}^{[+2]}  \right). 
\end{equation}
Using that $\bfT_{2,1}$ should not have a cut on the real line, i.e. $\bfT_{2,1} - \tilde{\bfT}_{2,1} = 0$, we find an equation on the $\Pf$ and $\mu_{12}$ which can be simplified to read
\begin{equation}
\label{eq:muconstraint1}
\tilde{\mu}_{12} - \mu_{12} = \Pf_1 \tilde{\Pf}_2 - \Pf_2 \tilde{\Pf}_1. 
\end{equation}
From the left band and the same analysis for $\bfT_{2,-1}$ we find 
\begin{equation}
\label{eq:muconstraint2}
\tilde{\mu}_{12} - \mu_{12} = \Pf^3 \tilde{\Pf}^4 - \Pf^4 \tilde{\Pf}^3,
\end{equation}
already revealing a relation between the upper and lower index $\Pf$ functions. Continuing up the upper band we impose the correct analyticity for all the other $\bfT$ functions (see appendix B.3 in \cite{Gromov:2014caa}, which leaves the following linear system
\begin{align}
\label{eq:linearsystem}
\frac{\reallywidetilde{\tilde{\Pf}^{\phantom{b}[2n]}_{a}}}{\tilde{\mu}_{12}} - 
\frac{\tilde{\Pf}^{[2n]}_{a}}{\mu_{12}} &=& - \Pf^{4 [2n]} \frac{\Pf_{a}\Pf^3 -  \Pft_{a}\Pft^3}{\mu_{12}\mut_{12}} + \Pf^{3 [2n]} \frac{\Pf_{a}\Pf^4 -  \Pft_{a}\Pft^4}{\mu_{12}\mut_{12}}, \quad &\text{ for } a =1,2\nn
\frac{\reallywidetilde{\tilde{\Pf}^{b[2n]}}}{\tilde{\mu}_{12}} - 
\frac{\tilde{\Pf}^{b [2n]}}{\mu_{12}} &=& +\Pf_1^{\phantom{4}[2n]} \frac{\Pf^{b}\Pf_2 -  \Pft^{b}\Pft_2}{\mu_{12}\mut_{12}} - \Pf_2^{\phantom{3}[2n]} \frac{\Pf^{b}\Pf_1 -  \Pft^{b}\Pft_1}{\mu_{12}\mut_{12}}, \quad &\text{ for } b =3,4,
\end{align}
where $n \in \Z\setminus\{0\}$ and the big tildes are applied after the shifts. We can reinterpret these equations as those expressing the analyticity of four new $\Pf$ functions $\Pf^1$, $\Pf^2$, $\Pf_3$ and $\Pf_4$ on all the $Z_n$ for $n\neq 0$, defined as follows: first consider $2 i\ad$-periodic solutions $\mu_{\alpha 3}$ and $\mu_{\alpha 4}$ (for $a = 1,2$) of the following Riemann-Hilbert problems
\be
\label{eq:RHproblemmu}
\frac{\mut_{a 4}}{\mut_{12}}-\frac{\mu_{a 4}}{\mu_{12}} = \frac{\Pft_a \Pft^3-\Pf_a \Pf^3}{\mu_{12}\mut_{12}} \quad \frac{\mut_{a 3}}{\mut_{12}}-\frac{\mu_{a 3}}{\mu_{12}} = \frac{\Pft_a \Pft^4-\Pf_a \Pf^4}{\mu_{12}\mut_{12}}. 
\ee
That these equations have a solution has to be considered separately for the undeformed and deformed case, but let us first finish our introduction of the $\Pf$ functions: given $\mu$ functions satisfying the above equations, we see that the four new $\Pf$ functions
\begin{equation}
\label{eq:newPf}
\begin{aligned}
\Pf^1 &\defeq -\frac{1}{\mu_{12}} \Pft_2 + \frac{\mu_{24}}{\mu_{12}} \Pf^4+ \frac{\mu_{23}}{\mu_{12}} \Pf^3, \\
\Pf^2 &\defeq + \frac{1}{\mu_{12}} \Pft_1 - \frac{\mu_{14}}{\mu_{12}} \Pf^4- \frac{\mu_{13}}{\mu_{12}} \Pf^3,
\end{aligned} \quad 
\begin{aligned}
\Pf_3 &\defeq  +\frac{1}{\mu_{12}} \Pft^4 - \frac{\mu_{23}}{\mu_{12}} \Pf_1+ \frac{\mu_{13}}{\mu_{12}} \Pf_2, \\
\Pf_4 &\defeq -\frac{1}{\mu_{12}} \Pft^3 - \frac{\mu_{24}}{\mu_{12}} \Pf_1+ \frac{\mu_{14}}{\mu_{12}} \Pf_2
\end{aligned}
\end{equation}
are all analytic apart from on $Z_0$ provided the linear system \eqref{eq:linearsystem} is satisfied for all $n\neq 0$. Since they are defined with short cuts this shows that all the eight $\Pf$ functions have only one $\hat{Z}_0$ cut. This analytic structure is illustrated in fig. \ref{fig:P}. 

\paragraph{Solving the Riemann-Hilbert problem.}\label{sec:solvingRHproblem} In order for the above construction to work we need to know that the newly introduced $\mu$ functions actually exist. For that we should show that the Riemann-Hilbert problem \eqref{eq:RHproblemmu} has a solution in both the undeformed and the deformed case. Note that these equations are of the form 
\be
\tilde{f}-f= g,
\ee
where in the deformed case $g$ is (anti)-periodic and indeed always has a solution in both cases 
\be
f(u) = \frac{1}{2\pi i} \int_{Z_0} K_{\ad}(u-v) g(v) dv,
\ee
modulo a regular $2i\ad$-periodic function. Interestingly, in the deformed case this regular function must be a constant due to its required real periodicity combined with Liouville's theorem, which colloquially put states that regular doubly-periodic functions are constants. The deformed kernel $K_{\ad}$ was defined in eqn.  \eqref{eq:RHdefkernel}, its undeformed counterpart in eqn. \eqref{eq:RHundkernel}. These kernels satisfy
\begin{align}
K_{\ad}(u+2\pi) &= - K_{\ad}(u), \nn
\lim_{\ad\rightarrow 0} \ad \cdot g K_{\ad} (g \ad u ) &= K_{1/g}(u),
\end{align}
i.e. $K_{\ad}$ is anti-periodic and has the expected undeformed limit. 
\\[5mm]
The solutions in the deformed case defined in this way are always anti-periodic in the real direction with anti period $2\pi$ and periodic in the imaginary direction with period $2 i\ad$, from which one can deduce that the newly introduced $\mu_{ab}$ will be anti-periodic when $\mu_{12}$ is periodic and vice versa. We see in both cases that $\check{\mu}_{ab}^{++} = \check{\mu}_{ab}$ for all $a,b$.\footnote{Note the check to emphasise that this is true for long cuts only.}

\paragraph{Combining the relations.} So far we have introduced eight $\Pf$ functions $\Pf_a$ and $\Pf^a$ with $a=1,\ldots,4$ and five $\mu$ functions $\mu_{12}$, $\mu_{1 3}$, $\mu_{23}$, $\mu_{14}$ and $\mu_{24}$, which are constrained by the equations \eqref{eq:muconstraint1}, \eqref{eq:muconstraint2}, \eqref{eq:RHproblemmu} and \eqref{eq:newPf}. We can introduce one more $\mu$ function to fill the upper triangle of a $4\times 4$ matrix as 
\begin{equation}
\label{eq:mu34definition}
\mu_{34} = \frac{1+\mu_{13}\mu_{24}-\mu_{14}\mu_{23}}{\mu_{12}}
\end{equation}
and subsequently finish filling the entire matrix by demanding anti-symmetry, i.e. \\$\mu_{ab} = -\mu_{ba}$ for all $a,b=1,\ldots, 4$. By construction all these $\mu_{ab}$ are $2i\ad$-periodic and satisfy 
\begin{equation}
\mut_{ab}-\mu_{ab} = \Pf_a \Pft_b -\Pft_a \Pf_b. 
\end{equation}
Moreover, the definition \eqref{eq:mu34definition} is now such as to impose that the Pfaffian of the matrix $\mu$ is equal to one:
\begin{equation}
\label{eq:Pfaffian}
\text{Pf} \left( \mu \right) \defeq \mu_{12}\mu_{34}-\mu_{13}\mu_{24}+\mu_{14}\mu_{23} = 1. 
\end{equation}
We can compactly write all these relations, as was first done in \cite{Gromov:2013pga}, in the following form: 
\be
\label{Pmusystem}
\boxed{
\mut_{ab}-\mu_{ab} = \Pf_a \Pft_b-\Pf_b \Pft_a, \quad \Pft_a = \mu_{ab} \Pf^b,\quad \Pf_a\Pf^a = 0,\quad  \text{Pf}(\mu) = 1,
}
\ee
called the \emph{$\Pf\mu$ system}. Note that in the construction in this section we only needed the algebraic relations between the different functions and their second sheet evaluations and their existence as guaranteed by the solution to the Riemann-Hilbert problem \eqref{eq:RHproblemmu}. That is, we did not explicitly use the real periodicity to find the constraints in \eqref{Pmusystem}. It is because of this that the $\Pf \mu$ system is identical for both the undeformed and the deformed case. 
\\[5mm]
By introducing the inverse of the $\mu$ matrix with upper indices as $\mu_{ab} \mu^{bc} = \delta^c_a$ we also find the upper-index form of the $\Pf \mu$ system
\be
\mut^{ab}-\mu^{ab} = -\Pf^a \Pft^b+\Pf_b \Pft^a, \quad \Pft^a = \mu^{ab} \Pf_b,\quad \Pf_a\Pf^a = 0,\quad  \text{Pf}(\mu) = 1.
\ee
\section{Deriving the $\Qf \omega$ system }
Next we construct the $\Qf\omega$ system from the $\Pf\mu$ system. Define the $4\times4$-matrix $U$ as
\be
U^b_a = \delta^b_a + \Pf_a \Pf^b
\ee
and consider the finite-difference equation
\be
X_a^- = U^b_a X_b^+
\ee
for the unknown functions $X_a$, $a=1,\cdots 4$. We can construct formal solutions of this equation which are analytic in the upper half-plane by the infinite product 
\be
X_a = \left(U^{[+1]}U^{[+3]}\cdots \right)^b_a \mathbf{C}_b
\ee
with $\mathbf{C}_b$ an arbitrary $2i\ad$-periodic vector. Note that $X_a$ inherits periodicity properties of the $\Pf_a$ and in the deformed case therefore is periodic by construction. Taking four independent vectors $\mathbf{C}_{b|i}$ labelled by $i$ we can build the matrix $M_{bi}$ and thus find four independent solutions
\be
\Qm_{a|i} = \left(U^{[+1]}U^{[+3]}\cdots \right)^j_a M_{ji}.
\ee
We can now define a set of $\Qf$ functions in the same way:
\be
\label{defQ}
\Qf_i = -\Pf^a \Qm_{a|i}^+ \text{ for Im}(u)>0,
\ee
since we are now working in the upper-half-plane-conventions (see section \ref{sec:switchingconventions}). An immediate consequence of these definitions is
\be
\label{QPF1}
\Qm_{a|i}^+-\Qm_{a|i}^- = \Pf_a \Qf_i.
\ee
Viewed as functions with long cuts the $\Qf$s only have one $\check{Z}_0$ cut, as follows by following the proof in \cite{Gromov:2014caa} verbatim. These $\Qf$s obey a system of equations quite similar to the $\Pf$ functions. To see this we first define the $\Qf$ analogue of the $\mu$ functions, called $\omega$: 
\\[5mm]
Consider the continuation of $\Qf$ through the long cut on the real axis,
\be
\label{defomega1}
\Qft_i = -\Pft^a \Qm_{a|i}^{\pm} = \Pf_b\mu^{ba}\Qm_{a|i}^{\pm},
\ee
where the shift direction in $\Qm_{a|i}^{\pm}$ is irrelevant as a simple application of the orthogonality condition $\Pf_a\Pf^a = 0$ will show. Define the $\Qf$ with upper indices as
\be
\Qf^i = \Pf_a \left(\Qm^{a|i}\right)^+,\quad \text{with } \Qm^{a|i} = \left(\Qm_{a|i}\right)^{-t},
\ee
with $M^{-t}$ indicating the transpose of the inverse of $M$. We can find that
\be
\left(\Qm^{a|i}\right)^- = V_b^a\left(\Qm^{a|i}\right)^+
\ee
where $V = U^{-1}$ is simply given by $V_b^a = \delta_b^a -\Pf_b \Pf^a$. This shows directly that $\Qf^i$ also has only one $\check{Z}_0$ cut and we can write
\be
\Pf_a = -\Qf^i \Qm_{a|i}^{\pm}.
\ee
This allows us to get rid of $\Pf$ in eqn. \eqref{defomega1} and find
\begin{align}
\label{defomega}
\Qft_i = -\Qf^j \Qm_{a|j}^{\pm} \mu_{ab}\Qm_{b|j}^{\pm} = -\Qf^j \omega_{ji}, \quad \text{where we define } \omega_{ji}=\Qm_{a|j}^{-}\mu_{ab}\Qm_{b|i}^{-},
\end{align}
such that $\omega$ is an antisymmetric matrix with long cuts playing the role analogous to $\mu$ in the $\Pf\mu$ system. Additionally we find that $\omega$ is $2i\ad$-periodic when viewed as a function with short cuts and that its discontinuity relation is
\be
\tilde{\omega}_{ij}-\omega_{ij} = \Qf_i \Qft_j -\Qft_i \Qf_j.
\ee
The final step involves setting the Pfaffian of $\omega$ to one: det$(\omega) = 1$, so from the defining equation of $\Qm_{a|i}$ we find that $\det \Qm$ is an $2i\ad$-periodic function. Since $\Qm$ is analytic in the upper half-plane, its $2i\ad$ periodicity implies that det$(\Qm)$ does not have any cuts. So we can rescale it without introducing new branch cuts  in our construction, by rescaling $\Qm_{a|i}$. This gives us det$(\omega)=1$ and combined with antisymmetry this implies Pf$(\omega)=1$ up to a sign, which we can fix by rescaling by $-1$ if necessary. This yields the \emph{$\Qf \omega$ system}
\be
\label{Qomegasystem}
\omegat_{ij}-\omega_{ij} = \Qf_i \Qft_j-\Qf_j \Qft_i, \quad \Qft_i = \omega_{ij} \Qf^j,\quad \Qf_i\Qf^i = 0,\quad  \text{Pf}(\omega) = 1,
\ee
valid for both the undeformed and deformed case. Finally, we can introduce the inverse of $\omega$ to complete the $\Qf\omega$ system with upper indices in analogy to the $\Pf \mu$ system in the previous section. 

\section{$QQ$ system}
From the objects defined so far we can now in fact define the entire $QQ$ system containing $256$ functions $\Qm_{A|I}$ with multi-indices $A$ and $I$: define the basic functions\footnote{The unimodularity constraint $\Qm_{1234|1234}=1$ follows from the $2i\ad$ periodicity of $\bfT_{1,0}$.}
\begin{align}
\Qm_{a|\emptyset} := \Pf_a, \quad \Qm_{\emptyset|i} := \Qf_i,\quad \Qm_{\emptyset|\emptyset}=\Qm_{1234|1234}=1
\end{align}
and generate the other $\Qm$s using the finite-difference equations known as the Pl\"{u}cker relations
\begin{align}
\Qm_{A|I}\Qm_{Aab|I} &= \Qm^+_{Aa|I}\Qm^-_{Ab|I}-\Qm^-_{Aa|I}\Qm^+_{Ab|I}, \nn
\Qm_{A|I}\Qm_{A|Iij} &= \Qm^+_{A|Ii}\Qm^-_{A|Ij}-\Qm^-_{A|Ii}\Qm^+_{A|Ij}, \nn
\Qm_{Aa|I}\Qm_{A|Ii} &= \Qm^+_{Aa|Ii}\Qm^-_{A|I}-\Qm^-_{Aa|Ii}\Qm^+_{A|I},
\end{align}
which find their origin in the theory of Grassmannians \cite{Kazakov:2015efa}. The $QQ$ system will not play a role in this thesis, but it is important for the theoretical embedding of the QSC in the mathematics literature as was analysed in e.g. \cite{Kazakov:2015efa}. 

\section{$H$ symmetry and reality}
The $QQ$ system derived above has a residual GL$(4)\times$GL$(4)$ gauge freedom dubbed \emph{$H$ symmetry}, that can be traced back to the Wronskian parametrisation in the previous chapter: all the equations are invariant under a transformation of the form
\be
Q_{A|I} \rightarrow \sum_{\substack{|B|=|A| \\ |J|=|I|}} \left(H_b^{\left[|A|-|I|\right]}\right)_A^B\left(H_f^{\left[|A|-|I|\right]}\right)_I^J Q_{B|J},
\ee
where $M_A^B = M_{a_1}^{b_1}M_{a_2}^{b_2}\cdots M_{a_{|A|}}^{b_{|B|}}$ and $H_b,H_f$ are arbitrary $u$-dependent GL$(4)$-matrices which are $2i\ad$-periodic and, in the the deformed case, also $2\pi$-(anti)-periodic. Note that due to the strict periodicity properties regular $H$-matrices are constant by Liouville's theorem, restricting the freedom of $H$ symmetry severely compared to the undeformed case. However, we will see that for most applications this amount of $H$ symmetry will suffice.

To analyse this a bit further, let us consider the conjugation properties of our basic functions $\Pf_a$,$\Qf_i$ and $\mu_{ab}$: it is not clear from our construction that we can ensure nice conjugation properties, but we do know that the energy extracted from the QSC should be real. This suggests that complex conjugation should be a symmetry of the QSC \cite{Gromov:2014caa}. Assuming this, we can pick up $\Pf_a$ and $\mu$ with nice reality properties using $H$ symmetry: for $\Pf_a$ the conjugation transformation then reads
\be
\bar{\Pf}_a = H_a^b \Pf_b,
\ee
where $H$ should be an invertible constant matrix to not change the analytic structure of the $\Pf_a$ functions. The newly defined $\Pf' = H^{1/2}\Pf$ will now be real functions. By multiplying the relevant $\Pf_a$ by $i$ we can actually pick the following conjugation rule for $\Pf_a$:
\be
\bar{\Pf}_a = (-1)^a \Pf_a,
\ee
which lead to real $\mathbb{T}$, consistent with our original parametrisation \eqref{eq:TsinPs}. We will nevertheless choose our $\Pf$ functions to be real.

\paragraph{Conjugation properties of $\mu$.} We can use the $\Pf \mu$ system to derive the conjugation properties of $\mu$: first we show that on $\Pf$ functions conjugation and continuation by the tilde commute, by assuming that all $\Pf$ functions can be expressed as a real function times a convergent series in the $x$ function. To show then that conjugation and continuation commute we show this property for the $x$ function, which in the undeformed case is simple since $x^{\text{und}}$ is real analytic:
\begin{equation}
\widetilde{\overline{x^{\text{und}}(u)}} = \widetilde{x^{\text{und}}\left(\overline{u}\right)} = \frac{1}{x^{\text{und}}\left(\overline{u}\right)} = \tilde{x}^{\text{und}}\left( \overline{u}\right) = \overline{\tilde{x}^{\text{und}}\left( u\right)}. 
\end{equation}
In the deformed case the $x$ function is \emph{not} real analytic, but we find nevertheless that conjugation and continuation of these functions commute:
\begin{align}
\widetilde{\overline{x(u)}} = \widetilde{\left( \frac{x(\bar{u})\xi+1}{x(\bar{u})+\xi} \right)} =  \frac{(x(\bar{u}))^{-1}\xi+1}{(x(\bar{u}))^{-1}+\xi}  =\left( \frac{x(\bar{u})\xi+1}{x(\bar{u})+\xi}\right)^{-1} =(\overline{x(u)})^{-1} = \overline{\tilde{x}(u)}.
\end{align}
Now we can use the $\Pf\mu$ equations \eqref{Pmusystem} to derive that
\be
\mu_{ab} = - (-1)^{a+b} \mu_{ab}.
\ee
For short cuts we find that $\bar{\hat{\mu}}_{ab} = -(-1)^{a+b}\hat{\mu}_{ab}$ or
\be
\label{mureality}
\overline{\hat{\mu}^{+}}_{ab} = -(-1)^{a+b}\hat{\mu}_{ab}^{+},
\ee
which is going to be useful in the next chapter.

\section{Conclusions}
Starting from the Wronskian parametrisation of the left and right $\hbT$ band we have in this chapter derived the $\Pf \mu$ system for the $\ads$ superstring and its $\eta$ deformation. From it we could derive the dual $\Qf \omega$ system and the $QQ$ system. The collection of these systems is what we call the \emph{quantum spectral curve}, which was our goal all along! The resulting equations are identical for the undeformed and deformed cases, which could have been suspected based on the similarities we have encountered along the way. We are not completely done yet though. At this point the only thing we know is that the deformed $\Pf$ and $\Qf$ functions are $2\pi$-periodic. In order to be able to find (deformed) string energies belonging to a particular state we need to understand which boundary conditions have to be imposed. This is where the undeformed and the deformed case are going to differ most strongly, as we will see in the next chapter.

%% file: solutionstate.tex
\section{Introduction}
In the previous chapter we finished our derivation of the $\Pf \mu$ system, which looks identical for both the undeformed and deformed case. This by itself could be viewed as an improvement over the original TBA-formulation, since the equations look a lot more manageable. In some sense these aesthetics are only superficial though, because none of the functions in the QSC are known in advance, in sharp contrast with the TBA equation which can immediately be solved due to our knowledge of the --arguably complicated -- TBA kernels. Of course we have always known that the QSC as presented in the previous chapter is not the complete description of the ($\eta$-deformed) $\ads$ spectral problem, we are missing boundary conditions. These boundary conditions are meant to make sure that for each state in our theory we find exactly one solution of the QSC such that we find a unique value for its energy. We can label states by their quantum numbers so we expect the boundary conditions to feature them in some way. The original TBA-equations only gave direct access to the ground state depending on the quantum number $J$ and it is therefore not clear whether the QSC we have derived can be applied to find the (deformed) string energies for other states as well. We will follow the approach laid out in \cite{Gromov:2013pga} to derive boundary data from the TBA equations depending on $J$, analysing the undeformed case before tackling the more subtle deformed case. We then generalise this to propose asymptotic dependence for all the $\Pf$ and $\Qf$ functions. We see how we, with a minimal set of assumptions, can constrain this dependence completely to yield a full proposal capable of fully solving the spectral problem for the ($\eta$-deformed) $\ads$ spectral problem. Finally we check these predictions at weak coupling, following \cite{Marboe:2014gma}.

\section{Information from TBA}
Our task is to find some property of the QSC functions $\Pf$, $\Qf$, $\mu$ and $\omega$ that depends on the quantum numbers of the state we would like to compute the energy of.\footnote{For good measure, let us mention here that in the undeformed case the AdS/CFT correspondence implies that this is the same as computing the anomalous dimension of the gauge theory operator dual to the string state. As the deformed string theory does not have a known dual we cannot make a similar claim for the deformed theory.} A priori, this could be anything, from pole residues or a value at some significant point to an integral over some significant domain. It turns out that the most natural guess is to look at the asymptotics of the QSC functions in some domain in the complex plane. As a starting point, we will show in this section that there is a ratio of $\mu$ functions that has simple behaviour in the limit $u\rightarrow \infty$ for the undeformed case and $u\rightarrow i\infty$ for the deformed case. This is based on a connection between the $\mu$ functions and the TBA-equations that was first analysed in \cite{Gromov:2011cx}. There are more modern approaches to finding asymptotics for the undeformed case, but because we would like to compare with the deformed case we opt to present this method.
\subsection{Connecting the $\mu$ functions to the TBA equations}
The ratio $Y^{(\alpha)}_+ / Y^{(\alpha)}_-$ does not depend on $\alpha$, as we see from the TBA equations \eqref{eq:TBAeqns}:
\be
\frac{Y_+^{(\alpha)}}{Y_-^{(\alpha)}} = \exp \left( \Lambda_P \star K^{Py}\right). 
\ee
In the following we will omit $\alpha$ and find an expression in the $\Pf\mu$ system that equals these two quantities. Starting from the $T$ system, writing $Y$s with short cuts we find
\be
\frac{Y_+}{Y_-} =  \frac{T_{1,0}T_{1,2}}{T_{0,1}T_{2,1}}\frac{T_{2,1}T_{2,3}}{T_{1,2}T_{3,2}} = \frac{\hbT_{1,0}\hbT_{2,3}}{\hbT_{0,1}\hbT_{3,2}}.
\ee
Not all the $\hbT$s here are part of the right band, so we will have to use the Hirota equation to continue the $\Pf\mu$ parametrisation into the center of the $T$ hook. Importantly, the $\mathbb{T}$s only satisfy the Hirota equation on the entire $T$ hook when viewed with \emph{long} cuts, so we will have to be careful here. We can reparametrise three of the $\hbT$s using simple, gauge specific relations:
\begin{align}
\hbT_{0,1} &= 1 \quad \hbT_{2,3} = \hbT_{1,1}^{[+3]}\hbT_{1,1}^{[-3]}, \nn
\mathbb{T}_{3,2} &= (-1)^{3\cdot 2} \bfT_{3,2} \left(\mu_{12}^{[2+3-1]}\right)^{3-2} = \bfT_{3,2}\mu_{12}^{[+4]},  \nn
\mathbb{T}_{2,3} &= (-1)^{2\cdot 3} \bfT_{2,3} \left(\mu_{12}^{[3+2-1]}\right)^{2-2}= \bfT_{3,2}.
\end{align}
Note that the gauge transformation from the $\bfT$s to the $\mathbb{T}$s is sensitive to the choice of cuts: the functions on both sides of the transformation agree close to the real axis, but after shifting things start to change since we have defined $\mu_{12}$ canonically with short cuts. Therefore, the safest thing to do is to start from the long cutted sheet (where things are naturally defined and we can use this equation everywhere) and translate the result into short cuts afterwards. We see here for example that we get inequivalent results if we had chosen the short-cutted $\mu_{12}$: in that case 
so
\be
\mathbb{T}_{3,2} = \mathbb{T}_{2,3}\mu_{12}^{[+4]} =  \mathbb{T}_{2,3}\mu_{12}
\ee
instead of $\hat{\mu}_{12}^{[+4]}$, which does not have a clear simplification on the short-cutted sheet. Now we find
\be
\frac{Y_+}{Y_-} = \frac{\hbT_{1,0}}{\mu_{12}}
\ee
and we only need to use the $\Pf\mu$ system to find $\hbT_{1,0}$. We will need to find $\hbT_{2,1}$ first:
\begin{align}
\mathbb{T}_{2,2}^+\mathbb{T}_{2,2}^- &= \mathbb{T}_{2,1}\mathbb{T}_{2,3}+\mathbb{T}_{1,2}\mathbb{T}_{3,2} = \mathbb{T}_{2,1}\mathbb{T}_{2,3}+\mathbb{T}_{1,2}\mathbb{T}_{2,3}\mu_{12} \nn
\mathbb{T}_{2,2}^+\mathbb{T}_{2,2}^- &= \mathbb{T}_{1,1}^{[+3]}\mathbb{T}_{1,1}^{[-3]}\hat{\mathbb{T}}_{1,1}^+\tilde{\hat{\mathbb{T}}}_{1,1}^- \nn
\mathbb{T}_{2,1} &= \frac{\mathbb{T}_{2,2}^+\mathbb{T}_{2,2}^--\mathbb{T}_{1,2}\mathbb{T}_{2,3}\mu_{12}}{\mathbb{T}_{2,3}} = 
\hat{\mathbb{T}}_{1,1}^+\tilde{\hat{\mathbb{T}}}_{1,1}^--\mathbb{T}_{1,2}\mu_{12},
\end{align}
where the most non-trivial step lies in the correct application of the hatted identity
\be
\hbT_{2,2} = \hbT_{1,1}^{[+2]}\hbT_{1,1}^{[-2]}.
\ee
This holds for \emph{short}-cutted $\mathbb{T}$'s, so application in the Hirota equation requires continuation. Now we can find $\mathbb{T}_{1,0}$:
\begin{align}
\mathbb{T}_{1,1}^+\mathbb{T}_{1,1}^- &= \mathbb{T}_{0,1}\mathbb{T}_{2,1}+\mathbb{T}_{1,0}\mathbb{T}_{1,2} = \mathbb{T}_{2,1}+\mathbb{T}_{1,0}\mathbb{T}_{1,2} \nn
\mathbb{T}_{1,0} &= \frac{\mathbb{T}_{1,1}^+\mathbb{T}_{1,1}^- -  \mathbb{T}_{2,1}}{\mathbb{T}_{1,2}} = \frac{\tilde{\hat{\mathbb{T}}}_{1,1}^+\mathbb{T}_{1,1}^- -  \left(\hat{\mathbb{T}}_{1,1}^+\tilde{\hat{\mathbb{T}}}_{1,1}^--\mathbb{T}_{1,2}\mu_{12}\right)}{\mathbb{T}_{1,2}} = \mu_{12}+\frac{\tilde{\hat{\mathbb{T}}}_{1,1}^+\mathbb{T}_{1,1}^-- \hat{\mathbb{T}}_{1,1}^+\tilde{\hat{\mathbb{T}}}_{1,1}^-}{\mathbb{T}_{1,2}}.
\end{align}
However, we also have to be careful with interpreting $\mathbb{T}_{1,1}^+$: the Hirota equations are valid only for long cuts, so we see that shifting $\mathbb{T}_{1,1}$ upwards depends on the cut structure: $\hbT_{1,1}^+$ is identically given by our expression in terms of $\Pf$s with the arguments shifted, but $\mathbb{T}_{1,1}^+$ has to be continued appropriately. We therefore find in terms of short cuts that
\begin{align}
&\hbT_{1,0} = \mu_{12}+\nn 
&\frac{\left( \Pf_1^{[+2]}\Pft_2-\Pft_1\Pf_2^{[+2]} \right)\left(\Pf_1\Pf_2^{[-2]}-\Pf_1^{[-2]}\Pf_2 \right)- \left(\Pf_1^{[+2]}\Pf_2-\Pf_1\Pf_2^{[+2]} \right)\left(\Pft_1\Pf_2^{[-2]}-\Pf_1^{[-2]}\Pft_2 \right)}{\Pf_1^{[+2]}\Pf_2^{[-2]}-\Pf_1^{[-2]}\Pf_2^{[+2]}} \nn
&= \mu_{12}+ \Pf_1\Pft_2-\Pft_1\Pf_2 = \tilde{\mu}_{12}.
\end{align}
This results in 
\be
\label{eq:mufromTBA1}
\frac{Y_+}{Y_-} = \frac{\tilde{\mu}_{12}}{\mu_{12}} = \frac{\mu_{12}^{++}}{\mu_{12}},
\ee
on the sheet with short cuts. 

\paragraph{Sensitivity to conventions.} An important aspect of this derivation is that it is sensitive to where we say long- and short-cutted versions of functions should coincide: slightly above or slightly below the real axis. In the derivation above we chose the upper-half-plane convention that functions agree just above the real axis. Let us see what changes if we take the functions to be equal just \emph{below} the real axis:
\begin{align}
\mathbb{T}_{2,2}^+\mathbb{T}_{2,2}^- &= \mathbb{T}_{1,1}^{[+3]}\mathbb{T}_{1,1}^{[-3]}\tilde{\hat{\mathbb{T}}}_{1,1}^+\hat{\mathbb{T}}_{1,1}^- \nn
\mathbb{T}_{2,1} &= \frac{\mathbb{T}_{2,2}^+\mathbb{T}_{2,2}^--\mathbb{T}_{1,2}\mathbb{T}_{2,3}\mu_{12}}{\mathbb{T}_{2,3}} = 
\tilde{\hat{\mathbb{T}}}_{1,1}^+\mathbb{T}_{1,1}^--\mathbb{T}_{1,2}\mu_{12},
\end{align}
since this time it is $\mathbb{T}_{2,2}^-$ that has to be continuated. Now we find
\begin{align}
\mathbb{T}_{1,1}^+\mathbb{T}_{1,1}^- &= \mathbb{T}_{0,1}\mathbb{T}_{2,1}+\mathbb{T}_{1,0}\mathbb{T}_{1,2} = \mathbb{T}_{2,1}+\mathbb{T}_{1,0}\mathbb{T}_{1,2} \nn
\mathbb{T}_{1,0} &= \frac{\mathbb{T}_{1,1}^+\mathbb{T}_{1,1}^- -  \mathbb{T}_{2,1}}{\mathbb{T}_{1,2}} = \frac{\mathbb{T}_{1,1}^+\tilde{\hat{\mathbb{T}}}_{1,1}^- -  \left(\tilde{\hat{\mathbb{T}}}_{1,1}^+\hat{\mathbb{T}}_{1,1}^--\mathbb{T}_{1,2}\mu_{12}\right)}{\mathbb{T}_{1,2}} \nn
&= \mu_{12}+\frac{\mathbb{T}_{1,1}^+\tilde{\hat{\mathbb{T}}}_{1,1}^-- \tilde{\hat{\mathbb{T}}}_{1,1}^+\mathbb{T}_{1,1}^-}{\mathbb{T}_{1,2}},
\end{align}
where this time it is $\mathbb{T}_{1,1}^-$ that has to be treated with care in the Hirota equation. Note first now that the equation relating $\mu_{12}$ to $\tilde{\mu}_{12}$ gets deduced from the absence of a discontinuity for $\bfT_{2,1}$ on $Z_0$ and this condition gets changed as a result of our choice, taking the form
\be
\mu_{12}-\tilde{\mu}_{12}= \Pf_1\Pft_2-\Pft_1\Pf_2,
\ee
which has a minus sign difference on the right-hand side with respect to the upper half plane version in eqn. \eqref{eq:muconstraint1}. We see that the expression we are after gets simplified in the same way as before:
\be
\mathbb{T}_{1,0}= \mu_{12}- \Pf_1\Pft_2+\Pft_1\Pf_2 = \tilde{\mu}_{12}.
\ee
So we see that ultimately the first equality in eqn. \eqref{eq:mufromTBA1} from which we deduce asymptotics of $\mu_{12}$ is not dependent on our choice.
\\[5mm]
Our choice of where to identify functions \emph{does} influence the analytic properties of $\mu_{12}$ though, thereby changing the second equality: By identifying in the upper half-plane we find on the short-cutted sheet that
\be
\mut_{12} = \mu^{[+2]}_{12},
\ee
whereas identifying in the lower half-plane results in 
\be
\mut_{12} = \mu^{[-2]}_{12},
\ee
as a simple inspection shows. So, although we found that 
\be
\frac{Y_+}{Y_-} = \frac{\tilde{\mu}_{12}}{\mu_{12}}
\ee
is true in both conventions, we see that in the upper-half-plane conventions 
\be
\frac{Y_+}{Y_-} = \frac{\mu_{12}^{++}}{\mu_{12}},
\ee
whereas in the lower-half-plane conventions 
\be
\frac{Y_+}{Y_-} = \frac{\mu_{12}^{--}}{\mu_{12}}. 
\ee
\subsection{Undeformed derivation}
The previous sections have shown that 
\be
\label{eq:mufromTBA}
\frac{\mu_{12}^{\pm \pm}}{\mu_{12}} = \exp \left( \Lambda_P \star K^{Py}\right) = \exp \left( \Lambda_P \star \left( K(v+i\ad  Q,u ) - K(v-i \ad Q,u )\right)\right),
\ee
with the sign depending on the chosen conventions. For now we will choose the lower-half-plane conventions, as the kernel $K$ was defined in that convention. We would like to consider this expression as $u\rightarrow \infty$. Coincidentally $x_s \rightarrow \infty$ as $u\rightarrow \infty$ and the entire $u$-dependence of the right-hand side of \eqref{eq:mufromTBA} is through the $x_s$ function, so we can just as well consider the expansion as $x_s\rightarrow \infty$: we find that
\be
K^{Py}(v,u) = \frac{1}{2\pi i}  \partial_v \left(  -\tilde{E}^Q - \frac{2 \tilde{p}_Q}{g x_s(u)} + \mathcal{O}\left(x_s^{-2}(u)\right) \right),
\ee
with the undeformed mirror momentum and energy defined in eqn. \eqref{eq:mirrormomentum}. From the TBA equations we know that these quantities express the total momentum and energy (see eqn. \eqref{eq:Energy}) when integrated with $\Lambda_P$, which we can use to simplify the convolution $\Lambda_P \star K^{Py}$:
\begin{align}
\label{eq:KPyundeformedexpanded}
\Lambda_P \star K^{Py} (u) &=\sum_P \int_{Z_0} dv \Lambda_P(v) \left(\frac{1}{2\pi}  \partial_v \left(   i \tilde{E}_Q + i \frac{2\tilde{ p}_Q}{g x_s(u)} + \mathcal{O}\left(x_s^{-2}(u)\right) \right) \right)\nn
&= -i P - 2 i \frac{E}{g x_s (u)} +\mathcal{O}\left(x_s^{-2}(u)\right) \nn
&=  -2 i \frac{ E}{g u} +\mathcal{O}\left(u^{-2}\right),
\end{align}
where we used the level-matching condition $P=0$ in the last line. So far so good, but what does this tell us about the asymptotics of $\mu_{12}$? Let us assume that as $u\rightarrow \infty$ $\mu_{12}$ has polynomial growth. To allow us to be precise about this we introduce the symbols $\simeq$ and $\sim$:\footnote{These have a deformed and undeformed meaning as usual, we introduce the undeformed meaning here. The deformed symbols have $u \rightarrow \pm i \infty$ depending on whether we are in the upper-(lower)-half-plane conventions.} we say $f\simeq g$ if and only if $\lim_{u\rightarrow \infty} f(u)/g(u) = 1$ and $f\sim g$ if there exists a constant $C\in \C$ such that $f \simeq C g$. For $\mu_{12}$ we assume that $\mu_{12} \sim u^M$. This implies that 
\be
\log \frac{\mu_{12}^{--}}{\mu_{12}} = -\frac{2iM}{g u}  +\bO{u^{-2}}
\ee
and after comparison with eqn. \eqref{eq:KPyundeformedexpanded} we find that $M = E$. So we find that $\mu_{12} \simeq u^E$. Can we do something similar in the deformed case?

\subsection{Now the deformed case}
The deformed case is more subtle: first off, it is not clear in which direction we should consider asymptotics, since taking the same limit $u\rightarrow \infty$ does not make sense on a cylinder. Really, the only two asymptotic directions available to us are up and down the cylinder, i.e. $u \rightarrow \pm i \infty$. In the lower-half-plane conventions taking the limit $u \rightarrow -i \infty$ makes a bit more sense, as it is there that $x_s \rightarrow \infty$. However, taking the limit parallel to the imaginary axis is tricky: even in the asymptotic regime the limit path is only a finite distance away from the (short) branch cuts of $\mu_{12}$. This might force us to include a prefactor in the asymptotic expression for $\mu_{12}$ to account for the influence of the branch cuts, spoiling the pure asymptotics as we saw them in the undeformed case.\footnote{The author thanks Vladimir Kazakov for insightful discussions concerning this point.} We still present the argument here to give the reader the chance to compare and see exactly where the problem arises. It is noteworthy that we are ultimately interested in the asymptotics of the $\Pf$ and $\Qf$ functions, which have only one cut on the real line. Their asymptotics in the limit $u\rightarrow \pm i \infty$ therefore are well-defined and do not suffer from the problem we rose for $\mu_{12}$. It would be great if we could find the correct asymptotics for the $\Pf$ and $\Qf$ functions without going through $\mu_{12}$. Possibly this can be done by approaching it from a different direction, for example in the framework outlined in \cite{Gromov:2017cja}. We hope to address this question again in the future. 
\\[5mm]
In the deformed case the identification of the $x_s$ function and the mirror energy and momentum are slightly different, see eqn. \eqref{eq:mirrormomentum}, yielding for the expansion of $K^{Py}$
\begin{equation}
K^{Py}(v,u) = \frac{1}{2\pi i}  \partial_v \left(  -i \ad \tilde{p}^Q - \tilde{E}_Q + \mathcal{O}\left(x_s^{-1}(u)\right) \right). 
\end{equation}
Following the same analysis we find
\begin{align}
\label{KPyexpansion}
\Lambda_P \star K^{Py} (u)= -i P + \ad E +\mathcal{O}\left(x_s^{-1}(u)\right) =  \ad E +\mathcal{O}\left(x_s^{-1}(u)\right),
\end{align}
where $E$ and $P$ here are the deformed string energy and momentum respectively and we used again the level-matching condition $P=0$. So far the analysis follows the undeformed one problem free. Our next step is to assume some asymptotic behaviour of $\mu_{12}$, which as mentioned could become problematic in the deformed case: it is to be expected that the finite distance to the branch cuts leads to some kind of contribution to the functional form of $\mu_{12}$ and in particular in its asymptotics. In short, an ansatz for the asymptotics of $\mu_{12}$ would be
\be
\mu_{12} \simeq m(u) e^{i u \alpha},
\ee
with $\simeq$ meaning that 
\begin{equation}
\lim_{u\rightarrow -i \infty}\frac{m(u) e^{i u \alpha}}{ \mu_{12}(u)} = 1,
\end{equation}
where the limit has to be taken avoiding branch cuts, i.e. here $|\text{Re}(u)| > \theta$ with $\alpha \in \C $ having positive real part.\footnote{Assuming $\alpha$ is imaginary or has a negative real part leads to a contradiction.} The function $m$ is expected to go to one when $u$ is taken infinitely far away from the branch cuts. We can now compute
\begin{align}
\log \frac{\mu_{12}^{--}}{\mu_{12}} \simeq 2 \alpha\ad \frac{m^{--}(u)}{m(u)} + \ldots,
\end{align}
which is no longer $u$ independent. One possible way out is to consider the case where $\alpha$ is irrational, such that $\mu_{12}$ is no longer periodic. Assuming $\mu_{12}$ is not periodic we can consider the limit $u\rightarrow -i\infty$ along the path with $|\text{Re}(u)| = |\text{Im}(u)| $, such that asymptotically $m$ goes to unity and we return to the pure asymptotics $\mu_{12} \sim e^{i u \alpha}$, with $f \sim g$ meaning that $f \simeq C g$ for some $C \in \C$. In this case, we can compare the two expansions to obtain $\alpha=E/2$, so that
\begin{equation}
\label{muasymp}
\mu_{12} \simeq m(u) e^{i u E/2 }. 
\end{equation}
We will use this result in the next section as one of the many checks that our proposal for the asymptotic behaviour of the $\Pf$ and $\Qf$ functions is consistent. Formally, it seems like we can only use this result in the special diagonal limit we just introduced, but this does not change much. 

\paragraph{Remark.} Note that this expression is very similar to its undeformed counterpart $\mu_{12} \sim u^{E}$, although the way we obtain this expression is subtly different. In the deformed case, we do not find energy and momentum as separate coefficients in the expansion of $K^{Py}$, but instead they come together already at the lowest order. This can be interpreted as a mixing of the conserved charges of the deformed theory as a result of the deformation. If we continue the expansion \eqref{KPyexpansion}, at second order we find an expression which manifestly gives the energy in the undeformed limit:
\begin{equation}
\frac{1}{x^{[-Q]}} - x^{[-Q]}-\frac{1}{x^{[+Q]}}+x^{[+Q]},
\end{equation}
even though in the deformed model this expression has no immediate physical interpretation.

\subsection{Comments on the periodicity of the deformed $\mu_{12}$}
\label{sec:muperiodicity}
It is important to note the factor of $1/2$ in the exponent in eqn. \eqref{muasymp}, as it implies that for odd integer $E$ the $\mu_{12}$ function is at most anti-periodic instead of periodic. This is exactly a possible situation at lowest order in perturbation theory, where quantum corrections play no role yet. Indeed, at lowest order the branch cuts of $\mu_{12}$ disappear -- and therefore the prefactor $m$ goes to unity -- as we send $\theta \rightarrow 0$. Using the regularity requirement for physical states as in the undeformed case, the only possible locations for poles are at the locations where the branch cuts disappear, at $u=2i \ad N$ for integer $N$, but an analysis along the lines of \cite{Marboe:2014gma} shows that at lowest order there are no poles at these points. Therefore $\mu_{12}$ is analytic at lowest order. Actually, we know that $\mu_{12}^+$ is real analytic, as we showed in eqn. \eqref{mureality}. As $\mu_{12}$ and $\check{\mu}_{12}$ coincide on the strip just below the real axis and $\check{\mu}_{12}$ is defined as the square root of a $2\pi$-periodic function, see eqn. \eqref{defmu}, both are at least $4\pi$-periodic just below the real axis. This allows us to write a Fourier expansion
\be
\mu_{12}(u) = \sum_{k\in \mathbb{Z}} a_k e^{ik u/2}
\ee
valid in the strip and using analyticity we can continue this expression anywhere in the complex plane. Combining this expansion with the known asymptotic behaviour around $u=-i \infty$ and the fact that $\mu_{12}^+$ is real we find\footnote{Using the lemma in appendix \ref{app:constrainingtrigonometricpolynomials}.} that $\mu_{12}$ is a (complex) \emph{trigonometric polynomial}: a \emph{trigonometric polynomial} is a function $f$ of the form
\begin{equation}
\label{eq:deftrigpol}
f(u) = \sum_{k=0}^N \left( a_k \sin( k u/2) + b_k \cos( k u/2) \right).
\end{equation}
If the coefficients are real (complex), $f$ is a real (complex) trigonometric polynomial. The \emph{order} of $f$ is denoted $N\in \N$.
\\[5mm]
The asymptotic behaviour directly enforces the periodicity properties: in the asymptotic region $\mu_{12}$ is $2\pi$-(anti)-periodic depending on the parity of $E$ and since $\mu_{12}$ is a trigonometric polynomial this property must hold everywhere. Interestingly, this means that for odd $E$ $\mu_{12}$ is not continuous on the cylinder. In short, we find that for $\theta=0$
\be
\label{eq:muperiod}
\mu_{12}(u) = (-1)^E \mu_{12}(u+2\pi).
\ee
This property can be interpreted as the periodic version of the property of ordinary odd-degree polynomials, having a different limit depending on whether we take $u \rightarrow \pm \infty$. Note that this property does not survive once we go to finite $\theta$: $\check{\mu}_{12}$ retains its periodicity from the lowest order, as it is manifestly a square root of a $2\pi$-periodic function. $\mu_{12}$, however, will lose its periodicity outside of the physical strip in favour of obeying its asymptotics, which for finite $\theta$ are generically no longer integer. Instead, $E$ becomes a measure of the jump of $\mu_{12}$ over the line Re$(u)=\pi$ very far in the lower half-plane. This is illustrated in fig. \ref{fig:mu}.

\section{Propagating to the rest of the QSC}
\label{sec:asymptotics}
\subsection{Switching to the upper-half-plane conventions}
Now that we have performed the analysis in the previous section we can switch permanently to the upper-half-plane conventions. We do this as before (see section \ref{sec:switchingconventions}), where now we have the additional changes
\begin{align}
\label{eq:uhpconventions2}
\tilde{\hat{\mu}}_{ab} = \hat{\mu}_{ab}^{--} \quad &\Longrightarrow \quad \tilde{\hat{\mu}}_{ab} = \hat{\mu}_{ab}^{++}, \nn
\mu_{12} \simeq m(u) e^{i u \frac{E}{2}} \quad &\Longrightarrow \mu_{12} \simeq m(u)e^{-i u \frac{E}{2}},
\end{align}
where now $\simeq$ means we consider the leading exponential term as $u\rightarrow i \infty$ avoiding branch cuts and the upper-half-plane $m$ does not have to coincide with the lower-half-plane $m$. 

\subsection{Anticipating the result}
Our actual aim in this chapter is to find whether the basic QSC functions $\Pf$ and $\Qf$ depend on the six quantum numbers \eqref{eq:quantumnumbers}: conjecturing that this is the case states that the QSC as we have derived it can be extended to describe any state in the system, provided we pick up the right ``boundary conditions'', in the form of asymptotics of the $\Pf$ and $\Qf$ functions. For the undeformed case these asymptotics were proposed in \cite{Gromov:2014caa} and we will discuss it alongside the deformed case. 
\\[5mm]
We conjecture that a state in the $\eta$-deformed model with given charges can be described in the QSC through the exponential large $iu$ asymptotics, similar to what we saw in the previous section. Writing $z= e^{-i u/2}$ for the deformed case and $z=u$ for the undeformed case we can write the asymptotics as
\be
\Pf_a \simeq A_a z^{-\tilde{M}_a}, \quad \Qf_i \simeq B_i z^{\hat{M}_i}\quad \Pf^a \simeq A^a z^{\tilde{M}_a}, \quad \Qf^i \simeq B^i z^{-\hat{M}_i},
\ee
where the deformed $\simeq$ was defined in the previous section. Note that the limit for both cases takes the form $z\rightarrow \infty$. The undeformed powers are given by
\begin{align}
\label{eq:undeformedcharges}
\tilde{M}^{\text{und}} &= \frac{1}{2}\left\{ J_1+J_2-J_3+2,J_1-J_2+J_3,-J_1+J_2+J_3+2,-J_1-J_2-J_3\right\}, \nn
\hat{M}^{\text{und}} &= \frac{1}{2}\left\{ \Delta-S_1 -S_2 +2 ,\Delta + S_1 +S_2,-\Delta-S_1 +S_2 +2 ,-\Delta+S_1 -S_2
\right\}.
\end{align}
whereas the deformed ones are 
\begin{align}
\label{charges}
\tilde{M} &= \frac{1}{2}\left\{ J_1+J_2-J_3+2,J_1-J_2+J_3,-J_1+J_2+J_3,-J_1-J_2-J_3-2\right\}, \nn
\hat{M} &= \frac{1}{2}\left\{ \Delta-S_1 -S_2 +2 ,\Delta + S_1 +S_2,-\Delta-S_1 +S_2 ,-\Delta+S_1 -S_2-2
\right\},
\end{align}
which follows the distribution of global charges of the classical spectral curve of the $\ads$ superstring. Note that the constant shifts in the powers of the asymptotics of $\Qf_i$ and $\Pf^a$ differ in both cases. 

\paragraph{Remark.} While our present choice of $z$ for the deformed case is the simplest function with asymptotic behaviour $e^{-i u/2}$, it is of course not unique. Moreover, it is not real, while this would be desirable given the reality of the $\Pf$ and $\Qf$ functions. It is in fact natural to consider $z=\sin u/2$. Although the undeformed limit of our asymptotics is ambiguous as it lives near the origin, we will see that formally taking the undeformed limit of $z=\sin u/2 \rightarrow \ad u/2 +\mathcal{O}\left(\ad^3\right)$ gives the correct undeformed power law asymptotics.

\subsection{Deriving the asymptotics}
\label{sec:derivingasymptotics}
We can deduce the above asymptotics from consistency of the QSC and the asymptotic behaviour we found for $\mu_{12}$. For the deformed case we can also use the fact that its asymptotics should limit to the undeformed asymptotics, whereas for the undeformed case we can use the knowledge of the classical spectral curve. It is most convenient to discuss these derivations separately. 

\paragraph{Undeformed asymptotics.} Starting from the fact that $\mu_{12} \sim u^E$ as $u\rightarrow \infty$\footnote{We use $\sim$ instead of $\simeq$ to allow for an normalisation constant.} we postulate that all the $\Pf$ and $\Qf$ functions have interesting asymptotics in this limit:\footnote{We refrain from putting an extra label ``und" on the quantities in this section, as this clutters the analysis immensely.}
\be
\Pf_a \simeq A_a u^{-\tilde{M}_a}, \quad \Qf_i \simeq B_i u^{\hat{M}_i-1}\quad \Pf^a \simeq A^a u^{\tilde{M}^a-1}, \quad \Qf^i \simeq B^i u^{-\hat{M}^i},
\ee
where all the constants $A,B, \tilde{M},\hat{M}$ are still completely unrelated. The shifts by $-1$ could be viewed as purely a convenience, but we will see that they constitute a prime difference between the undeformed and the deformed cases.
\\[5mm]
We can use the equation \eqref{QPF1} to find the asymptotics of $\mathcal{Q}_{a|i}$: writing its asymptotics as
\be
\mathcal{Q}_{a|i} \simeq C_{a|i}u^{d_{a|i}}
\ee
we find from the asymptotic analysis, replacing the shift $\pm i/g$ by $1+\tfrac{i}{g} \partial_u$ as we let $u\rightarrow \infty$, of \eqref{QPF1} that 
\begin{equation}
C_{a|i} = -\frac{ig}{2} \frac{ A_a B_i}{d_{a|i}}, \quad d_{a|i} = -\tilde{M}_a + \hat{M}_i. 
\end{equation}
Doing a direct power counting on the asymptotic regime of equation \eqref{defQ} shows that $\tilde{M}^a = \tilde{M}_a$, whereas the coefficients yield the equation
\begin{equation}
\frac{ig}{2} \frac{A_a A^a}{d_{a|j}} =-1 \qquad \text{ for every } j.
\end{equation}
This equation can be solved for the products $A_{a_0}A^{a_0}$ where no summation is implied and $a_0 = 1, \ldots, 4$. The solution reads
\be
\label{eq:undeformedA0}
A_{a_0}A^{a_0} = -\frac{ig}{2}\frac{\prod_j \left( \tilde{M}_{a_0} - \hat{M}_{j} \right)}{\prod_{b\neq a_0} \left( \tilde{M}_{a_0} - \tilde{M}_{b} \right)}, \quad \text{ for } a_0 =1, \ldots, 4.
\ee
Performing a similar analysis on 
\be
\Pf_a = -\Qf^i \Qm_{a|i}^{\pm}
\ee
we find that $\hat{M}^i = \hat{M}_i$ and a similar expression to eqn. \eqref{eq:undeformedA0} for the product $B_{j_0}B^{j_0}$. 
\\[5mm]
We can also find asymptotics for $\omega$ and $\mu$: Since $\omega^{++} = \omega$ we find that $\omega\simeq 1$, as this is the only $2i/g$-periodic function with power-like asymptotics . Choosing the right $\Qf_i$ basis we can choose it to be such that it is anti-symmetric with $\omega_{12}=1=\omega_{34}$ and all other elements vanishing. To derive $\mu$ asymptotics we analyse eqn. \eqref{defomega}. Using the definition of $\omega^{ij}$ and $\Qm^{a|i} = -\Qm_{a|i}^{-t}$ we can invert these relations to obtain
\be
\mu^{ab} = (\Qm^{a|i})^-(\Qm^{b|j})^-\omega_{ij},
\ee
which we can use directly to find the asymptotics of $\mu$ in terms of $\tilde{M}_a$ and $\hat{M}_i$. This defines the asymptotics of all the important functions in the yet unknown powers $\tilde{M}$ and $\hat{M}$. Before we continue, let us first see what happens in the deformed case. 

\paragraph{Deformed asymptotics.} This time we postulate that all the $\Pf_a$ and $\Qf_i$ have interesting asymptotics as we send $u\rightarrow i \infty $ outside the branch cut strip:
\be
\Pf_a \simeq A_a z^{-\tilde{M}_a}, \quad \Qf_i \simeq B_i z^{\hat{M}_i}\quad \Pf^a \simeq A^a z^{\tilde{M}^a}, \quad \Qf^i \simeq B^i z^{-\hat{M}^i},
\ee
where we refrain from interpreting the powers for now and do not impose a relation between the four sets of powers. Note that the shifts of $-1$ are absent. Now, again using eqn. \eqref{QPF1}
\be
\label{QPQ2}
\mathcal{Q}_{a|i}^+-\mathcal{Q}_{a|i}^- = \Pf_a \Qf_i
\ee
to deduce asymptotics for $\Qm_{a|i}$, and writing the asymptotics as
\be
\mathcal{Q}_{a|i} \simeq C_{a|i}z^{d_{a|i}}
\ee
we find that
\be
C_{a|i}\left(e^{\ad d_{a|i}/2}- e^{- \ad d_{a|i}/2} \right)z^{d_{a|i}} \simeq A_a z^{-\tilde{M}_a}B_i z^{\hat{M}_i},
\ee
which when comparing powers leads to $d_{a|i} = -\tilde{M}_a + \hat{M}_i$. It is in this analysis that the shift of $-1$ is no longer necessary: in the undeformed case the shift compensated the action of the derivative, but its $\eta$-deformed analogue does not require this compensation. We also find
\be
C_{a|i} = \frac{A_a B_i }{2\sinh \ad \frac{d_{a|i}}{2}}.
\ee
Performing the same analysis on eqn. \eqref{defQ} we find by comparing the powers that $\tilde{M}^a = \tilde{M}_a$. Comparing coefficients in eqn. \eqref{defQ} we find
\be
\label{eq:Acoefficients}
\frac{A_a A^ae^{ \ad \frac{d_{a|i}}{2}} }{2\sinh \left(\ad \frac{d_{a|i}}{2}\right)} =-1 \qquad \text{ for every } i.
\ee
Note that due to eqn. \eqref{QPQ2} we could have written eqn. \eqref{defQ} with a minus shift and comparing the coefficients of the asymptotics of that equation leads to a slightly different formula:
\be
\label{eq:Acoefficientsminus}
\frac{A_a A^ae^{-\ad \frac{d_{a|i}}{2}} }{2\sinh \left(\ad \frac{d_{a|i}}{2}\right)} =-1 \qquad \text{ for every } i.
\ee
The solutions of these equations coincide only when
\be
\label{eq:sumconstraint}
\sum_{a}\tilde{M}_a = \sum_{i}\hat{M}_i.
\ee
This is an interesting consequence of the deformation: in the undeformed case the effect of the shift is not leading and therefore unseen in this analysis, requiring us to find a different way to argue that eqn. \eqref{eq:sumconstraint} should hold (which in fact it does). Solving these equations again we find
\be
\label{eq:A0A0}
A_{a_0}A^{a_0} =2\frac{\prod_{j}\sinh\left( \ad \frac{\tilde{M}_{a_0}- \hat{M}_{j}}{2}\right)}{\prod_{b\neq a_0} \sinh\left( \ad \frac{ \tilde{M}_{a_0}- \tilde{M}_{b}}{2}\right)} \qquad \text{ for } a_0=1,\cdots,4.
\ee
Performing a similar analysis on
\be
\label{derB}
\Pf_a = -\Qf^i \Qm_{a|i}^{\pm},
\ee
we find that $\hat{M}^i = \hat{M}_i$ and that
\be
\label{eq:B0B0}
B^{j_0} B_{j_0}  =-2\frac{\prod_{a}\sinh\left(\ad \frac{\hat{M}_{j_0}- \tilde{M}_{a}}{2}\right)}{\prod_{j\neq j_0} \sinh\left(\ad  \frac{ \hat{M}_{j_0}- \hat{M}_{j}}{2}\right)}\qquad\text{ for } j_0=1,\cdots,4.
\ee
The rest of the analysis, finding the asymptotics of $\omega$ and $\mu$ is completely analogous to what we described in the previous section. 

\subsection{Dependence of $M$ on the global charges.} Our remaining task now is to find the dependency of the $M$'s on the global charges. In order to deduce this we should consider the following set of constraints on the asymptotics:
\begin{itemize}
\item The $\mu_{12}$ asymptotics have a specified form as follows from the TBA.
\item The sum of $\tilde{M}_a$ and the $\hat{M}_i$ should be equal, see eqn. \eqref{eq:sumconstraint}.\footnote{That this is true in the undeformed case follows from the comparison with the classical spectral curve \cite{Gromov:2014caa}.}
\item We should find the appropriate Bethe equations at weak coupling, which we analyse for the deformed case in section \ref{sec:weakcoupling} for the $\mathfrak{sl}_2$ sector.
\item In the left-right symmetric sector where $J_2=J_3=S_2=0$,\footnote{This is the sector where the left and right $Y$ functions and associated $T$ functions are (assumed to be) equal, not to be confused with the left and right \emph{bands} of the $T$ hook. In this sector the non-zero charges are $J$ and $S_1 =S$.} we get constraints on the powers. Namely, $ \Pf^4 = \Pf_1$ and $\Pf^3=\Pf_2$ implies that in the deformed case
\begin{align}
\tilde{M}_1=-\tilde{M}_4,\quad \tilde{M}_2=-\tilde{M}_3,\quad \hat{M}_1=-\hat{M}_4,\quad \hat{M}_2=-\hat{M}_3,
\end{align}
whereas in the undeformed case
\begin{align}
\tilde{M}_1^{\text{und}} &=-\tilde{M}_4^{\text{und}}+1,\quad \tilde{M}_2^{\text{und}}=-\tilde{M}_3^{\text{und}}+1, \nn
\hat{M}_1^{\text{und}}&=-\hat{M}_4^{\text{und}}+1,\quad \hat{M}_2^{\text{und}}=-\hat{M}_3^{\text{und}}+1,
\end{align}
thus constraining their dependence on the other charges and on constants.
\item We can use $H$ symmetry of the $QQ$ system to make sure that all the $\Pf$ and $\Qf$ have different asymptotics and we can order them such that asymptotically
\be
|\Pf_1|<|\Pf_2|<|\Pf_3|<|\Pf_4| \text{ and } |\Qf_2|> |\Qf_1| > |\Qf_4|>|\Qf_3|,
\ee
implying bounds on the asymptotics.
\item For the deformed case we know that in the undeformed limit we should obtain the asymptotics from the undeformed case.
\end{itemize}
The quickest way by far to obtain the complete asymptotics for the undeformed case is to note that the functions $\Pf_a$ and $\Qf_i$ should be related to the known classical spectral curve of the $\ads$ superstring computed in \cite{Beisert:2005bm}. Comparing these quantities directly yields the equations \eqref{eq:undeformedcharges}. 

\subsubsection{Deriving the deformed charges.}\label{sec:obtainingdeformedcharges} Obtaining the deformed charges is less straightforward, simply because the classical spectral curve is not known for the $\eta$-deformed model. We therefore proceed by imposing the constraints listed in the previous paragraph and additionally impose that the dependence on the charges is linear and that the $\Pf$ and the $\Qf$ asymptotics, depend only on the $J_i$, and only on $\Delta$ and the $S_i$, respectively. We can then use the $\mu_{12}$ asymptotics to find the dependence on $J_1$ and $\Delta$ (since $E=\Delta-J$) and  constrain the dependence on $S_1$. From the ordering of $\Qf$ functions and the sum constraint we then find that also the $S$ dependence is exactly as in the undeformed case.\footnote{We follow the undeformed case and use the ABA diagram, see appendix C in \cite{Gromov:2014caa}.} At this point it seems clear that the only way to ensure consistency with the undeformed asymptotics is to let the dependence on the other charges be the same as in the undeformed case. This leaves only a freedom to add constants, and at this point the asymptotics can be written using four independent constants $\alpha,\beta,\gamma,\delta$:
\begin{align}
\tilde{M} &= \frac{1}{2}\left\{ J_1+J_2-J_3+2\alpha,J_1-J_2+J_3-2\beta,-J_1+J_2+J_3+2\beta,-J_1-J_2-J_3-2\alpha\right\}, \nn
\hat{M} &= \frac{1}{2}\left\{ \Delta-S_1 -S_2 +2\gamma ,\Delta + S_1 +S_2+2\delta,-\Delta-S_1 +S_2 -2\delta ,-\Delta+S_1 -S_2-2\gamma
\right\}.
\end{align}
The fact that at weak coupling we should find the $\mathfrak{sl}_2$ Bethe equations forces that $\beta=0$, as follows from the derivation in the next section. A final constraint from the $\mu_{12}$ asymptotics is that $\alpha=\gamma+\delta$.
\\[5mm]
We can fix the constants by comparison with the undeformed case. In the undeformed limit our asymptotics for the $\Pf$s and $\Qf$s are proportional to some power of $\ad$ and therefore diverge or vanish. More precisely, for some $N_i>0$
\begin{align}
z^{-\tilde{M}_{1,2}} \rightarrow \bO{c^{-N_{1,2}}},\quad z^{\tilde{M}_{3,4}} \rightarrow \bO{c^{N_{3,4}}}.
\end{align}
In the left-right symmetric sector, where $A^1=A_4$, we find that the product of the coefficients $A_1A_4$ goes as $\bO{c}$, implying at least one of the coefficients vanishes in the undeformed limit. From these constraints we see that, if the undeformed limit at this level is regular, only the $\Pf_{1,2}$ asymptotics can have a finite undeformed limit, whereas the leading $\Pf_{3,4}$ asymptotics necessarily vanish. Hence the subleading term must become leading in the undeformed limit.\footnote{Since the $\Pf_{3,4}$ are auxiliary variables anyway, their asymptotics are fixed by consistency of the QSC, hence uniquely in terms of the well-defined undeformed limit of the $\Pf_{1,2}$.} From this reasoning we find the following comparison of the undeformed and deformed asymptotics:
\begin{align}
\tilde{M}_{1,2} = \tilde{M}^{\text{und}}_{1,2},\quad \tilde{M}_{3,4}  = \tilde{M}^{\text{und}}_{3,4}- 1.
\end{align}
This comparison holds precisely when $\alpha= 1$, and $\beta=0$ as found independently above. Performing a similar analysis for the $\Qf_i$ (where the roles of $\Qf_{1,2}$ and $\Qf_{3,4}$ are reversed) we find that
\begin{align}
\hat{M}_{1,2}-1 = \hat{M}^{\text{und}}_{1,2}-1,\quad \hat{M}_{3,4}  = \hat{M}^{\text{und}}_{3,4}-1,
\end{align}
which holds precisely when $\gamma = 1$ and $\delta=0$. This completely fixes the asymptotics to be as in eqs. \eqref{charges}. Note that in this interpretation the charges do not change while taking the undeformed limit, only the auxiliary $\tilde{M}$ and $\hat{M}$ jump.
\section{Putting together the results}
For the readers convenience let us summarise all the results. We went through the complete derivation of the quantum spectral curve for the $\ads$ superstring and its $\eta$ deformation. The final form of the $\Pf \mu$- and the $\Qf \omega$-system is the same for both models and are
\be
\mut_{ab}-\mu_{ab} = \Pf_a \Pft_b-\Pf_b \Pft_a, \quad \Pft_a = \mu_{ab} \Pf^b,\quad \Pf_a\Pf^a = 0,\quad  \text{Pf}(\mu) = 1
\ee
and 
\be
\omegat_{ij}-\omega_{ij} = \Qf_i \Qft_j-\Qf_j \Qft_i, \quad \Qft_i = \omega_{ij} \Qf^j,\quad \Qf_i\Qf^i = 0,\quad  \text{Pf}(\omega) = 1.
\ee
The $\Pf$ functions have one $\hat{Z}_0$ cut (see fig. \ref{fig:P}), the $\mu$ functions have a ladder of short cuts (see fig. \ref{fig:mu}). The $\Qf$ functions on the other hand have one $\check{Z}_0$ cut and the associated $\omega$ have a ladder of long cuts. The deformed $\Pf$ and $\Qf$ functions are defined on a (cover of a) cylinder with radius one, the situation for the deformed $\mu_{ab}$ is more complicated: at zero coupling $\theta$ $\mu_{ab}$ is either periodic or anti-periodic, but for finite coupling this property is generically lost, depending on whether we choose long or short cuts. When viewed as a function with short cuts, however, $\hat{\mu}_{ab}$ generically are only (anti-)periodic on the upper half of the physical strip, where they coincide with the long-cutted $\check{\mu}_{ab}$. Outside of this region the $\mu_{ab}$ lose their periodicity properties at non-zero coupling to reflect the fact that the deformed string energy $E$ is not integer. This can be interpreted as the trigonometric version of the phenomenon in the undeformed case that $\mu_{ab}$ develops a branch point at infinity for non-zero coupling.
\\[5mm]
Although we started from the ground-state TBA-equations we ended up with a proposed set of boundary conditions for the $\Pf$ and $\Qf$ functions that can potentially be used for any state: with $z=u$ the undeformed asymptotics are 
\be
\Pf_a \simeq A_a z^{-\tilde{M}_a}, \quad \Qf_i \simeq B_i z^{\hat{M}_i}\quad \Pf^a \simeq A^a z^{\tilde{M}_a}, \quad \Qf^i \simeq B^i z^{-\hat{M}_i},
\ee
whereas the deformed asymptotics, with $z= e^{-i u/2}$, are 
\be
\Pf_a \simeq A_a z^{-\tilde{M}_a}, \quad \Qf_i \simeq B_i z^{\hat{M}_i}\quad \Pf^a \simeq A^a z^{\tilde{M}_a}, \quad \Qf^i \simeq B^i z^{-\hat{M}_i},
\ee
where $\simeq$ is defined as\footnote{We use the upper-half-plane conventions here.}
\begin{align}
\text{undeformed case:}&\quad  f \simeq g \, \text{ if and only if } \, \lim_{u\rightarrow \infty} f(u)/g(u) = 1, \nn
\text{deformed case:}& \quad f \simeq g \, \text{ if and only if } \, \lim_{u\rightarrow i \infty} f(u)/g(u) = 1.
\end{align}
The undeformed powers are given by
\begin{align}
\tilde{M}^{\text{und}} &= \frac{1}{2}\left\{ J_1+J_2-J_3+2,J_1-J_2+J_3,-J_1+J_2+J_3+2,-J_1-J_2-J_3\right\}, \nn
\hat{M}^{\text{und}} &= \frac{1}{2}\left\{ \Delta-S_1 -S_2 +2 ,\Delta + S_1 +S_2,-\Delta-S_1 +S_2 +2 ,-\Delta+S_1 -S_2
\right\}.
\end{align}
whereas the deformed ones are 
\begin{align}
\tilde{M} =\tilde{M}^{\text{und}} - \{0,0,1,1\}, \quad \hat{M} = \hat{M}^{\text{und}} - \{0,0,1,1\}.
\end{align}
When supplemented with the regularity condition that none of the QSC functions should have any poles these boundary conditions allow for a complete solution of the undeformed case either perturbatively at weak coupling \cite{Marboe:2014gma,Marboe:2017dmb} or numerically at finite coupling \cite{Gromov:2015wca}. Further developments in the $\eta$-deformed case suggest the same holds for this case, but it is too early to make definite statements here. Lastly, due to the presence of trigonometric functions instead of rational functions as in the undeformed case we can rightfully call the $\eta$-deformed QSC to be a \emph{trigonometric} QSC. 

\section{Checking using weak coupling}
\label{sec:weakcoupling}
We can check the derived quantum spectral curve for the $\eta$-deformed model and asymptotics by comparing to the Bethe-Yang equations in \eqref{eq:BetheYang}. We will zoom in on the $\mathfrak{sl}_2$ sector of the theory, similar to the analysis done for the undeformed case in \cite{Marboe:2014gma}. In this sector only three of the global charges ($J_1 = J$, $S_1=S$ and $\Delta = J+S +\mathcal{O}\left(\theta^2\right)$) are non-zero. The Bethe-Yang equations in this sector are
\be
\label{eq:BY2}
\left(\frac{1}{q} \frac{x_j^++\xi}{x_j^-+\xi}\right)^J = \prod_{k\neq j}^{S}\left(\frac{x_j^+-x_k^-}{x_j^--x_k^+} \right)^{-1}  \frac{1-\frac{1}{x_j^+ x_k^-}}{1-\frac{1}{x_j^- x_k^+}}
\ee
and, using 
\be
S(u)=\sin\frac{u}{2}
\ee
as defined in eqn. \eqref{eq:Sfunction} and $S_j \defeq S\left(u_j\right)$ we can perform the $\theta \rightarrow 0$ limit to reduce eqn. \eqref{eq:BY2} to
\begin{equation}
\label{XXZBethe}
\left( \frac{S^+_j}{S^-_j}\right)^J= \prod_{k\neq j}^{S}\frac{S^{--}(u_j-u_k)}{S^{++}(u_j-u_k)},
\end{equation}
which are the $\mathfrak{sl}_2$-\textsc{xxz} Bethe equations. As in the undeformed case we expect to find these equations also from the QSC at zero coupling, by associating the roots $u_j$ with zeroes of $\mu_{12}^+$. These roots can then be associated to the exact Bethe roots in the TBA description as zeroes of $Y_{1,0}+1$, since zeroes of $\mu_{12}^+$ imply zeroes of $Y_{1,0}+1$.\footnote{See the discussion in section 4.5 of \cite{Gromov:2014caa} or \cite{Gromov:2009zb}.} In particular we will see how this puts restrictions on the asymptotics defined in the previous section.
\\[5mm]
We start by expanding the $\Pf\mu$ system at lowest order: using $H$ symmetry we can set $A_1=\mathcal{O}\left( \theta^2\right)$, such that $\Pf_1$ vanishes at lowest order. This splits the $\Pf\mu$ system into two parts as in the undeformed case \cite{Marboe:2014gma} and using some algebra we obtain the following $TQ$-like equation for $Q=\mu_{12}^+$:
\be
-T Q +\frac{1}{\left(\Pf_2^-\right)^2}Q^{--}+\frac{1}{\left(\Pf_2^+\right)^2}Q^{++}=0,
\ee
where $T$ is given by the following rational function of $\Pf_a$s:
\be
\label{eq:defT}
T = \frac{\Pf_3^-}{\Pf_2^-}-\frac{\Pf_3^+}{\Pf_2^+}+\frac{1}{\left(\Pf_2^-\right)^2}+\frac{1}{\left(\Pf_2^+\right)^2}.
\ee
As long as we can make sure that $T$ is pole free, we can use the usual philosophy to obtain an equation on the zeroes of the function $Q$: at each zero $u_k$ of $Q$ we must have
\be
\label{wcbe1}
\left(\frac{\Pf_2^+}{\Pf_2^-}\right)^2(u_k) = - \frac{Q^{--}}{Q^{++}}(u_k).
\ee
To further analyse this, we first focus on $\mu_{12}$: as argued in section \ref{sec:muperiodicity}, $\mu_{12}$ is a $4\pi$-periodic analytic function at lowest order. 
\\[5mm]
Taking into account the asymptotics from eqs. \eqref{eq:uhpconventions2} (which in this sector read $\mu_{12}\simeq z^S$ after setting $m=1$) and the reality condition \eqref{mureality} we restrict the Fourier series on $\mu_{12}^+$ using the lemma proven in appendix \ref{app:constrainingtrigonometricpolynomials}, showing it is a real trigonometric polynomial, i.e. of the form
\be
Q(u) = \sum_{k=1}^S \left( a_k \sin(ku/2) + b_k \cos(k u/2)\right),
\ee
with real coefficients. Restricted to the complex strip with real part $[0,4\pi[$, $Q$ has $2S$ zeroes. Using Louiville's theorem we can prove easily that
\be
Q(u) \sim \prod_{k=1}^{2S} S \left( \frac{u-u_k}{2} \right),
\ee
where the $u_k$ are the zeroes of $Q$. The periodicity of $Q$, cf. eqn. \eqref{eq:muperiod},
\be
\label{req1}
Q(u+2\pi) = (-1)^S Q(u),
\ee
relates the roots in a simple way, as it is equivalent to the following statement about the zeroes:
\be
\{u_k\}_{1\leq k \leq 2S} = \{u_k+2\pi \mod 4\pi \}_{1\leq k \leq 2S}.
\ee
This allows us to rewrite
\be
Q(u)  \sim \prod_{k=1}^{S} S(u-u_k),
\ee
giving us the right-hand side of eqn. \eqref{wcbe1}. We can also analyse $\Pf_2$: at lowest order its branch cut vanishes and leaves a possible pole at zero. After factoring out this pole by multiplying with an appropriate power of the factor $S(u)$, $\Pf_2$ is a real analytic function with a convergent Fourier series. Using its asymptotics and reality we can use the lemma in appendix \ref{app:constrainingtrigonometricpolynomials} to restrict the series to be a trigonometric polynomial. In fact, assuming as in the undeformed case that the pole at the origin has order $J/2$, the trigonometric polynomial trivialises to a constant, leaving us with
\be
\Pf_2(u) \sim S(u)^{-J/2}.
\ee
This implies that $T$ of eqn. \eqref{eq:defT} is indeed pole free, and if we combine this result with the above we find indeed that eqn. \eqref{wcbe1} produces the $\mathfrak{sl}_2$-\textsc{xxz} Bethe equations \eqref{XXZBethe}. This shows that the asymptotics we proposed in the previous section are consistent with the Bethe-Yang equations.

\section{Intermezzo: why was the road this long?} A very tempting question to ask is why it was not possible to use the undeformed QSC more directly to find its deformed analogue: already on the level of the TBA equations it is clear that the analytic structure of the two cases are very similar: both have square root branch cuts, which come in two types and both deal with functions which are otherwise well-behaved: there are few -- if any -- singularities. So why was there no shortcut? 
\\[5mm]
As a first naive guess one would use that, since the exponential map is entire, it is possible to map the complex plane of the undeformed case to the cylinder without destroying any analytic properties of the functions involved. This map would change the location of the branch cut from the real line to the unit circle, but that might not be the biggest issue: more problematic is that it is clear that such a map would not result in the right asymptotics, meaning asymptotics that would ultimately yield the $\mathfrak{sl}_2$-\textsc{xxz} Bethe equations. In fact the match with these equations provided us with a very helpful guideline in finding the correct description for the deformed asymptotics.
\\[5mm]
A second option would be to consider a complete uniformisation of the QSC by considering everything on the $x$ plane instead of the $u$ plane. This makes sense for the one-cutted $\Pf$ and $\Qf$ functions, which indeed can be expanded in a convergent $x$-series at least in a neighbourhood of the real line (see appendix \ref{app:constrainingtrigonometricpolynomials}). Whether such series are sufficient for $\mu$ is not clear though: remember that in the deformed case the behaviour of $\mu_{12}$ is very particular and for irrational energies seemingly can not be put on a finite cover of the cylinder. Since the deformed $x$-function is $2\pi$-periodic it seems unlikely that it is possible to describe the QSC in terms of $x$ only. 
\\[5mm]
Finally a more hand-wavy argument comes from the observation that the $\eta$-deformed QSC is to the undeformed QSC as the \textsc{xxz} spin chain is to the \textsc{xxx} spin chain: indeed, their weak coupling limits yield \textsc{xxz} and \textsc{xxx} Bethe equations respectively and just like the spin chains the $\eta$-deformed string theory carries a $q$-deformed version of the undeformed symmetry algebra. Now, since we know that even these simpler models exhibit fundamentally different behaviour\footnote{For example, the excitation spectrum of the \textsc{xxz} is only similar to the \text{xxx} spectrum for one regime of the deformation parameter.}, it seems unlikely that their big string-theory brothers would in some sense be the same. 
\\[5mm]
Lastly, let us mention that based on the similar representation theory one could have argued that the QSC equations should be the same in the two cases, although we lack the knowledge to perform this argument precisely (and are not sure whether the required theory has already been developed rigorously). Then, based on the previous paragraph, the asymptotics could be bootstrapped from consistency with the weak coupling expansion and the undeformed limit. This procedure is quite subtle though, as we have seen in the previous section and is therefore difficult to perform without a firmer understanding of the underlying functions. Moreover, it would not be clear that the resulting QSC actually describes the spectrum of the $\eta$-deformed model, as there would be no direct connection between the quantum numbers in the $S$-matrix description and the QSC. 

\section{Implications from mirror symmetry}
\label{sec:mirrordualityandlimit}
As we already explored in chapter \ref{chap:etadeformed}, the family of $\eta$-deformed strings is actually closed under the associated double Wick rotation. Namely, taking $\theta \rightarrow \pi - \theta$ at fixed $\ad$ is equivalent to a double Wick rotation. To explicitly match the parametrisation this should be combined with a shift $u \rightarrow u + \pi$. To match the labelling of states and charges as in e.g. \cite{Arutyunov:2007tc}, we should moreover interchange the charges $J_{2,3} \leftrightarrow S_{1,2}$ and a re-identify the string circumference $J=J_1$, in terms of the mirror length $R=\ad J$ \cite{Arutynov:2014ota}. Note that the shift of $\theta$ and $u$ interchanges the $x$ functions $x_s$ and $x_m$ of eqs. (\ref{eq:xsfunction} \ref{eq:xmfunction}) as it should.\footnote{To be precise, $x_s$ becomes $-x_m$ and vice versa. This relative sign is inconsequential and can be avoided by considering $\theta \rightarrow \theta + \pi$ instead \cite{Arutynov:2014ota} -- the model is invariant under $\theta \rightarrow - \theta$ -- but then the inequivalent models would be parametrised by $\theta \in [-\pi/2,0] \cup [\pi/2,\pi]$.} As such, the $\eta$-deformed quantum spectral curve at $\theta=\theta_0$ not only describes the spectrum of the these models at $\theta_0$, but also their thermodynamics at $\pi-\theta_0$. Interestingly, this shift of theta actually exchanges the (analytic properties of) the $\Pf\mu$ and $\Qf\omega$ systems. Of course, to consider strict thermodynamics, in this second picture one does not want to add ``string excitations'' to the mirror model, meaning we would consider all charges except the energy $\Delta$ and the mirror length $R$ to be zero in the QSC asymptotics. We should also take into account that we started by computing Witten's index -- $\mbox{Tr}\left((-1)^F e^{-\beta H}\right)$ -- in the mirror theory, and to undo this and get back to the standard free energy, we should add $i\pi$ chemical potentials for the $y$ particles of the TBA. In the undeformed case it is well known that such chemical potentials introduce particular exponentially decay in the QSC asymptotics. It is an interesting question to understand the appropriate generalisation of these asymptotics in the deformed context. Of course, once this is understood it should be simple to add general chemical potentials to the partition function.
\\[5mm]
In this context it is particularly interesting to consider the ``undeformed'' mirror limit $\theta\rightarrow\pi$,\footnote{At the level of the sigma model this is in some sense a maximal deformation limit, as discussed in section \ref{sec:maximaldeformationlimit}. Algebraically speaking, it is a contraction limit \cite{Pachol:2015mfa}.} because there we describe the spectral problem of the mirror model which is the light-cone gauge-fixed version of a string sigma model itself \cite{Arutyunov:2014cra,Arutyunov:2014jfa}, as well as the thermodynamics of the undeformed $\ads$ string which have recently been explicitly related to the Hagedorn temperature \cite{Harmark:2017yrv}. To concretely take this limit, we shift $u\rightarrow u + \pi$, rescale $u \rightarrow 2 \ad u$, and consider the limit $\ad \rightarrow 0^+$ with $\theta = 2 \arccos(2 g \sinh \ad$), see section \ref{sec:spectrumandthermodynamics}. In this limit, the cut structure of the QSC is as expected: the $\Pf$ and $\mu$ functions now have \emph{long} cuts with branch points $\pm 2g$,\footnote{These are the branch points in the conventions of \cite{Gromov:2014caa}; we have tacitly continued our discussion in the TBA conventions, where the branch points are located at $\pm 2$.} while the $\Qf$ and $\omega$ functions have short cuts. Beyond this, the discussion of asymptotics immediately follows the one for the regular undeformed limit as discussed in section \ref{sec:obtainingdeformedcharges}, up to the above mentioned interchange of charges.\footnote{Formally, we are still considering asymptotics around $u \rightarrow i \infty$, not crossing cuts. In the undeformed (mirror) limit we can move around at infinity and consider $u\rightarrow \infty$ instead, since the $\Pf$ and $\Qf$ functions have no obstructing cuts. More concretely, the analysis of for instance the $\mu$ asymptotics from the undeformed string and mirror TBA is unaffected by choice between $i \infty$ and $\infty$.} In other words, up to some state relabelling, the QSC for the spectrum of the mirror model is obtained by simply flipping the branch cut structure. This is formally equivalent to exchanging the undeformed $\Pf\mu$ and $\Qf\omega$ systems. These same QSC equations can also be used to efficiently compute the Hagedorn temperature in the setup of \cite{Harmark:2017yrv}. The only required modification is in the prescribed asymptotics, which should be by exponential decay of the form $e^{-\pi u}$ to account for the above-mentioned difference between Witten's index and the regular free energy, and a possible shift of the energy charge due to the finite difference between the $J$ charge and the classical scaling dimension which are taken to infinity in the spectral problem and Hagedorn temperature problem respectively. This question is under investigation by the authors of \cite{Harmark:2017yrv}.

%% file: conclusions.tex
In this thesis we have derived the quantum spectral curve for the $\eta$-deformed superstring in a language that allowed us to simultaneously review the construction of the QSC for the $\ads$ superstring. The resulting description of the spectral problem is arguably much simpler than the starting point, the TBA equations. As discussed in detail in chapter \ref{chap:etadeformed}, the $\eta$-deformed model is not directly a string itself, and strictly speaking the deformed QSC equations describe the spectrum of the classically light-cone gauge fixed model, a natural one-parameter generalisation of the gauge fixed $\ads$ superstring. Although the deformation does not affect the form of the QSC equations, it shows up in the underlying analytic structure. As such, the $\eta$-deformed QSC can be thought of as a trigonometrisation of the rational (undeformed) QSC describing the spectrum of $\mathcal{N}=4$ super Yang-Mills theory. Indeed, rather than living on a plane with cuts, the deformed QSC functions live on a cylinder with cuts, or an appropriate cover in some cases. 
\\[5mm]
We started from the very beginning, at the definition of the undeformed $\ads$ superstring. From its Poisson structure we reviewed how the $\eta$ deformation can be introduced. Its spectral problem can be approached in the same way as was done for the undeformed superstring: we discussed how one can find the exact $S$-matrix for the undeformed theory using integrability. In the language of Hopf algebras we discussed how the $\eta$ deformation, which can be viewed as a quantum deformation, establishes itself on the $S$-matrix level by constructing the relevant quantum group. From the representation theory of the quantum group the $S$ matrix followed.  We reviewed how the mirror trick allowed for the usage of the TBA program to treat the spectral problem, culminating in the ground-state TBA equations. These formed the starting point of our work. 
\\[5mm]
We rigorously derived this QSC starting from the $\eta$-deformed TBA equations via the associated analytic $Y$ and $T$ systems. By an appropriate use of notation this derivation simultaneously discusses how to derive the undeformed QSC from the undeformed TBA-equations. The analysis showed that most of the objects admit a natural trigonometrisation, such as integration contours, but also trigonometric solutions to finite-difference equations, thus allowing for a simultaneous treatment of both constructions.
\\[5mm]
To describe arbitrary excited states we proposed asymptotics for the QSC functions depending on the quantum numbers of such states. We reviewed the undeformed case where power-law asymptotics are specified for large real values of the spectral parameter. In the deformed case this asymptotic direction is not available, and we instead considered exponential asymptotics for large imaginary spectral parameter. In particular we found that this asymptotic prescription can nevertheless be smoothly linked to the power law prescription for the undeformed model. The precise dependence of the asymptotics on the quantum numbers follows from the classical spectral curve in the undeformed case, in the deformed case we used that this dependence should be consistent with the structure of the QSC as well as with both the weak coupling and the undeformed limit. The QSC equations together with these asymptotics form a system whose solutions encode the energy of the states of the underlying string theories. This is very similar to the usual description of energies in quantum-integrable models using Bethe equations: the equations are well-behaved, but finding explicit solutions might not be easy. 

\section*{Future directions} Now that we have derived the $\eta$-deformed QSC there is a whole host of possible further directions of study.

\paragraph{Finding deformed string energies.} Of course, the prime usage of the deformed QSC is the computation of string energies for the $\eta$-deformed string theory. The spectrum of the deformed theory should interpolate from the undeformed string spectrum to the undeformed mirror theory spectrum, each extreme having a string theory interpretation. The deformation is expected to lift degeneracies in the spectrum, which could yield interesting results. For example, the type of numbers appearing in the weak coupling expansion of the $\ads$ string energies have been under 
investigation, see. e.g. \cite{Leurent:2013mr}, yielding only so-called multiple zeta values. Finding whether these numbers get deformed -- and if so how-- would be intriguing. 
\\[5mm]
Perturbatively we can try to approach solving the QSC at weak coupling through the algorithm proposed in \cite{Marboe:2014gma,Marboe:2017dmb}. A particular challenge here is solving the trigonometric Bethe equations or Q system that would kick-start the procedure. We have started to analyse the solution for the $\eta$-deformed Konishi state: this state has non-zero quantum numbers $L=S=2$ and its $Q$ function
\be
Q(u) \sim 4\cosh (\ad) \cos(u) - \sqrt{4 \cosh(2c) +5} +1
\ee
is a deformation of the undeformed $Q$-function $Q(u) = u^2-\tfrac{1}{12}$. Continuing this analysis to find the energy of the Konishi state would be a good test case from where to generalise. Finding solutions at strong coupling would be very exciting, but since this also has not been achieved in the undeformed case this seems a difficult challenge. A numerical solution such as is available for the undeformed case \cite{Gromov:2015wca} seems more approachable, as it is based on techniques similar to those in the perturbative solution. 

\paragraph{$\eta$-deformed classical spectral curve.} To perform a direct check of the asymptotics we proposed for the $\eta$-deformed QSC it would be useful to construct the classical spectral curve for the $\eta$-deformed model, and contrast it with our QSC and asymptotics, similarly to how this was done in the undeformed case. This might also unravel more details about the relation between the classical and quantum spectral curves, as the currently known examples of known pairs of spectral curves are all of rational type and derive from the $\ads$ case. 

\paragraph{Studying thermodynamics.} As mentioned in section \ref{sec:mirrordualityandlimit}, it would also be very interesting to use the trigonometric QSC to look at the thermodynamics of our models, in particular the undeformed string. Here it would be great to understand how to incorporate chemical potentials in the $\eta$-deformed QSC, analogous to how this was done in the undeformed case \cite{Kazakov:2015efa} at least for purely imaginary chemical potentials corresponding to twists. In the undeformed case, a purely imaginary chemical potential $i \beta = \log \text{x}$ introduces exponential asymptotics of the form $\text{x}^{-i u}$. As power law behaviour naturally became exponential behaviour in the deformed case, it is not immediately clear how these twist exponentials should be modified. There are two reasons to believe these exponentials might not need modification at all. First, multiplying the $\Pf$ functions by similar exponentials -- $\text{x}^{-iu/2\ad}$ -- would precisely introduce the correct twists in the weak coupling Bethe equations, cf. Section \ref{sec:weakcoupling}. Second, from the perspective of the light-cone gauge-fixed string sigma model, certain twists do not translate to chemical potentials but instead affect the level matching condition, cf. e.g. \cite{deLeeuw:2012hp}. Setting $P=\beta$ in the discussion surrounding eqn. \eqref{KPyexpansion}, would introduce exactly a term of the form $\text{x}^{-iu/2\ad}$ in $\mu_{12}$. Adding such exponentials raises one immediate concern however, as the frequency of this exponential, $\beta/2c$, is not a multiple of $\pi$ for generic $\beta$ or $c$ and it would therefore dramatically affect the surface on which our equations are defined. It would be great to fully develop our understanding of such chemical potentials or twists in the QSC. Once this is understood, it would be interesting to consider the Hagedorn temperature computation of \cite{Harmark:2017yrv} in our deformed setting, and interpolate from the string to the ``mirror'' Hagedorn temperature.

\paragraph{Other deformed QSCs.} We saw that the $\eta$-deformed $\ads$ string ``contains'' many other integrable deformations of the $\ads$ string, fitting into the class of homogeneous Yang-Baxter deformations \cite{vanTongeren:2015soa,Osten:2016dvf,Kawaguchi:2014qwa,Matsumoto:2014gwa,Hoare:2016wsk,Borsato:2016pas}. Namely, by considering particular coupled infinite boost and $q\rightarrow1$ limits, it is possible to extract many homogeneous Yang-Baxter deformations of $\ads$ at the level of the sigma model action \cite{Hoare:2016hwh}, including for instance the gravity dual of canonical non-commutative SYM. It would be very interesting to see whether such boosts can be implemented directly in the QSC, to thereby extract the QSC and in particular its asymptoticity data that describe these models which often have an interesting AdS/CFT interpretation \cite{vanTongeren:2015uha,vanTongeren:2016eeb}. Of course it would also be interesting to directly investigate the QSC description of homogeneous Yang-Baxter models, beyond the basic Cartan-twisted ones. The one loop spectral problem for the simplest non-Cartan twisted model has recently been investigated in \cite{Guica:2017mtd}, indicating non-trivial asymptotics for the QSC. Of course, it would be great to use the general algebraic structure of twisted models to formulate a general Yang-Baxter deformed QSC.

\paragraph{The root of unity case.} Another interesting direction to investigate is the model defined by the exact $q$-deformed S-matrix for $q$ a root of unity, i.e. $q=e^{i \pi/k}$, instead of real.  The representation theory of quantum algebras is quite different when $q$ is a root of unity, which will presumably reflect itself in the QSC -- as it does in the TBA \cite{Arutyunov:2012zt,Arutyunov:2012ai} -- from both an algebraic as well as an analytical perspective, thereby providing a great test case for the QSC formalism. The number of $Y$ functions in this case is finite to begin with and the $Y$-system equations take a different form. Also the $T$ system has more complicated interactions than the Hirota equations. Our investigations have revealed that already at the level of the analytic $Y$-system difficulties arise: it is not straightforward how to find a set of discontinuity equations that correctly reproduce the ground-state TBA-equations, as the naive approach yields seemingly increasingly complicated expressions, although possible hidden symmetry of the $Y$-system equations might simplify this. Also, the additional overall real periodicity that we got in our real $q$ case becomes imaginary periodicity in the root of unity case, and this periodicity is generically not compatible with the periodicity we would expect for $\mu$. However, with the explicit knowledge of the trigonometric QSC one could try to approach the problem from two sides, by first proposing a root-of-unity QSC and then tracing how to obtain it from an analytic $Y$- and $T$-system.

\paragraph{Elliptic QSC.} Finally, even more speculatively, one might wonder whether it is possible to find an elliptic deformation of the $\ads$ string, or at least its exact S-matrix. The resulting theory should constitute the ``\textsc{xyz}" cousin of the $\ads$ (\textsc{xxx}) and $\eta$-deformed (\textsc{xxz}) string theories, following the integrability folkore. It is at this moment unclear what the corresponding QSC would look like, as there seem to be some major obstructions, most importantly this one: the QSC as currently formulated makes heavy use of large $u$ asymptotics, which does not seem to allow for further compactification of the spectral plane.\footnote{A possible way out would be to formulate the asymptotics in a different way. For example, for the undeformed one-cutted $\Pf$ functions we can find their asymptotics using a suitable contour integral around the cut, thereby no longer needing an infinite direction. This was observed by Till Bargheer in private communication.} It thus seems that an elliptic QSC, if it exists, will have a considerably different structure.

%% file: xfunctions.tex
In the spectral problem for $\ads$ and its deformations a central role is played by the $x$ functions, first introduced in section \ref{sec:xfunctions}. They are the basic building block of the $S$ matrix and the TBA kernels, specify the crossing equation and its solution and can ultimately be used to represent solutions to the QSC as convergent series in this function (see appendix \ref{app:constrainingtrigonometricpolynomials}). To aid the reader we collect all the relevant definitions of the $x$ functions in this appendix. We will stick to the conventions used to derive the TBA equations, $Y$ system and $T$ system, i.e. to the lower-half-plane conventions as introduced in \ref{sec:shortcuttedfunctions}. 
\section{Characterisation}
The $x$ functions are solutions to the shortening condition for the fundamental representations of the manifest symmetry algebra of the $S$ matrix, $\mathfrak{psu}_q(2|2)^{\otimes 2}$ (with $q=1$ for the undeformed case) when viewed as a functional equation: the deformed shortening condition is
\begin{equation}
e^{\ii u} =  \frac{x(u) + \frac{1}{x(u)} + \xi + \frac{1}{\xi}}{\frac{1}{\xi} - \xi},
\end{equation}
where
\begin{equation}
\xi = -\frac{\ii}{2} \frac{ h \left(q-q^{-1}\right)}{\sqrt{1-\tfrac{h^2}{4}\left(q-q^{-1}\right)^2}} = \ii \tan \frac{\theta}{2},
\end{equation}
such that the only free parameter in the shortening condition is $\theta \in (-\pi,\pi]$. Due to the left-hand side we can restrict our analysis to the cylinder with $-\pi\leq $Re$(u)<\pi$. The undeformed shortening condition is much simpler and can be obtained by taking the undeformed limit: rescaling $u\rightarrow \ad u$ and then sending $\ad\rightarrow 0$ we obtain
\begin{equation}
\label{eq:undeformedshorteningcondition}
x(u) + \frac{1}{x(u)} =u.
\end{equation}
Both these shortening conditions obviously have more than one solution: whenever $x$ is a solution, so is $1/x$. Moreover, a solution $x(u)$ can only have a zero when the left-hand side diverges, which happens when Im$(u)\rightarrow - \infty$ in the deformed case and $u\rightarrow \infty$ in the undeformed case. It is also fairly obvious that as a function on the complex plane $x$ has two branch points at $\pm \theta$ ($\pm 2$ in the undeformed case). We will consider only long and short branch cuts originating from those branch points. Finally, just like in our discussion about analytic continuation in section \ref{sec:analyticalcontinuation} the two solutions $x(u)$ and $1/x(u)$ are related by analytic continuation. 
\\[5mm]
We can characterise the solutions of the shortening condition that we find most convenient uniquely by the following set of properties:
\begin{itemize}
\item it has a short or long cut between the branch points,
\item on the complement of the branch cut it is an entire function,
\item it behaves as $x(u-i\e) \simeq u$ as $u\rightarrow \infty$ in the undeformed case\footnote{Choosing here $u+i\e$ yields the $x$ functions in the upper-half-plane conventions.} and as $x(u) \simeq e^{iu}$ as $u \rightarrow -i\infty$ in the deformed case.
\end{itemize}
This singles out the following solutions: the undeformed $x$-functions are
\begin{align}
\label{eq:appundeformedx}
x^{\text{und}}_s(u) = \frac{u}{2} \left( 1 + \sqrt{1-\frac{4}{u^2}}\right), \nn
x^{\text{und}}_m(u) = \frac{1}{2} \left( u - \ii \sqrt{4-u^2}\right),
\end{align}
whereas the deformed ones are given by
\begin{align}
\label{eq:appdeformedx}
x_s(u) &= -\ii \csc \theta \left( e^{\ii u } -\cos \theta -\left( 1-e^{iu}\right) \sqrt{\frac{\cos u - \cos \theta}{\cos u - 1}}\right),\nn
x_m(u) &= -\ii \csc \theta \left( e^{\ii u } -\cos \theta +\left( 1+e^{iu}\right) \sqrt{\frac{\cos u - \cos \theta}{\cos u + 1}}\right). 
\end{align}
\section{Properties}
\paragraph{Conjugation.}
The undeformed $x$-functions are both real analytic and as such conjugation of these functions is simple: for both $x_s$ and $x_m$ we find
\begin{equation}
\overline{x^{\text{und}}(u)} = x^{\text{und}}\left(\overline{u}\right). 
\end{equation}
The deformed case is more complicated: 
\begin{equation}
\overline{x_s(u)} = \frac{x_s\left( \overline{u}\right)+\xi}{x_s\left( \overline{u}\right)\xi +1}, \qquad \overline{x_m(u)} = \frac{x_m\left( \overline{u}\right)\xi+1}{x_m\left( \overline{u}\right) +\xi}.
\end{equation}
\paragraph{Continuation.} 
By construction both sets of $x$ functions coincide on the lower half-plane and are inverse on the upper half-plane, following the lower-half-plane conventions. Following the definition of the tilde continuation this implies that in both cases
\begin{equation}
\widetilde{x^{\text{und}}(u)} = \frac{1}{x^{\text{und}}\left(u\right)}, \qquad \widetilde{x(u)} = \frac{1}{x(u)},
\end{equation}
making continuation of functions expressed through the $x$ functions fairly easy.

\paragraph{Commutativity of conjugation and continuation.}
An interesting and non-trivial property that survives deformation is the fact that conjugation and continuation using the tilde of the $x$ functions commutes. In the undeformed case this is quite trivial, as
\begin{equation}
\widetilde{\overline{x^{\text{und}}(u)}} = \widetilde{x^{\text{und}}\left(\overline{u}\right)} = \frac{1}{x^{\text{und}}\left(\overline{u}\right)} = \tilde{x}^{\text{und}}\left( \overline{u}\right) = \overline{\tilde{x}^{\text{und}}\left( u\right)}. 
\end{equation}
In the deformed case the $x$ function is \emph{not} real analytic, but we find nevertheless that conjugation and continuation of these functions commute:
\begin{align}
\widetilde{\overline{x(u)}} = \widetilde{\left( \frac{x(\bar{u})\xi+1}{x(\bar{u})+\xi} \right)} =  \frac{(x(\bar{u}))^{-1}\xi+1}{(x(\bar{u}))^{-1}+\xi}  =\left( \frac{x(\bar{u})\xi+1}{x(\bar{u})+\xi}\right)^{-1} =(\overline{x(u)})^{-1} = \overline{\tilde{x}(u)}.
\end{align}
This simplifies the analysis of the $\Pf$ functions.

%% file: definitionsandconventions.tex
\section{Deformed kernels}
\label{App:kernelsandenergies}
The kernels that appear in the $\eta$-deformed TBA-equations \eqref{eq:TBAeqns} are defined as follows:
\be
\label{eq:basickernel}
K_M (u) = \frac{1}{2\pi i} \frac{d}{du} \log S_M(u) = \frac{1}{2\pi} \frac{ \sinh M\ad}{\cosh M\ad- \cos u},
\ee
with
\be
\label{eq:sinkernel}
S_{Q}(u) = \frac{S^{[-Q]}}{S^{[+Q]}} = \frac{\sin \frac{1}{2} (u-i Q \ad)}{\sin \frac{1}{2} (u+i Q \ad)},
\ee
using our definition in eqn. \eqref{eq:Sfunction}. This now defines for us
\be
K_{MN} (u) = K_{M+N} (u)+ K_{|M-N|} (u)+2\sum_{j=1} ^{\min(M,N)-1} K_{|M-N|+2j} (u).
\ee
The other kernels are defined directly from the scattering matrices
\bea
S_-^{yQ}(u,v) &=& q^{Q/2} \frac{x(u) - x^+(v) }{x(u) - x^-(v) }\sqrt{\frac{x^+(v)}{x^-(v)}}, \\
S_+^{yQ}(u,v) &=& q^{Q/2} \frac{1/x(u) - x^+(v) }{1/x(u) - x^-(v) }\sqrt{\frac{x^+(v)}{x^-(v)}}, \\
S_-^{Qy}(u,v) &=& q^{Q/2} \frac{x^-(u) - x(v) }{x^+(u) - x(v) }\sqrt{\frac{x^+(u)}{x^-(u)}} ,\\
S_+^{Qy}(u,v)& =& q^{Q/2} \frac{x^-(u) - 1/x(v) }{x^+(u) - 1/x(v) }\sqrt{\frac{x^+(u)}{x^-(u)}},\\
S_{xv}^{QM}(u,v) &=& q^{Q} \frac{x^-(u) - x^{+M}(v) }{x^+(u) - x^{+M}(v) } \frac{x^-(u) - x^{-M}(v) }{x^+(u) - x^{-M}(v) }\frac{x^+(u)}{x^-(u)} \prod_{j=1}^{M-1} S_{Q+M-2j}(u-v) , \qquad \quad
\eea
where $x=x_m$, $x^{\pm}(v) = x(v\pm i Q \ad)$, $x^{\pm M}(v) = x(v\pm i M \ad)$. They are given by
\bea
\label{eq:TBAkernels2}
K_{xv}^{QM}(u,v) &=& \frac{1}{2\pi i} \frac{d}{du} \log S_{xv}^{QM} (u,v), \\
K_{vwx}^{MQ}(u,v) &=& -\frac{1}{2\pi i} \frac{d}{du} \log S_{xv}^{QM} (v,u), \\
K_{\beta}^{Qy}(u,v) &=& \frac{1}{2\pi i} \frac{d}{du} \log S_{\beta}^{Qy} (u,v), \\
K_{\beta}^{yQ}(u,v) &=& \beta \frac{1}{2\pi i} \frac{d}{du} \log S_{\beta}^{Qy} (v,u),
\eea
and finally the dressing phase kernel is defined as
\be
K^{PQ}_{\mathfrak{sl}(2)}(u,v) = \frac{1}{2\pi i} \frac{d}{du} \log S^{PQ}_{\mathfrak{sl}(2)},
\ee
where the dressing phase was defined in \eqref{eq:sl2dressingphase}. The natural dressing factor for the mirror model that appears in the TBA equations is
\begin{equation}
\Sigma_{12} = \frac{1-\frac{1}{x_1^+ x_2^-}}{1-\frac{1}{x_1^- x_2^+}} \sigma_{12},
\end{equation}
defined by appropriate analytic continuation of the above objects. Fused to describe bound state scattering we have \cite{Arutyunov:2012ai}\footnote{Note that the definition of $\Sigma$ in our real $q$ case \cite{Arutynov:2014ota} is slightly different from the one of \cite{Arutyunov:2012ai} for the $|q|=1$ case. The present definition is natural from the point of view of mirror duality. We effectively go back to the $|q|=1$ conventions in our definition of $K^{\Sigma}_{QM}$ below, to get kernels and TBA equations analogous to the undeformed model.}
\begin{align}
-i\log\Sigma^{QM}(y_1,y_2) =& \, \Phi(y_1^+,y_2^+)-\Phi(y_1^+,y_2^-)-\Phi(y_1^-,y_2^+)+\Phi(y_1^-,y_2^-) \label{eq:improveddressingphase} \\ &\, -\frac{1}{2}\left(\Psi(y_1^+,y_2^+)+\Psi(y_1^-,y_2^+)-\Psi(y_1^+,y_2^-)-\Psi(y_1^-,y_2^-)\right) \nonumber \\
&\, +\frac{1}{2}\left(\Psi(y_{2}^+,y_1^+)+\Psi(y_{2}^-,y_1^+)-\Psi(y_{2}^+,y_1^-)
-\Psi(y_{2}^-,y_1^-) \right) \nonumber \\
&\, - i \log \frac{i^Q \, \Gamma_{q^2}\left(M - \frac{i}{2\ad}(u(y_1^+) - v(y_2^+))\right)}{i^{M} \Gamma_{q^2}\left(Q+ \frac{i}{2\ad}(u(y_1^+) - v(y_2^+))\right)}\frac{1- \frac{1}{y_1^+y_2^-}}{1-\frac{1}{y_1^-y_2^+}}\sqrt{\frac{y_1^+ + \xi}{y_1^-+\xi} \frac{y_2^- + \xi}{y_2^+ + \xi}} \nonumber \\
& \, + \frac{i}{2} \log q^{Q-M} e^{-i(Q+M-2)(u-v)}\nonumber,
\end{align}
where
\begin{equation}
\label{eq:Psi}
\Psi(x_1,x_2) \equiv i \oint_{\mathcal{C}} \frac{dz}{2 \pi i} \frac{1}{z-x_2} \log  \frac{\Gamma_{q^2} (1+\frac{i}{2\ad}(u_1-u(z)))}{\Gamma_{q^2} (1-\frac{i}{2\ad}(u_1-u(z)))} \, ,
\end{equation}
where we have denoted $x^{-1}$ by $u$, as is common in the literature. We define the associated integration kernel as
\begin{equation}
K^{\Sigma}_{QM}(u,v) = \frac{1}{2\pi i} \frac{d}{du} \log\Sigma^{QM}(y_1(u),y_2(v)) \sqrt{\tfrac{y_1^+}{y_1^-}\tfrac{y_1^-+\xi}{y_1^+ + \xi} \tfrac{y_2^-}{y_2^+}\tfrac{y_2^+ + \xi}{y_2^- + \xi}}.
\end{equation}
where we added a factor that reduces to one in the undeformed limit to simplify expressions below. These expressions are necessary to derive the simplified expression for the dressing phase contribution to the simplified TBA-equations, as discussed in appendix \ref{app:dressing}.
\\[5mm]
Some important identities for the kernels are
\bea
\label{kernelids}
K_{Qy}(u,v) &\defeq& K_{-}^{Qy}(u,v)-K_{+}^{Qy}(u,v)  = K(u+ i Q \ad,v) -K(u- i Q\ad,v) \\
K_{yQ}(u,v) &\defeq& K_{-}^{yQ}(u,v)+K_{+}^{yQ}(u,v)  = K(u,v- i Q \ad) -K(u,v+ i Q \ad) \\
K_Q(u-v) &=& K_{-}^{yQ}(u,v) - K_{+}^{yQ}(u,v) = K_{-}^{Qy}(u,v)+K_{+}^{Qy}(u,v)  ,\qquad  \qquad \qquad 
\eea
with the kernel
\be
K(u,v) = \frac{1}{2\pi i} \frac{d}{du} \log\frac{x(u) - 1/x(v) }{x(u) - x(v) }.
\ee
We can define the similar kernel 
\be
\hat{K}(u,v) = \vartheta\left(|u|-\theta \right) \frac{1}{2\pi i} \frac{d}{du} \log\frac{x(u) - 1/x_s(v) }{x(u) - x_s(v) },
\ee
where $\vartheta$ is the Heaviside theta-function, and the associated 
\be
\hat{K}_{M}(u,v)  = \bar{K}(u+ i M \ad,v) +\bar{K}(u- i M \ad,v).
\ee
The mirror energy and momentum are defined through the mirror $x$-functions as follows: 
\be
\label{eq:mirrormomentum}
 \tilde{p}^Q = \frac{i}{\ad}\log \left( q^Q \frac{x^+}{x^-}\frac{x^-+\xi}{x^++\xi}\right), \quad \tilde{E}_Q= -\log \left( \frac{1}{q^Q}\frac{x^++\xi}{x^-+\xi}\right). 
\ee

\section{Undeformed quantities.}
\label{app:undeformedkernelsandenergies}
After letting $q\rightarrow 1$, taking $x= x^{\text{und}}_m$ (defined in \eqref{eq:appundeformedx}) and replacing the shift distance $\ad \rightarrow 1/g$ the kernels defined above are the undeformed TBA-kernels. In particular this means we use $S^{\text{und}}$ instead of $S$. The undeformed mirror energy and momentum are given by
\begin{equation}
\label{eq:undeformedmirrorenergyandmomentum}
\tilde{p}_{\text{und}}^{Q} = g x\left( u -\frac{ \ii Q}{g} \right)-g x\left( u +\frac{ \ii Q}{g} \right)+\ii Q, \qquad \tilde{E}^{\text{und}}_Q= \log \left( \frac{x\left(u-iQ/g\right)}{x\left(u+iQ/g\right)}\right). 
\end{equation}
Let us nevertheless mention the undeformed version of $S_Q$ (defined in eqn. \eqref{eq:sinkernel}) and $K_M$ (defined in eqn. \eqref{eq:basickernel}) explicitly:
\begin{equation}
\label{eq:undeformedsinkernel}
S_Q^{\text{und}}(u) = \frac{u-i Q /g}{u+i Q /g}, \qquad K_Q^{\text{und}}(u) = \frac{1}{\pi} \frac{g Q}{Q^2 + g^2 u^2}.  
\end{equation}
Finally, the undeformed dressing kernel follows from the description above: in particular the contour $\mathcal{C}$ turns into the unit circle, undoing the rescaling and shifting caused by the deformation.

%% file: dressingphase.tex
An important role in the derivation of the analytic $Y$-system from the TBA equations is played by the dressing-phase kernel $ K_{\mathfrak{sl}(2)}^{PQ}$. Its explicit form is very convoluted, making its analysis tricky. In order to derive the analytic $Y$-system we need to know the discontinuities that arise because of the presence of the dressing phase. A particularly simple form of the dressing phase that divulges this information was derived in \cite{Cavaglia:2010nm} for the undeformed case, based on kernel identities found in \cite{Arutyunov:2009ux}. For the deformed case we need a similar simple form, which we will derive in this appendix. 
\\[5mm]
The dressing phase largely drops out of the simplified TBA-equations obtained by acting with $(K+1)^{-1}$ as explained in section \ref{sec:simpTBA}. Stripping off an extra $\s$ of the dressing-phase term in the simplified TBA-equations \eqref{eq:simpTBA} by acting with $s^{-1}$ defined in eqn. \eqref{eq:sinvdef}, we are interested in\footnote{Let us mention again that we have changed notation compared to the paper \cite{Klabbers:2017vtw}, where this derivation first appeared: in this thesis we use a hat to denote functions with short cuts throughout, whereas in \cite{Klabbers:2017vtw} the short-cutted kernels were denoted with a check.}
\begin{equation}
\label{eq:Ksigmadef}
\hat{K}^\Sigma_Q(u,v)  = \lim_{\epsilon\rightarrow 0^+}\left(K^{\Sigma}_{Q1} (u,v+i \ad - i \epsilon) + K^{\Sigma}_{Q1} (u,v-i \ad + i \epsilon)\right) - K^{\Sigma}_{Q2}(u,v),
\end{equation}
which vanishes for $|v|< \theta$ \cite{Arutyunov:2012ai}. Here we will need more detailed properties of this kernel for $|v|>\theta$. We can express this kernel in terms of simpler kernels already appearing in the TBA, similarly to how this was done in the undeformed case \cite{Arutyunov:2009ux}.
\\[5mm]
Let us work at the S matrix rather than kernel level, with $\hat{\Sigma}_Q$ denoting the relevant S matrix. There are no particular subtleties in taking the $\epsilon \rightarrow 0$ limit and direct evaluation gives
\begin{align*}
-i \log \hat{\Sigma}_Q = & \, \Phi(y_1^-,x)- \Phi(y_1^-,1/x)- \Phi(y_1^+,x)+ \Phi(y_1^+,1/x)\\
& \, + \frac{1}{2} \left( \Psi(y_1^-,x)- \Psi(y_1^-,1/x)+ \Psi(y_1^+,x)- \Psi(y_1^+,1/x)\right)\\
&\, + \Psi(x,y_1^+)-\Psi(x,y_1^-)\\
& \, - i \log i^Q \tfrac{\, \Gamma_{q^2}\left(\tfrac{Q}{2} - \frac{i}{2\ad}(u-v)\right)}{\Gamma_{q^2}\left(\tfrac{Q}{2}+ \frac{i}{2\ad}(u - v)\right)}\frac{1- \frac{1}{y_1^+ x}}{1-\frac{x}{y_1^-}}\sqrt{\tfrac{y_1^+}{y_1^-} \tfrac{1}{x^2}}\\
& \, + \frac{i}{2} \log q^{Q} e^{-i(Q-2)(u-v)}
\end{align*}
where $x=x(v - i \e)$ and $\Psi$ and $\Phi$ were defined in \eqref{eq:Psi} and \eqref{eq:Phi}.
\\[5mm]
Let us now rewrite the contour integrals in the $\Phi$ and $\Psi$ terms as integrals over the interval $[-\theta, \theta]$. The counter-clockwise contour $\mathcal{C}$ can be described by $x(u)$ as $u$ runs from $\theta$ to $-\theta$ followed by $1/x(u)$ as it runs from $-\theta$ to $\theta$. This means that for any function $f(z)$ invariant under inverting its argument,
\begin{equation*}
\oint_{\mathcal{C}} \frac{dz}{2 \pi i} \frac{1}{z-y} f(z) = \int_{\hat{Z}_0}\frac{dt}{2\pi i} \frac{d x(t)}{dt} \left(\frac{1}{x(t)-y} + \frac{1}{x(t)^2} \frac{1}{\frac{1}{x(t)}-y} \right)f(x(t)),
\end{equation*}
with $\hat{Z}_0$ defined in eqs. \eqref{eq:Zs}. We recognise this combination of $x$ functions as
\begin{equation*}
\frac{\partial}{\partial t} \log S(s,t), \qquad S(s,t) = \frac{x(s)-x(t)}{x(s)-1/x(t)},
\end{equation*}
such that, with $v=u(y)$,
\begin{equation*}
\oint_{\mathcal{C}} \frac{dz}{2 \pi i} \frac{1}{z-y} f(z) = -\int_{\hat{Z}_0} \frac{dt}{2\pi i} \left(\frac{\partial}{\partial t} \log S(v,t)\right) f(x(t)).
\end{equation*}

With this identity, the first line above results in
\begin{align*}
&\partial_u \Delta \Phi \equiv \partial_u \left(\Phi(y_1^-,x)- \Phi(y_1^-,1/x)- \Phi(y_1^+,x)+ \Phi(y_1^+,1/x)\right)  \\
 & \qquad =  2 i \int_{\hat{Z}_0} \frac{dt_1}{2\pi i} \int_{\hat{Z}_0} \frac{dt_2}{2\pi i} \frac{\partial}{\partial u} \frac{\partial}{\partial t_1} \log S_{Qy}(u,t_1) \frac{\partial}{ \partial t_2} \log \bar{S}(t_2,v) \log S^{[2]}_{q\Gamma}(t_1,t_2),
\end{align*}

where\footnote{Note that we use the label $q\Gamma$ for readability, though on the right-hand side it is really $\Gamma_{q^2}$ which appears.}
\begin{equation*}
\log S^{[Q]}_{q\Gamma}(t_1,t_2) = \log   \frac{\Gamma_{q^2}(\frac{Q}{2}-\frac{i}{2\ad}(t_1-t_2))}{\Gamma_{q^2}(\frac{Q}{2}+\frac{i}{2\ad}(t_1-t_2))},
\end{equation*}
and we recall that $S_{Qy}(u,v) = S(u+ i \ad Q,v)/S(u - i \ad Q,v)$ which naturally comes out of the $y_1^\pm$ terms,\footnote{Our sign conventions for kernels differ from \cite{Arutyunov:2009ux} at this and other points.} and note that for Im$(s)<0$
\begin{equation*}
\frac{\partial}{\partial t} \log \frac{x(s)-x(t)}{x(s)-1/x(t)}\frac{1/x(s)-1/x(t)}{1/x(s)-x(t)} = 2\frac{\partial}{\partial t} \log \frac{x(t) - x(s)}{x(t) - 1/x(s)} = - 2 \frac{\partial}{ \partial t} \log \bar{S}(t,s),
\end{equation*}
which naturally comes out of the $x$ terms. Integrating by parts in the $t_1$ integral, noting that $K_{Qy}$ vanishes at $\pm \theta$, and taking out factors of $2 \pi i$ to define kernels, we end up with
\begin{equation*}
\frac{1}{2\pi}\frac{\partial}{\partial u} \Delta \Phi =  2 K_{Qy} \,\hat{\star}\, K_{q\Gamma}^{[2]} \,\hat{\star}\, \bar{K},
\end{equation*}
where $K_{q\Gamma}^{[Q]}(u) = \tfrac{1}{2\pi i} \tfrac{d}{du} \log  S_{q\Gamma}^{[Q]}(u)$. For the first line of $\Psi$ terms we similarly find
\begin{align*}
& \Delta \Psi \equiv \frac{1}{2}\left(\Psi(y_1^-,x)- \Psi(y_1^-,1/x)+ \Psi(y_1^+,x)- \Psi(y_1^+,1/x) \right) \\
& \qquad = - i \int \frac{dt}{2\pi i}  \frac{\partial}{ \partial t} \log \bar{S}(t,v)  \left(\log S^{[2]}_{q\Gamma}(u-i \ad Q,t)+\log S^{[2]}_{q\Gamma}(u+i \ad Q,t)\right).
\end{align*}
We now notice that
\begin{align*}
S^{[2]}_{q\Gamma}(u-i \ad Q,t) S^{[2]}_{q\Gamma}(u+i \ad Q,t)  = &\, \frac{\Gamma_{q^2}(1+\frac{Q}{2}-\frac{i}{2\ad}(u-t))}{\Gamma_{q^2}(1+\frac{Q}{2}+\frac{i}{2\ad}(u-t))}\frac{\Gamma_{q^2}(1-\frac{Q}{2}-\frac{i}{2\ad}(u-t))}{\Gamma_{q^2}(1-\frac{Q}{2}+\frac{i}{2\ad}(u-t))}\\
= &\, (-1)^{Q-1} e^{i(Q-1)(u-t)} S^{[Q+2]}_{q\Gamma}(u,t) S^{[Q]}_{q\Gamma}(u,t),
\end{align*}
where in the second equality we used the defining property of the $q$-gamma function $Q-1$ times in the second term, the resulting product of ratios of trigonometric functions cancelling pairwise up to a phase. By the same property
\begin{equation*}
S^{[Q+2]}_{q\Gamma}(u,t) =S^{[Q]}_{q\Gamma}(u,t)/(e^{i(u-t)}S_Q(u-t))
\end{equation*}
We hence find
\begin{equation*}
\frac{1}{2\pi} \frac{\partial}{\partial u}  \Delta \Psi = (2 K_{q\Gamma}^{[Q]}-K_Q)\, \hat{\star}\,\bar{K} + \frac{Q-2}{2\pi} \star \bar{K}.
\end{equation*}
The next line gives, upon integration by parts again
\begin{equation*}
\frac{1}{2\pi} \frac{\partial}{\partial u} \left(\Psi(x,y_1^+) - \Psi(x,y_1^-)\right) = K_{Qy} \,\hat{\star}\, K_{\Gamma_{q^2}}^{[2]}.
\end{equation*}
For the second to last line, analogously to the undeformed case we directly find
\begin{equation*}
\frac{1}{2\pi i} \frac{\partial}{\partial u} \log i^Q \tfrac{\, \Gamma_{q^2}\left(\tfrac{Q}{2} - \frac{i}{2\ad}(u-v)\right)}{\Gamma_{q^2}\left(\tfrac{Q}{2}+ \frac{i}{2\ad}(u - v)\right)}\frac{y_1^+- \frac{1}{x}}{y_1^- - x}\sqrt{\tfrac{y_1^-}{y_1^+}x^2} = K_{q\Gamma}^{[Q]}(u-v) - \frac{1}{2} \bar{K}_Q(u,v) - \frac{1}{2} K_Q(u-v).
\end{equation*}
Finally, the last line precisely cancels the constant term in $\frac{1}{2\pi} \frac{\partial}{\partial u}  \Delta \Psi$ above, as
\begin{equation*}
1 \star \bar{K} = -\frac{1}{2}.
\end{equation*}
Putting everything together, we find
\begin{equation*}
\hat{K}_Q^\Sigma =   2 K_{Qy} \,\hat{\star}\, K_\Gamma^{[2]} \,\hat{\star}\, \bar{K} + (2 K_{\Gamma_{q^2}}^{[Q]}-K_Q)\,\hat{\star}\, \bar{K} + K_{Qy} \,\hat{\star}\, K_{q\Gamma}^{[2]}+ K_{q\Gamma}^{[Q]} - \frac{1}{2} \bar{K}_Q - \frac{1}{2} K_Q
\end{equation*}

To simplify this, we note that the $\bar{K}$ and $K_Q$ kernel satisfy\footnote{To see this, extend the integration to run from $-\pi$ to $\pi$ a little above and a little below the real axis -- hence the factors of $1/2$ -- and then note that $K_Q(u-v)$ has poles at $u=v \pm Q i \ad$, and $\bar{K}(u,v)$ has ones at $u=v\pm i \epsilon$.}
\begin{equation*}
K_Q \,\hat{\star}\, \bar{K} = \frac{1}{2} \bar{K}_Q - \frac{1}{2} K_Q ,\qquad 1 \star \bar{K} = -\frac{1}{2}.
\end{equation*}
Similarly
\begin{equation}
\label{eq:Kqgammaidentity}
K^{[Q]}_{q\Gamma}\,\hat{\star}\, \bar{K}  = - \frac{1}{2} K^{[Q]}_{q\Gamma}+  \frac{1}{2}  \sum_{N=0}^\infty\bar{K}_{Q+2N},
\end{equation}
which we will encounter again later. Note that this expression is manifestly compatible with the relation $K^{[Q+2]}_{q\Gamma}  = K^{[Q]}_{q\Gamma} - K_Q -1$ encountered above. Using these identities we can finally simplify $\hat{K}_Q^\Sigma$ to
\begin{align*}
\hat{K}_Q^\Sigma & \, = \sum_{N=1}^{\infty} \left(K_{Qy}\,\hat{\star}\, \bar{K}_{2N} + \bar{K}_{Q+2N}\right).
\end{align*}

\paragraph{Contribution in the TBA.} The contribution of the improved mirror dressing factor to $\Delta$ appearing in the TBA equations is given by
\begin{equation}
\Delta^\Sigma(u) = 2 \sum_Q L_Q \star \hat{K}^\Sigma_Q (u).
\end{equation}
This function has a short cut on the real line. We would like to continue it from the upper half-plane to have long cuts. Using the simplified expression for $\check{K}_Q^\Sigma$ discussed in the previous section, we can simplify the result by following appendix C of \cite{Cavaglia:2010nm} with appropriately adapted integration contours, to arrive at
\begin{equation}
\begin{aligned}
\label{eq:simplifieddressing}
\Delta^\Sigma(u^*) & \, = \sum_\alpha \log Y^{(\alpha)}_- \star_{\gamma_x} \sum_{N=1}^{\infty} \bar{K}_{2N}(u^*) \\
& \, = \sum_\alpha \oint_{\gamma_x} \log Y^{(\alpha)}_-(z) \left(\sum_{N=1}^{\infty} K(z + 2 i N \ad, u) + K(z - 2 i N \ad, u)\right).
\end{aligned}
\end{equation}
This result is exactly the same in the undeformed case. We will use it in section \ref{sec:discdelta} to derive the discontinuity equations for $\Delta$, which governs the behaviour of $Y_Q$ functions. We will also use it later in section \ref{sec:restoringdressing} to rederive the $Y_Q$ TBA-equation.

%% file: Ysystemcomputations.tex
\section{Deriving a local discontinuity equation for $Y_{\pm}$}
We are looking for a local discontinuity equation that we can impose on the $Y$-system equations instead of the non-local discontinuity \eqref{eq:Y-nonlocal}. As an example of how these computations are carried out in practice we show it in detail here. We compute (simultaneously for the deformed and undeformed case)
\be
\left[\log \frac{Y_-}{Y_+} \right]_{\pm 2N}(u),
\ee
with $N \in \mathbb{N}$. We search for an expression for $\log \frac{Y_-}{Y_+} (u+2N \ad i )$, by examining the integral expression for this quantity:
\be
\log \frac{Y_-}{Y_+} (u) = -\int_{Z_0} \Lambda_P(v) K_{Py} (v,u) = -\int_{Z_0} \Lambda_P(v)\left(K(v+ i P \ad,u) -K(v- i P \ad,u)\right).
\ee
This integral is only computable if the poles at the branch points $\pm \theta$ are smoothed out in the product of the $Y$ functions with the kernels. The only remaining poles of the integrand then sit at $v \pm i P \ad = u$. Choosing Im$(u) >0$, we see that when we start shifting $u$ exactly one of the two poles crosses the integration contour and only those kernels for which $|P|< 2N$ contribute. We see that

{
\bea
\log \frac{Y_-}{Y_+} (u\pm 2N\ad i) &=& -\int_{Z_0} \Lambda_P(v) K_{Py} (v,u\pm 2N\ad i) \nn
&+& \sum_{P=1}^{2N}\oint_{v=u\pm i(2N-P)\ad} dv \Lambda_P(v)K(v\pm iP\ad,u\pm 2N\ad i) 
\eea
}

\noindent such that we find 
\be
\log \frac{Y_-}{Y_+} (u\pm 2N\ad i) = -\int_{Z_0} \Lambda_P(v) K_{Py} (v,u \pm 2N\ad i) - \sum_{P=1}^{2N}\Lambda_P(u\pm i(2N-P)\ad).
\ee
Combined with $\Lambda_P\in \mA_P$ this implies that the discontinuity at the lines $\pm 2N\ad i$ is given by 
\be
\left[\log \frac{Y_-}{Y_+} \right]_{\pm2N}(u) = -\sum_{P=1}^{2N} \left[\Lambda_P\right]_{\pm(2N-P)}(u) 
= -\sum_{P=1}^{N} \left[\Lambda_P\right]_{\pm(2N-P)}(u).
\ee

\section{Deriving the TBA equations for $Y_{(v)w}$ functions from the analytic $Y$-system}
\label{app:Yvwdiscs}
\begin{figure}
\centering
\begin{subfigure}{7.5cm}
\includegraphics[width=7.5cm]{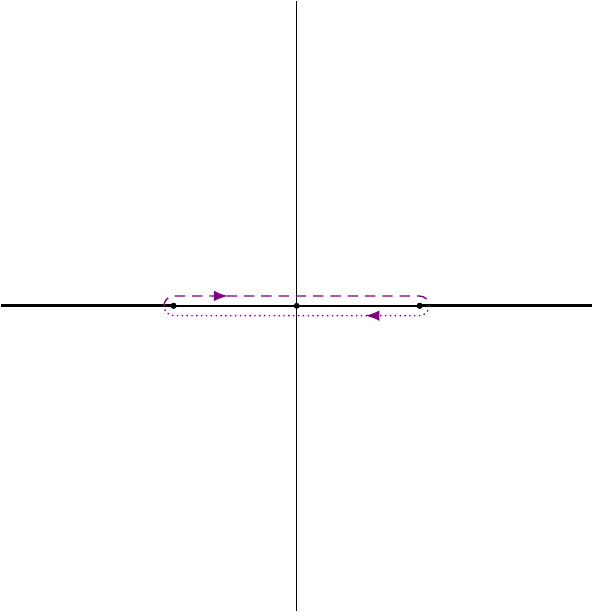}
\subcaption{}
\end{subfigure} \quad 
\begin{subfigure}{6cm}
\includegraphics[width=6cm]{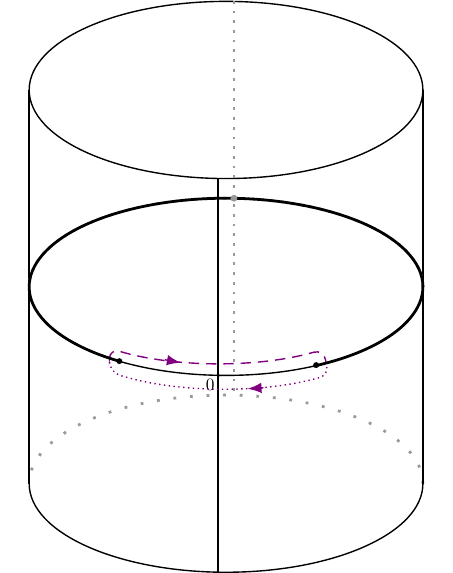}
\subcaption{}
\end{subfigure}
\caption{The contour $\gamma_0$ in the undeformed (a) and in the deformed (b) case. The dotted line indicates that the lower half of the contour runs on the second sheet.}
\label{fig:gamma0}
\end{figure}

We present a full derivation of $Y_{(v)w}$ TBA-equations (see section \ref{sec:rederiveYvwTBA} in the main text for a summary of this derivation). 
\\[5mm]
We can write the left-hand side of the TBA equation as a contour integral as
\be
\label{eq:appYvwcontour}
\log Y_{M|(v)w}(u) =\sum_{\tau}\sum_{l=0}^{\infty}  \int_{Z_0}  \frac{dz}{2\pi i} H(z+2\ii \tau N \ad -u) \left[ \log Y_{M|(v)w} \right]_{\pm 2M+l}.
\ee
We use the $Y$-system equations to find the necessary discontinuity equations. In particular we will derive discontinuities corresponding to the plus sign in \eqref{eq:appYvwcontour}, the ones with a minus sign follow straightforwardly. Taking logs of the $Y$-system equations \eqref{eq:Ysystem} for $Y_{M|vw}$ we find for $M\geq 2$
\be
\log Y_{M|vw}^+ +\log Y_{M|vw}^- =\sum_{N=1}^{\infty} A_{MN} \Lambda_{N|vw}-\Lambda_{M+1},
\ee
with $A_{MN}$ as in \eqref{eq:AMJ}. Taking discontinuity brackets at $M+2l-1$ yields
\bea
\label{hulpje1v}
& &\left[ \log Y_{M|vw} \right]_{M+2l} =  \sum_{N=1}^{\infty} A_{MN} \left[L_{N|vw}\right]_{M+2l -1} +  \left[\log Y_{M-1|vw}\right]_{M+2l-1} \nn 
&+& \left[\log Y_{M+1|vw}\right]_{M+2l-1} -\left[\Lambda_{M+1}\right]_{M+2l-1} - \left[\log Y_{M|vw}\right]_{M+2(l-1)}. 
\eea
This is the rule that we can use repeatedly to bring the discontinuity containing $\log Y_{Q|vw}$ with $Q<M$ closer and closer to the real line. Applying it once yields
\bea 
& &\left[ \log Y_{M|vw} \right]_{M+2l} = \nonumber \\
&=& \sum_{N=1}^{\infty}  A_{MN} \left[L_{N|vw}\right]_{M+2l -1} + \sum_{N=1}^{\infty}  A_{M-1,N} \left[L_{N|vw}\right]_{M-1+2l -1} -\left[ \log Y_{M|vw} \right]_{M+2l-2}\nonumber \\ 
&+&  \left[\log Y_{M-2|vw}\right]_{M+2l-2} +\left[\log Y_{M|vw}\right]_{M+2(l-1)} + \left[\log Y_{M+1|vw}\right]_{M+2l-1} \nn
&-& \left[\log Y_{M-1|vw}\right]_{M+2l-3}
-\left[\Lambda_{M+1}\right]_{M+2l-1} - \left[\Lambda_{M}\right]_{M-1+2l-1},\nn 
\eea
so we can conclude that 
\bea 
& &\left[ \log Y_{M|vw} \right]_{M+2l} -\left[\log Y_{M+1|vw}\right]_{M+2l-1} =  \sum_{Q=M-1}^M \sum_{N=1}^{\infty}  A_{QN} \left[L_{N|vw}\right]_{Q+2l -1} \nn
&+&  \left[\log Y_{M-2|vw}\right]_{M+2l-2}-\left[\log Y_{M-1|vw}\right]_{M+2l-3}-\sum_{Q=M}^{M+1} \left[\Lambda_Q\right]_{Q+2l-2}.
\eea
Repeated application now leads to the result
\bea
\label{rep1v}
& &\left[ \log Y_{M|vw} \right]_{M+2l} -\left[\log Y_{M+1|vw}\right]_{M+2l-1} = \sum_{Q=2}^M \sum_{N=1}^{\infty}  A_{QN} \left[L_{N|vw}\right]_{Q+2l -1}\nn
&+&  \left[\log Y_{1|vw}\right]_{1+2l}-\left[\log Y_{2|vw}\right]_{2l}-\sum_{Q=3}^{M+1} \left[\Lambda_Q\right]_{Q+2l-2}.
\eea
Plugging in the Y-system equation for $Y_{1|vw}$ gives us
\bea
\label{rep2v}
& &\left[ \log Y_{M|vw} \right]_{M+2l} -\left[\log Y_{M+1|vw}\right]_{M+2l-1} = \nonumber \\
&=&\sum_{Q=1}^M \sum_{N=1}^{\infty}  A_{QN} \left[L_{N|vw}\right]_{Q+2l -1}-  \left[\log Y_{1|vw}\right]_{1+2l-2} +\left[\Lambda_--\Lambda_+ \right]_{2l}-\sum_{Q=2}^{M+1} \left[\Lambda_Q\right]_{Q+2l-2}.\nn
\eea
This equation is also valid for $M=1$ such that we can continue to treat all the $Y_{M|vw}$ simultaneously. Using the line \eqref{rep1v} with first term $\left[ \log Y_{M+1|vw} \right]_{M+2l-1}$ (So $M\rightarrow M+1,l\rightarrow l-1$) gives 
\bea
\left[ \log Y_{M+1|vw} \right]_{M+2l-1} = \left[\log Y_{M+2|vw}\right]_{M+2l-2}  +\sum_{Q=2}^{M+1} \sum_{N=1}^{\infty}  A_{QN} \left[L_{N|vw}\right]_{Q+2l -3}\nn
+  \left[\log Y_{1|vw}\right]_{2l-1}-\left[\log Y_{2|vw}\right]_{2l-2}-\sum_{Q=3}^{M+2} \left[\Lambda_Q\right]_{Q+2l-4}.
\eea
Note the existence of a term $\left[\log Y_{1|vw}\right]_{2l-1}$ in \eqref{rep2v} and also in the last line. When combining the statements, these terms will cancel. We can repeatedly do this to end up with
\bea
\label{resvw1}
& &\left[ \log Y_{M|vw} \right]_{M+2l} = \left[ \log Y_{M+l+1|vw} \right]_{M+l-1} -\sum_{J=1}^{l}\sum_{Q=1+J}^{M+J} \left[\Lambda_Q\right]_{Q+2l-2J} \nn
&+&\sum_{J=0}^l\sum_{Q=1+J}^{M+J} \sum_{N=1}^{\infty}  A_{QN} \left[L_{N|vw}\right]_{Q+2l -1-2J} + \left[\Lambda_--\Lambda_+\right]_{2l} -\left[\log Y_{l+1|vw}\right]_{l-1}, \quad 
\eea
where the first and last term vanish except when $l=0$ because of the fact that the functions in the brackets are analytic on the respective strips. The triple sum can be put in a more suggestive form:
\bea
\label{longder1}
& &\sum_{J=0}^l\sum_{Q=1+J}^{M+J} \sum_{N=1}^{\infty}  A_{QN} \left[L_{N|vw}\right]_{Q+2l -1-2J} \nn
&=& \sum_{J=1}^l\sum_{Q=1+J}^{M+J}  \left( \left[L_{Q+1|vw}\right]_{Q+2l -1-2J}+\left[L_{Q-1|vw}\right]_{Q+2l -1-2J}\right)+  \sum_{Q=1}^{M} \sum_{N=1}^{\infty}  A_{QN} \left[L_{N|vw}\right]_{Q+2l -1} \nn
&=&  \sum_{J=1}^l \sum_{Q=J+1}^{M+J-1}  \left[L_{Q|vw}\right]_{Q+2l-2J}  +  \sum_{J=0}^{l-1} \sum_{Q=2}^{M+1}  \left[L_{Q+J|vw}\right]_{Q+2l-J-2} +  \sum_{Q=1}^{M}  \left[L_{Q+l+1|vw}\right]_{Q+l-1} \nn
&+& \sum_{Q=1}^{M-1} \left[L_{Q|vw}\right]_{Q+2l } +  \sum_{J=1}^l   \left[L_{J|vw}\right]_{2l-J}\nn
&=& \sum_{J=1}^l \left( 2 \sum_{Q=1}^{M-1}  \left[L_{Q+J|vw}\right]_{Q+2l-J} + \left[L_{M+J|vw}\right]_{M+2l-J}\right)+\sum_{Q=1}^{M-1} \left[L_{Q|vw}\right]_{Q+2l } \nn 
&+& \sum_{J=1}^l   \left[L_{J|vw}\right]_{2l-J}.
\eea
This leads to the result (where we have now included the discontinuities below the real axis)
\bea
& &\left[ \log Y_{M|vw} \right]_{(M+2l)\tau} =  \left[\Lambda_--\Lambda_+  \right]_{2l\tau } -\delta_{l,0}\left[\Lambda_-\right]_0 -\sum_{J=1}^{l}\sum_{Q=1+J}^{M+J} \left[\Lambda_Q\right]_{(Q+2l-2J)\tau }\nn
&+& \sum_{J=1}^l \left( 2 \sum_{Q=1}^{M-1}  \left[L_{Q+J|vw}\right]_{(Q+2l-J)\tau} + \left[L_{M+J|vw}\right]_{(M+2l-J)\tau}\right)\nn &
+&\sum_{Q=1}^{M-1} \left[L_{Q|vw}\right]_{(Q+2l)\tau } 
 +  \sum_{J=1}^l   \left[L_{J|vw}\right]_{(2l-J)\tau}.
\eea
We can define the discontinuities in terms of $D$ functions as in \cite{Cavaglia:2010nm}, by defining them as
\bea
& &D^{M|vw}_{(M+2l)\tau}(u) =\left(\Lambda_--\Lambda_+ \right)(u+2l\ad \tau i) +\sum_{Q=1}^{M-1} L_{Q|vw}(u+i(Q+2l)\ad\tau) \nn
&+& \sum_{J=1}^l \left( 2 \sum_{Q=1}^{M}  L_{Q+J|vw}(u+i(Q+2l-J)\ad\tau) + L_{M+J|vw}(u+i(M+2l-J)\ad\tau)\right)\nn
&+&  \sum_{J=1}^l   L_{J|vw}(u+i(2l-J)\ad\tau)-\sum_{J=1}^{l}\sum_{Q=1+J}^{M+J} \Lambda_Q(u+i(Q+2l-2J)\ad \tau).
\eea
We have the following identity for these functions:
\bea
\label{Didentityv}
D^{M|vw}_{(M+2l)\tau }(u) - D^{M|vw}_{(M+2l-2)\tau}(u+2\tau i \ad ) &=& L_{l|vw}(u+\tau li \ad)+L_{M+l|vw}(u+\tau (M+l)i \ad ) \nn
&+& 2 \sum_{Q=l+1}^{M-1+l}  L_{Q|vw}(u+\tau Q\ad i) -\sum_{Q=1+l}^{M+l} \Lambda_Q(u+\tau Q \ad i), \nn
\eea
Now we can plug this into our expression for $\log Y_{vw}$ \eqref{eq:appYvwcontour} (or \eqref{eq:contourintegralYvw} in the main text)
\bea
\label{92v}
&=&\sum_{\tau}\sum_{l=0}^{\infty} \int_{Z_0}  \left[\log Y_{M|vw}\right]_{\tau (M+2l)}(z) H(z+\tau (M+2l)-u) \nn
&=&\sum_{\tau}\sum_{l=0}^{\infty} \int_{Z_0} \left(\left[D_{(M+2l)\tau }^{M|vw}\right]_{0}(z) -\delta_{l,0}\left[\Lambda_-\right]_0  \right) H(z+\tau (M+2l) -u) \nn
&=& - \left(\int_{Z_0+i\e}-\int_{Z_0-i\e}\right)\left(\Lambda_-(z) H(z+M \ad i -u) + \Lambda_-(z) H(z-M \ad i -u)\right) \nn
&+& \sum_{\tau}\sum_{l=0}^{\infty} \int \left(\left[D_{(M+2l)\tau}^{M|vw}\right]_{0}(z) \right) H(z+\tau (M+2l)\ad i-u) .
\eea
Using the recursion relation for $D$ functions we find that 
\bea
\label{93v}
& & \sum_{\tau}\sum_{l=0}^{\infty} \int \left(\left[D_{(M+2l)\tau}^{M|vw}\right]_{0}(z) \right) H(z+\tau (M+2l)\ad i -u) \nn
&=&  - \sum_{\tau}\sum_{l=1}^{\infty}\tau  \int_{Z_0 -i\tau \e } \left(D_{(M+2l)\tau}^{M|vw}(z) -D_{(M+2l-2)\tau}^{M|vw}(z+2\tau \ad i )\right) H(z+\tau (M+2l)\ad i- u) \nn
&-&\sum_{\tau} \tau \int_{Z_0 -i\tau \e }\left(\left(\Lambda_--\Lambda_+ \right) + \sum_{Q=1}^{M-1} L_{Q|vw}(z+\tau i \ad Q )\right) H(z+\tau M \ad i-u).
\eea
Combining all the terms with $\Lambda_{\pm}$ from \eqref{92v} and \eqref{93v} gives:
\bea
&-& \left(\int_{Z_0+i\e}-\int_{Z_0-i\e}\right)\left(\Lambda_-(z) H(z+M \ad i -u) + \Lambda_-(z) H(z-M \ad i -u)\right) \nn
&-&\sum_{\tau} \tau \int_{Z_0 -i\tau \e }\left(\Lambda_--\Lambda_+ \right) H(z+\tau M \ad i-u).
\eea
Using the relation
\bea
\int_{Z_0 \pm i \e} (r -\tilde{r}) f = \oint_{\gamma_0} r f\pm \oint_{\gamma_x} r f,
\eea
for an analytic function $f$, a function $r$ with square root branch cuts and the contours $\gamma_0$ and $\gamma_x$, which are defined in fig. \ref{fig:gamma0} and fig. \ref{fig:gammax} respectively, we can rewrite this expression for $Y_-$'s as 
\bea
\left(\Lambda_--\Lambda_+\right) \hat{\star} K_M. 
\eea
Collecting all the remaining terms and plugging in the relation from \eqref{Didentityv} into the first line leads to
\bea
\label{derivvw1}
& &\sum_{Q=1}^{M-1} L_{Q|vw}\star K_{M-Q} - \sum_{\tau}\sum_{l=1}^{\infty}\tau  \int_{Z_0 -i\tau \e }H(z+\tau (M+2l)\ad i-u)\cdot  \nn
& &\left(L_{l|vw}(z+\tau li \ad)+L_{M+l|vw}(z+\tau (M+l)i \ad) +2 \sum_{Q=l+1}^{M-1+l}  L_{Q|vw}(z+\tau Qi \ad)\right. \nn
&-& \left. \sum_{Q=1+l}^{M+l} \Lambda_Q(z+\tau Q\ad i)\right).
\eea
These terms excluding the $\Lambda_Q$ terms in the last line together form the $K_{PQ}$ kernel:
\bea
&+& \sum_{Q=1}^{M-1} L_{Q|vw}\star K_{M-Q}- \sum_{\tau}\sum_{l=1}^{\infty}\tau  \int_{Z_0 -i\tau \e }H(z+\tau (M+2l)\ad i-u)\cdot  \nn
& &\qquad \left(L_{l|vw}(z+\tau li \ad)+L_{M+l|vw}(z+\tau (M+l)i \ad) +2 \sum_{Q=l+1}^{M-1+l}  L_{Q|vw}(z+\tau Qi \ad)\right) \nn
&=& \sum_{l=1}^{\infty} L_{l|vw} \star \left( 2\sum_{Q=1}^{\min(M,l)-1} K_{|M-l|+2Q} +  K_{|M-l|} +  K_{M+l} \right) = \sum_{l=1}^{\infty} L_{l|vw} \star K_{lM}.\quad 
\eea
The $\Lambda_Q$ term in \eqref{derivvw1} can be simplified as well:
\bea
& & \sum_{\tau}\sum_{l=1}^{\infty}\tau  \int H(z+\tau (M+2l)\ad i-u)\sum_{Q=1+l}^{M+l} \Lambda_Q(z+\tau Q\ad i) \nn
&=& \sum_{\tau}\sum_{l=1}^{\infty}\tau  \int H(z+\tau (M+2l-Q)\ad i-u)\sum_{Q=1+l}^{M+l} \Lambda_Q(z) \nn
&=& - \sum_{l=1}^{\infty} \sum_{Q=1+l}^{M+l} \Lambda_Q\star K_{M+2l-Q} \nn
&=& -\sum_{Q=1}^{\infty}  \sum_{l=\max(M-Q,0)}^{M-1}  \Lambda_{Q+1}\star K_{Q+1+2l-M},
\eea
after shifting $l\rightarrow l+Q-M$. Our total right-hand side for the the TBA-equation of $Y_{M|vw}$ now reads
\bea
\sum_{N=1}^{\infty} L_{N|vw} \star K_{NM} + \left(\Lambda_--\Lambda_+\right) \hat{\star} K_M-\sum_{Q=1}^{\infty}  \sum_{l=\max(M-Q,0)}^{M-1}  \Lambda_{Q+1}\star K_{Q+1+2l-M}. 
\eea
Rewriting the $\Lambda_{\beta}$'s into $L_{\beta}$'s and using the TBA equation for $Y_{\beta}$ on the resulting $\log Y_{\beta}$-terms yields
\bea
& &\sum_{N=1}^{\infty} L_{N|vw} \star K_{NM} + \left(L_--L_+\right) \hat{\star} K_M\nn
&-&\sum_{Q=1}^{\infty} L_Q\star K_{Qy}\hat{\star}K_M-\sum_{Q=1}^{\infty}  \sum_{l=\max(M-Q,0)}^{M-1}  \Lambda_{Q+1}\star K_{Q+1+2l-M}. 
\eea
Gathering all the terms depending on $Y_Q$'s gives us
\bea
-\sum_{Q=1}^{\infty} L_Q\star \left( K_{Qy}\hat{\star}K_M+(1-\delta_{Q,1}) \sum_{l=\max(M-Q+1,0)}^{M-1}  K_{Q+2l-M}\right)
\eea
and using the fact that the terms in the brackets sum up to exactly $K_{xv}^{QM}$ we find the TBA-equation for $Y_{vw}$:
\bea
\log Y_{M|vw}^{(\alpha)} = L_{N|vw}^{(\alpha)} \star K_{NM} +\left(L_-^{(\alpha)}-L_+^{(\alpha)}\right) \hat{\star} K_M-\Lambda_Q \star K_{xv}^{QM}.
\eea

\section{Deriving discontinuities for $Y_Q$ }
\label{app:YQdiscs}
For the derivation of the $Y_Q$ TBA-equation in section \ref{sec:rederivingYQ} from the analytic $Y$-system we need to derive discontinuities of the form
$$
\left[ \log Y_{M} \right]_{\pm (2M+l)},
$$
where $M,l\in \mathbb{N}$. We will derive them from the $Y$-system in a spirit very similar to the corresponding derivation in the $Y_{(v)w}$ case.\footnote{This means in particular that we will leave the dependence on the sign $\pm$ implicit in the derivation, restoring it in the final result.}
\\[5mm]
Taking logs on both sides of the $Y$-system equations \eqref{Ysys:Q} gives us
\begin{align}
\log Y_Q^+ + \log Y_Q^- &= -\sum_{Q'}A_{QQ'} L_{Q'} + \sum_{\alpha} L_{vw|Q-1}^{(\alpha)} \mbox{ with } Q\geq 2 \nn
\log Y_1^+ + \log Y_1^- &=-\sum_{Q'}A_{QQ'} L_{Q'} + \sum_{\alpha} L_-^{(\alpha)},
\end{align}
with $A$ as defined in \eqref{eq:AMJ}. We will leave the $\alpha$-summation implicit from now on. Taking discontinuity brackets at $M+2l-1$ for $M\geq 2$ leads to 
\bea
\label{YQh1}
\left[ \log Y_{M} \right]_{M+2l} &=& -\sum_{N=1}^{\infty} A_{MN} \left[L_{N}\right]_{M+2l -1} + \left[L_{vw|M-1}^{(\alpha)}\right]_{M+2l -1} - \left[ \log Y_{M} \right]_{M+2l-2} \nonumber \\
&=& -\sum_{N=1}^{\infty} A_{MN} \left[\Lambda_{N}\right]_{M+2l -1} +  \left[\log Y_{M-1}\right]_{M+2l-1} + \left[\log Y_{M+1}\right]_{M+2l-1} \nonumber\\ &-& \left[\log Y_{M}\right]_{M+2(l-1)}	+ \left[L_{vw|M-1}^{(\alpha)}\right]_{M+2l -1}
\eea
Using the second line of \eqref{YQh1} first yields: 
\bea 
& &\left[ \log Y_{M} \right]_{M+2l} = \nonumber \\
&=& -\sum_{N=1}^{\infty}  A_{MN} \left[\Lambda_{N}\right]_{M+2l -1} - \sum_{N=1}^{\infty}  A_{M-1,N} \left[\Lambda_{N}\right]_{M-1+2l -1} -\left[ \log Y_{M} \right]_{M+2l-2}\nonumber \\ 
&+&  \left[\log Y_{M-2}\right]_{M+2l-2} +\left[\log Y_{M}\right]_{M+2(l-1)} + \left[\log Y_{M+1}\right]_{M+2l-1}-\left[\log Y_{M-1}\right]_{M+2l-3} \nn
&+& \sum_{Q=M-1}^M \left[L_{vw|Q-1}^{(\alpha)}\right]_{Q+2l -1},
\eea
so we can conclude that 
\bea 
\left[ \log Y_{M} \right]_{M+2l} &=&  \left[\log Y_{M+1|w}\right]_{M+2l-1} -\sum_{Q=M-1}^M \sum_{N=1}^{\infty}  A_{QN} \left[\Lambda_{N}\right]_{Q+2l -1}+  \left[\log Y_{M-2}\right]_{M+2l-2}\nn
&-& \left[\log Y_{M-1}\right]_{M+2l-3} + \sum_{Q=M-1}^M \left[L_{vw|Q-1}^{(\alpha)}\right]_{Q+2l -1}.
\eea
Applying this step repeatedly gives
\bea
\label{rep1Q}
& &\left[ \log Y_{M} \right]_{M+2l} = \left[\log Y_{M+1}\right]_{M+2l-1}  \nonumber \\
&-&\sum_{Q=2}^M \sum_{N=1}^{\infty}  A_{QN} \left[\Lambda_{N}\right]_{Q+2l -1}+  \left[\log Y_{1}\right]_{1+2l}-\left[\log Y_{2}\right]_{2l}+\sum_{Q=1}^{M-1} \left[L_{vw|Q}^{(\alpha)}\right]_{Q+2l}.
\eea
Plugging in the Y-system equation for $Y_{1}$ \eqref{Ysys:Q} gives us
\bea
\label{rep2Q}
& &\left[ \log Y_{M} \right]_{M+2l} -\left[\log Y_{M+1}\right]_{M+2l-1} = \nonumber \\
&=&-\sum_{Q=2}^M \sum_{N=1}^{\infty}  A_{QN} \left[\Lambda_{N}\right]_{Q+2l -1}-  \left[\log Y_{1}\right]_{1+2l-2} -\left[\Lambda_{2}\right]_{2l}+\left[\L^{(\alpha)}_-\right]_{2l}+\sum_{Q=2}^{M} \left[L_{vw|Q-1}^{(\alpha)}\right]_{Q+2l-1} \nonumber \\
&=& -\sum_{Q=1}^M \sum_{N=1}^{\infty}  A_{QN} \left[\Lambda_{N}\right]_{Q+2l -1}-  \left[\log Y_{1}\right]_{2l-1}+\left[\L^{(\alpha)}_-\right]_{2l}+\sum_{Q=1}^{M-1} \left[L_{vw|Q}^{(\alpha)}\right]_{Q+2l}
\eea
Using the line \eqref{rep1Q} with first term $\left[ \log Y_{M+1|vw} \right]_{M+2l-1}$ (So $M\rightarrow M+1,l\rightarrow l-1$) gives 
\bea
\label{rep3Q}
& &\left[ \log Y_{M+1} \right]_{M+2l-1} = \left[\log Y_{M+2}\right]_{M+2l-2}  \nonumber \\
&-&\sum_{Q=2}^{M+1} \sum_{N=1}^{\infty}  A_{QN} \left[\Lambda_{N}\right]_{Q+2l -3}+  \left[\log Y_{1}\right]_{2l-1}-\left[\log Y_{2}\right]_{2l-2}+\sum_{Q=3}^{M+1} \left[L_{vw|Q-1}^{(\alpha)}\right]_{Q+2l-3}.\qquad
\eea
Note the existence of a term $\left[\log Y_{1}\right]_{2l-1}$ in \eqref{rep2Q} and also in \eqref{rep3Q}. When combining the statements, these terms will cancel. We can repeatedly do this to end up with
\bea
& &\left[ \log Y_{M} \right]_{M+2l} = \left[ \log Y_{M+l+1} \right]_{M+l-1} +\sum_{J=1}^{l+1}\sum_{Q=J}^{M+J-2} \left[L_{vw|Q}^{(\alpha)}\right]_{Q+2l-2J+2} \nn
&-&\sum_{J=0}^l\sum_{Q=1+J}^{M+J} \sum_{N=1}^{\infty}  A_{QN} \left[\Lambda_{N}\right]_{Q+2l -1-2J} + \left[L_-^{(\alpha)}\right]_{2l} -\left[\log Y_{l+1}\right]_{l-1},
\eea
where the first and last term usually vanish because of the fact that the functions in the brackets are analytic on the respective strips (except at the very important exception $l=0$, for which the last term does contribute). Note the similarities with the derivation of for vw-functions by comparing with \eqref{resvw1}. An interesting difference is the upper bound on the $J$ summation for vw functions: in the previous section (when we treated the vw TBA-equation) we found that the last term $J=l+1$ was trivial, but due to the structure of the relevant $Y$-system equations ($\Lambda_{Q+1}$ vs $L_{vw|Q-1}$) the situation is different here: we really do need to write the $J=l+1$ term. We find that, while being careful about the implications of changing $M+2l\rightarrow \pm (M+2L)$,
\begin{align}
\left[ \log Y_{M} \right]_{M+2l} &= \sum_{J=1}^{l+1}\sum_{Q=1+J}^{M+J-1} \left[L_{vw|Q-1}^{(\alpha)}\right]_{Q+2l-2J+1} \nn
&-\sum_{J=0}^l\sum_{Q=1+J}^{M+J} \sum_{N=1}^{\infty}  A_{QN} \left[\Lambda_{N}\right]_{Q+2l -1-2J} + \left[L_-^{(\alpha)}\right]_{2l} -\delta_{l,0}\left[\log Y_{1}\right]_{-1} \nn
\left[ \log Y_{M} \right]_{-(M+2l)}& = \sum_{J=1}^{l+1}\sum_{Q=1+J}^{M+J-1} \left[L_{vw|Q-1}^{(\alpha)}\right]_{-(Q+2l-2J+1)} \nn
&-\sum_{J=0}^l\sum_{Q=1+J}^{M+J} \sum_{N=1}^{\infty}  A_{QN} \left[\Lambda_{N}\right]_{-(Q+2l -1-2J)} + \left[L_-^{(\alpha)}\right]_{-2l} -\delta_{l,0}\left[\log Y_{1}\right]_{1}.
\end{align}
The $Y$-system also provides us with a way to rewrite $\left[\log Y_{1}\right]_{-1}$: taking brackets at $0$ gives us
\be
\label{Y1property}
\left[\log Y_{1}\right]_{1} + \left[\log Y_{1}\right]_{-1} = \left[L_-^{(\alpha)} \right]_0.
\ee
The similarity with the w- and vw-case can be exploited once more to rewrite the triple sum term. Taking expression \eqref{longder1} and replacing $L_{N|(v)w}\rightarrow -\Lambda_N$ we find directly that we can write the discontinuities for $l\geq 1$ as
\begin{align}
&\left[ \log Y_{M} \right]_{\tau(M+2l)} = \sum_{J=1}^{l+1}\sum_{Q=1+J}^{M+J-1} \left[L_{vw|Q-1}^{(\alpha)}\right]_{\tau(Q+2l-2J+1)}  + \left[L_-^{(\alpha)}\right]_{\tau 2l} -\delta_{l,0}\left[\log Y_{1}\right]_{-\tau 1} \nn &-  \sum_{J=1}^l   \left[\Lambda_{J}\right]_{\tau(2l-J)}
-\sum_{J=1}^l \left( 2 \sum_{Q=1}^{M-1}  \left[\Lambda_{Q+J}\right]_{\tau(Q+2l-J)} + \left[\Lambda_{M+J}\right]_{\tau(M+2l-J)}\right)-\sum_{Q=1}^{M-1} \left[\Lambda_{Q}\right]_{\tau(Q+2l) } .
\end{align}
Now, defining the functions $D_{\pm(M+2l)}^Q$ as follows for $l\geq 0$
\begin{align}
D_{\tau (Q+2l)}^Q(u) &= \sum_{J=1}^{l+1}\sum_{M=J}^{Q+J-2} L_{vw|M}^{(\alpha)}(u+\tau(M+2l-2J+2)i\ad)  + L_-^{(\alpha)}(u+2\tau l i \ad)   \nn
&-\sum_{J=1}^l \left( 2 \sum_{M=1}^{Q-1}  \Lambda_{M+J}(u+\tau(M+2l-J)i\ad) + \Lambda_{Q+J}(u+\tau(Q+2l-J)i\ad)\right)\nn
&-\sum_{M=1}^{Q-1} \Lambda_{Q}(u+\tau(M+2l)i\ad) -  \sum_{J=1}^l  \Lambda_{J}(u+\tau(2l-J)i\ad)
\end{align}
and find that they obey the following recursion:
\begin{align}
&D_{\tau (M+2l)}^Q(u) - D_{\tau (M+2l-2)}^Q(u +\tau 2 i\ad ) = \nn
&= \sum_{Q=2+l}^{M+l} L_{vw|Q-1}^{(\alpha)}(u+\tau(Q-1)i\ad)-   \Lambda_{l}(u+\tau li\ad) \nn
&-\left( 2 \sum_{Q=1}^{M-1}  \Lambda_{Q+l}(u+\tau(Q+l)i\ad) +\Lambda_{M+l}(u+i \tau (M+l)\ad)\right) \nn
&=\sum_{Q=l+1}^{M+l-1} L_{vw|Q}^{(\alpha)}(u+\tau Q i\ad)-   \Lambda_{l}(u+\tau li\ad) - 2 \sum_{Q=1+l}^{M-1+l}  \Lambda_{Q}(u+\tau Qi\ad) -\Lambda_{M+l}(u+\tau (M+l)i \ad).
\end{align}
This is similar to the corresponding expression in \cite{Cavaglia:2010nm} except for the lower bound on $Q$ for the vw-terms. The discontinuity of $Y_Q$ can now be written as follows for all $l\geq0$:
\begin{align}
\left[ \log Y_{M} \right]_{\tau(M+2l)}(u) &= \left[D_{\tau (M+2l)}^Q\right]_0 (u) -\delta_{l,0}\left[\log Y_{1}\right]_{-\tau 1}(u). 
\end{align}
This will allow us to continue the derivation of the $Y_Q$ TBA-equation in section \eqref{sec:rederiveYQ}. 

%% file: constrainingtrigpol.tex
In this appendix we will consider a few technical lemmas that are necessary for the correctness of the argumentation in the main text. We will show that $\Pf$ functions can be expanded in terms of the $x$ function and that periodic analytic functions with exponential growth must be trigonometric polynomials.

\section{$\Pf$ functions can be written as convergent $x$-series}
To have a better grip on the QSC functions and for example be able to actually construct solutions it is useful to find nice representations of the $\Pf$ functions. In \cite{Marboe:2014gma} extensive use is made of the fact that is possible to write $\Pf$ functions as convergent series on the $x$ plane. In this section we present an argument why this is so.
\paragraph{Undeformed case.} $\Pf$ functions have short cuts and it seems natural to expect an expansion in the $x_s$ function, which we will denote simply by $x$ from now on. As a function $x: \C \rightarrow \C$ it is not surjective, its image is given by 
\begin{equation}
\C_{>1} \defeq \left\{z \in \C \, | \, |z| >1   \right\}.
\end{equation}
Therefore the inverse $x^{-1} : \C_{>1} \rightarrow \C$ exists and admits an extension to the unit circle $|z| =1$. Now let $\Pf_x : \C \rightarrow \C$ be defined as $\Pf_x = \Pf \circ x^{-1}$, where $\Pf$ is any $\Pf$ function. This function is meromorphic on $\C_{>1}$ by construction and has a cut on the unit circle. One can continue this function into the unit disk, where it will have more cuts. In the $u$ language, the inside of the unit disk is the second sheet of $\Pf$, where there is an infinite ladder of short cuts. Since $\Pf$ has polynomial asymptotics we find that $\Pf$ must have polynomial asymptotics in $x$, i.e. $\Pf(u) \simeq u^M$ implies $\Pf_x(x) \simeq x^M$ as $x\rightarrow \infty$. This shows that $\Pf_x$ is meromorphic with polynomial growth as $x\rightarrow \infty$. If we assume $\Pf$ is pole-free on the $u$ plane, following the regularity requirement, $\Pf_x$ is analytic on the $\C_{>1}$ and it follows, using the usual polynomial version of the lemma in the next section, that 
\begin{equation}
\label{eq:Pinxexpanded}
\Pf_x (x) = \sum_{k=-M}^{\infty} \frac{c_k}{x^k}. 
\end{equation}
A particularly useful property of this expression is that analytic continuation of $\Pf_x$ can be done rather explicitly by inverting the $x$ functions, although the resulting series usually does not converge everywhere.

\paragraph{Deformed case.} In the deformed case the $u$ domain takes the form of a cylinder. $\Pf$ functions are periodic or anti-periodic, and in the latter case multiplication by an anti-periodic prefactor such as $e^{iu/2}$ makes it periodic. We will consider only this case. The deformed $x$-function is surjective as a function $x:i \R \times [-\pi,\pi) \rightarrow \C_{>\mathcal{C}}$, where now 
\begin{equation}
\C_{>\mathcal{C}} = \left\{ z\in \C \, | \, |z|^2 + \left(z^*-z \right)\xi >1  \right\},
\end{equation}
i.e. the part of the complex plane outside of the contour $\mathcal{C}$ as introduced in \eqref{eq:contourC}. As before we can invert $x$ and extend it to the boundary of its domain $\mathcal{C}$ to define $\Pf_x : \C_{>\mathcal{C}} \rightarrow \C$ in exactly the same way as in the undeformed case: $\Pf_x = \Pf \circ x^{-1}$. Since $\Pf$ has exponential asymptotics $\Pf \simeq e^{-iuM}$ and $x\simeq e^{-iu}$ in the deformed case this yields $\Pf_x \simeq x^M$ as $x\rightarrow \infty$, as in the undeformed case. Therefore we come to the conclusion that also in this case one can expand $\Pf_x$ as in \eqref{eq:Pinxexpanded}. 
\\[5mm]
Note that, although this result holds for all our one-cutted functions (using the $x_m$ function for the $\Qf$), we cannot say the same about $\mu$. Therefore, it is not possible to treat the entire QSC on the $x$ plane alone. 
\section{Characterising analytic periodic functions with exponential growth} A crucial point in the analysis of the quantum spectral curve is that analytic functions on the complex plane with polynomial growth as $u\rightarrow \infty$ are quite special: one can prove that they should be polynomials. It seems only natural to assume that when suitably adapted, a similar statement should also hold for periodic functions. This turns out to indeed be true: we prove in this chapter a lemma that states that every function with exponential growth as $u\rightarrow i\infty$ is a trigonometric polynomial. This slightly technical lemma is useful in finding solutions of the $\eta$-deformed quantum spectral curve, but we use it in particular for the derivation of the $\mathfrak{sl}_2$-\textsc{xxz} Bethe equations in section \ref{sec:weakcoupling}.

\paragraph{Lemma.} Let $f: i \R \times [-\pi,\pi) \rightarrow \C$ be analytic and have infinitely many non-zero coefficients $a_n$ on the positive side (thus for positive $n$) of its Fourier series
\be
f(u) = \sum_{n\in \Z} a_n e^{-inu}.
\ee
Then $f$ grows superexponentially, i.e. 
\be 
\lim_{u \rightarrow i \infty} f(u) e^{iu M} 
\ee
diverges for all $M \in \N$. 
\\[5mm]
\textbf{Proof.} For all $m\geq 0$ and for all $r>0$
\begin{align}
|a_m|&= \left| \frac{1}{2\pi} \int_{Z_0} dz e^{imz}f(z)  \right|= \left| \frac{1}{2\pi} \int_{Z_0} dz e^{im(z+ir)}f(z+ir)  \right| \nn
&= \frac{1}{2\pi} e^{-rm} \left| \int_{Z_0} dz e^{imz}f(z+ir)  \right| \leq e^{-rm} M(r),
\end{align}
where $M(r) = \max_{z\in S^1+ir} |f(z)|$. If $M$ grows exponentially there exists an $n'$ such that $|M(r)| \leq e^{r n'}$ for large enough $r$. Now take any $n_*>n'$ such that $a_{n_*}\neq 0$, we see from the computation that 
\be 
|a_{n_*}| \leq e^{r\left(n'-n_*\right)} \text{for all } r,
\ee
a contradiction. So $M$ must grow superexponentially and so does $f$. \hfill\ensuremath{\blacksquare}
\\[5mm]
We see that by contraposition the lemma states that every function that grows exponentially must have only finitely many non-zero entries on the positive side of its Fourier series. Repeating the same lemma for the negative side then leads to the conclusion that analytic functions $f$ for which there exist $M_{\pm}$ such that 
\begin{equation}
\lim_{u \rightarrow \pm i \infty} f(u) e^{iu M_{\pm}} 
\end{equation}
converge must be trigonometric polynomials. For the function in section \ref{sec:weakcoupling} we use the asymptotics to constrain the Fourier series on one side and its reality to transfer this to the other side as well. This allows for the usage of the lemma. 